\documentclass[fleqn,usenatbib]{mnras}
\usepackage{natbib}
\usepackage[T1]{fontenc}
\usepackage{ae,aecompl}
\usepackage{graphicx}
\usepackage{amsmath}
\usepackage{amssymb}
\usepackage{siunitx}
\usepackage{multirow}
\usepackage[table,xcdraw]{xcolor}
\usepackage[utf8]{inputenc}
\usepackage[T1]{fontenc}
\usepackage{multicol}
\usepackage{pdfpages}

\title[Spectral modelling of jellyfish galaxies]{Spatially resolved self-consistent spectral modelling of jellyfish galaxies from MUSE with FADO: trends with mass and stripping intensity}

\author[Azevedo et al.]{Gabriel M. Azevedo$^{1}$
\thanks{gabriel.azevedo@ufrgs.br},
Ana L. Chies-Santos$^{1}$ \thanks{ana.chies@ufrgs.br},
Rogério Riffel$^{1,2}$
\thanks{riffel@ufrgs.br},
Jean M. Gomes$^{3}$,
\newauthor
Augusto E. Lassen$^{1}$, 
João P. V. Benedetti$^{1}$,
Rafael S. de Souza$^{4}$,
Quanfeng Xu$^{5,6}$
\\
$^{1}$Instituto de F\'isica, Universidade Federal do Rio Grande do Sul (UFRGS), Av. Bento Goncalves, 9500, Porto Alegre, RS, Brazil\\
$^{2}$ Instituto de Astrof\'\i sica de Canarias, Calle V\'\i a L\'actea s/n, E-38205 La Laguna, Tenerife, Spain\\
$^{3}$Instituto de Astrof\'isica e Ciências do Espaço, Universidade do Porto, CAUP, Rua das Estrelas, PT4150-762 Porto, Portugal\\
$^{4}$Centre for Astrophysics Research, University of Hertfordshire, College Lane, Hatfield, AL10~9AB, UK\\
$^{5}$Shanghai Astronomical Observatory, Chinese Academy of Sciences, 80 Nandan Rd., Shanghai 200030, China\\
$^{6}$School of Mathematics and Computer Science, Yunnan Minzu University, 2929 Yuehua Street, Kunming, 650500, China\\
}

\date{Accepted 2023 May 25. Received 2023 May 17; in original form 2023 April 10}

\usepackage{newtxtext,newtxmath}

\begin{document}
\label{firstpage}
\pagerange{\pageref{firstpage}--\pageref{lastpage}}
\maketitle

\begin{abstract}
We present a spatially resolved stellar population analysis of 61 jellyfish galaxies and 47 control galaxies observed with ESO/MUSE attempting to understand the general trends of the stellar populations as a function of the stripping intensity and mass. This is the public sample from the GASP programme, with $0.01 < z < 0.15$ and $8.9 <\log(M_{\star}/M_{\odot}) < 12.0$. We apply the spectral population synthesis code {\sc fado} to fit self-consistently both the stellar and nebular contributions to the spectra of the sources. We present 2D morphological maps for mean stellar ages, metallicities, gas-phase oxygen abundances, and star formation rates for the galaxies with Integrated Nested Laplace Approximation ({\sc inla}), which is efficient in reconstructing spatial data of extended sources. We find that ``extreme stripping'' and ``stripping'' galaxies are typically younger than the other types. Regarding stellar and nebular metallicities, the ``stripping'' and ``control passive'' galaxies are the most metal-poor. Based on the phase space for jellyfish cluster members we find trends in ages, metallicities, and abundances with different regions of the diagram. We also compute radial profiles for the same quantities. We find that both the stripping and the stellar masses seem to influence the profiles, and we see differences between various groups and distinct mass bins. The radial profiles for different mass bins present relations already shown in the literature for undisturbed galaxies, i.e., profiles of ages and metallicities tend to increase with mass. However, beyond $\sim0.75$ effective radius, the ages of the most massive galaxies become similar to or lower than the ages of the lower mass ones.
\end{abstract}

\begin{keywords}
galaxies: evolution -- galaxies: interactions -- galaxies: stellar content -- galaxies: individual: jellyfish galaxies
\end{keywords}

\section{Introduction}

A galaxy's environment plays an essential role in determining its morphology and star formation. The ram pressure stripping \citep[RPS,][]{Gunn&Gott72} is a phenomenon that changes both in dense environments. This effect occurs when a galaxy falls into denser regions of the cluster/group and has its gas removed by the ram pressure caused by the gas from the intracluster medium (ICM) when it is sufficient to overcome the gravitational potential of the galaxy. Stripping has already been observed in all sorts of galaxies: spirals \citep{stripping_spiral_virgo,candidates,vestige,roman2019}, ellipticals \citep{elipticalfireball}, dwarfs \citep{dwarf_fireball}, and even ring galaxies \citep{gaspV}.

The most extreme cases of galaxies undergoing RPS are called jellyfish galaxies. Their name is due to the tail-like structures of stripped gas resembling the tentacles of the namesake sea animal. The gas extends in the opposite direction of the galaxy infall \citep[see][and references therein]{review_jellyfish}. Although pieces of evidence of the tails of gas in such stripped galaxies were observed for the first time in 1984 \citep{first_obs_RPS}, it was just in the last years that astronomers have been dedicating special attention to those objects, creating large catalogues and commanding surveys dedicated to jellyfish galaxies. Such objects have been observed with photometry and spectroscopy in the optical, UV, IR, and radio \citep{budhies, gaspI, vestige, UVITrps, roman2019, LoTSS-I}. Large amounts of atomic and ionised gas have been found in the tails and disk of these galaxies \citep{budhies, gaspI, virgo,  gaspXIII, HI_JO206, roman2019, HI_JO204, molecular_JW100, HI_JO206eJO201}. For instance, recently over 70 jellyfish candidates were identified in the A901/2 multi-cluster system \citep{roman2019}, 95 were detected in low-redshift ($z<0.05$) clusters \citep{LoTSS-I}, and 60 were observed in low-redshift ($z<0.05$) groups \citep{LoTSS-II}, and over 200 at intermediate-redshift ($0.2<z<0.9$) systems \citep{durret21,durret22}. A great sample of jellyfish candidates at redshifts 0.04 < z < 0.07 was observed and described in \cite{wings}. From the initial selection, 94 were further observed by the GAs Stripping Phenomena in galaxies with MUSE (GASP) survey \citep{gaspI}, an integral-field spectroscopic survey that uses the Multi-Unit Spectroscopic Explorer (MUSE) spectrograph in the Very Large Telescope (VLT).

Previous works suggest that systems with cluster mergers can trigger RPS events \citep{owers12,McPartland16}, thus enhancing the formation of jellyfish galaxies. For example, the multi-cluster system  A901/2, formed by four subsystems, presents a large population of stripped galaxies that seem to be preferentially near boundary regions where diffuse gas moving along with the subcluster and diffuse gas from the remnant of the system meet \citep{roman2019,ruggiero19,roman2021}. In such regions, the significant change in the speed increases the ram pressure up to a factor of $\sim$1000 \citep{ruggiero19}. But this type of galaxy forms even in lower mass groups (down to $10^{12.5}\, M_{\odot}$), although less frequently than in clusters \citep{LoTSS-II}. Cluster jellyfish galaxies present {an enhanced global star formation rate} compared to other star-forming galaxies in the same environment \citep{enhancedSFR, roman2019, LoTSS-I}. At the same time, group jellyfish galaxies do not suggest such enhancement and seem to be stripped more slowly than those in clusters \citep{oman21,LoTSS-II}.

The stripping is more efficient in high-mass clusters and low-mass galaxies, but the most extreme cases are massive galaxies \citep{gaspIX}. Both observations and simulations indicate that the jellyfish galaxies are recent infallers in clusters, with an infall time between $\sim$ 1 and 3 Gyr \citep{gaspIX, illustris}. The fate of these transforming galaxies is still a matter of debate. It is believed that the RPS can turn spirals and irregulars into S0's and ellipticals by removing the gas of the galaxy to the point of quenching it \citep{disktoS0}. Besides transforming to S0s, spirals may also become less concentrated (lower Sérsic indices, \citealt{roman2021}). Observations of a jellyfish galaxy falling along the line of sight \citep{gaspII, gaspXVIII} and the existence of truncated disks among jellyfish candidates \citep{spectacular_virgo,gaspIII} support the idea that the stripping is outside-in (i.e. occurs first in the outer regions of the galaxy). It has also been proposed that part of the great amount of ultra-compact dwarfs (UCD) and intracluster globular clusters (GC) at low redshift clusters can originate from those \ion{H}{ii} regions that are stripped from the tails of jellyfish galaxies \citep{gaspXIII}.

The star formation enhancement in cluster jellyfish galaxies occurs likely due to shock waves produced by the ram pressure \citep{gaspXXX}. As a result, \ion{H}{ii} clumps are formed in the disc and the tails, extending up to 80\,kpc from the galactic centre. The clumps in both disk and tails follow similar M$_{gas}$--M$_{\star}$, L$_{H\alpha}$--$\sigma$ and SFR--M$_{gas}$ relations \citep{gaspXIII}, where M$_{\star}$ is the stellar mass, M$_{gas}$ is the gas mass, L$_{H\alpha}$ is the H$_{\alpha}$ luminosity, $\sigma$ is the velocity dispersion of the gas, and SFR is the Star Formation Rate. In addition, the gas kinematics in the tails tend to follow the gas kinematics at the galaxy disc \citep{gaspI, gaspIV, gaspXV}. Some galaxies undergoing RPS are found to have a much larger amount of H$_2$ than undisturbed galaxies, suggesting that the stripping is causing conversion of \ion{H}{i} into H$_2$ so that it could be one of the causes of star formation enhancement \citep{molecular_JW100, HI_into_H2}. The SFR depends on various parameters of the galaxies and host clusters, but \cite{SFR_tails} found some general trends: galaxies with lower SFR in the tails are found at relatively large clustercentric distances (from $0.4$ up to $1.8R_{200}$) and galaxies with large SFR in the tails (> 0.25 M$_{\odot}yr^{-1}$
) that are moving at great speed ($>2\sigma_{cl}$) in the innermost regions of the clusters ($<0.5R_{200}$) are massive and hosted in low-mass clusters. Also, the RPS occurs preferentially at intermediate clustercentric distances in massive clusters ($\sigma_{cl} > 900$ km/s) and at lower distances in intermediate ($600$ km/s $< \sigma_{cl} < 900$ km/s) and low-mass ($\sigma_{cl} < 600$ km/s) clusters.

The study of some cases of strongly stripped galaxies finds a decrease in the metallicity of the star-forming clumps in the tails, along with their distance to the stellar disk, which supports a scenario where the metal-poor ICM is cooled and mixed to the metal-rich stripped gas \citep{met_decrease_tail}. Also, the stripped galaxies seem to present an extended X-ray emission component that follows the interstellar medium (ISM) and is compatible with the thermal cooling of a warm plasma. This suggests that the ICM is being cooled on the surface of the ISM and eventually can be accreted, triggering episodes of star formation \citep{gasp_termo}. Besides those aspects, a strong magnetic field parallel to the direction of the tails may be a critical factor in allowing in situ star formation in these tails, preventing the cold gas clouds from exchanging heat and momentum with their surroundings \citep{gasp_magnetic}.

Most of the literature results are based on deriving the SFR of those galaxies using their emission lines fluxes. However, by doing this, only the recent star formation is probed \citep[e.g.][]{Kennicutt+98}. Aiming to investigate deeper the star formation in jellyfish galaxies, we performed a self-consistent analysis of the stellar population and nebular emission using the spectral synthesis code {\sc fado} \citep{fado} in 113 moderate-quality (signal-to-noise ratio > 2 in the \ion{H}{$\alpha$}) public datacubes downloaded from the ESO archive from the GASP survey \citep{gaspI}.

The data and sample selection are explained in \autoref{section2}, along with more details about the sample. The stellar population synthesis code, other tools and models used in this work, are presented in \autoref{section3}. In \autoref{section4}, we show the results of the synthesis and the respective analysis. In \autoref{section5}, we discuss the relation of the synthesis results with the environment, the masses, and the intensity of the stripping. Finally, in \autoref{section6}, we summarise our work and present our conclusions. In the supplementary material, we present 2D morphology maps of mean stellar ages and metallicities, H$_{\alpha}$ emission, surface density star formation rates ($\Sigma_{\mathrm{SFR}}$), and oxygen abundances for the galaxies of the sample. Throughout this work we used the following cosmology: $H_0 = 67\,km\,s^{-1}Mpc^{-1}$, $\Omega_M = 0.3$, $\Omega_{\Lambda} = 0.7$ \citep{planck}.

\section{Data}
\label{section2}

In this work, we analyse a sample of 113 public datacubes of galaxies from the GASP survey \citep{gaspI}. The reduced data of all the galaxies from GASP \href{http://www.eso.org/sci/publications/announcements/sciann17080.html}{DR1} and \href{https://archive.eso.org/cms/eso-archive-news/second-data-release-of-the-large-programme--dissecting-gas-strip.html}{DR2} were downloaded from the ESO Archive. The data from GASP \citep{gaspI} was obtained using the spectrograph MUSE \citep{muse} on VLT, covering the wavelength range of 4650 -- 9300 $\si{\angstrom}$ with a spectral sampling of 1.25 $\si{\angstrom}$, spectral resolution of $\sim 2.6\, \si{\angstrom}$, FoV of 1 arcmin$^2$, angular sampling of 0.2 arcsec, and spatial resolution of 0.4 arcsec, so it is capable of observing regions of the galaxies distant from the disk and with good resolution.

The GASP parent sample encompasses three other surveys covering low-redshift environments ($0.4 \lesssim z \lesssim 0.7$). Galaxies in clusters are selected from WINGS and OMEGAWINGS \citep{wings, omegawings}, and galaxies within groups and from the field are selected from PM2GC catalogue \citep{pm2gc}. The galaxies in this work whose names begin with ``JW'' or ``JO'' refer to WINGS and OMEGAWINGS galaxies, thus the ones in clusters, while the others refer to groups or field galaxies.

To ensure minimum reliability in the synthesis and filter very low-quality spectra we fit only the spaxels with a signal-to-noise ratio (SNR) > 2 in the \ion{H}{$\alpha$} line. To compute the SNR, we first calculate the ratio between the flux and the standard deviation arrays in an $8\,\si{\angstrom}$ wavelength window blue and redwards of the rest-frame $H_{\alpha}$ emission line and evaluate the mean. With these criteria, we ensure that we fit both disk and tails of the jellyfish. To improve the SNR in the continuum and simultaneously avoid a significant loss of spatial resolution, we have binned the spaxels 2x2 (sum 4 spaxels in 1). By doing such binning, the spatial resolution becomes 0.8 arcsec, which is lower than the usual lower limit of the atmospheric seeing ($\gtrsim 0.8$ arcsec). In our efforts to enhance the signal-to-noise ratio (SNR), we employed both Butterworth filtering and Voronoi binning techniques. Unfortunately, these methods did not yield satisfactory results. The application of Voronoi binning led to a substantial spatial degradation in the tails of the galaxies. At the same time, the implementation of Butterworth filtering resulted in a significant increase of SNR solely within the central regions of the galaxies.

We used the dust maps from \cite{schlegel} and the CCM reddening law with $R_v = 3.1$ \citep{CCM,odonnell} to correct the fluxes according to the extinction caused by the dust from the Milky Way. We computed the redshifts using the difference between the measured wavelengths of prominent emission lines and their respective rest-frame wavelengths.

Following \cite{gaspIX} we classified the disturbed galaxies into different groups based on their \ion{H}{$\alpha$} emission. The ``extreme stripping'' (ES) are the most extreme cases of jellyfish and have the longest tails; ``stripping'' (S) galaxies are the ones with ongoing evident stripping, but more moderate than the ES ones; ``post-stripping'' (PS) are galaxies with active star formation in the inner regions but have a truncated disk in the outskirts; and ``?'' are the ones for which the cause of the disturbance is unknown. Our sample consists of 11 ES, 46 S, 4 PS, and 5 ``?''. Additionally, the control sample is formed of 47 galaxies, of which 36 are star-forming (CSF) and 11 are passive (CP). The control sample is formed of galaxies that do not present clear signs of stripping nor \ion{H}{$\alpha$} emission beyond the disk \citep{enhancedSFR,gaspXXVII}.

\section{Stellar Population Synthesis}
\label{section3}

Implementing stellar population synthesis (SPS) techniques on a vast sample of datacubes presents a complex challenge. By design, these algorithms function on an individual spectrum, necessitating spectra extraction from each desired spaxel within a datacube before executing the SPS separately. This process requires substantial data storage capacity and considerable computational time. MUSE cubes typically range between 3 and 6 GB in size, with SPS outputs for a single cube potentially exceeding 10 GB. In addition to the substantial CPU memory required, synthesizing an entire galaxy on a personal computer takes approximately a whole day. The analysis was performed on a machine with the following specifications: CPU - 2.45GHz AMD 7763 (128C/256T); GPU - NVIDIA A100 (80GB); OS - Ubuntu Linux 18.04 64-bit; RAM - 4 TB. By employing this system, it was possible to apply the spectral synthesis technique to a substantially increased quantity of spectra simultaneously, resulting in a reduction of the total processing time to approximately one week.

\subsection{Spectral Fitting Code}

In order to implement the SPS in the MUSE datacubes, we used the code \href{http://spectralsynthesis.org/fado.html}{{\sc fado}} \citep[Fitting Analysis using Differential evolution Optimisation,][]{fado}. {\sc fado} uses a genetic differential evolution optimisation algorithm to fit the spectra with the additional constraint of the nebular emission (continuum plus lines), computed with the flux of various emission lines (H$\alpha$, H$\beta$, [\ion{N}{ii}], [\ion{S}{ii}], [\ion{O}{iii}], etc).  This constraint is critical for the synthesis of star-forming galaxies \citep{fado2} since it enables the computation of nebular spectra (in addition to the stellar one) and the number of young stars (t < 20 Myr) necessary to produce the measured gas ionisation. The inclusion of the nebular component makes \textsc{fado} a very suitable SPS code to analyse the stellar populations in the tails and disks of the jellyfish galaxies, which are the main sites of star formation and present many \ion{H}{ii} regions. During the stripping the SFR of these galaxies is known to increase throughout its whole extent \citep{enhancedSFR}, suggesting the ubiquitous presence of young stellar populations along the galactic structure.

\subsection{The base of elements}

In this work, we used a base of elements comprising 68
Simple Stellar Populations (SSPs) from \cite{BC03}. The SSPs were computed using the `Padova 1994' \citep{padova94-1,padova94-2,padova94-3,padova94-4,padova94-6}, and \cite{schonberner83,koster&schonberner86,winget+87,vassiliadis&wood93,vassiliadis&wood94} evolutionary tracks and the STELIB stellar library \citep{stelib} extended with BaSeL library \citep{bessel89,bessel91,fluks94,allard&hauschildt95,kurucz95,rauch02} for the UV. We have adopted a \cite{chabrier} Initial Mass Function (IMF), with lower and upper mass cutoffs equal to 0.1 $M_{\odot}$ and 100 $M_{\odot}$, respectively.

We selected SSPs with 17 different ages ranging from 1 Myr to 13 Gyr and 4 metallicities. This base has a 3.0\,\r{A} Full-Width Half Maximum (FWHM) spectral resolution in the wavelength range encompassed by our data. The base has spectra for young populations down to 1 Myr, which is fundamental to describing the underlying continuum of star-forming galaxies. The solar metallicity adopted by those models is $Z_{\odot} = 0.02$.
The SSPs selection follows the methodology described in \cite{dametto2014}. In short, we have computed the differences in the spectra of two populations with consecutive ages and solar metallicity. The most significant differences appear for populations younger than 5 Myr; we choose 1, 2.09, and 3 Myr as representative ages in that range.

We make such a selection because using the entire base of \cite{BC03} is redundant for the synthesis. Populations with similar ages have very similar theoretical spectra; with the noise addition, they are nearly indistinguishable. Additionally, using too many SSPs increases the computational time enormously without improving the fit accuracy. The final counts with populations with 17 ages (t = 1.00, 2.09, 3.02, 5.01, 10.00, 25.12, 50.00, 101.50, 321.00, 508.80, 718.70 Myr and 1.02, 2.00, 3.00, 5.00, 10.00, 13.00 Gyr) and 4 metallicities (Z = $\frac{1}{200}Z_{\odot}, \frac{1}{5}Z_{\odot}, 1Z_{\odot}, \frac{5}{2}Z_{\odot}$), which have 7 representative ages for young populations (t < 100 Myr), 7 representative ages for intermediate populations (100 Myr < t < 5 Gyr), and 3 for old populations (t $\geq$ 5 Gyr). 

\subsection{Synthesis post-processed by-products} 

To enable quick statistical analysis of large galaxy samples, reducing the synthesis output into a few quantities characterising the overall stellar and nebular component of a galaxy is useful. We will next define the most commonly used in the analysis of SPS results. These secondary best-fit quantities include the first moments of the distribution and from emission lines luminosity computed with {\sc fado}.

For each spaxel, the mean stellar ages are defined in two ways \citep{starlight,fado}. The mean age weighted by light can be expressed as:
\begin{equation}
    <\log t>_L = \sum_{j=1}^{N_{\star}} x_j \log(t_j)\,,
\end{equation}
meanwhile, the mean age weighted by mass is expressed as:
\begin{equation}
    <\log t>_M = \sum_{j=1}^{N_{\star}} \mu_j \log(t_j)\,,
\end{equation}
where $t_j$ is the age of the j-th SSP from the base measured in years, $x_j$ is the fraction with which it contributes to light, $\mu_j$ is the fraction with which it contributes to mass, and $N_{\star}$ is the number of SSPs in the base. Similarly, the mean stellar metallicities are defined as:
\begin{equation}
    <Z>_L = \sum_{j=1}^{N_{\star}} x_j Z_j\,,
\end{equation}
\begin{equation}
    <Z>_M = \sum_{j=1}^{N_{\star}} \mu_j Z_j\,,
\end{equation}
where $Z_j$ is the metallicity of the j-th SSP.

Besides the stellar metallicities, it is equally interesting to visualise the gas-phase metallicities in our sample. For that, we computed the gas-phase oxygen abundances of the ISM through the calibration of \cite{dopita16}:
\begin{equation}
\log(O/H) + 12 = 8.77 + \log([\ion{N}{ii}]/[\ion{S}{ii}]) + 0.264\times \log([\ion{N}{ii}]/H_{\alpha})  \,,
\end{equation}
where [\ion{N}{ii}] is the flux of the [\ion{N}{ii}]$\lambda$ 6548 \r{A} + [\ion{N}{ii}]$\lambda$ 6583 \r{A} lines and [\ion{S}{ii}] is the flux of the [\ion{S}{ii}]$\lambda$ 6716 \r{A} + [\ion{S}{ii}]$\lambda$ 6731 \r{A} lines.

We also compute the recent star formation rates via SPS, as done in \cite{asari07,riffel20}:
\begin{equation}
SFR(t) = \frac{dM_{\star}^c}{dt} \approx \frac{\Delta M_{\star}^c}{\Delta t} = \frac{\sum_{j_i}^{j_f}M_{\star,j}^c}{\Delta t}\,,
\end{equation}
where j refers to the j-th population in the base, and $M_{\star}^c$ is the total mass converted to stars in the time interval $\Delta t$. This is different from the mass locked inside stars, which must be corrected for the mass returned to the ISM by stellar evolution. {\sc fado} computes the total $M_{\star}^c$ and we can then use the coefficients $\mu_j$ to find the mass converted into stars in a given time. In this work, we chose $\Delta t = 25.12$ Myr. As presented by \cite{asari07,riffel20}, SFRs computed via synthesis with $\Delta t$ around 20 and 25 Myr show a strong correlation to the SFRs computed via $H_{\alpha}$ emission.

Adopting a uniform dust screen model, the attenuation in the V band is estimated from the Balmer decrement via the expression: 
\begin{equation}
A_V = \frac{2.5R_V}{k(\lambda_{H_{\alpha}})-k(\lambda_{H_{\beta}})}\times\log\Bigg[\Bigg(\frac{H_{\alpha}}{H_{\beta}}\Bigg)_{int}\times\Bigg(\frac{H_{\beta}}{H_{\alpha}}\Bigg)_{obs}\Bigg]\,,
\end{equation}
where $(H_{\alpha}/H_{\beta})_{int}$ is the intrinsic ratio between the H$_{\alpha}$ and H$_{\beta}$. Assuming the case B recombination with electronic density $n_e=10^2\,$cm$^{-3}$ and temperature $T_e=10000\,$K, $(H_{\alpha}/H_{\beta})_{int} = 2.87$ \citep{osterbrock}. It is worth mentioning that this assumption is widely used for normal star-forming galaxies and HII regions. $(H_{\alpha}/H_{\beta})_{obs}$ is the observed ratio and $k(\lambda)$ is the attenuation function, where we assume the Calzetti reddening law \citep{Calzetti}. $k(\lambda) = A_{\lambda}/E(B-V)$, where A$_{\lambda}$ is the extinction in the wavelength $\lambda$ and $E(B-V) = A_V/R_V$. For these calculations, we adopted the extinction factor $R_V=3.1$ \citep{CCM,odonnell}.

\section{Results}
\label{section4}

Performing SPS on datacubes provides detailed spatially resolved information about the SFH of the galaxy. For a given spectrum in the IFU cube, {\sc fado} computes the fractional contribution of each SSP in the base. The fractional contribution can be computed as light or mass fractions since the model spectral energy distribution (SED) for an SSP, such as in \cite{BC03}, is normalised by one solar mass of that population. In this work, we only performed the synthesis in the spectra with SNR(H$_{\alpha}$) $>$ 2. That said, we then computed the mean stellar ages and metallicities of each spaxel, using both the light (light-weighted) and stellar mass fractions (mass-weighted). Moreover, massive stars (e.g., O \& B) are much more luminous but less numerous than less massive ones. In addition, they are short-lived (of the order of a few Myr). Therefore, the luminosity of stellar populations containing a small number of high-mass stars can easily dominate the total luminosity of a galaxy in a given spectral band (UV and optical), thus yielding a high young stellar population light-fraction contribution. On the other hand, stellar mass fractions are usually dominated by low mass long-lived stars (e.g., K \& M) because they are more numerous \citep[see][for a review]{IMF}. Thus, weighting the stellar populations by mass tends to be skewed towards older populations, typically resulting in higher fractions of these populations. Because of this difference, it is interesting to present our measurements in both ways.

\begin{figure}
\centering
\includegraphics[width=\linewidth]{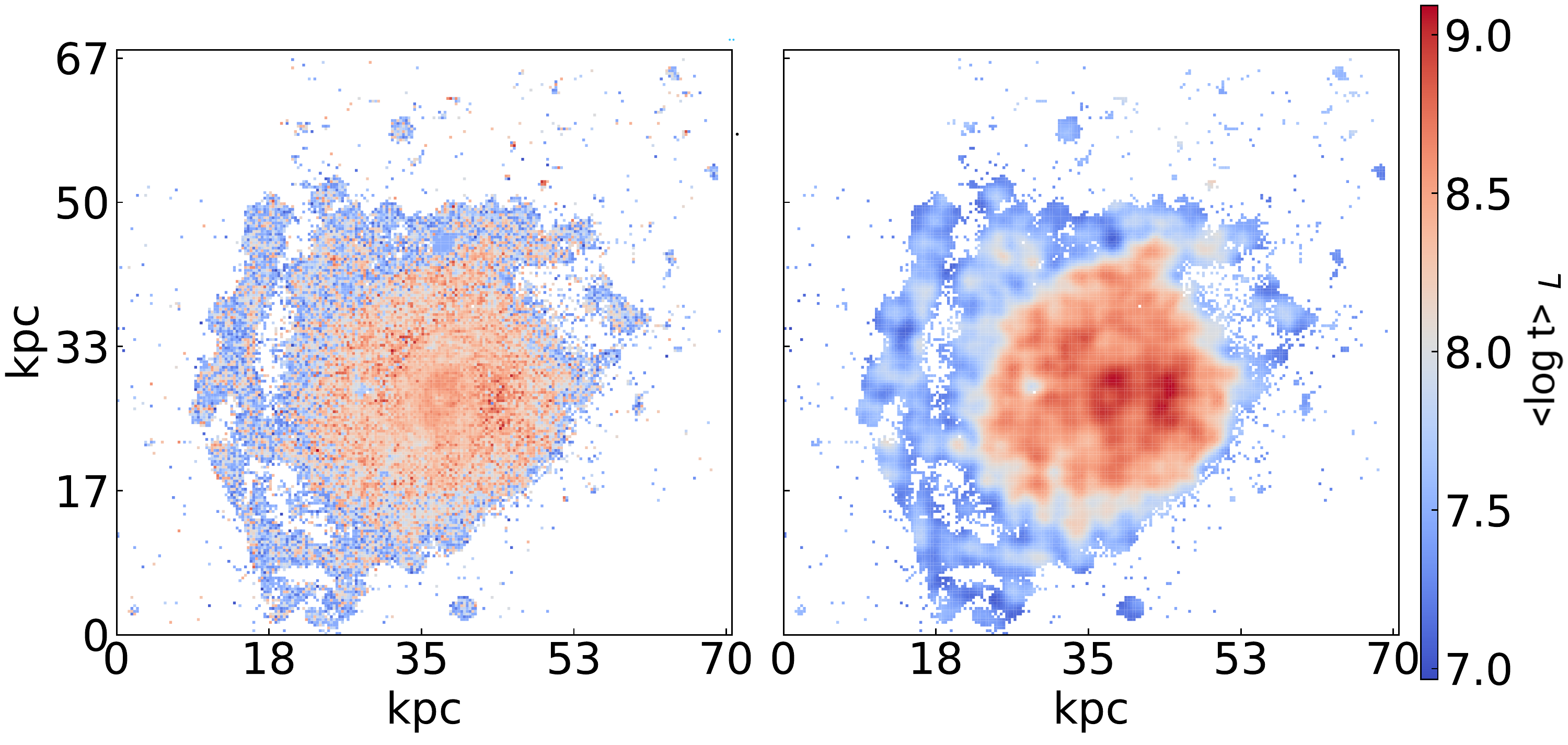}
\caption{Left: 2D morphological map of $<\log t>_L$ (mean stellar age light-weighted) for the galaxy JO85, located at z$=$0.0356 in the Abell2589 cluster. Right: Same map for the same galaxy but now applying the Integrated Nested Laplace Approximation ({\sc inla}), which reveals and highlights previously less apparent substructures (left panel).}
\label{inla}
\end{figure}

To better visualise the 2D morphological maps of mean stellar ages and metallicities, as well as the maps of other quantities, we used the \href{https://www.r-inla.org/}{Integrated Nested Laplace Approximation} ({\sc inla}), an alternative method to Markov Chain Monte Carlo for fast Bayesian inference. As shown in \cite{inla}, this method is efficient in reconstructing spatial data of extended sources (such as our 2D maps) considering the spatial correlation. In other words, it smoothes the results of the spectral synthesis considering the spatial correlation between neighbouring pixels, since it is not physically accurate that close regions have considerably different values for the same quantity. \autoref{inla} exemplifies the difference between a 2D map with and without the implementation of {\sc inla}. 

\begin{figure}
\centering
\includegraphics[width=\linewidth]{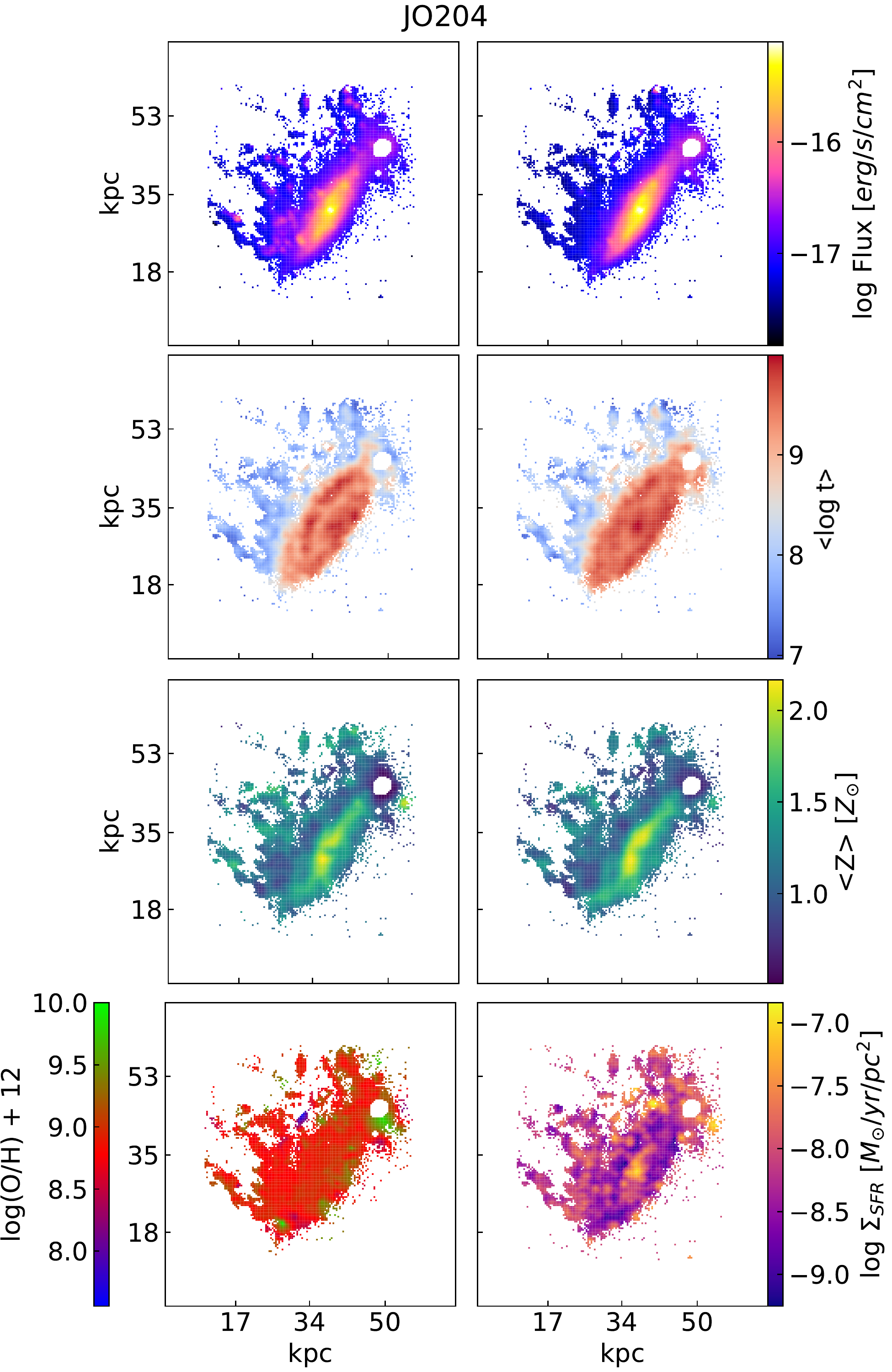}
\caption{This is an example of 2D morphological maps for the jellyfish galaxy JO204 located at a redshift of $z=0.0549$ in the Abell957x cluster. First row: maps of fluxes for H$_{\alpha}$+[N$_{II}$] lines (left) and for 6530--6600 \AA continuum window (right). Second row: maps of $<\log t>_L$ (left) and $<\log t>_M$ (right). Third row: maps of $<Z>_L$ (left) and $<Z>_M$ (right). Fourth row: oxygen abundances (left) and star formation rate surface density (right). {\sc inla} was implemented in all maps.}
\label{JO113_maps}
\end{figure}

For the 113 galaxies in our sample, we constructed 2D maps of mean stellar ages and metallicities weighted by light and by mass, gas-phase oxygen abundances, and SFR surface densities ($\Sigma_{SFR}$), which is the ratio of SFR to the area in pc$^{2}$. \autoref{JO113_maps} presents the maps for the galaxy JO204, located at z$=0.0424$ in Abell957x cluster, as an example. The extra material section presents figures similar to \autoref{JO113_maps} for the rest of the sample. The maps seem to be ``clumpy'', due to jellyfish galaxies' clumpy star formation nature. Most of the stellar age maps show that the most central regions are older than the outskirts and tails. The mean stellar and gas metallicities do not present any clear pattern. Neither mean stellar ages nor metallicities match precisely the structures seen in the $H_{\alpha}$ and continuum maps. As discussed further in \autoref{radial}, it is a trend of the sample that the mean ages decrease with radius, but the radial profiles of stellar and gas metallicities are more varied.

In order to analyse the distribution of star formation through time, we define the quantities t$_{50L}$ and t$_{50M}$, which represent the look-back time at which 50 per cent of the stellar light and mass, respectively, was formed in a galaxy. Half of the mass present in stars today in a galaxy is formed by populations with ages $\geq$t$_{50M}$, and half of its light is emitted by populations with ages $\geq$t$_{50L}$. Similarly, we introduce t$_{90L}$ and t$_{90M}$, which, like t$_{50L}$ and t$_{50M}$, represent the lookback time at which 90 per cent of the light and stellar mass, respectively, was formed in a galaxy. The range of those quantities for different groups in the sample, as well as mean ages, are listed in \autoref{tabela}. Given the large spread in such values for the same group, we see no clear distinction between the types of galaxies.

\begin{table*}
\centering
\caption{$<\log t>_L$, $<\log t>_M$, t$_{50L}$, t$_{90L}$, t$_{50M}$ and t$_{90M}$ for all the groups in our sample.}
\begin{tabular}{lcccccc}
\hline
Type    &
$<\log t>_L\,(\log yr)$   &
$t_{50L}$ &
$t_{90L}$ &
$<\log t>_M\,(\log yr)$   &
$t_{50M}$ &
$t_{90M}$  \\ \cline{1-7}
ES  & 7.2 - 9.1 & 50 Myr - 1 Gyr   & 3 Myr - 321 Myr  & 7.6 - 9.8 & 50 Myr - 13 Gyr & 5 Myr - 1 Gyr    \\
S   & 6.7 - 9.5 & 5 Myr - 13 Gyr   & 5 Myr - 1 Gyr    & 6.7 - 10.1  & 5 Myr - 13 Gyr  & 5 Myr - 13 Gyr   \\
PS  & 7.7 - 8.9 & 50 Myr - 1 Gyr   & 50 Myr - 101 Myr & 7.7 - 9.8 & 50 Myr - 10 Gyr & 50 Gyr - 1 Gyr   \\
?   & 6.7 - 9.0 & 5 Myr - 2 Gyr    & 5 Myr - 509 Myr  & 6.7 - 9.8 & 5 Myr - 10 Gyr  & 5 Myr - 1 Gyr    \\
CSF & 7.6 - 9.4 & 50 Myr - 10 Gyr  & 3 Myr - 509 Myr  & 7.6 - 10.1 & 50 Myr - 13 Gyr & 25 Myr - 13 Gyr  \\
CP  & 8.1 - 9.9 & 321 Myr - 13 Gyr & 3 Myr - 3 Gyr    & 8.8 - 10.1 & 1 Gyr - 13 Gyr  & 321 Myr - 13 Gyr \\
\hline
\end{tabular}%
\label{tabela}
\end{table*}

\subsection{Intensity of stripping}
\label{j-class}

Intending to explore the relation of stellar populations of our jellyfish galaxy candidates with the intensity of the stripping, we classified the galaxies into groups following \cite{gaspIX} classification mentioned in \autoref{section2}. Based on the \ion{H}{$\alpha$} emission, we separated the galaxies into ``extreme stripping'', ``stripping'', ``post-stripping'' and ``unknown'', which encompass respectively 11, 46, 4, and 5 galaxies, plus a control sample of 47 galaxies (36 star-forming and 11 passive). We should keep in mind the difficulties that the projection may imply in the identification and classification of such objects. For instance, if a jellyfish galaxy with very long tails is moving along the line-of-sight, tails and disk overlap in the observations, and it is difficult for us to estimate the real length of the tails or even to realise that galaxy is a jellyfish at all. Simulations estimate that approximately 30 per cent of the existing jellyfish galaxies are not identified as such because of projection effects \citep{illustris}.

\begin{figure}
\centering
\includegraphics[width=0.7\linewidth]{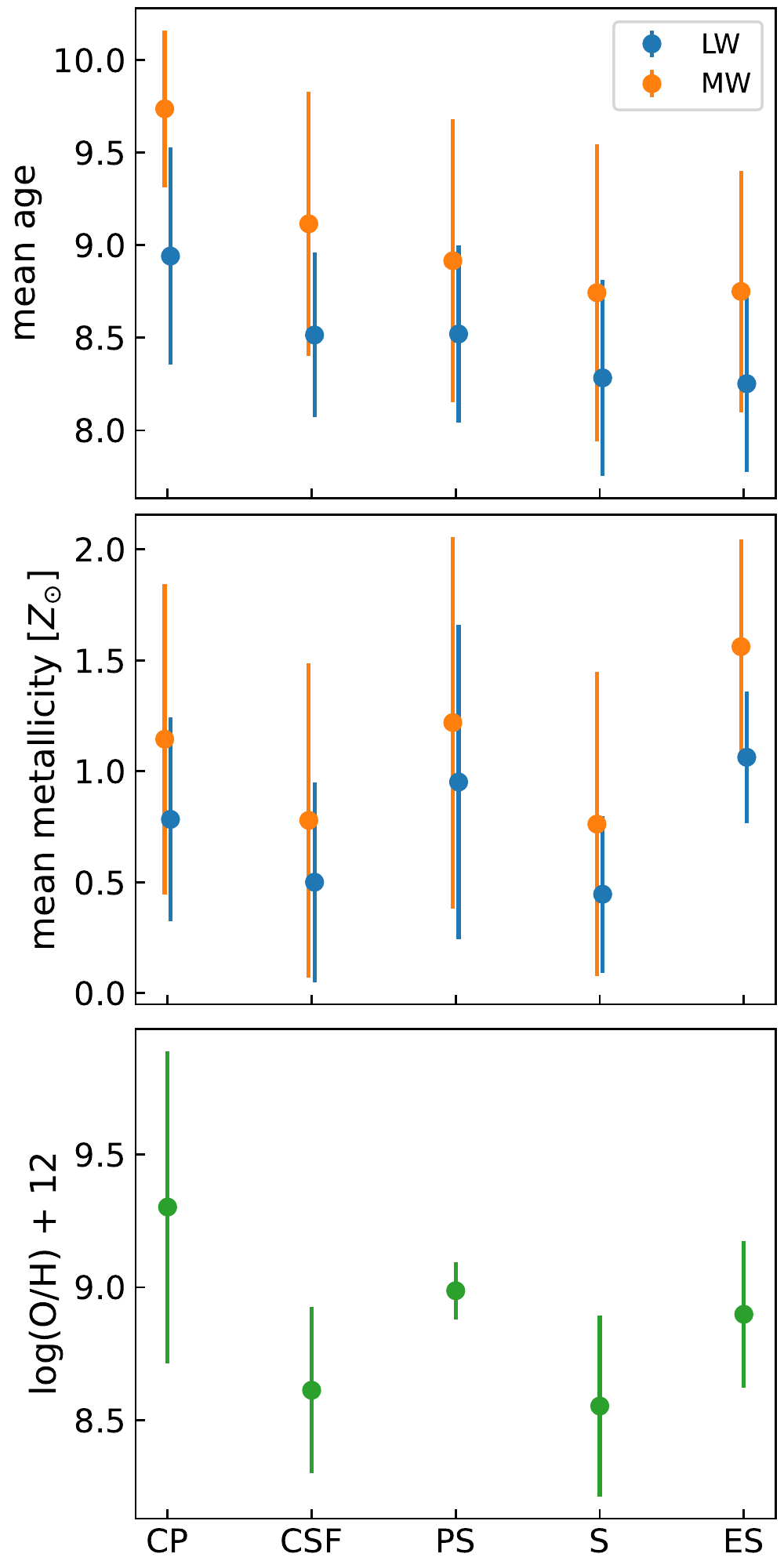}
\caption{Mean stellar ages (top), mean stellar metallicities (middle), and nebular oxygen abundance (bottom) are grouped by J-class. J5 represents the group of the most stripped jellyfish candidates, while J1 represents those galaxies with slight asymmetries.}
\label{Jclass}
\end{figure}

Therefore we calculate the average of the mean stellar ages and metallicities of each class, as well as the gas-phase mean oxygen abundances, computed as the arithmetic average of $<\log t>_L$, $<\log t>_M$, $<Z>_L$, $<Z>_M$ and $\log(O/H) + 12$ of the galaxies of each group, shown in \autoref{Jclass}. The mean values (stellar age, metallicity, and oxygen abundance) for one galaxy are taken from the synthesis of its integrated spectrum, and the average of the groups as the numerical average of the values of objects in each group. The average mean stellar ages of ES and S galaxies tend to be lower than the other groups, which is consistent with the fact that the stripping enhances star formation along all the galactic radii. However, the uncertainties (standard deviations) pose a challenge in arriving at definitive conclusions.

As computed by \cite{SFR_tails}, the most extreme jellyfish galaxies have a much higher SFR in the tails compared to those with more moderate stripping. Despite this increase in SFR in the tails for higher classes, we do not find any significant difference in the global sSFR of the groups. We computed the mean and standard deviation of the logarithm of masses and sSFR of each group, presented in \autoref{sSFR}. The sSFRs of ES and S galaxies are similar to the CSF ones. Those three have in fact sSFRs higher than PS and CP groups, which encompass galaxies completely or partially quenched. However, the uncertainties in the measurement of those two groups are the largest. Although there seems to be no very clear difference in the global sSFRs, differences arise between the groups when investigating their radial profiles of sSFRs (see \autoref{radial}).

We do not observe a correlation between the intensity of stripping and the stellar or gas-phase metallicities. The ES group seem to be the one with higher metallicities, but both PS and CP groups are more metal-rich than the S galaxies. Again it is difficult to draw strong conclusions from this figure given the uncertainties. The nebular oxygen abundance shows a pattern resembling the stellar metallicities, with the difference that the CP group is the one with higher abundances, although it is also the one with the largest standard deviation. 

\begin{table}
\centering
\caption{Mass, SFR, and sSFR for the galaxies in our sample, as well as the J-classes.}
\begin{tabular}{lcc}
\hline
Group  & log($M_{\star}/M_{\odot}$)  & log(sSFR [$yr^{-1}$]) \\ \cline{1-3}
ES     & 10.5$\pm$0.6 & -10$\pm$1 \\
S      &  9.9$\pm$0.7 & -10$\pm$1 \\
PS     & 10.4$\pm$0.4 & -13$\pm$4 \\
Unkown &  9.3$\pm$0.8 &  -9$\pm$1 \\
CSF    & 10.0$\pm$0.6 & -10.3$\pm$0.8 \\
CP     & 10.3$\pm$0.6 & -13$\pm$4 \\
\hline
\end{tabular}
\label{sSFR}
\end{table}

\section{Discussion}
\label{section5}

In this section, we explore the impact of strip intensity, orbit within the host cluster, and the total stellar mass on stellar population properties (such as ages and metallicities), nebular characteristics, and SFR. First, we investigate our sample's mean stellar ages and metallicities, grouping them accordingly to different stripping stages (see \autoref{j-class}). Then, we examine the scatter of stellar ages and metallicities across different orbits (phase space diagram \ref{phasespace}). Furthermore, we construct 1D radial profiles for each galaxy using light and mass-weighted stellar ages and metallicities, gas-phase oxygen abundances, and specific star formation rates (sSFR). Finally, we compute the median profiles for different bins of stellar mass and groups of stripping intensity.

\subsection{Phase space}
\label{phasespace}

One further analysis interesting to explore is the phase space diagram, which is the galaxy's line-of-sight velocity normalised by the cluster velocity dispersion ($v/\sigma$) versus the projected clustercentric distance normalised by $R_{200}$ \footnote{Clustercentric radius, which encompasses a region with a mean density equal 200 times the critical density of the Universe}. The values of $r/R_{200}$ and $v/\sigma$ for the galaxies of our sample were taken from \cite{SFR_tails}, which contains the values for 54 galaxies in clusters only, where $R_{200}$ were available. It is important to stress that the values for velocities and radius in the phase space are projected measurements, possibly propagating great uncertainties. Velocity and distance are 3D quantities, but we observe only radial velocities and radii projected on the plane of the sky. These projected values work similarly to lower limits.

\begin{figure}
\centering
\includegraphics[width=\linewidth]{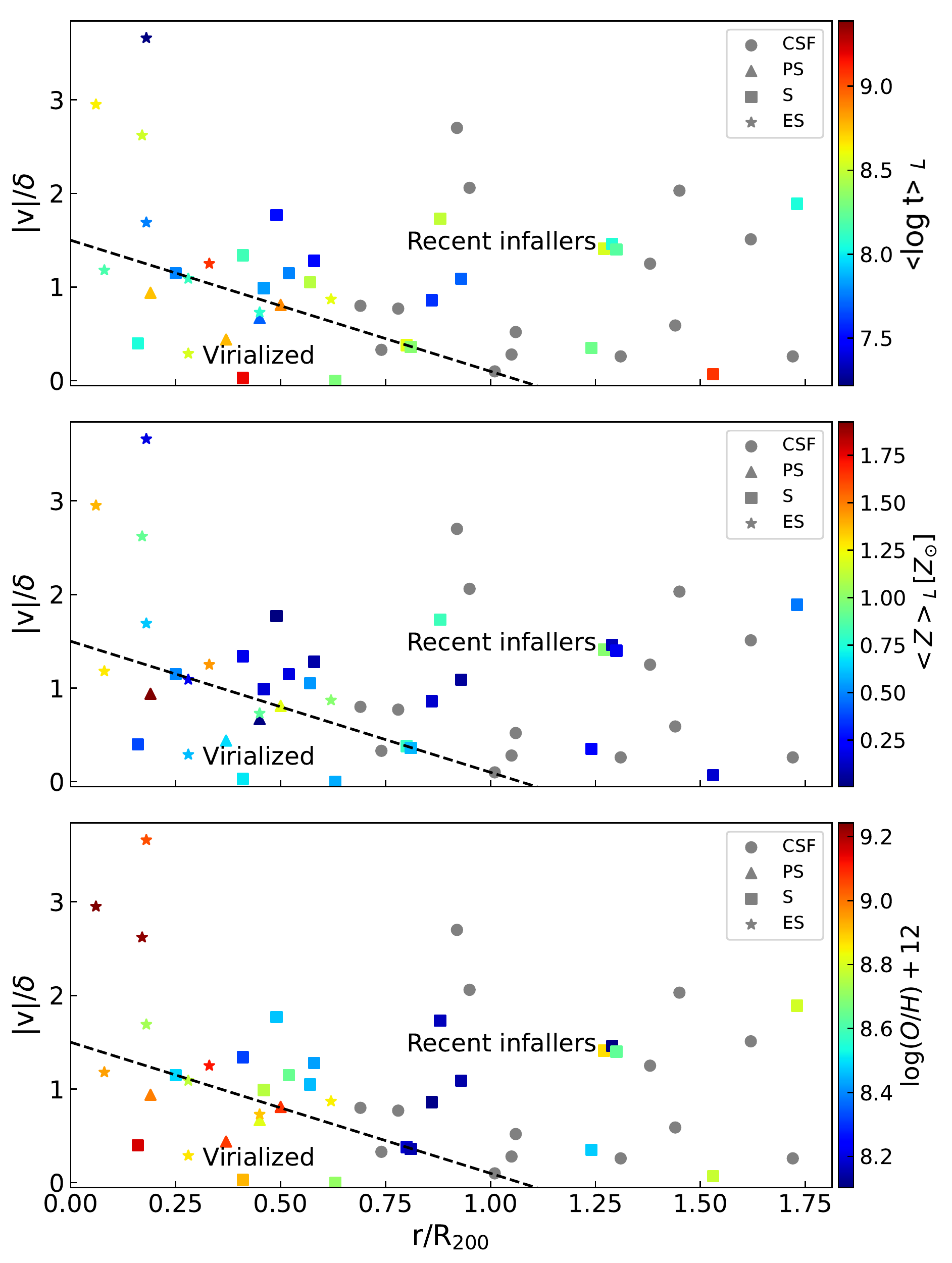}
\caption{Phase space diagram of our sample, presenting the radial velocity vs clustercentric distance, colour-coded by light-weighted age (top panel), by light-weighted metallicity (middle panel), and gas-phase oxygen abundances (bottom panel). The dashed line encompasses the approximated region where one expects most galaxies to be virialized. The grey dots are galaxies from the control sample.}
\label{phase-space}
\end{figure}

Nevertheless, it is worth verifying whether there are possible age and metallicity differences with this environmental probe. \autoref{phase-space} shows two phase-space panels, where $<\log t>_L$ and $<Z>_L$ are plotted in the colour dimension. The region in the lower-left corner is the approximate region where one expects that most galaxies are virialized \citep{mahajan-phase-space, budhies, rhee_phase-space, gaspIX}. Outside it and to the right is the region of recent infallers, where most galaxies are in their first infall or in the ``back-splash'' (i.e. previously entered the cluster but are now beyond the virial radius). This segregation is based on the time since infall (which is defined as the time since the galaxy crossed R$_{200}$ for the first time). The virialized region is dominated by galaxies that entered the cluster $\gtrsim$ 4 Gyr, meanwhile the outside of the virialized region of the phase space, the dominance is for those that entered $\lesssim$ 3 Gyr \citep{rhee_phase-space}.

We computed the medians of ages and metallicities for the jellyfish galaxies in the different regions of the diagram, adopting the median absolute deviation as the uncertainties and not considering control galaxies. We have found that the jellyfish in the virialized region have median ages equal to $8.3\pm0.3$ and $9.0\pm0.5$, for light- and mass-weighted ages respectively. On the other hand, the median ages of the remaining jellyfish are $8.1\pm0.4$ and $8.4\pm0.5$. There is no significant difference between the ages of galaxies inside and outside the virialized region when weighted by light. But when weighted by mass the values present a difference larger than their uncertainties, with the ages in the virialized region being higher.

Compared to ages, the stellar metallicities, on the other hand, seem to have a clearer difference in the medians. The virialized region presents a median of $(0.6\pm0.2)\,Z_{\odot}$ and $(1.0\pm0.6)\,Z_{\odot}$, for light- and mass-weighted ages, while we have $(0.2\pm0.2)\,Z_{\odot}$ and $(0.2\pm0.1)\,Z_{\odot}$ for the recent infallers region. This indicates that the galaxies in the virialized region are more metal-rich than the remaining. For the nebular oxygen abundances, we see a difference between the median of the regions, with the median of $\log(O/H) +12 = (8.9\pm0.1)$ for the virialized region and equal to $(8.6\pm0.2)$ for the recent infallers. The phase space for gas abundances clearly has galaxies with lower values outside the virialized region, however, there are three galaxies with velocities above $2\,\sigma$ and positions below $0.25\,R_{200}$ that have $\log(O/H) + 12 \gtrsim 9$ that are increasing the median. The differences in the values of stellar metallicities are much clearer than in ages and gas abundances. A possible bias involved in this analysis is that mean ages and oxygen abundances involve logarithmic scales, while stellar metallicities are in linear scales. Also, at the same time that the metallicities present the clearest segregation between the regions of the phase space, it is the less reliable result in the synthesis, given that our base has only 4 values for it.

\begin{figure}
\centering
\includegraphics[width=\linewidth]{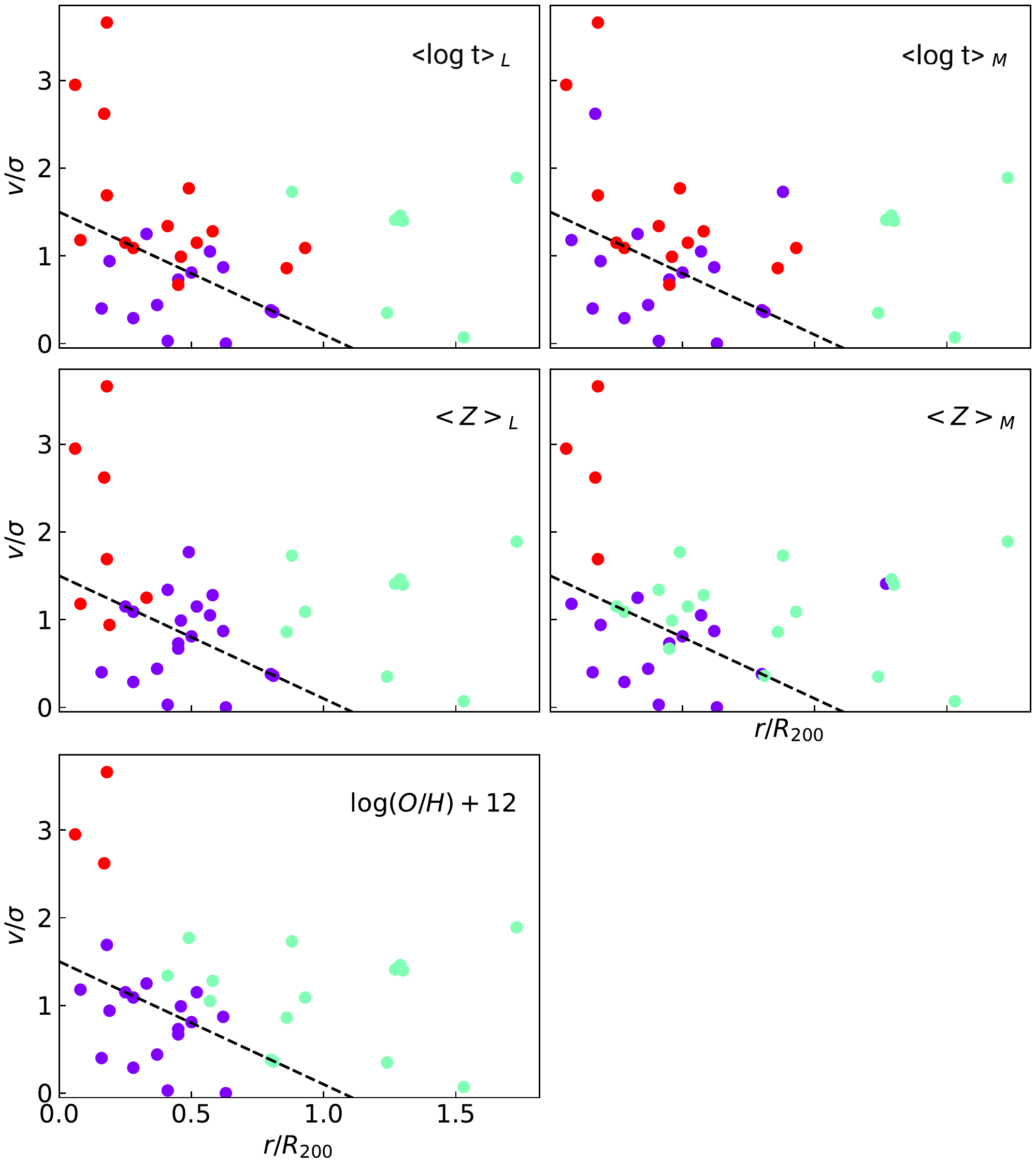}
\caption{K-means clustering method applied in the phase space for cluster members jellyfish galaxies. The first row contains the segregation based on stellar age, the second row the stellar metallicity, and the third row the nebular oxygen abundance. The dashed line encompasses the approximated region where one expects most galaxies to be virialized. Each colour represents a ``k-group'' and their means are listed in \autoref{kmean_tab}.}
\label{kmean}
\end{figure}

In addition to this analysis, we applied a k-mean clustering algorithm \citep{k-means} in the phase space (see \autoref{kmean}). This algorithm computes k centroids (we used k=3 in this case) in a given parameter space and groups the points based on the closest centroid. The parameters we used are the position in phase space plus a measurement of ages or metallicity. This way it is easier to identify if some specific region of the phase space has typically greater or lesser values of stellar ages, metallicities, and nebular abundances. For example, the red group in the stellar ages evaluated k-means, which is dominated by galaxies with $r/R_{200} < 1$ and $v/\sigma \gtrsim 1$, has a mean age lower than the purple and green groups. On the other hand, we see that the red group in the $<Z>_L$ segregation has a higher mean metallicity than the others, while the green group in the $<Z>_M$ present a lower mean metallicity than the remaining. This leads us to the conclusion that the galaxies outside the virialized region and with $r/R_{200}$ are generally the most metal-poor, which is in accordance with \autoref{phase-space}. The segregation in the gas-phase oxygen abundance is even more clear since the 3 galaxies with the higher velocity (red group) have higher mean values, followed by the purple group dominated by galaxies in the virialized region, and at last, the green group dominated by galaxies outside the virialized region and with $r/R_{200} \gtrsim 0.5$. The mean values of each group are listed in \autoref{kmean_tab}, and computing the medians instead of the means does not change significantly the results.

\begin{table}
\centering
\caption{Means of clustering groups}
\resizebox{\linewidth}{!}{%
\begin{tabular}{lccccc}
\hline
Group  &   $<\log t>_L$   &   $<\log t>_M$   &   $<Z>_L$   &   $<Z>_M$   & $\log(O/H)+12$\\ \cline{1-6}
Purple & 8.6$\pm$0.3 & 9.2$\pm$0.4 & 0.5$\pm$0.3 & 1.4$\pm$0.4 & 8.9$\pm$0.2   \\
Green  & 8.4$\pm$0.3 & 8.7$\pm$0.5 & 0.4$\pm$0.3 & 0.2$\pm$0.3 & 8.4$\pm$0.3   \\
Red    & 7.9$\pm$0.4 & 7.9$\pm$0.4 & 1.1$\pm$0.5 & 1.1$\pm$0.7 & 9.2$\pm$0.1   \\
\hline
\end{tabular}%
}
\label{kmean_tab}
\end{table}

\subsection{Analysis of radial profiles}
\label{radial}

\begin{figure}
\centering
\includegraphics[width=\linewidth]{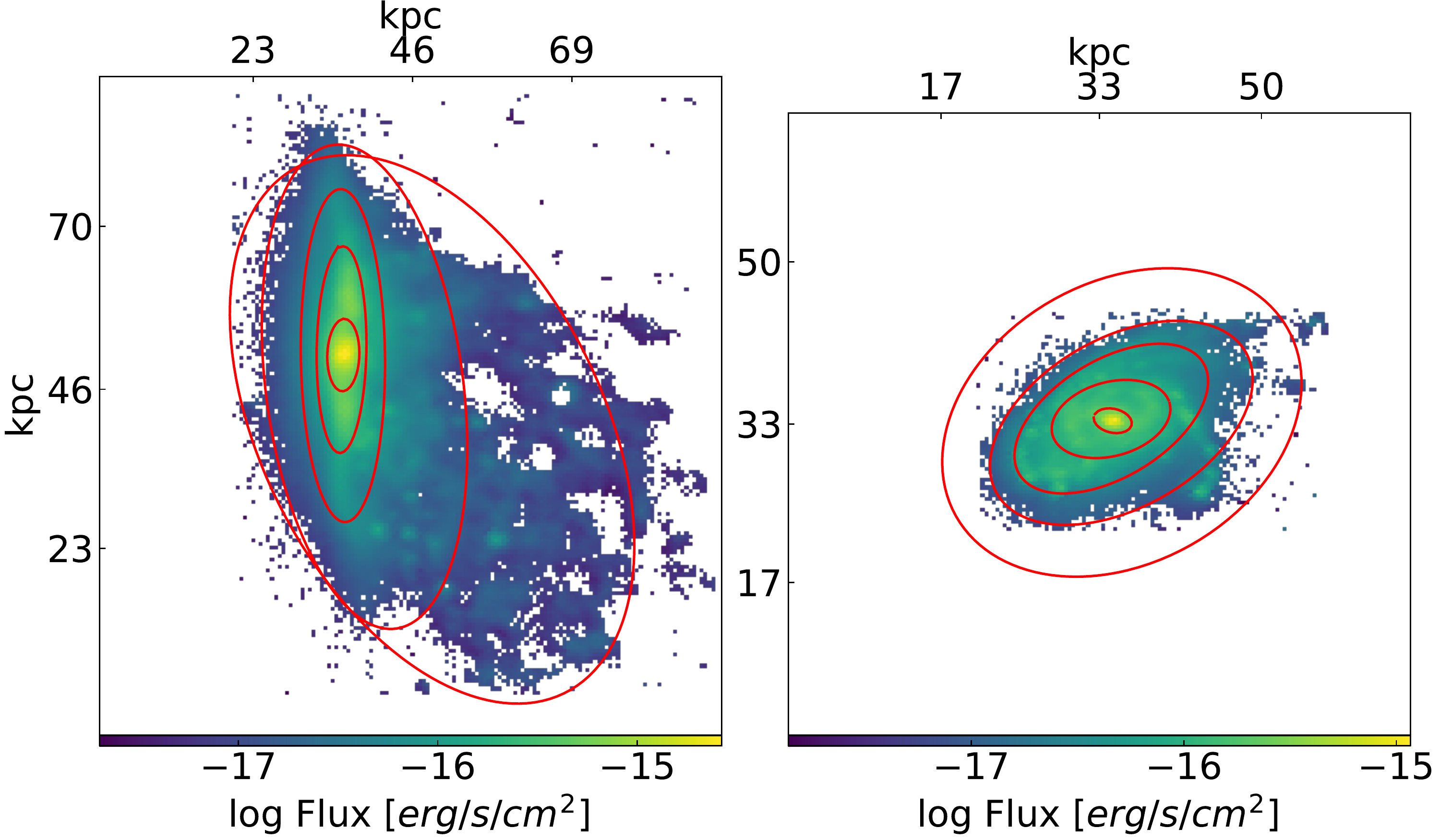}
\caption{This figure shows logarithmic continuum flux maps for two galaxies, an ``extreme stripping'' (ES) galaxy on the left and a ``stripping'' (S) galaxy on the right. The classification was determined based on the $H_\alpha$ emission and the presence of tentacle-shaped substructures. The red curves are elliptical isophotes obtained with {\sc photutils}, which effectively fits the optical continuum flux maps for the galaxies in our sample within a featureless wavelength window of 6530--6600 \AA.}
\label{isophotes}
\end{figure}

Jellyfish galaxies are disturbed and asymmetric. Still, we can fit elliptical isophotes using an iterative method described by \cite{isophotmethod87} that fits considerably well the continuum flux maps for the galaxies in our sample in the featureless wavelength window 6530--6600 \AA. In order to measure elliptical isophotes, we use the Python library \href{https://photutils.readthedocs.io/en/stable/}{\sc photutils} \citep{photutils}, an \href{https://www.astropy.org/}{\sc astropy} package \citep{astropy} for photometry. \autoref{isophotes} shows examples of computed isophotes for a ``stripping'' and a ``extreme stripping'' galaxy.

\begin{figure*}
\centering
\includegraphics[width=0.9\linewidth]{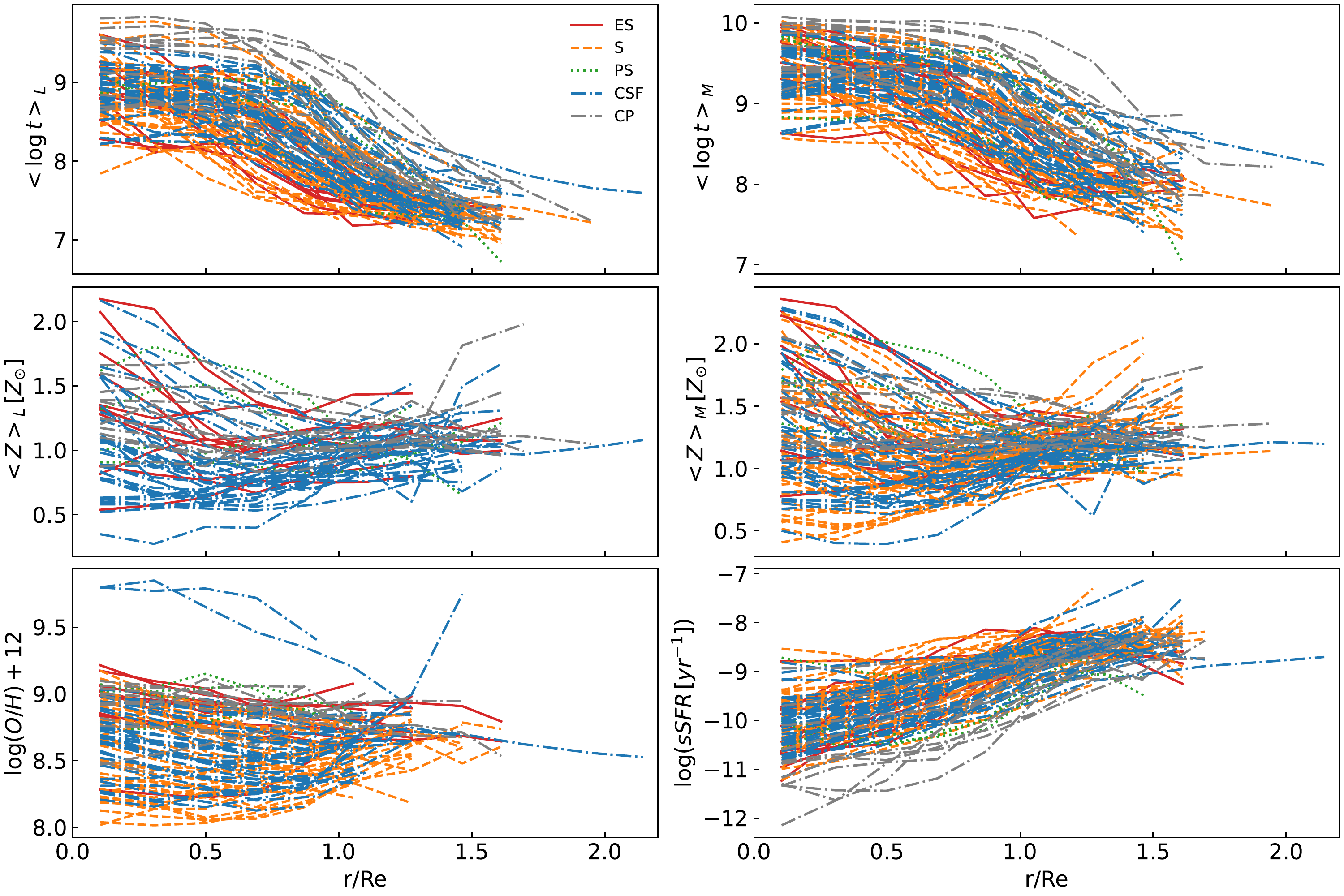}
\caption{Radial profiles of the jellyfish and control galaxies. The first row shows the mean ages, and the second one shows the mean stellar metallicities, both light-weighted (left) and mass-weighted (right). The third row shows the gas metallicities. The red, orange, green, blue, and grey lines denote respectively the ``extreme stripping'', ``stripping'', ``post-stripping'', ``control star-forming'' and ``control passive'' galaxies.}
\label{radial_profiles}
\end{figure*}

By computing the median ages and metallicities for the spaxels between two ellipses, we can obtain radial profiles for stellar ages, stellar metallicities, and gas metallicities. We consider the radius the semi-major axis of the ellipses since the centre and eccentricity of the isophotes can vary largely for the most asymmetric and clumpy galaxies. The radius is normalised by the semi-major axis of the isophote, which contains 90 per cent of the total light of the galaxy (measured in the continuum maps), i.e., the effective radius ($\,R_e$). \autoref{radial_profiles} presents the radial profiles for the individual galaxies in our sample, both light and mass-weighted. We also compute the median profiles for each type of galaxy (ES, S, PS, CSF, and CP). Disregarding spaxels dominated by AGN/LINER-like emission does not change the curves considerably.

 The mass-weighted median age curves are higher than the light-weighted ones, which is expected and also observed in the 2D maps since the stellar masses are dominated by older populations and light is dominated by younger populations. There is a significant age scatter in the profiles of different galaxies. In general, the age profiles decrease with radius and converge around $1.5\,R_e$. Considering light-weighted profiles, the ES group has the lowest mean ages along almost all radii, followed by S and CSF galaxies. The PS group presents similar ages to the other star-forming groups, but it stays approximately constant until $\sim0.9\,R_e$, when it starts decreasing and gets very close to the profile of the CP group. On the other hand, when considering mass-weighted ages, the ES profile is close to the PS profile and higher than the S and CSF ones until $0.5\,R_e$, but after $\sim0.75\,R_e$ it presents the lowest values again.

Stellar and nebular metallicities also present great spread in the individual galaxy profiles. Mass-weighted stellar metallicities seem to be altogether greater than light-weighted ones, as already shown in \autoref{Jclass}. The scatter in stellar metallicity and gas-phase oxygen abundance is relatively larger than the scatter in ages. Until almost the entire effective radius the S and CSF groups present the lowest stellar metallicities and oxygen abundances. The PS stellar metallicities profile is the largest between $\sim0.5\,R_e$ and $\sim0.75\,R_e$, however, the ES mass profile is more metal-rich in the central regions, presenting a rapid decrease. Both S and CSF metallicities and oxygen abundance profiles start decreasing and then increase along the effective radius. The increase present in the ES, S, and CSF profiles $0.75\,R_e$ is consistent with a chemical enrichment caused by the increased SFR in the outskirts and tails. It is worth noting that the PS group is the less populated one, formed by only 4 galaxies. Indeed, there is a point missing in the gas metallicity profile of the PS group, because we were not able to compute the abundances for distances much larger than the effective radius, except for one of the galaxies. This problem arises most likely due to the low SNR of emission lines in the galactic outskirts and due to quenched disks.

We computed radial sSFR for our galaxies dividing the median SFR by the median mass in each annulus. In general, the sSFRs increase for larger radius, except for the PS profile, which has an initial decrease until $\sim0.5\,R_e$. The CP galaxies present the lowest sSFR for radii lower than $\sim0.75\,R_e$, where it becomes very close or even higher than the profile of the PS galaxies. The median sSFR of the remaining three groups are very close to each other for most radii, and also larger than the other two groups.

\begin{figure*}
\centering
\includegraphics[width=0.9\linewidth]{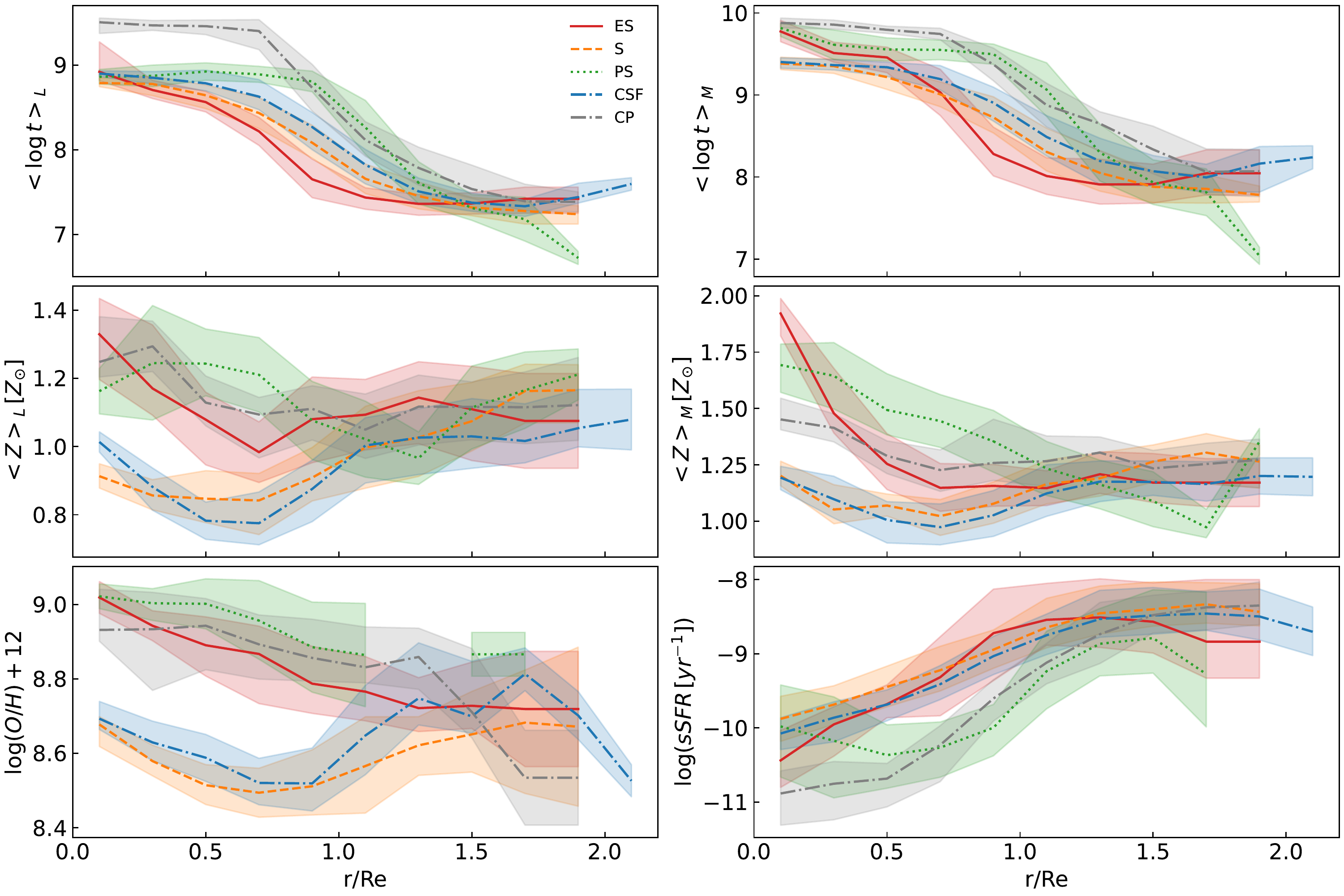}
\caption{Median radial profiles of each group of jellyfish and control galaxies. The first row shows the mean ages, and the second one shows the mean stellar metallicities, both light-weighted (left) and mass-weighted (right). The third row shows the gas metallicities. The red, orange, green, blue, and grey lines denote respectively the ``extreme stripping'', ``stripping'', ``post-stripping'', ``control star-forming'' and ``control passive'' galaxies.}
\label{radial_profiles_median}
\end{figure*}

\begin{figure*}
\centering
\includegraphics[width=0.9\linewidth]{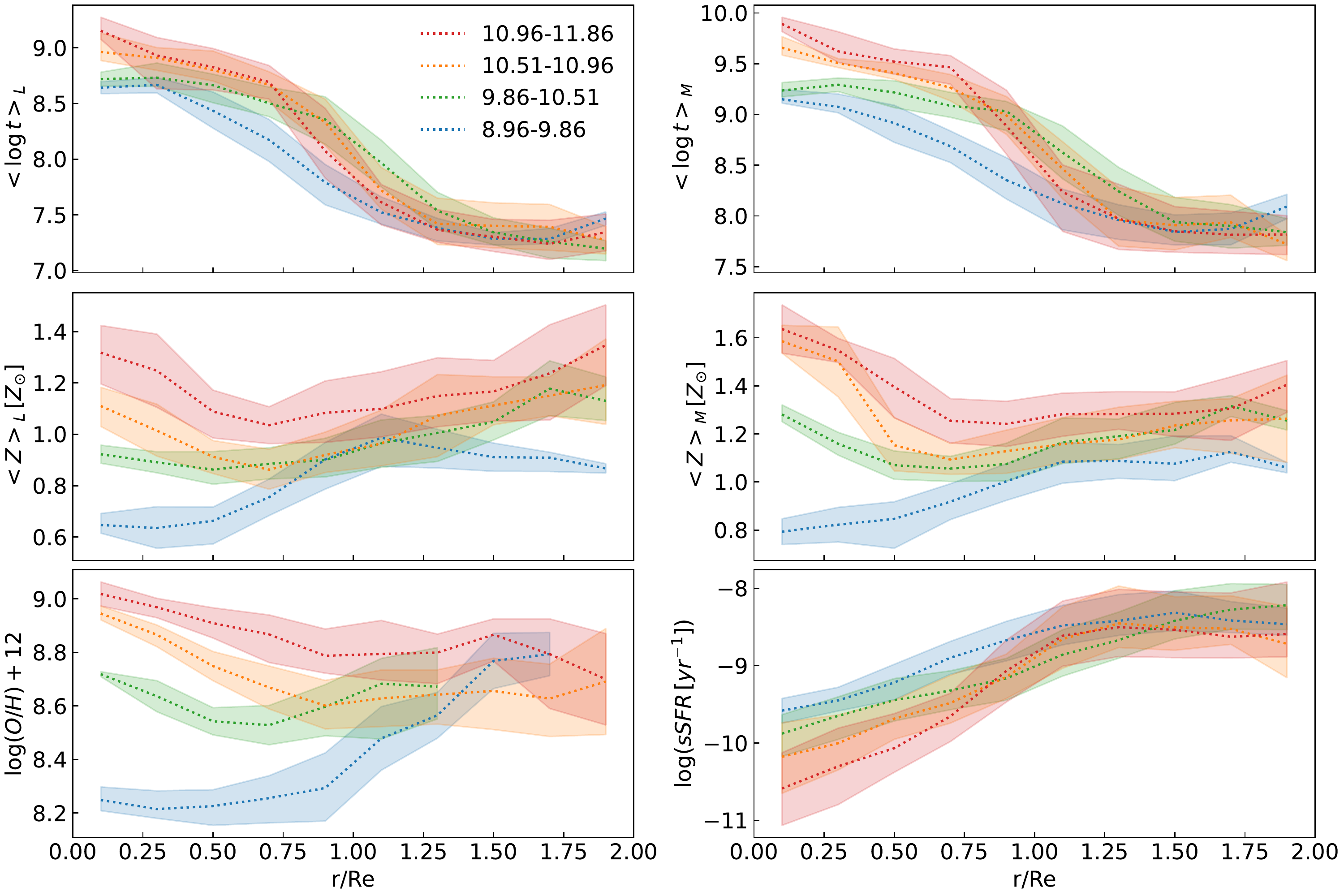}
\includegraphics[width=0.9\linewidth]{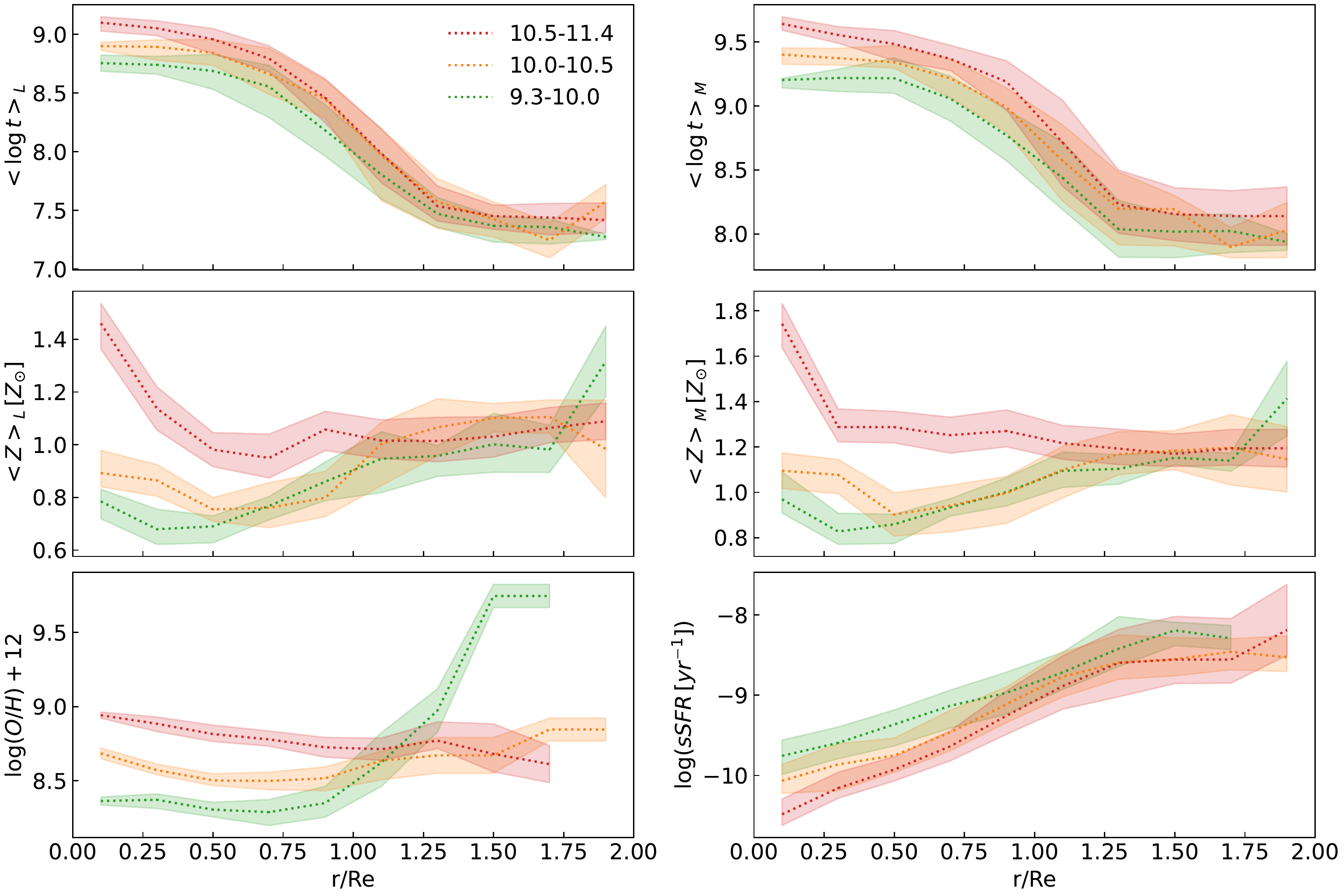}
\caption{Radial profiles of the jellyfish (top) and control star-forming (bottom) galaxies grouped in bins of mass. The first row presents the logarithmic median ages, the second presents the median stellar metallicities, and the third row presents the radial profiles of oxygen abundance and sSFR. The left panels are light-weighted, while the right panels are weighted by mass.}
\label{radial_profiles_mass}
\end{figure*}

To further investigate the relation between radial profiles and stellar masses, we divided the 61 jellyfish galaxies in our sample into four stellar mass bins, with 15 or 16 galaxies each, and then computed the median profiles for each bin. Those bins are $\log(M_{\star}/M_{\odot}) =$ 8.96-9.86, 9.86-10.51, 10.51-10.96, and 10.96-11.86. \autoref{radial_profiles_mass} shows the radial profiles of the four bins. For both light and mass-weighted ages the more massive bin is the largest in the central regions but decreases more rapidly than the others, especially after $\sim0.75\,R_e$. In almost all the effective radii the second most massive bin present ages higher than the third one, and for greater radii the relation is inverted. The age profiles of the less massive bin stay the lowest for more than $1\,R_e$ and after $\sim1.25\,R_e$ the profiles converge. We also show, for comparison, the profiles for control star-forming galaxies, although we divide this subsample into only three mass bins, given that it is less populated than the jellyfish galaxies.

The patterns found here are somewhat different from previous studies in the literature for undisturbed galaxies. Segregation in mass is also seen in \cite{gonzalez14}, who computed the radial profiles of mean ages for galaxies in the CALIFA survey. That work also separated their sample into bins of mass, and both the whole sample and the sub-samples of disk-dominated galaxies and spheroidal-dominated galaxies present the same pattern of ages increasing with mass, especially for $r\lesssim1\,R_e$. \cite{parikh21} analysed a sample of MaNGA galaxies and found resembling results, although the less massive bin in their sample of late-type galaxies is not amidst the younger ones. \cite{peng10} shows that the fraction of red galaxies and the quenching efficiency increase with stellar mass and environment density for zCOSMOS and SDSS galaxies. \cite{mobasher09} finds that the $\Sigma_{SFR}$ decrease with increasing mass for COSMOS objects and according to \cite{gallazzi05} the stellar ages grow with mass for SDSS galaxies, suggesting that the most massive systems have the most of their star formation activity at earlier epochs. 

In our work, although such relation of ages increasing with mass is maintained until $\sim0.8\,R_e$, our most massive bins present a steeper decrease, and after such value and do not keep having the largest ages. On top of that, the profiles of the two most massive bins are very close to each other, especially for the light-weighted values. A possible explanation for this more rapid decay in ages is that the ES galaxies and also the ones with the longest tails between the ``stripping'' group are usually between the most massive objects \citep{gaspIX}. The more massive galaxies are expected to have larger gas reservoirs to form very long tails without having all of their gas stripped and without being quenched. Because a great number of galaxies among the most massive ones are jellyfish with long tails, the stellar ages decrease greatly once the fitted ellipses reach the outskirts and tail of those objects.

The stellar metallicities show a different separation between the bins. The most massive bin seems to be the most metal-rich along all radii. Until at least $\sim0.6\,R_e$ the metallicities increase with mass, but for radii larger than that the second and third most massive bins become very similar. The profile of the less massive bin is the more metal-poor for almost all radii, except around $\sim1\,R_e$. There seems to be a correlation between mass and stellar metallicity for jellyfish galaxies, especially for radii $<1\,R_e$. \cite{parikh21} and \cite{lian18} also study the stellar metallicity radial profiles of MaNGA galaxies and find that Z grows with stellar mass. That is exactly the case in our sample, so such a result is somewhat expected.

Regarding nebular metallicities, there is also a clear dependency on oxygen abundance and mass. For almost all the effective radii the higher the mass is, the larger the median gas-phase metallicity. For radii $\gtrsim1\,R_e$ the uncertainties increase greatly and the curves start converging. For both stellar and nebular metallicities, the less massive bin profiles increase up to $1\,R_e$ before flattening, while the others decrease or are approximately constant with effective radius. In the case of oxygen abundance, it increases the steepness of the growth after $\sim1\,R_e$. These results are in accordance with the mass-metallicity relation \citep{mzr}. Galaxies with lower masses have lower gas-phase metallicities than galaxies with higher masses. The reason is that shallow potentials are less effective in retaining the metals in the ISM, which are expelled by the galactic winds.

In the case of the sSFR radial profiles, a trend with mass becomes apparent, where higher mass bins present lower values of sSFR for radii $\lesssim0.75\,R_e$. After such value, the profiles of the two most massive bins become very close to each other and higher than the profile of the third most massive bin. This is very similar to what happens in the age profiles. At the same radius where the median ages decrease more rapidly, the sSFR also grows at a higher rate. Such growth of the sSFR is also probably due to the high SFR in the tails of the most prominent cases of jellyfish, which are among the most massive galaxies in the sample. Similar trends in stellar mass were also found in \cite{belfiore18} for MaNGA main sequence galaxies and in \cite{coenda19} for CALIFA late-type galaxies, i.e, the sSFR increase for decreasing stellar mass. However, our sample follows such a trend only until $\sim0.75\,R_e$.

Our results, combined with previous works, indicate that both intensities of stripping and stellar masses influence the ages, metallicities, gas-phase abundances, and sSFRs of the jellyfish galaxies. While the median age profiles of S galaxies decay at a rate similar to CSF galaxies but are lower, the ES galaxies can have older populations in the inner regions, but the ages decrease more rapidly in the outskirts and tails. The PS galaxies have stellar ages approximately constant with effective radius and are typically older than other star-forming galaxies. ES and PS galaxies are generally more metal-rich than S and CSF galaxies, for both stellar and nebular metallicities. The patterns in the sSFRs of those groups follow the same relation as in the ages, only where the ages decrease, the sSFRs increase. The radial profiles for different mass bins present relations, in a certain way, already shown in the literature for undisturbed galaxies. However, after $\sim0.75\,R_e$ the ages and sSFRs of the most massive bins differ from the expected.

\section{Summary and conclusions}
\label{section6}
In this work, we perform a spatially resolved stellar population synthesis in jellyfish galaxies taking into account nebular emission. We used a large sample of jellyfish galaxies and undisturbed comparative ones from GASP with the MUSE IFS. Our sample comprises 61 jellyfish galaxies, 47 control galaxies (both star-forming and quenched), and 5 objects with asymmetries that seem to be caused by phenomena other than RPS. We implement the stellar population synthesis on the spectra with SNR$>2$ in the H$_{\alpha}$ emission line using the code {\sc fado} with a base of 68 SSPs from \cite{BC03}. {\sc fado} allows us to synthesise spectra under self-consistent boundary conditions based on measurements of emission lines and computation of nebular emission, resulting in more reliable results for star-forming regions, which involve ionising populations (t $\lesssim$ 20 Myr). Our main results and conclusions are summarised below:

\begin{itemize}
    \item We created 2D morphological maps with {\sc inla} of mean stellar ages and metallicities, H$_{\alpha}$ emission, SFR surface densities, and gas-phase oxygen abundances for all galaxies.
    \item From the stellar population synthesis, we were able to derive SFHs. We observe that the control and ``post-stripping'' galaxies reach a fraction of light and mass close to 1 sooner than ``stripping'' and ``extreme stripping'' galaxies, reflecting the recent star formation enhancement due to RPS.
    \item Computing the means of $<\log t>$, $<Z>$ and $\log(O/H) + 12$ of galaxies in each type of intensity of stripping, we have found that ``stripping'' and ``extreme stripping'' are typically younger than the other groups, although the scatter in the values is considerably large. The stellar populations of ``extreme stripping'' galaxies are generally the most metal-rich, followed by ``post-stripping'' and ``control passive'' galaxies, and then by ``stripping'' and ``control star-forming'' ones. The relations between gas-phase abundances are very similar to the ones between stellar metallicities, except the ``control passive'' group presents the highest abundances.
    \item Based on the phase space diagram and k-means clustering algorithm for jellyfish cluster members: galaxies with $r/R_{200} < 1$ and $v/\sigma \gtrsim 1$ have a mean age lower than the others; galaxies outside the virialised region and with $r/R_{200} > 0.5$ are generally the most metal-poor; excluding the 3 galaxies with $v/\sigma > 2$ and $r/R_{200} < 0.5$, galaxies inside the virialised region have higher nebular oxygen abundances than the ones outside it.
    \item We produce radial profiles for stellar ages and metallicities, nebular oxygen abundances, and sSFRs, by fitting elliptical isophotes to the continuum maps of the galaxies. In addition to the individual galaxy profiles, we also computed the median curves grouping the galaxies based on the intensity of stripping and stellar masses.
    \item The age profiles of ``stripping'' galaxies decay at a rate similar to ``control star-forming'' galaxies but are lower. The ``extreme stripping'' galaxies can have older populations than ``stripping'' and ``control star-forming'' groups in the inner regions, but the ages decrease more rapidly after $\sim0.75\,R_e$, probably due to the long tails of star-forming regions. ``Post-stripping'' are typically older than other star-forming galaxies along almost all radii.
    \item ``Extreme stripping'' galaxies and ``post-stripping'' galaxies are generally more metal-rich than ``stripping'' and ``control star-forming'' galaxies, for both stellar and nebular metallicities.
    \item The radial profiles for different mass bins present relations, in a certain way, already shown in the literature for undisturbed galaxies, i.e., profiles of ages and metallicities tend to increase with mass, and sSFRs decrease with mass \citep{gallazzi05, mobasher09, gonzalez14, belfiore18, lian18, coenda19, parikh21}. However, after $\sim0.75\,R_e$, the ages of the most massive galaxies become similar to or lower than the ages of lower mass galaxies.
\end{itemize}

Jellyfish galaxies prove themselves to be curious objects and wonderful systems to deepen our understanding of galaxy evolution. In this work, we perform a comprehensive spatially resolved stellar population synthesis on the large public sample of jellyfish galaxies from GASP, attempting to extract general trends. Our analysis found interesting aspects about the influence of ram pressure stripping and stellar mass on the spatial distribution of stellar populations and nebular content on such objects.

\section*{Data Availability}
The data underlying this article are available from the ESO archive.  (\url{https://archive.eso.org/scienceportal/home}).

\section*{Acknowledgements}
We thank the referee, Florence Durret, for useful comments that have helped improve the paper.
This study was developed as part of an undergraduate research project and later completed during the first part of the PhD of GMA funded by Conselho Nacional de Desenvolvimento Científico e Tecnológico (CNPq) and Coordenação de Aperfeiçoamento de Pessoal de Nível Superior (CAPES Proj. 0001) and the Programa de Pós-Graduação em Física (PPGFis) at UFRGS. ACS acknowledges funding from the CNPq and the Rio Grande do Sul Research Foundation (FAPERGS) through grants CNPq-11153/2018-6, CNPq-314301/2021-6, FAPERGS/CAPES 19/2551-0000696-9. ACS was partially supported by the Chinese Academy of Sciences (CAS) President's International Fellowship Initiative (PIFI) through grant 2021VMC0005. RR acknowledges CNPq (Proj. 311223/2020-6,  304927/2017-1 and 400352/2016-8), FAPERGS (Proj. 16/2551-0000251-7 and 19/1750-2) and CAPES. RR also acknowledges support from the Fundaci\'on Jes\'us Serra and the Instituto de Astrof{\'{i}}sica de Canarias under the Visiting Researcher Programme 2023-2025 agreed between both institutions, from the ACIISI, Consejer{\'{i}}a de Econom{\'{i}}a, Conocimiento y Empleo del Gobierno de Canarias and the European Regional Development Fund (ERDF) under grant with reference ProID2021010079, and the support through the RAVET project by the grant PID2019-107427GB-C32 from the Spanish Ministry of Science, Innovation and Universities MCIU. This work has also been supported through the IAC project TRACES, which is partially supported through the state budget and the regional budget of the Consejer{\'{i}}a de Econom{\'{i}}a, Industria, Comercio y Conocimiento of the Canary Islands Autonomous Community. J.M.G. thanks FCT, which supported this work via Fundo Europeu de Desenvolvimento Regional (FEDER) through COMPETE2020 - Programa Operacional Competitividade e Internacionalização (POCI) through the research grants UID/FIS/04434/2019, UIDB/04434/2020 and UIDP/04434/2020 and an FCT-CAPES Transnational Cooperation Project. J.M.G. is supported by the DL 57/2016/CP1364/CT0003 contract and acknowledges the previous support by the fellowships CIAAUP-04/2016-BPD in the context of the FCT project UID/FIS/04434/2013 and POCI-01-0145-FEDER-007672, and SFRH/BPD/66958/2009 funded by FCT and POPH/FSE (EC). JPVB acknowledges financial support from CNPq, CAPES (Proj. 0001), and the Programa de Pós-Graduação em Física (PPGFis) at UFRGS. We acknowledge the use of HAL9000 hosted at SHAO. We thank Mi Chen, Fernanda Roman-Oliveira, Nicolas Dullius Mallmann and Marina Trevisan for helpful discussions. 

\bibliographystyle{mnras}
\bibliography{refs} 

\section*{Suplementary material}

In this section, we present 2D maps similar to \autoref{JO113_maps} for the remaining galaxies in the sample.

\begin{figure}
\centering
\includegraphics[width=\linewidth]{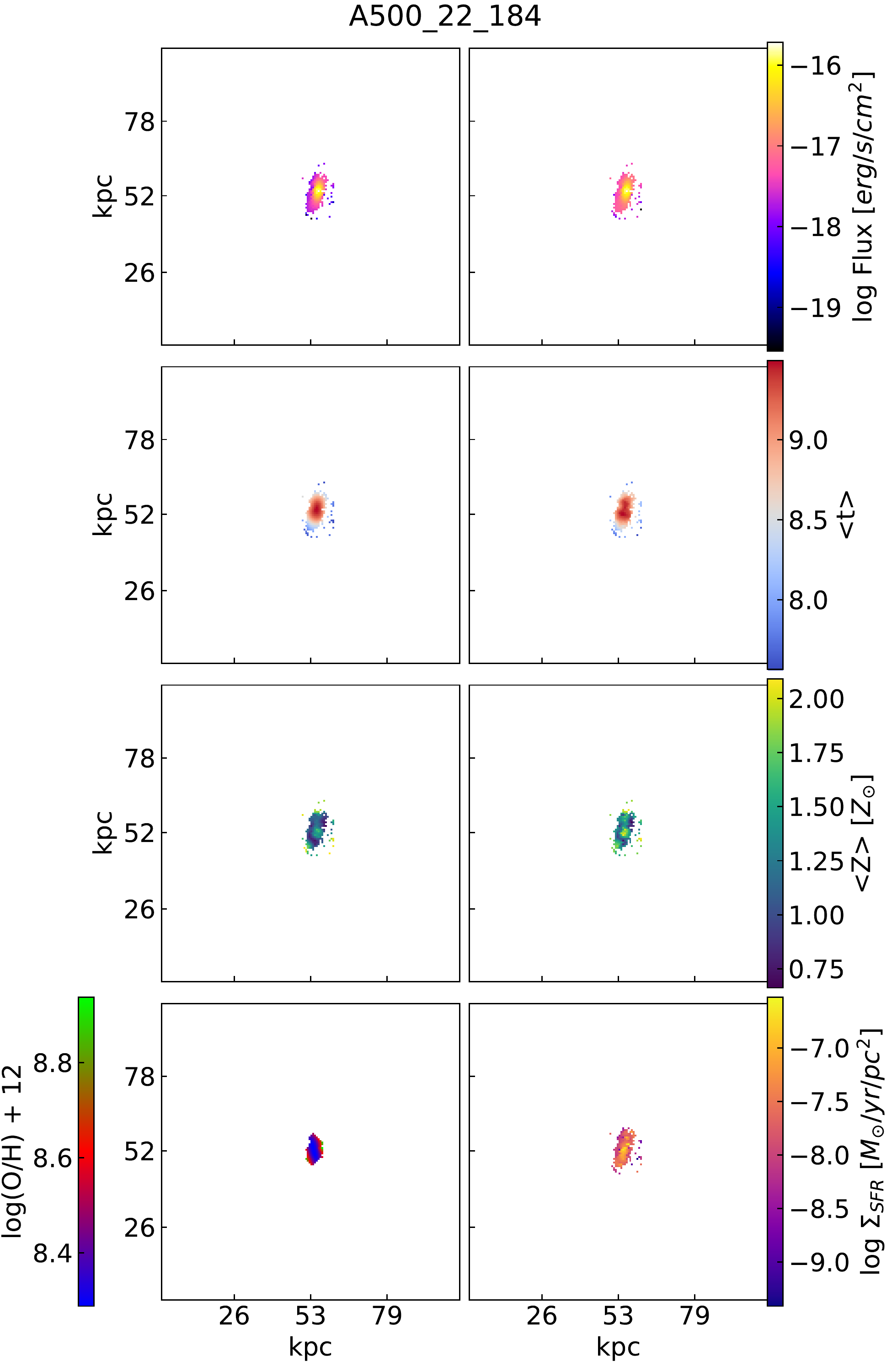}
\end{figure}

\begin{figure}
\centering
\includegraphics[width=\linewidth]{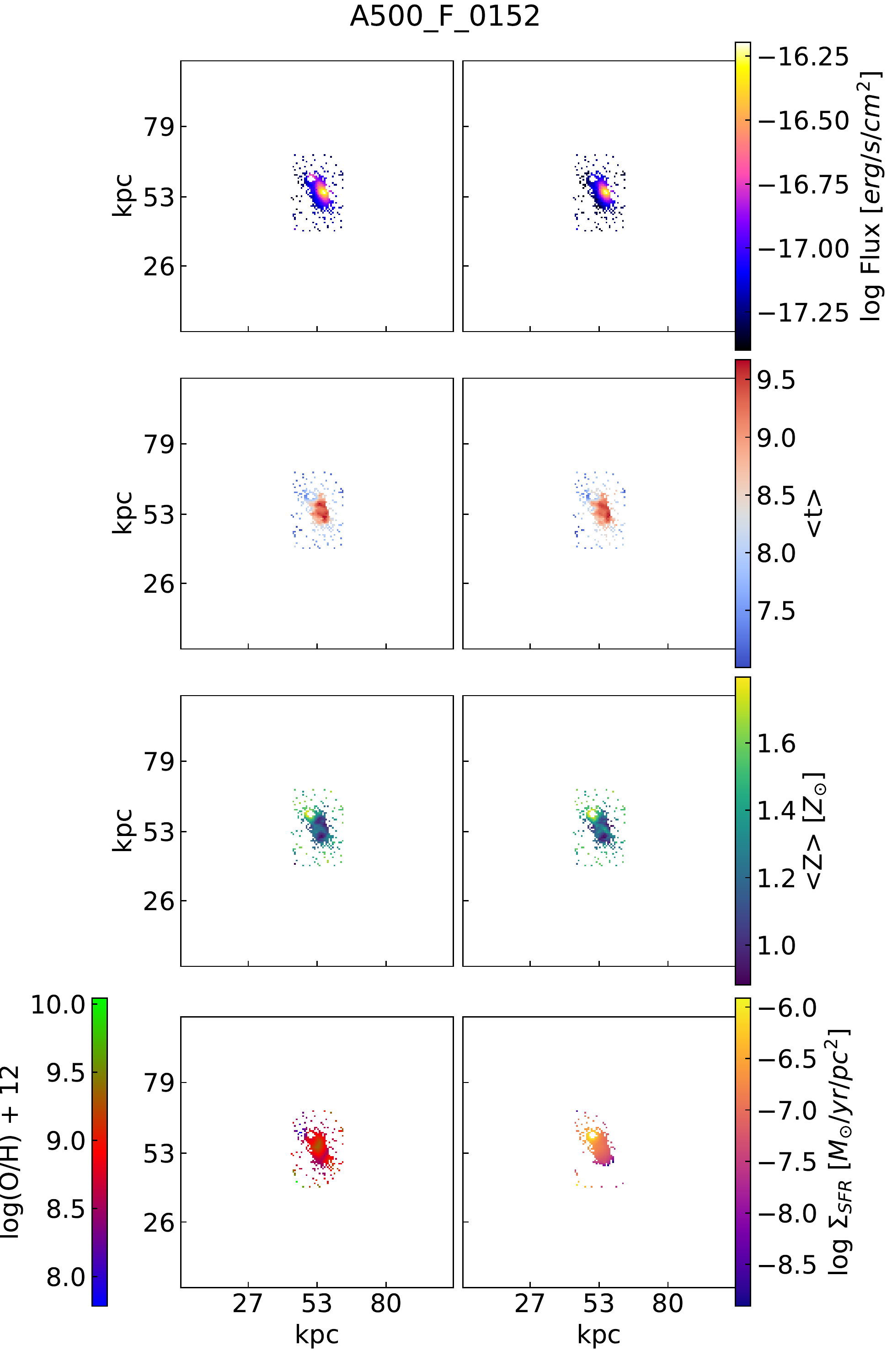}
\end{figure}

\begin{figure}
\centering
\includegraphics[width=\linewidth]{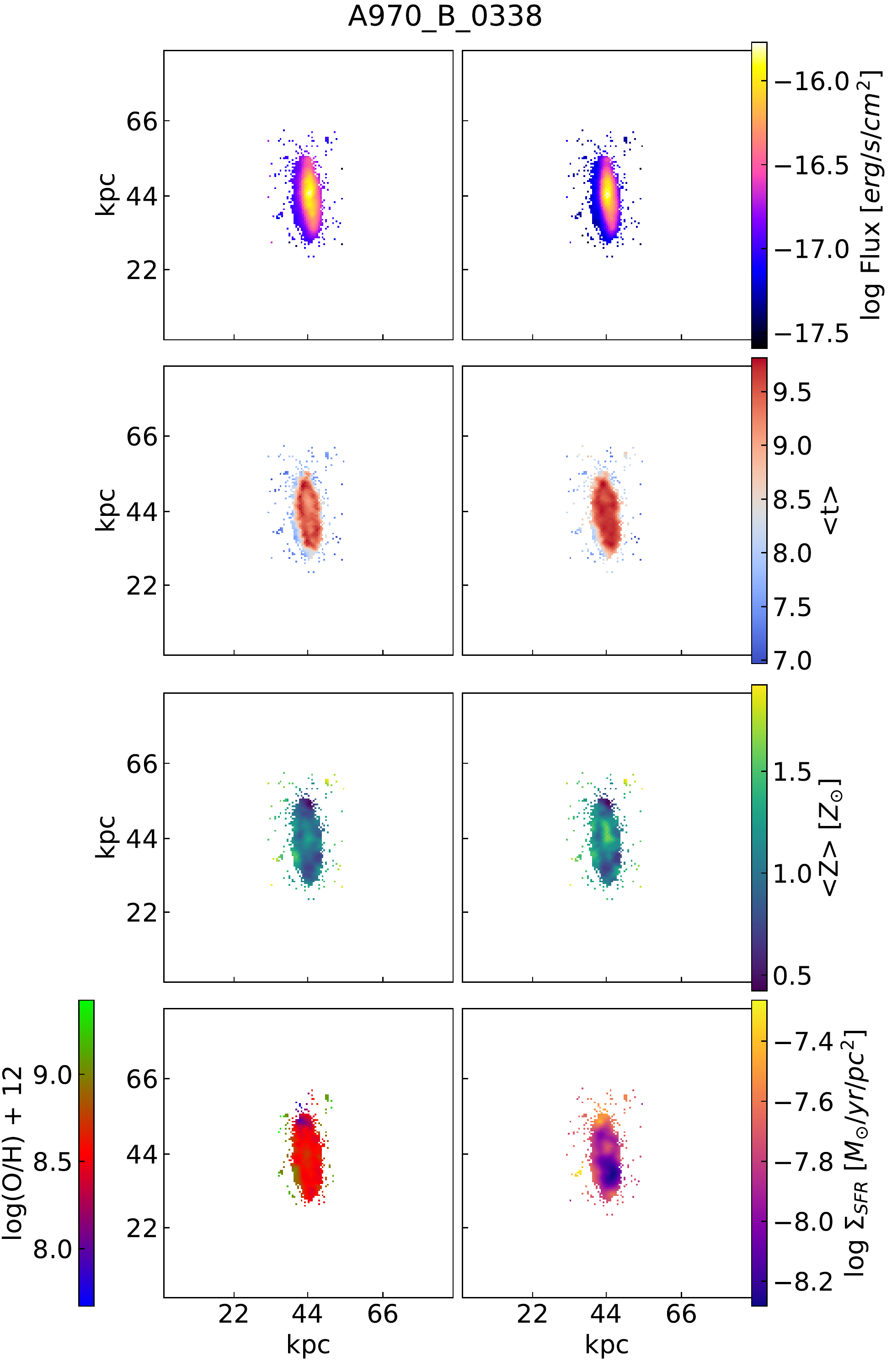}
\end{figure}

\begin{figure}
\centering
\includegraphics[width=\linewidth]{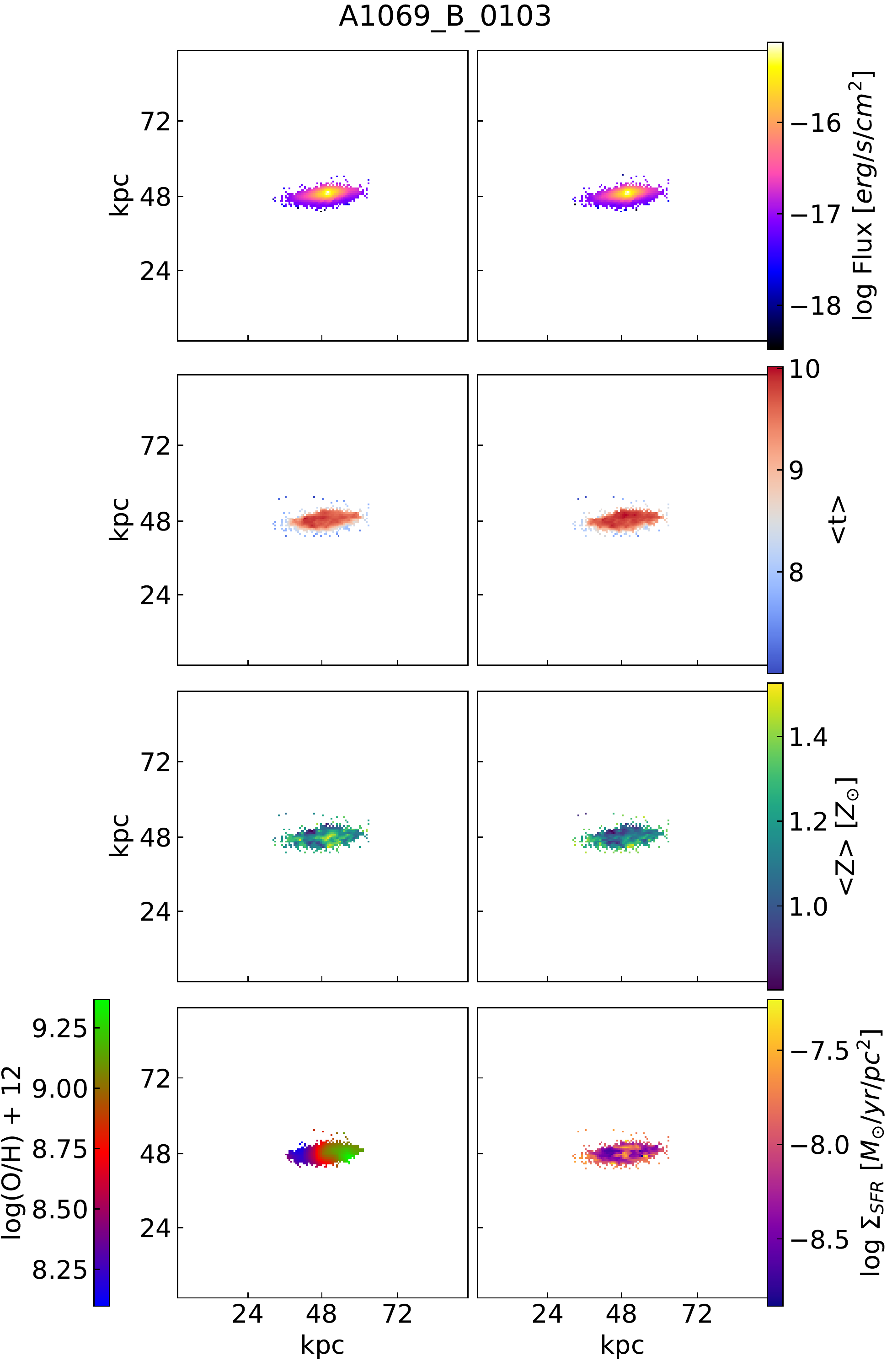}
\end{figure}

\begin{figure}
\centering
\includegraphics[width=\linewidth]{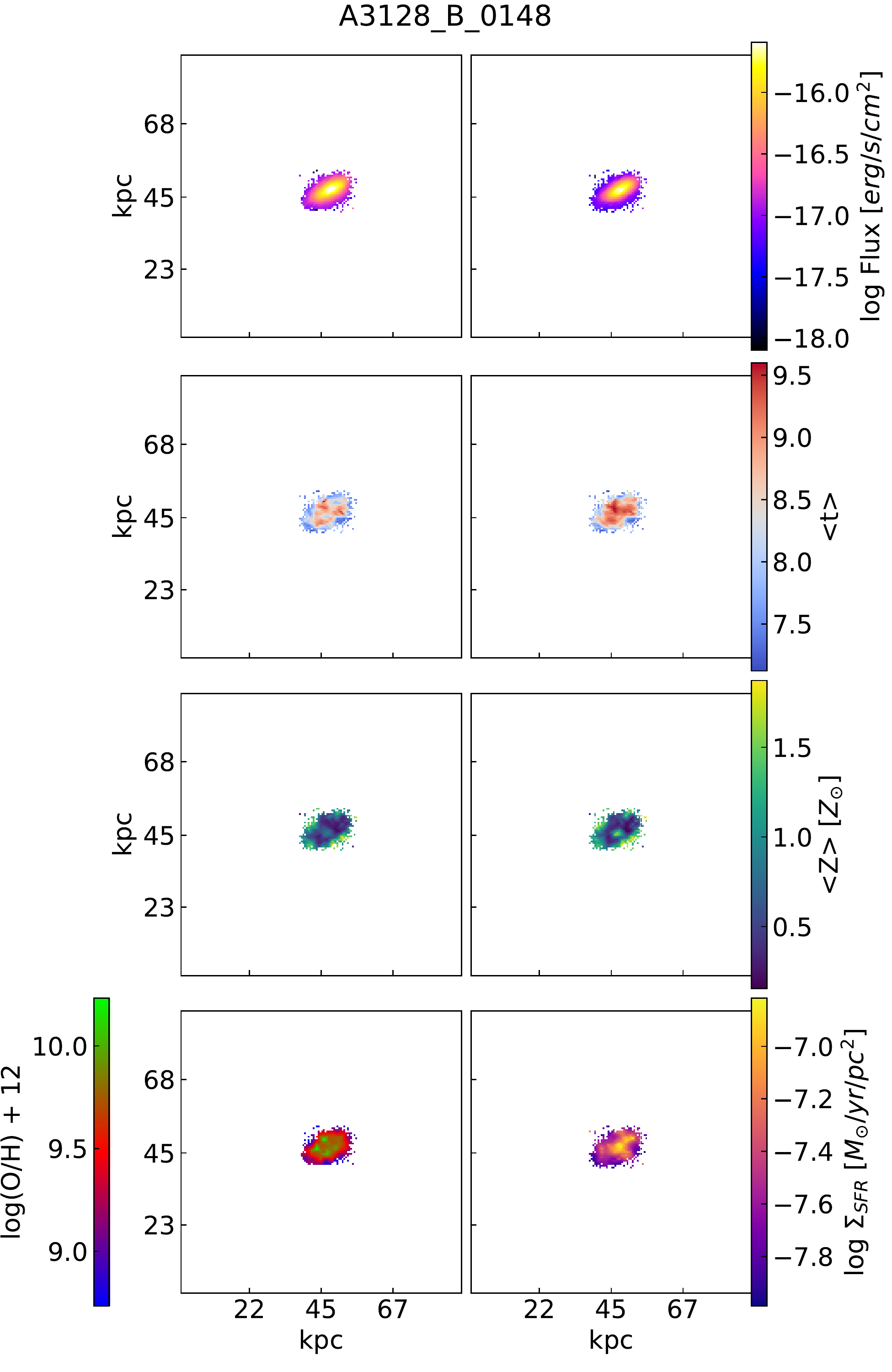}
\end{figure}

\begin{figure}
\centering
\includegraphics[width=\linewidth]{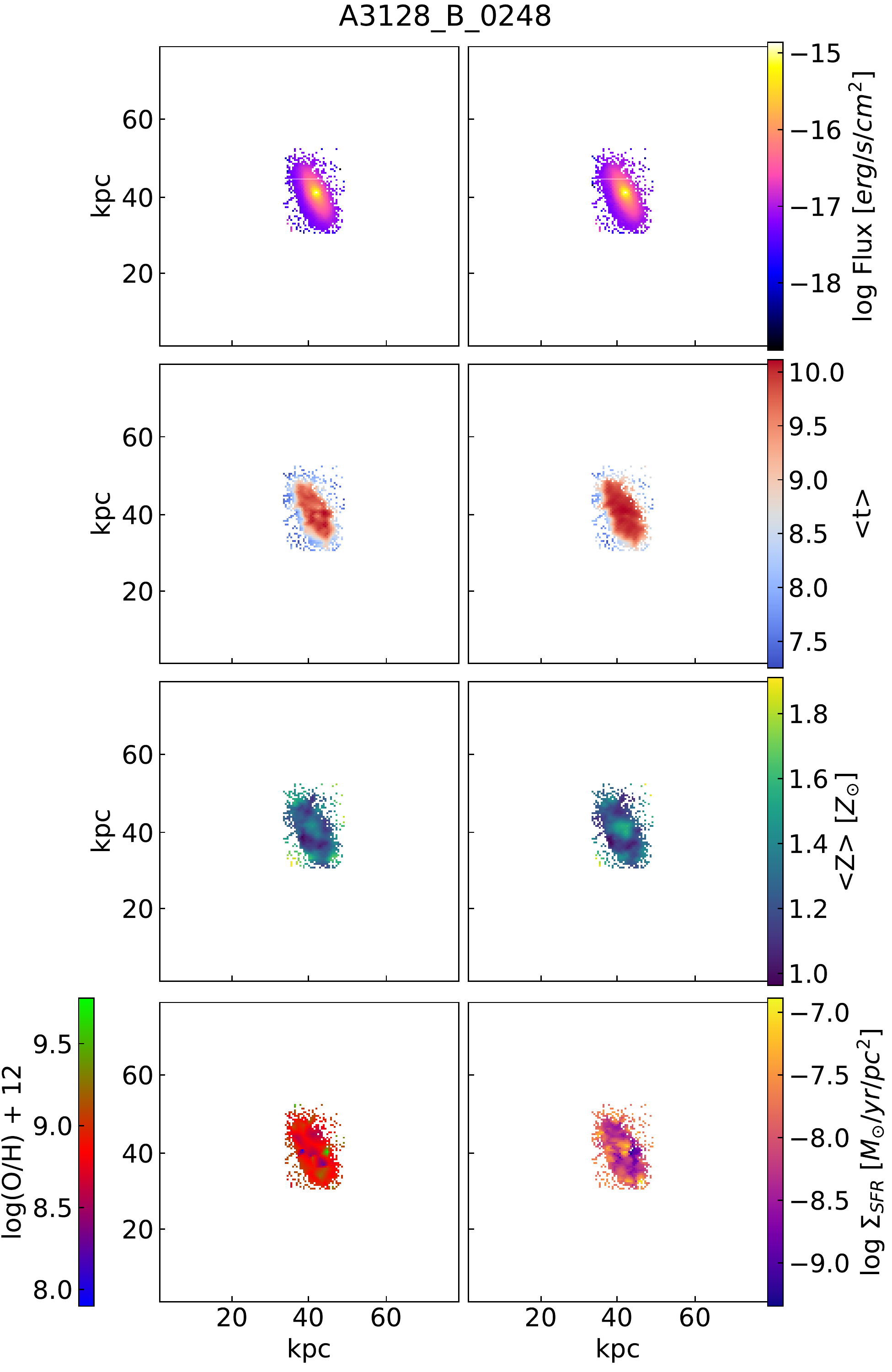}
\end{figure}

\begin{figure}
\centering
\includegraphics[width=\linewidth]{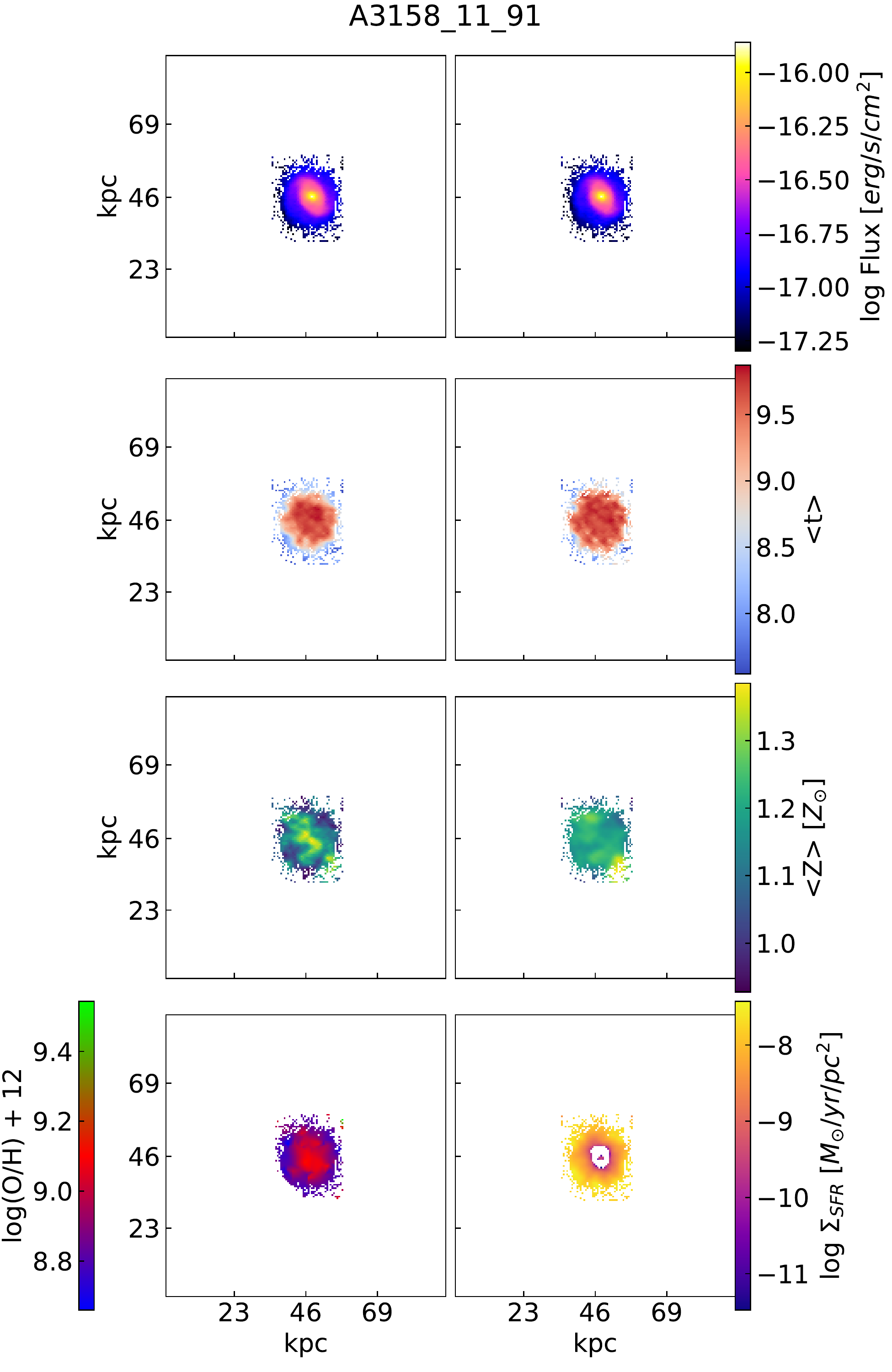}
\end{figure}

\begin{figure}
\centering
\includegraphics[width=\linewidth]{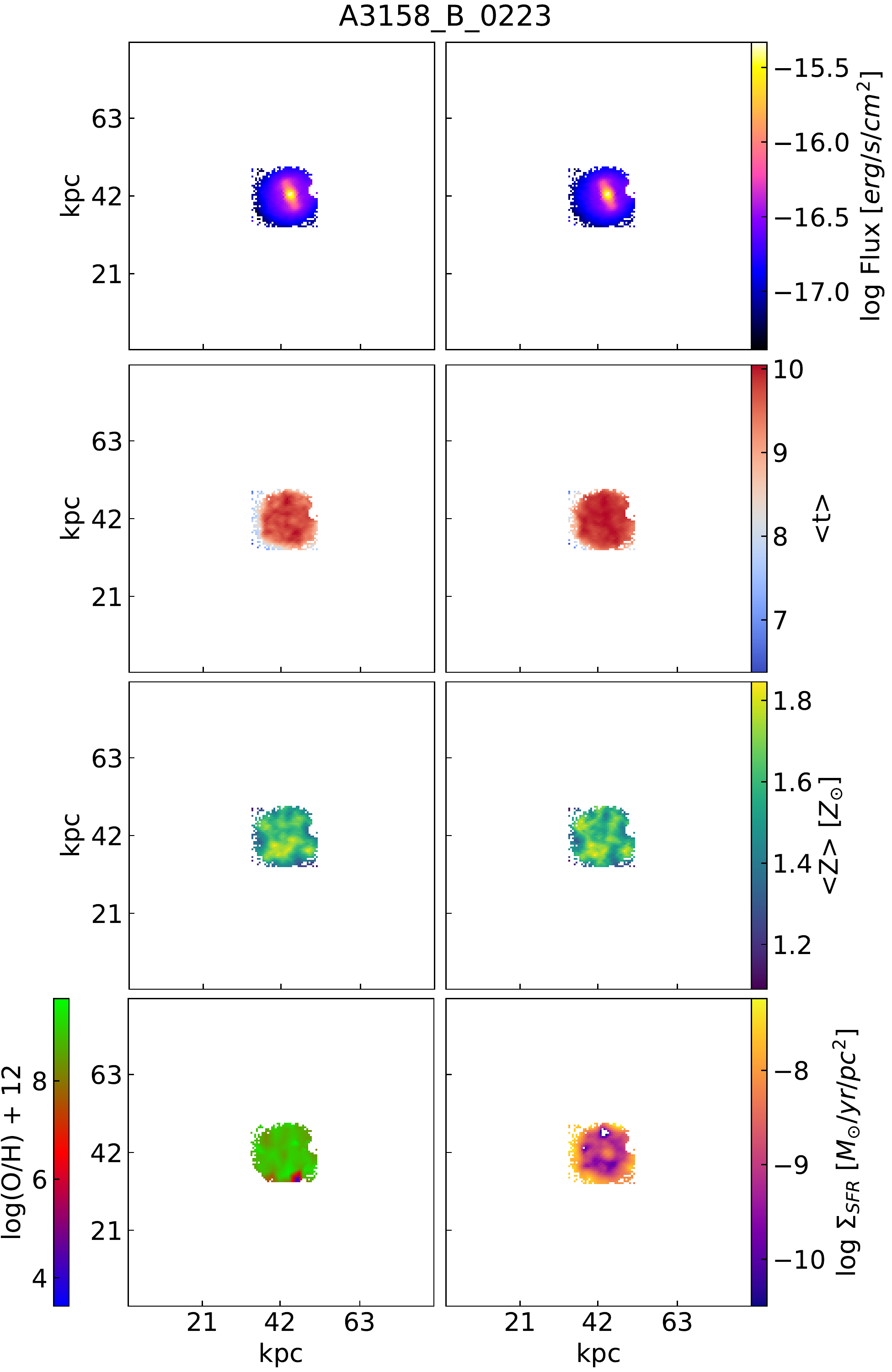}
\end{figure}

\begin{figure}
\centering
\includegraphics[width=\linewidth]{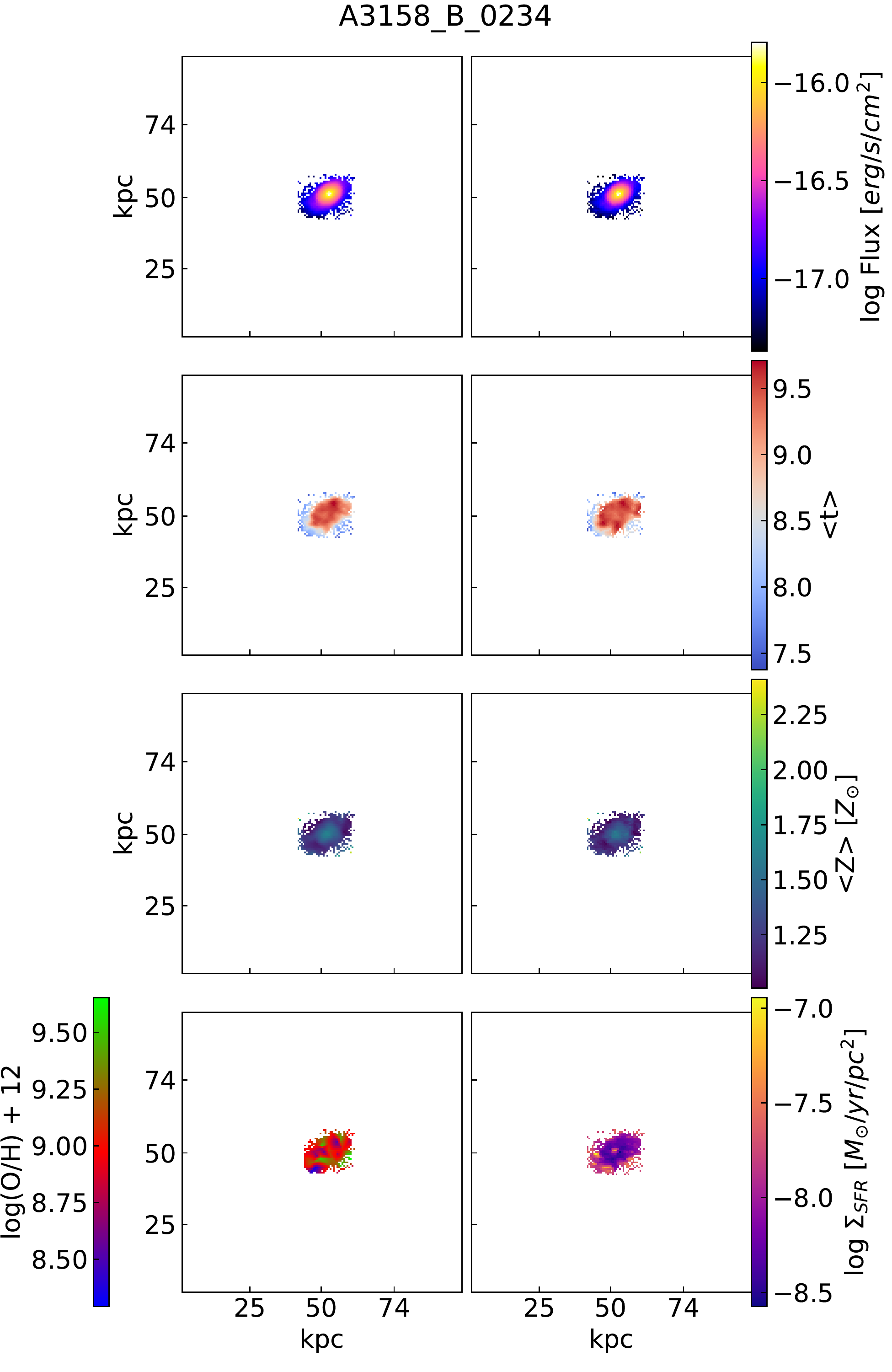}
\end{figure}

\begin{figure}
\centering
\includegraphics[width=\linewidth]{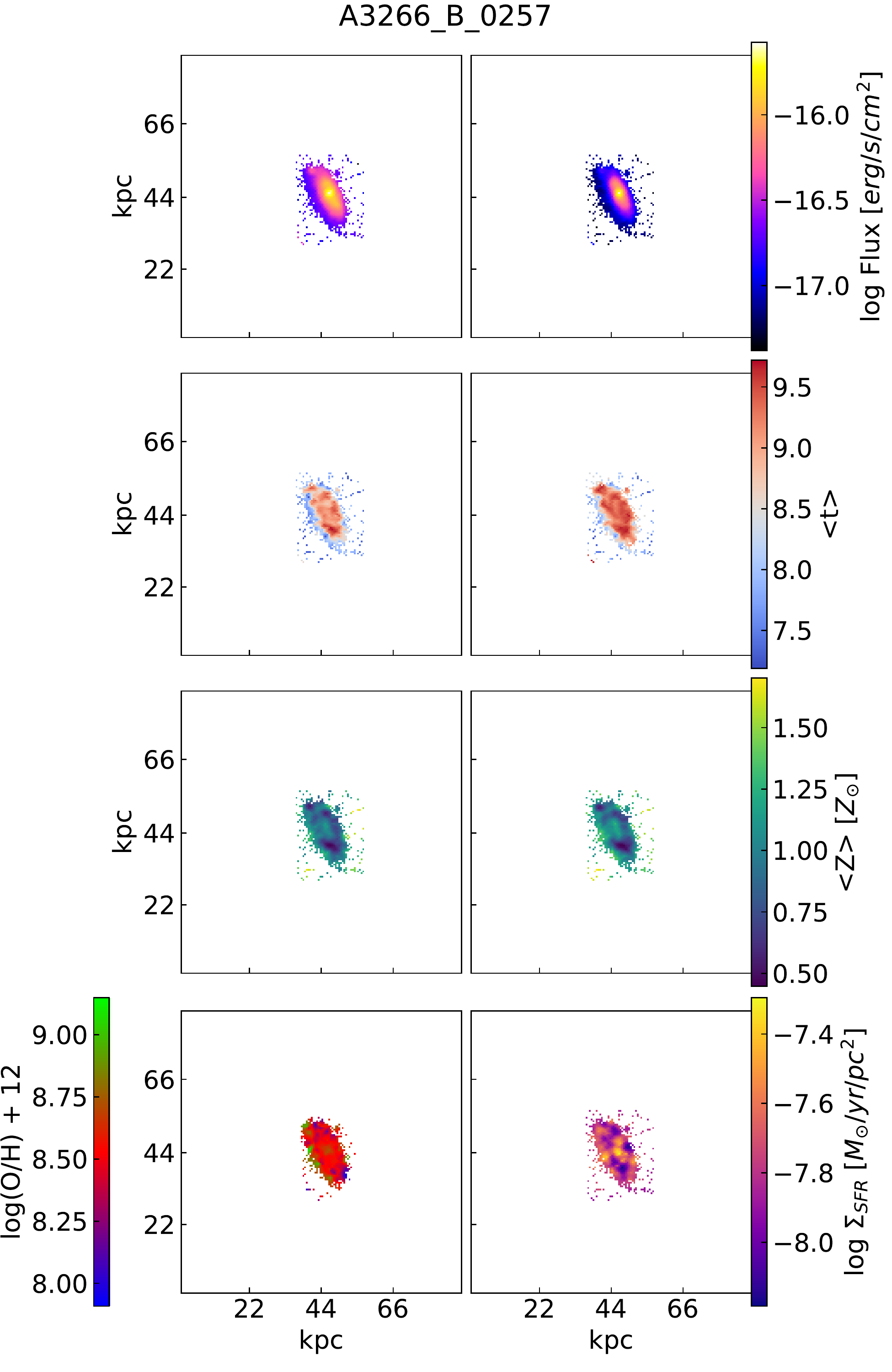}
\end{figure}

\begin{figure}
\centering
\includegraphics[width=\linewidth]{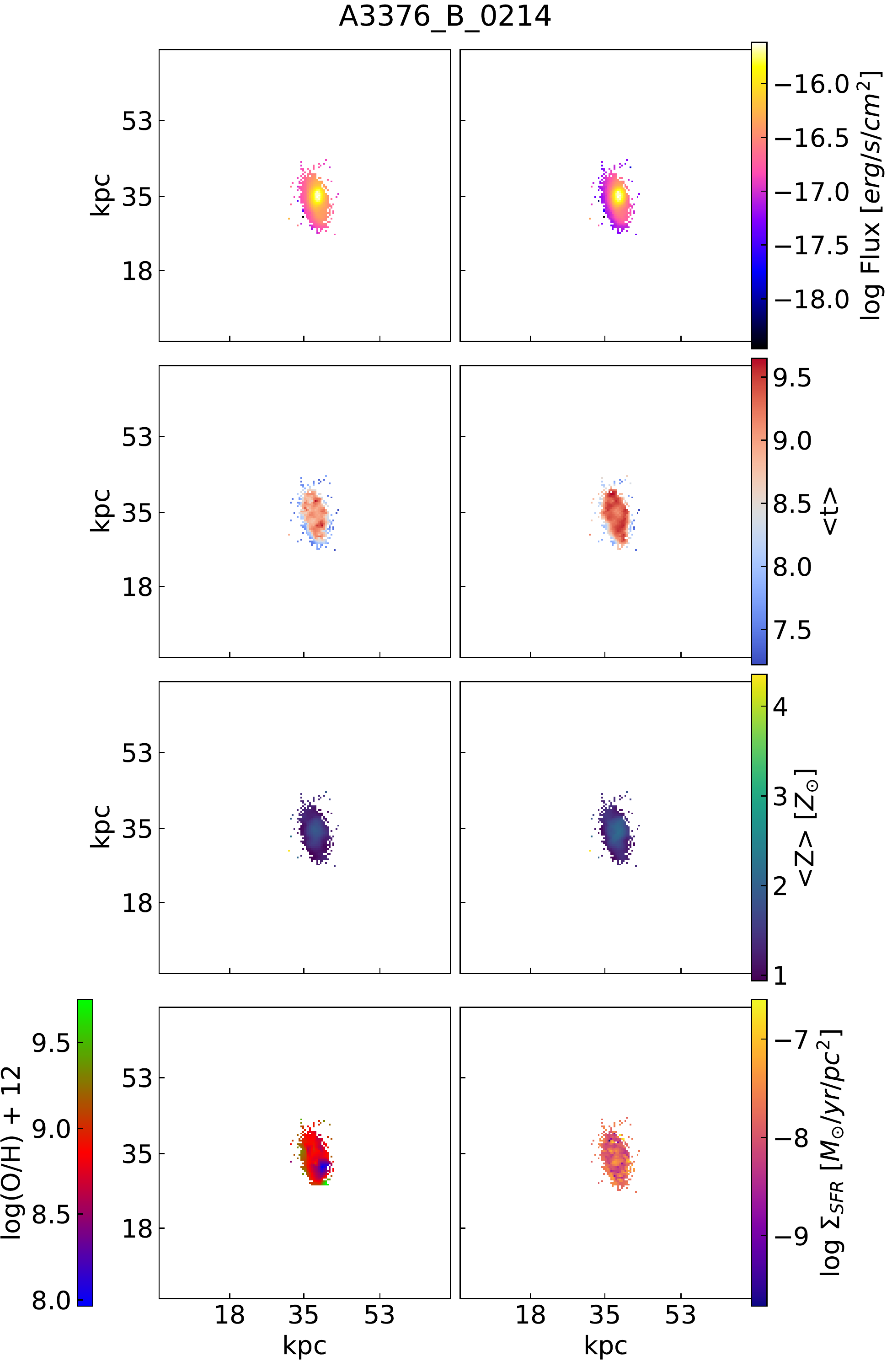}
\end{figure}

\begin{figure}
\centering
\includegraphics[width=\linewidth]{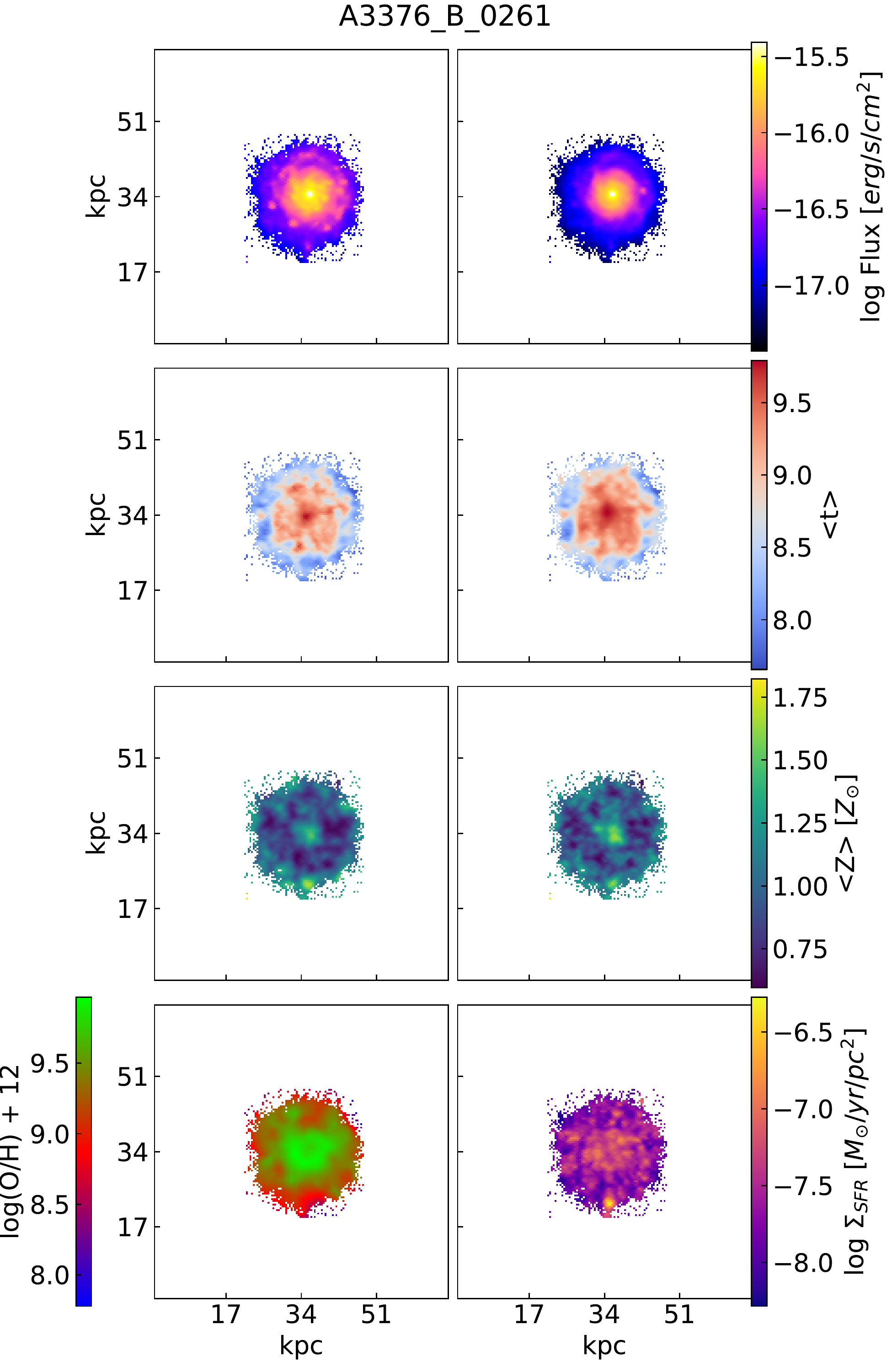}
\end{figure}

\begin{figure}
\centering
\includegraphics[width=\linewidth]{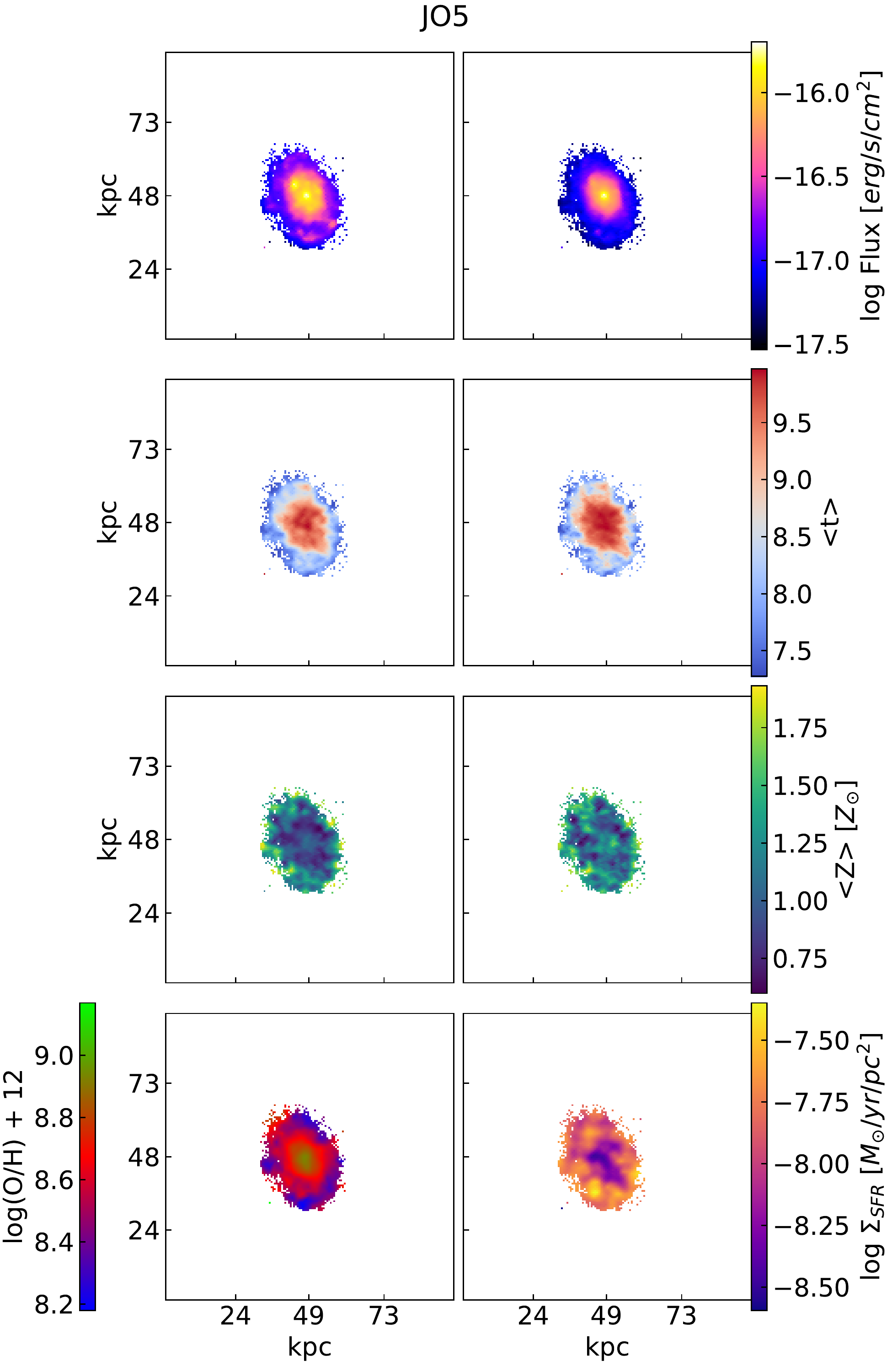}
\end{figure}

\begin{figure}
\centering
\includegraphics[width=\linewidth]{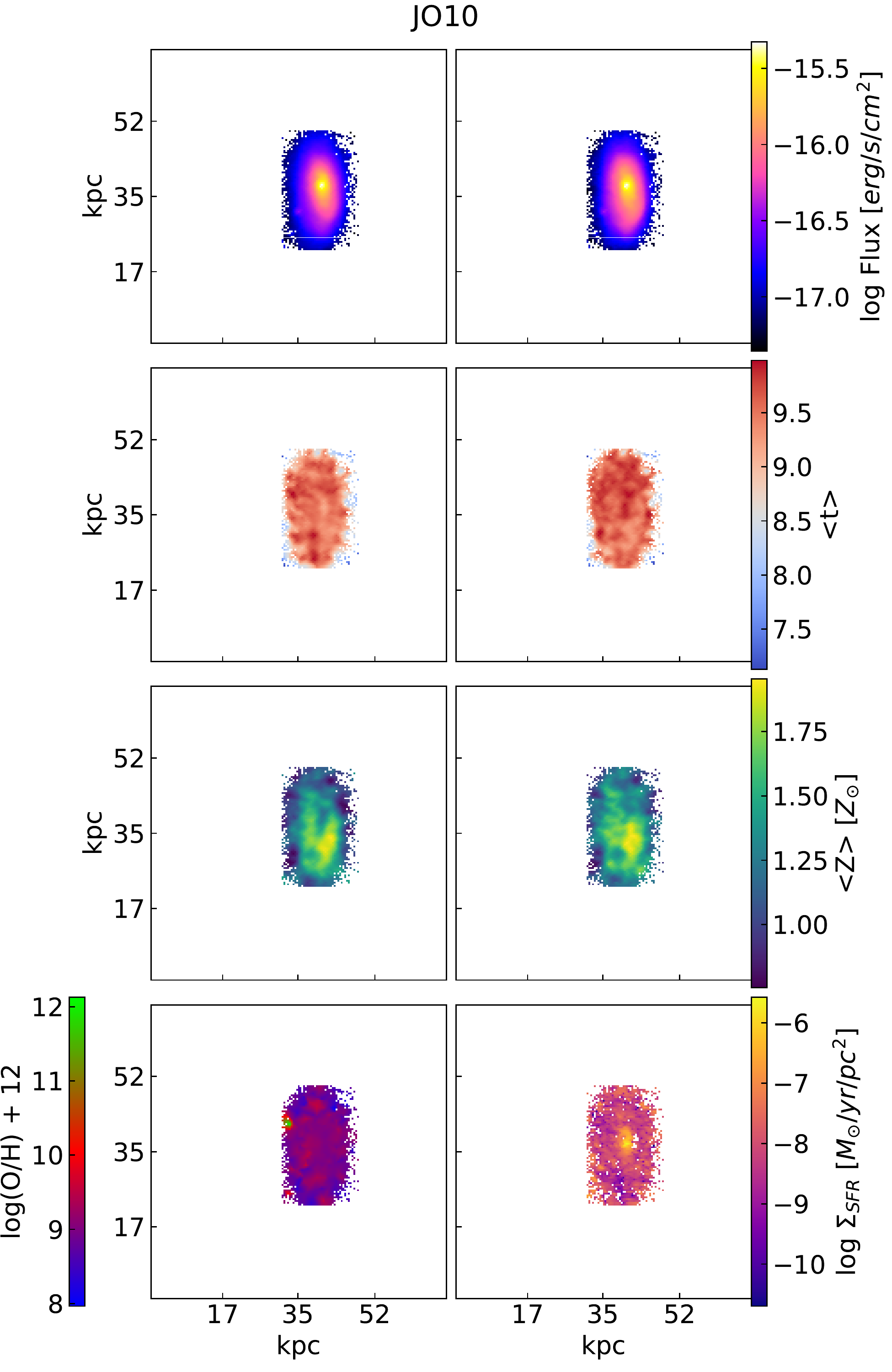}
\end{figure}

\begin{figure}
\centering
\includegraphics[width=\linewidth]{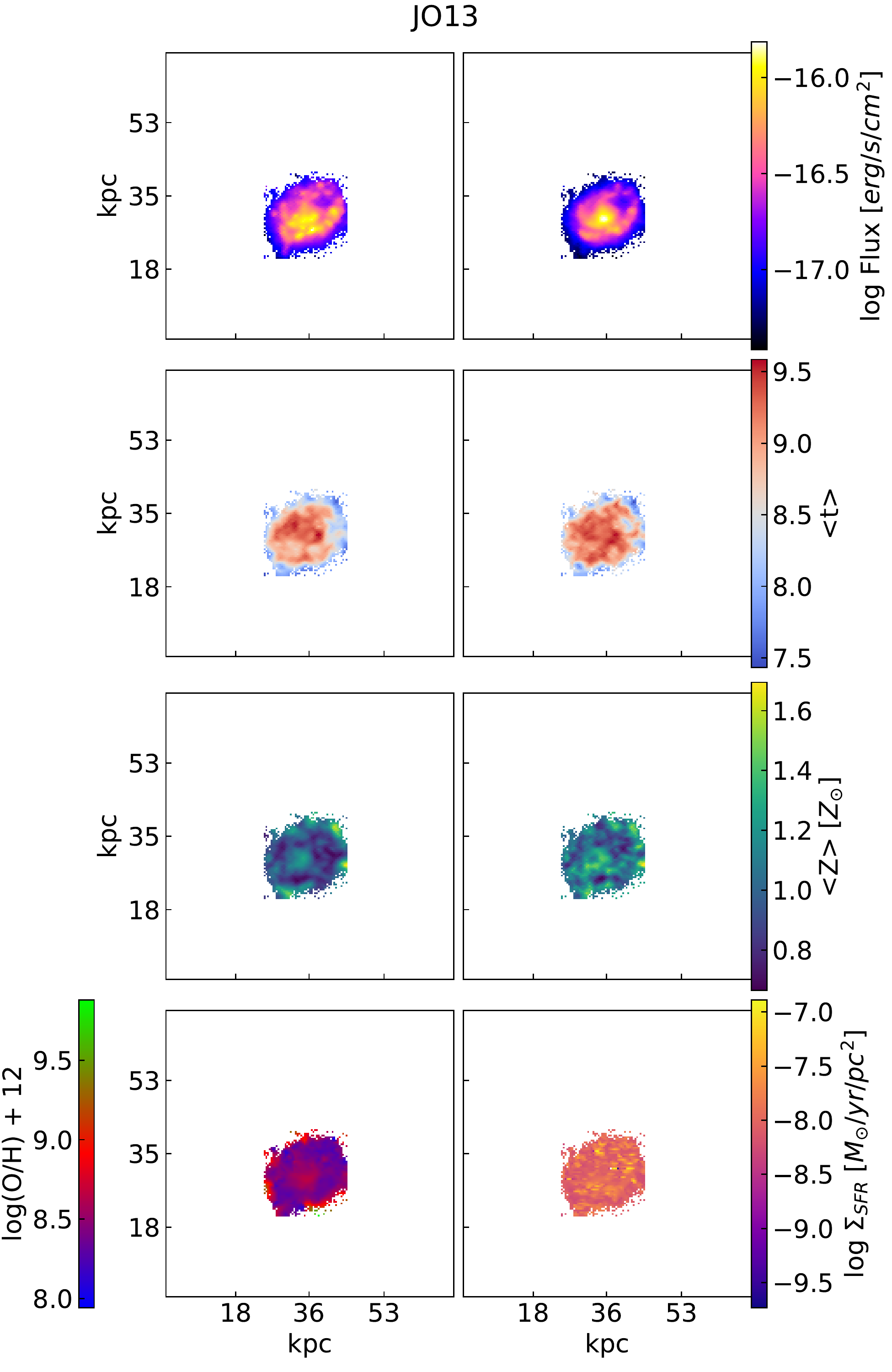}
\end{figure}

\begin{figure}
\centering
\includegraphics[width=\linewidth]{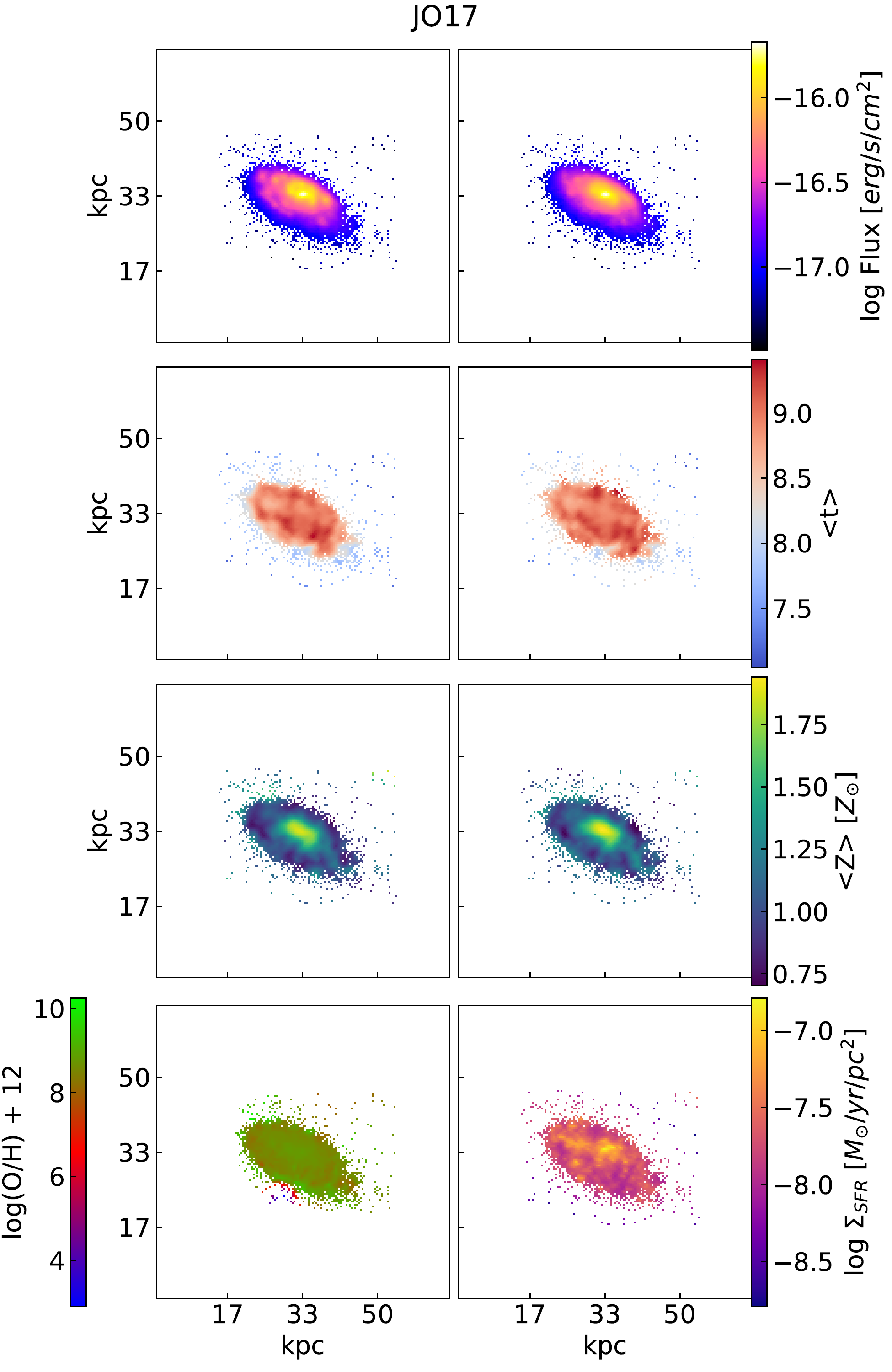}
\end{figure}

\begin{figure}
\centering
\includegraphics[width=\linewidth]{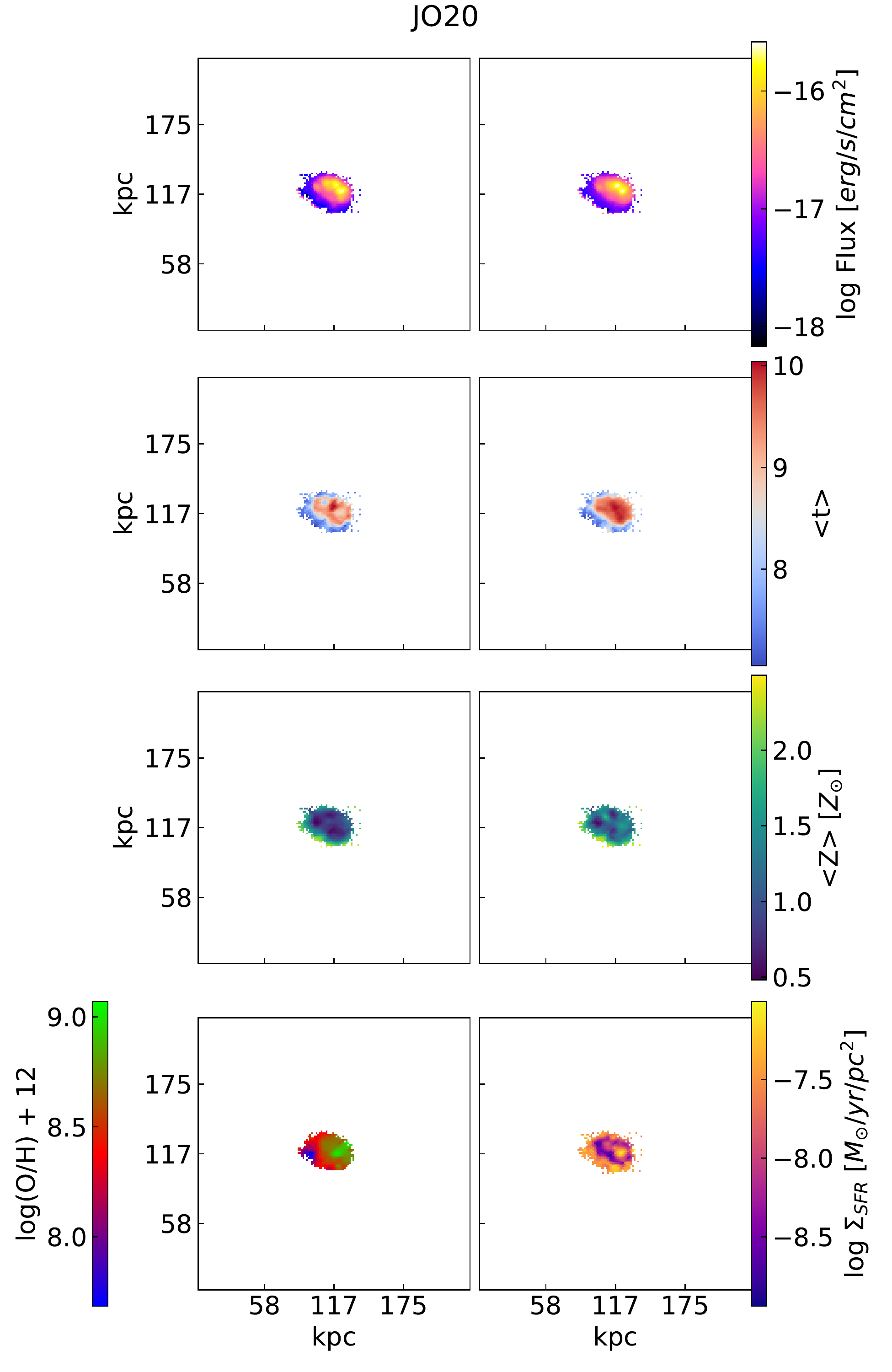}
\end{figure}

\begin{figure}
\centering
\includegraphics[width=\linewidth]{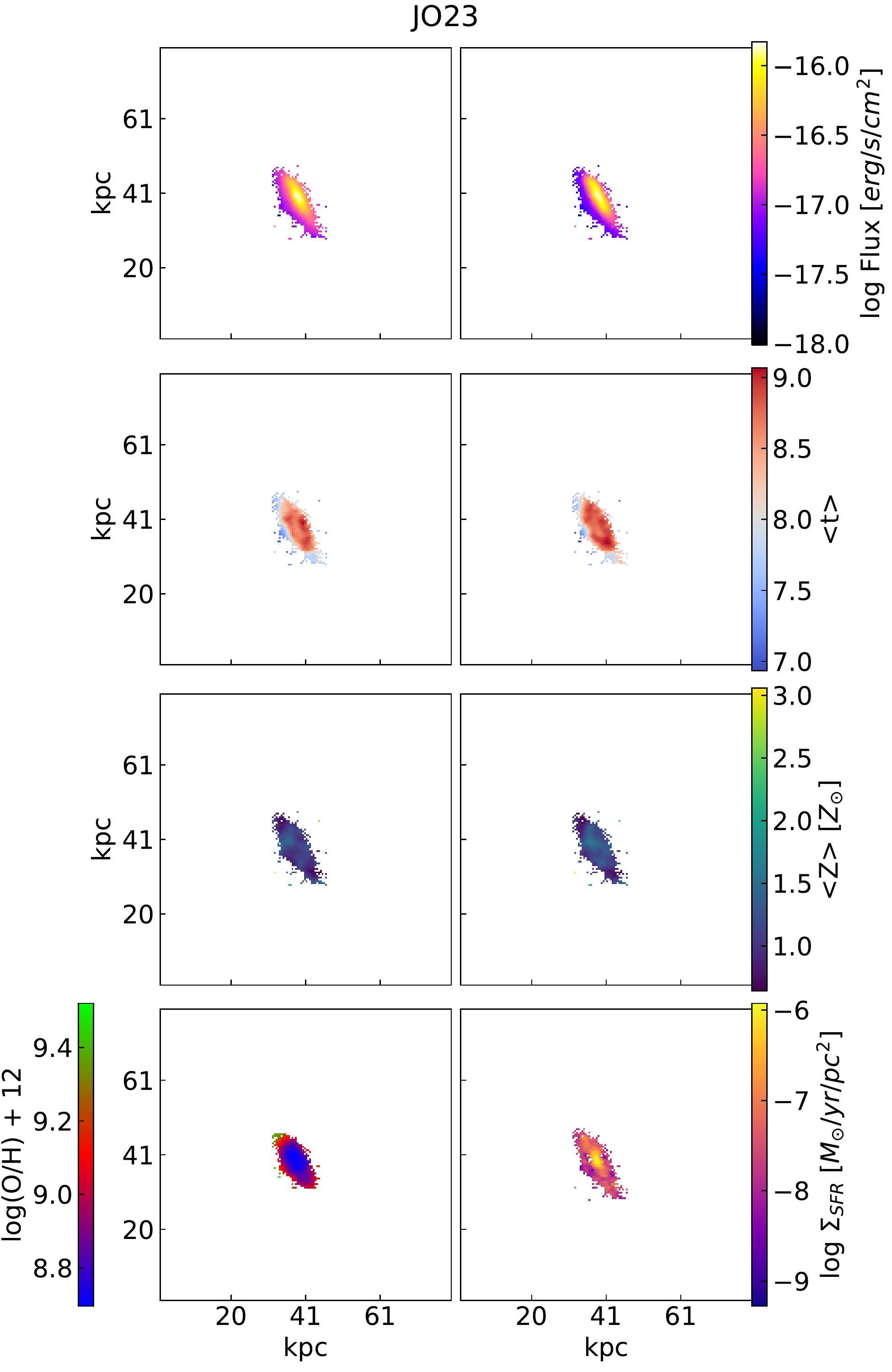}
\end{figure}

\begin{figure}
\centering
\includegraphics[width=\linewidth]{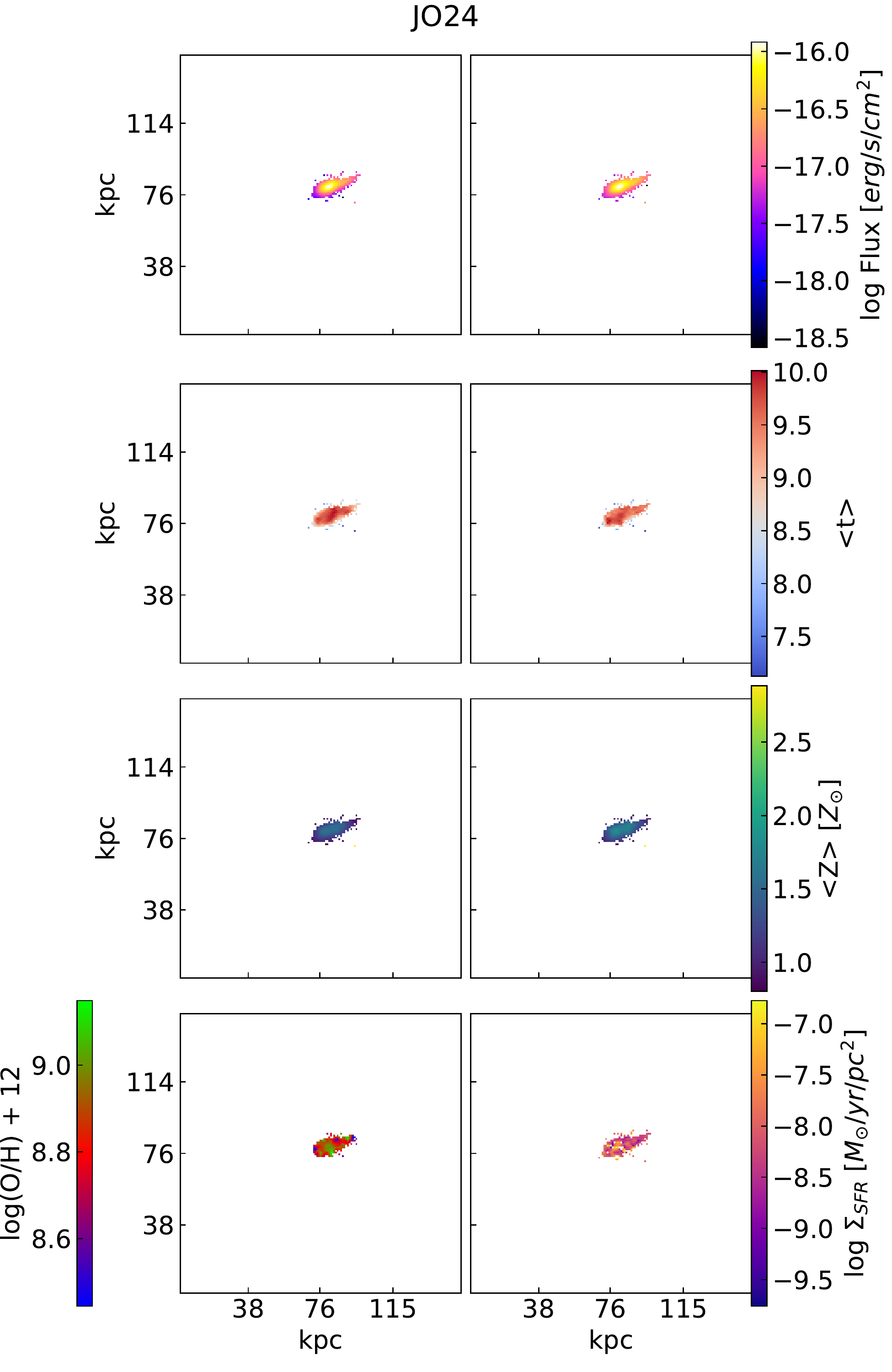}
\end{figure}

\begin{figure}
\centering
\includegraphics[width=\linewidth]{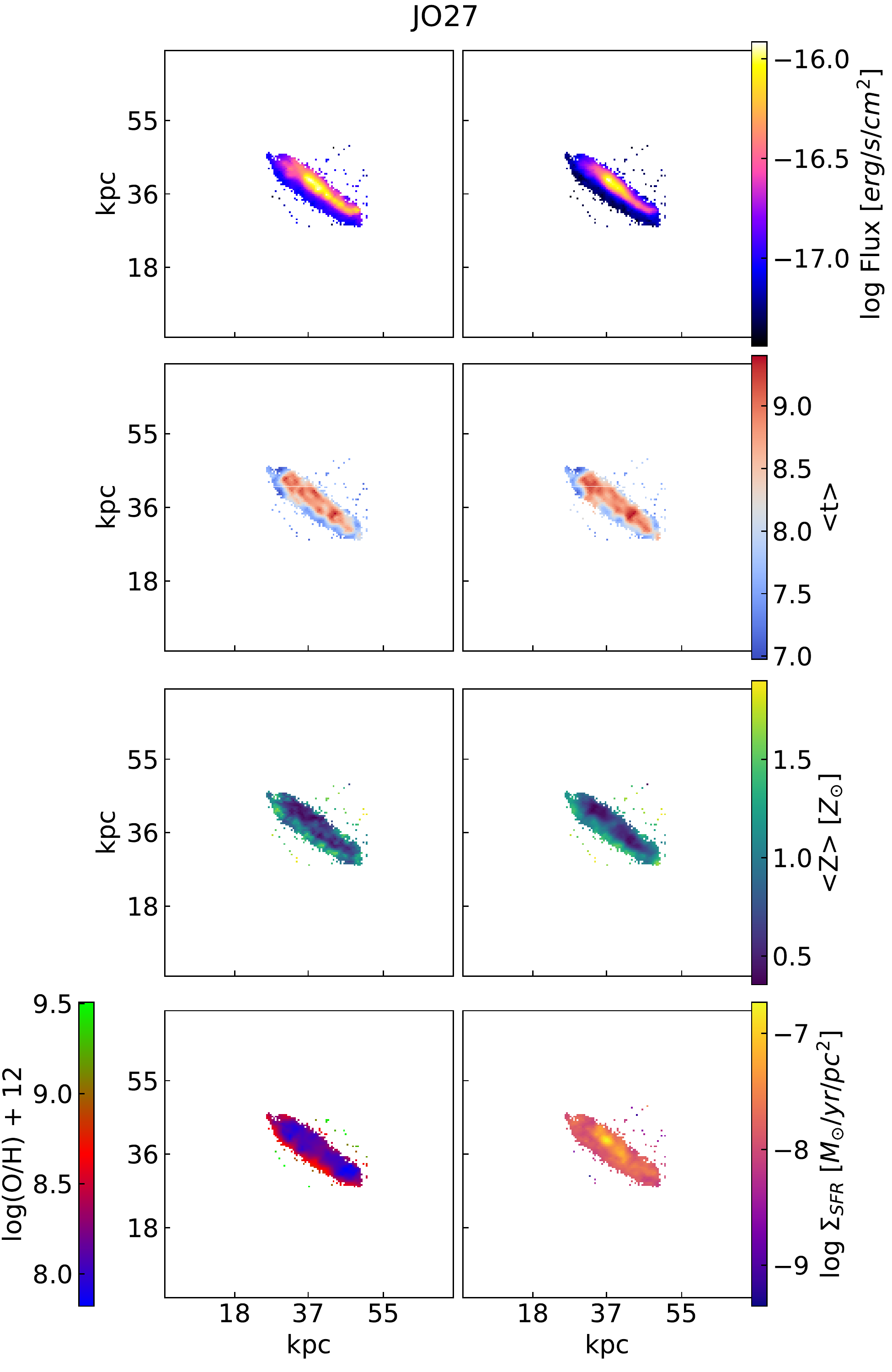}
\end{figure}

\begin{figure}
\centering
\includegraphics[width=\linewidth]{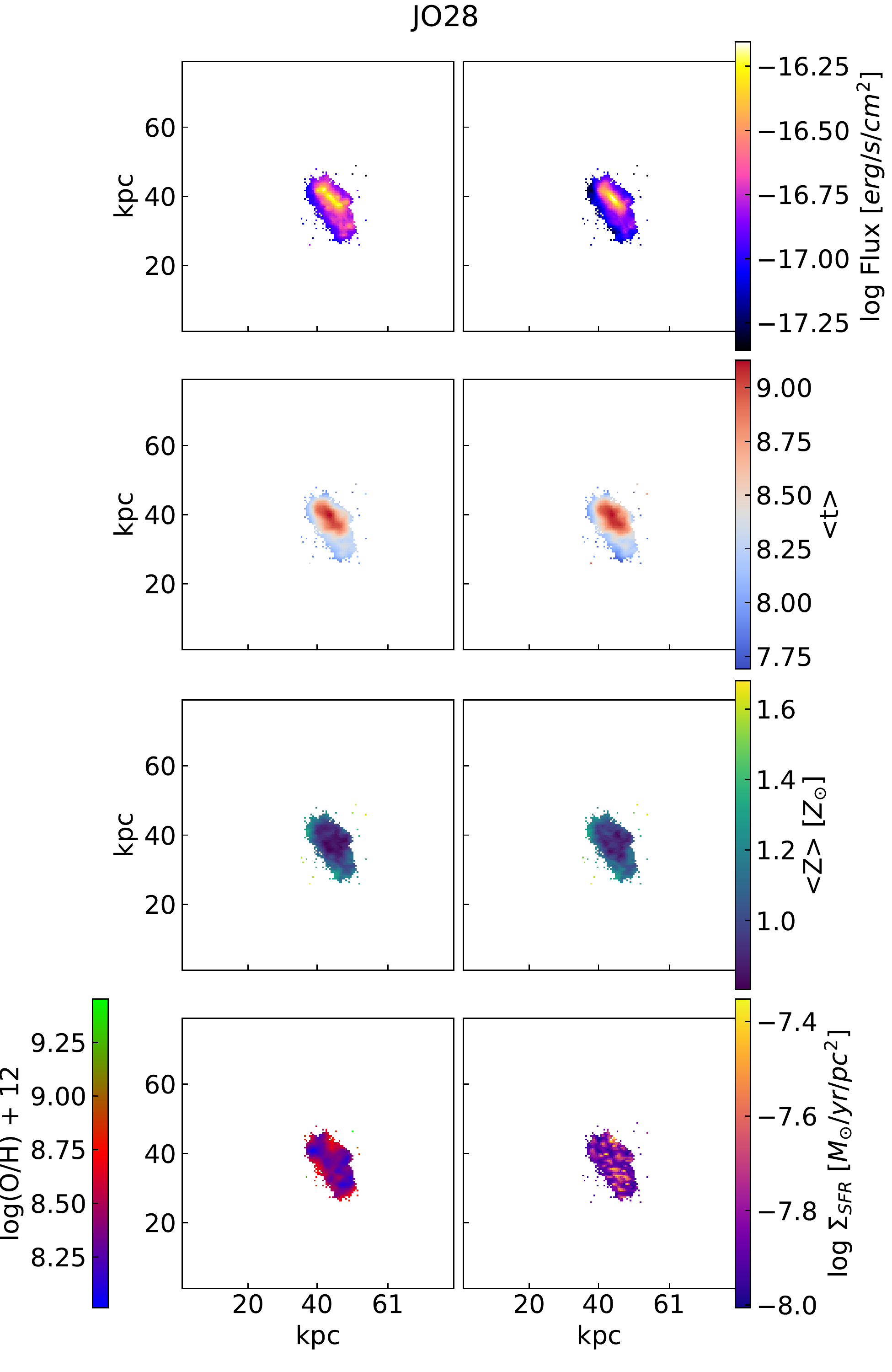}
\end{figure}

\begin{figure}
\centering
\includegraphics[width=\linewidth]{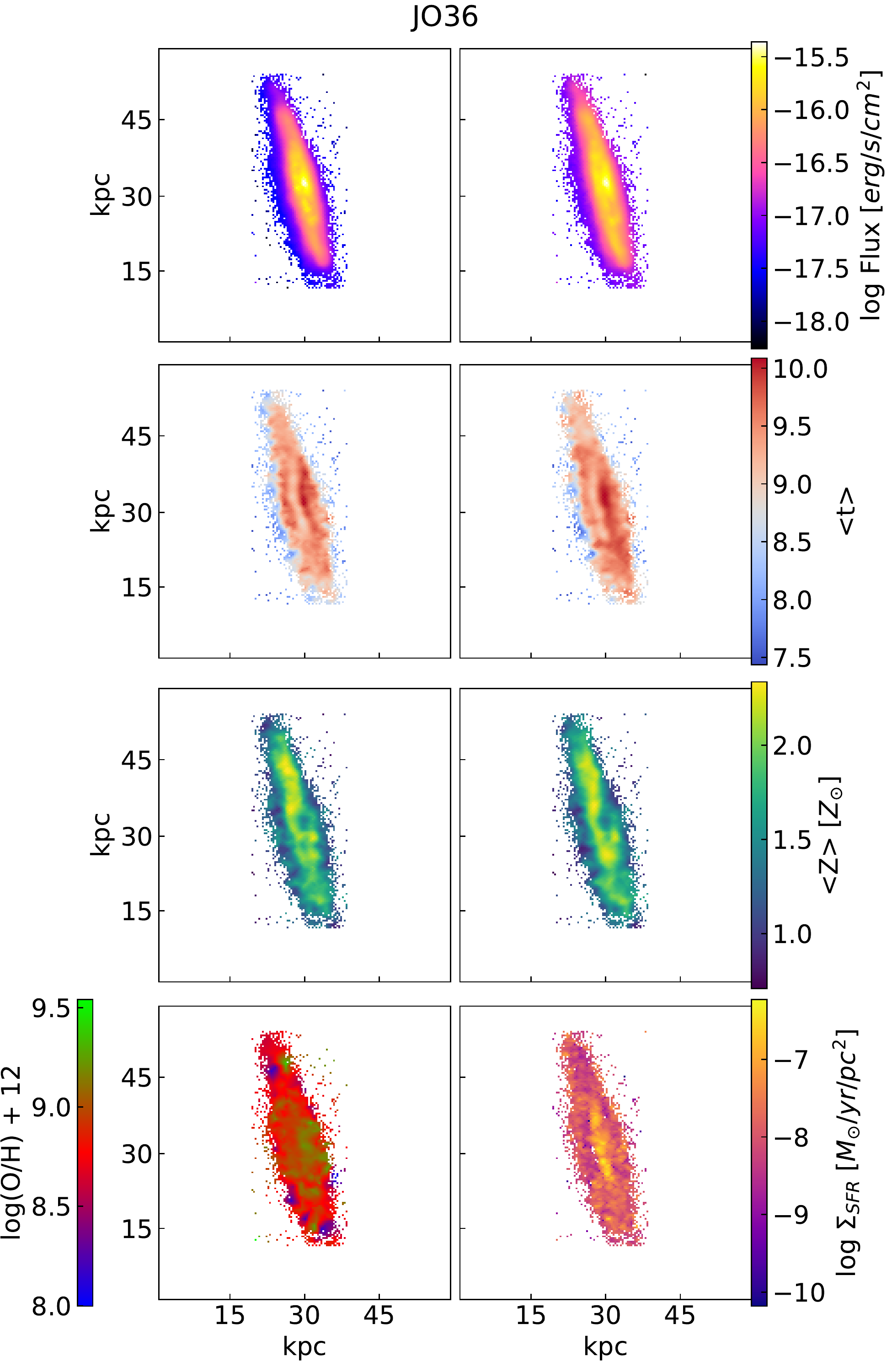}
\end{figure}

\begin{figure}
\centering
\includegraphics[width=\linewidth]{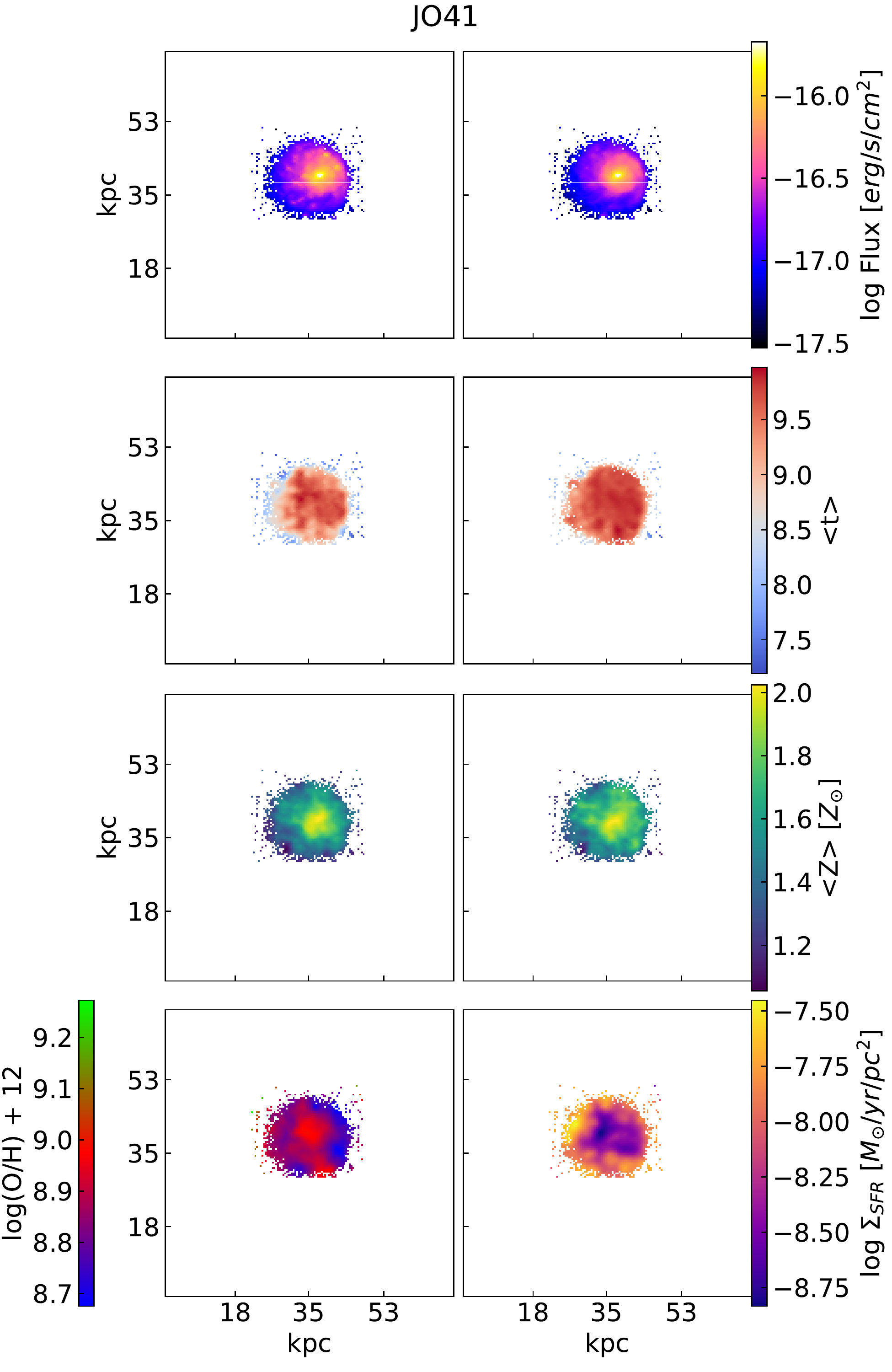}
\end{figure}

\begin{figure}
\centering
\includegraphics[width=\linewidth]{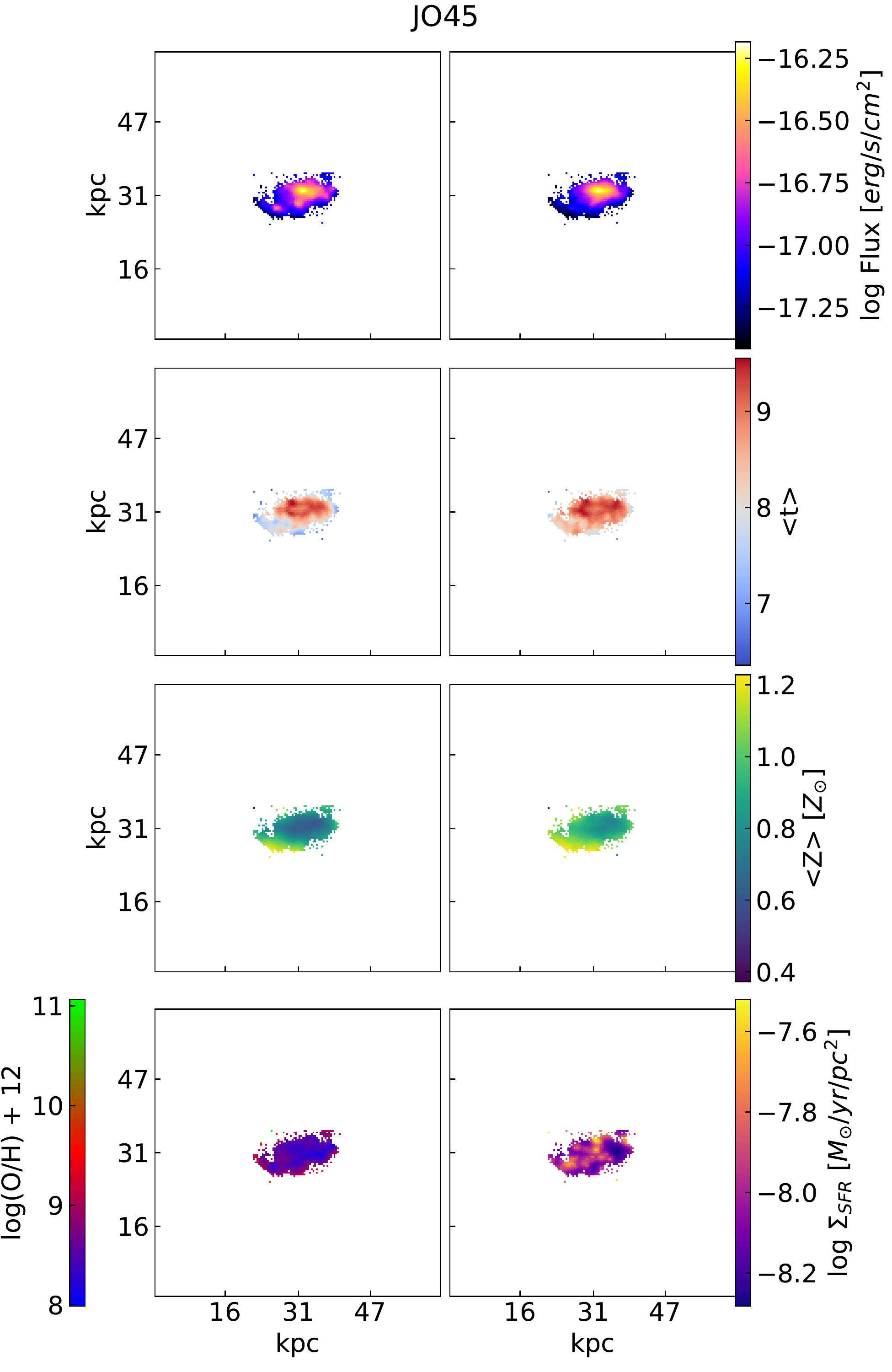}
\end{figure}
\begin{figure}
\centering
\includegraphics[width=\linewidth]{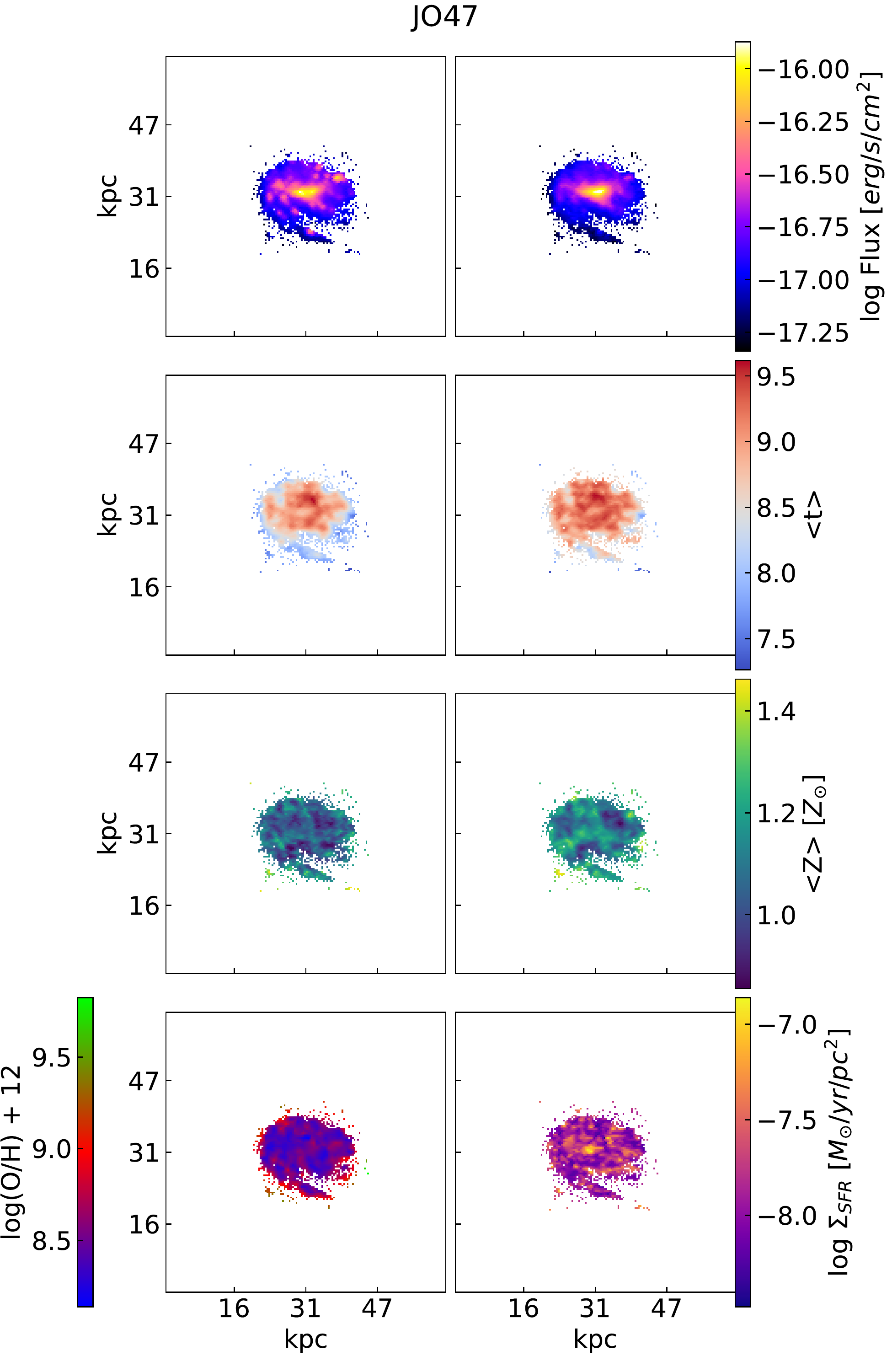}
\end{figure}

\begin{figure}
\centering
\includegraphics[width=\linewidth]{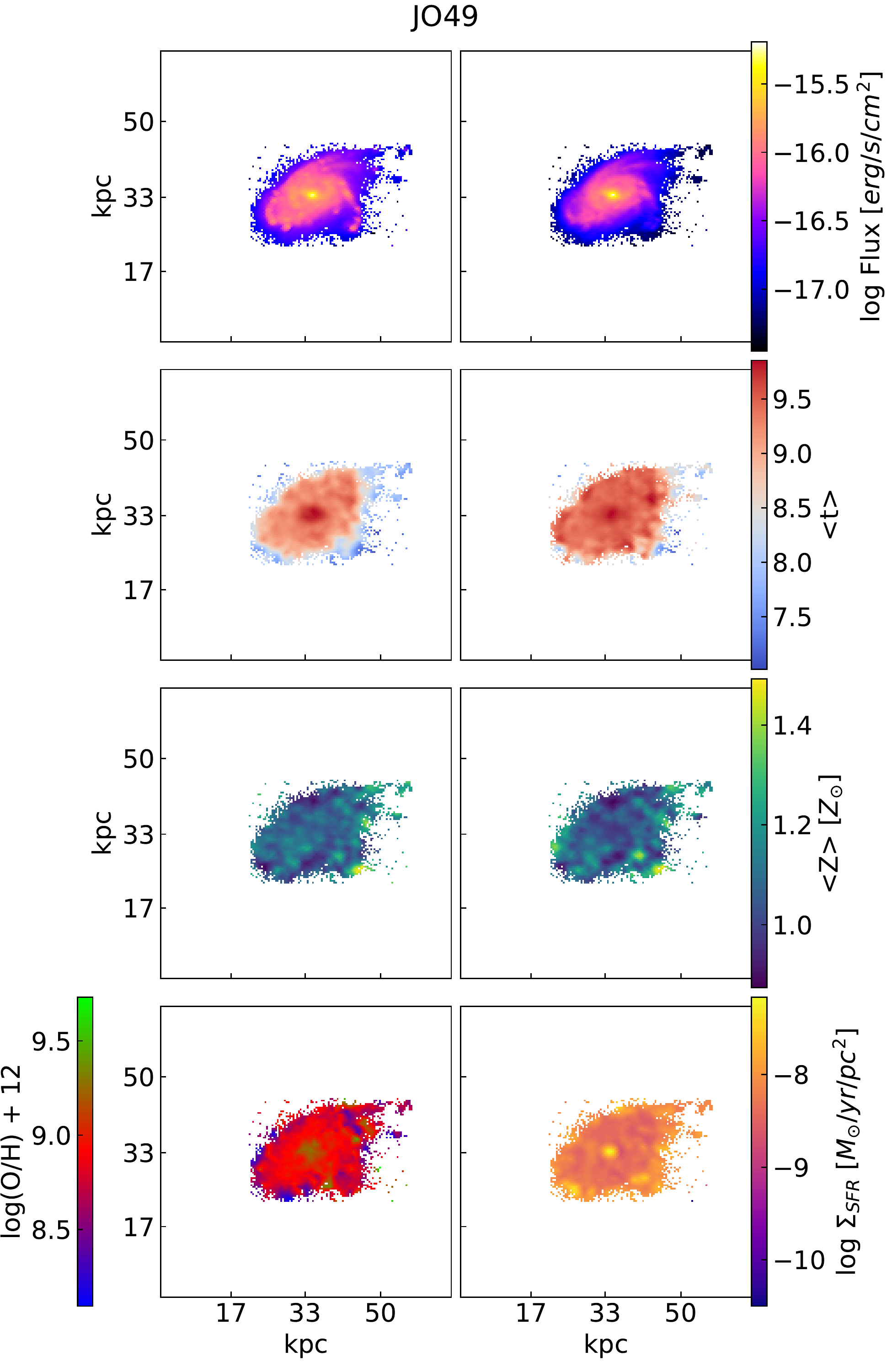}
\end{figure}

\begin{figure}
\centering
\includegraphics[width=\linewidth]{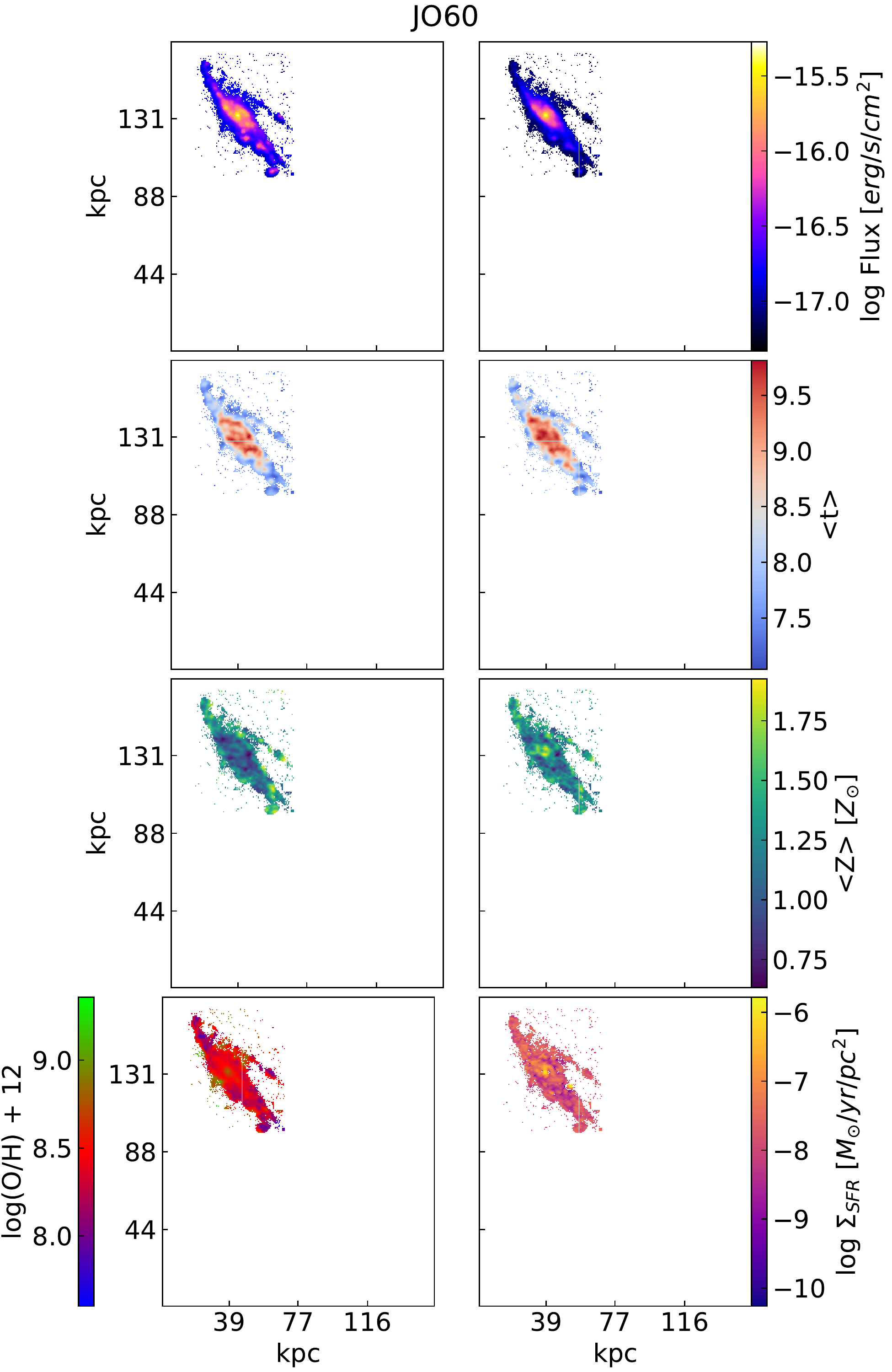}
\end{figure}

\begin{figure}
\centering
\includegraphics[width=\linewidth]{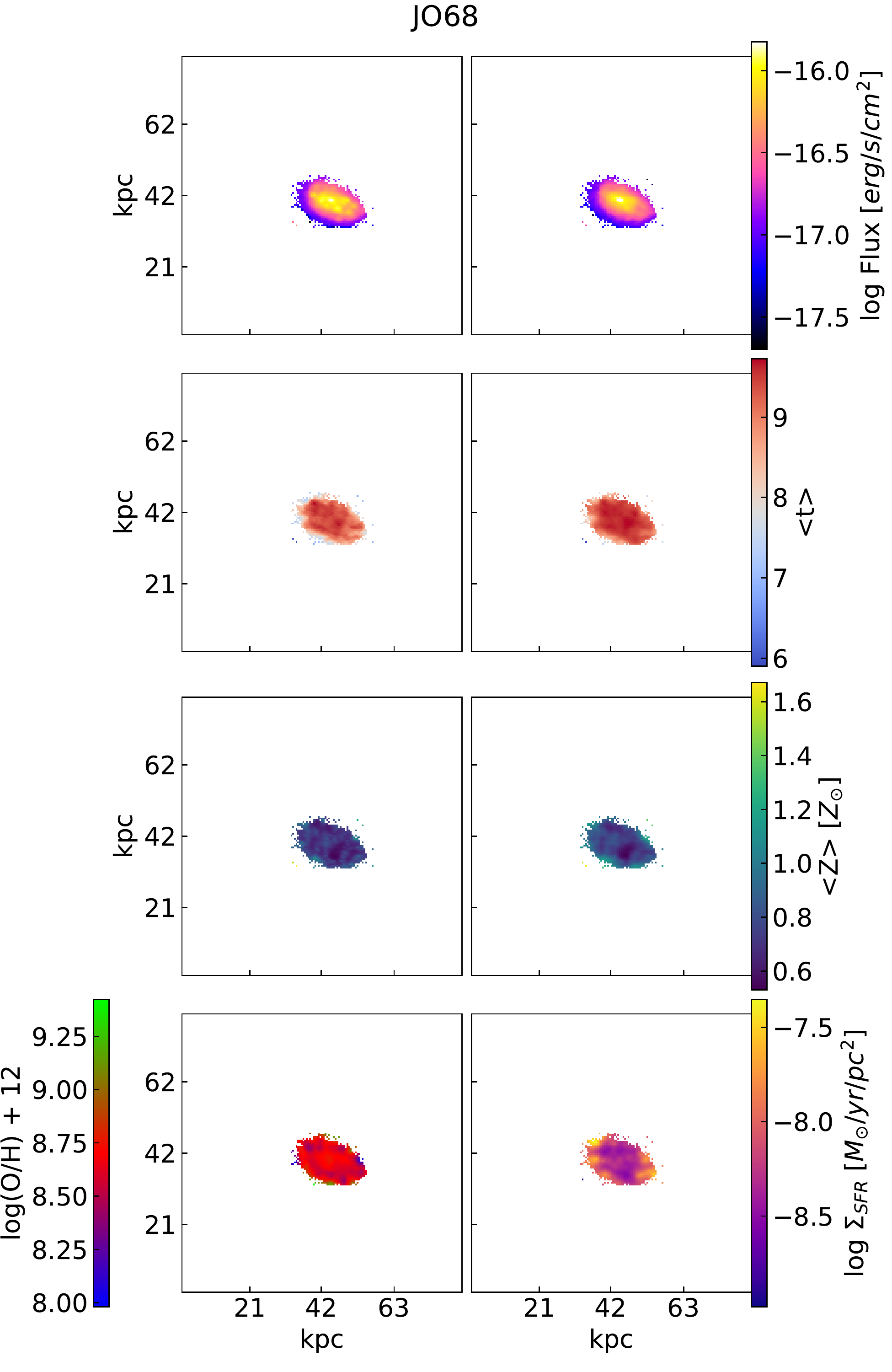}
\end{figure}

\begin{figure}
\centering
\includegraphics[width=\linewidth]{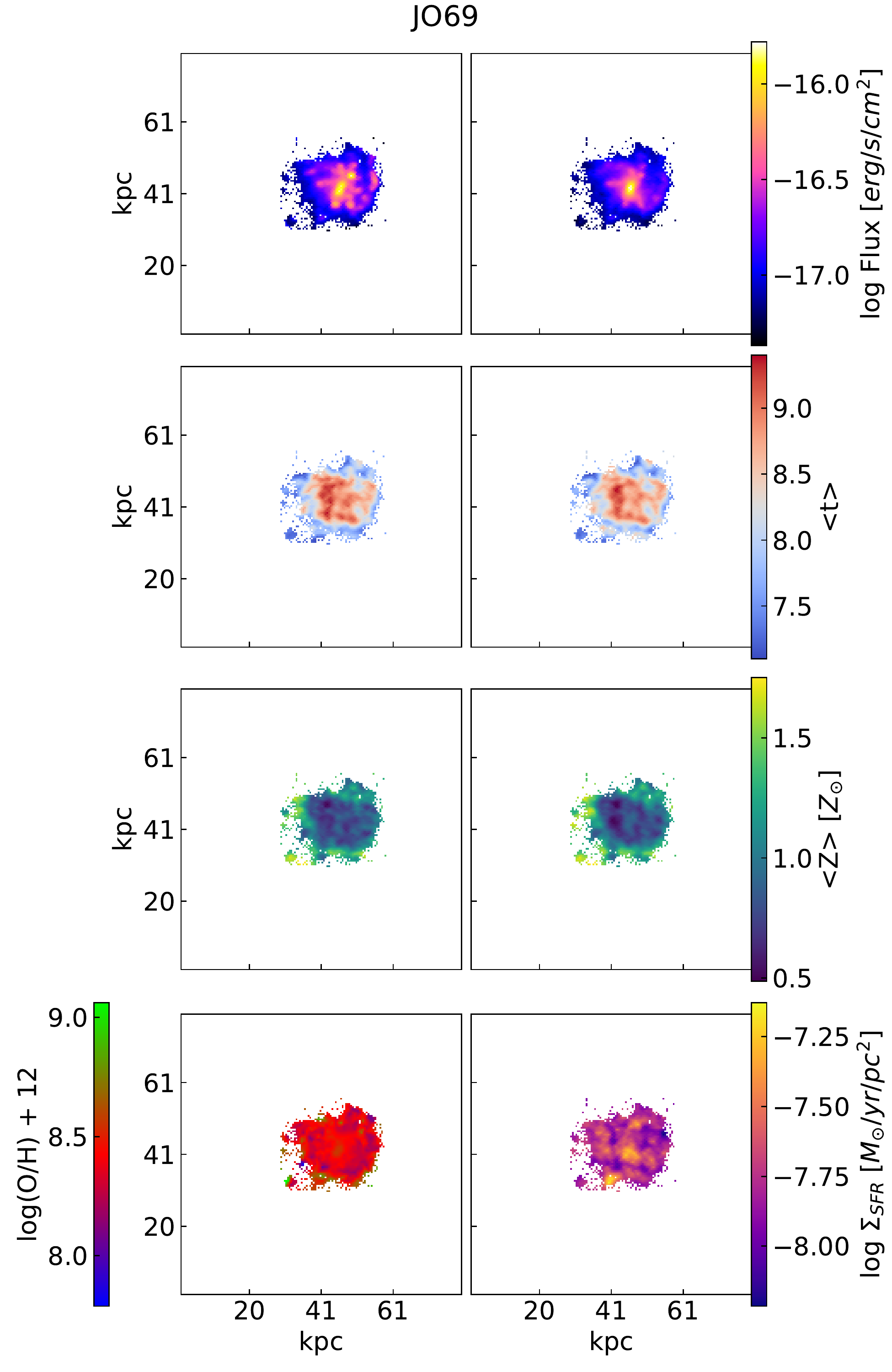}
\end{figure}

\begin{figure}
\centering
\includegraphics[width=\linewidth]{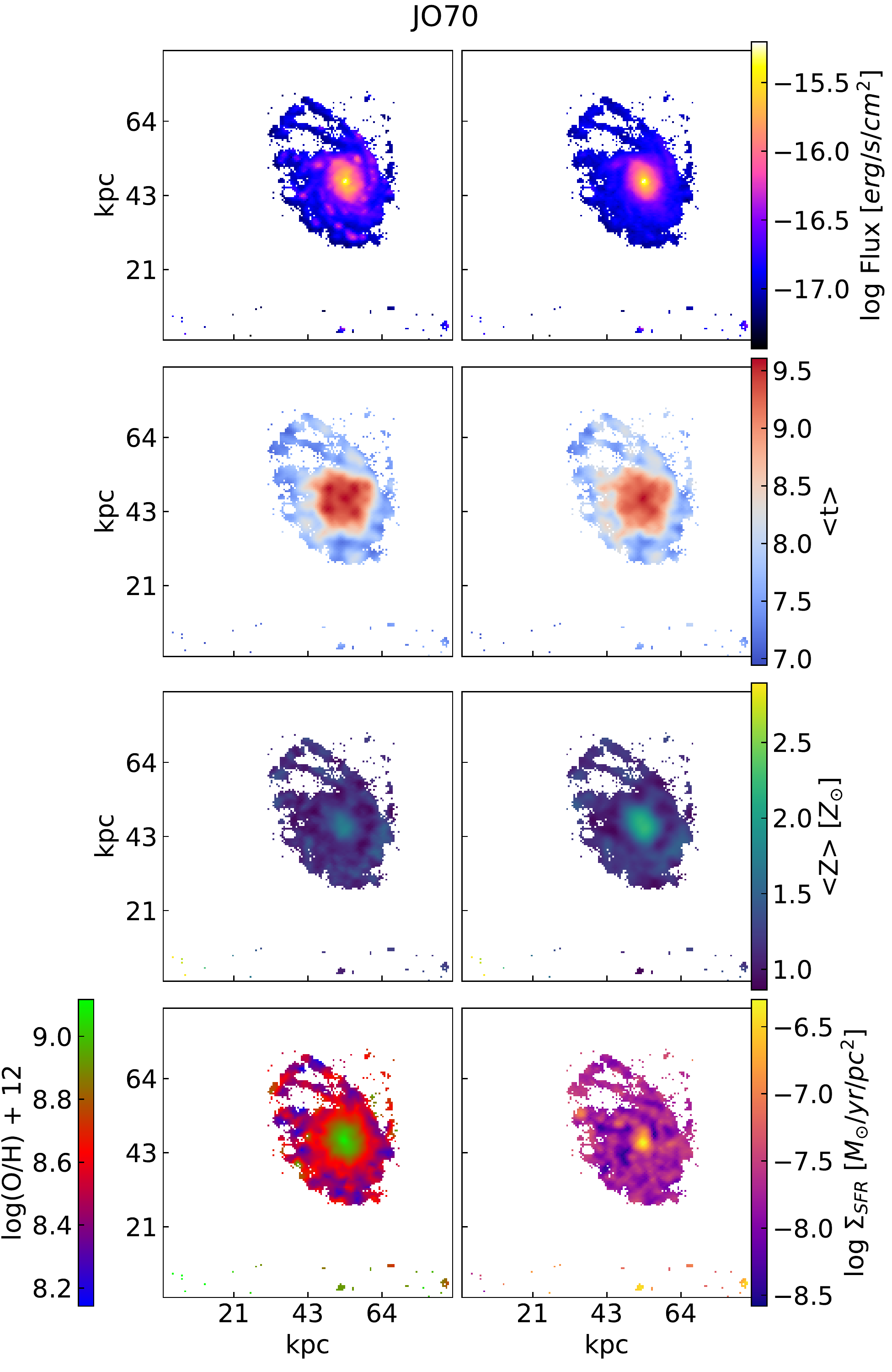}
\end{figure}

\begin{figure}
\centering
\includegraphics[width=\linewidth]{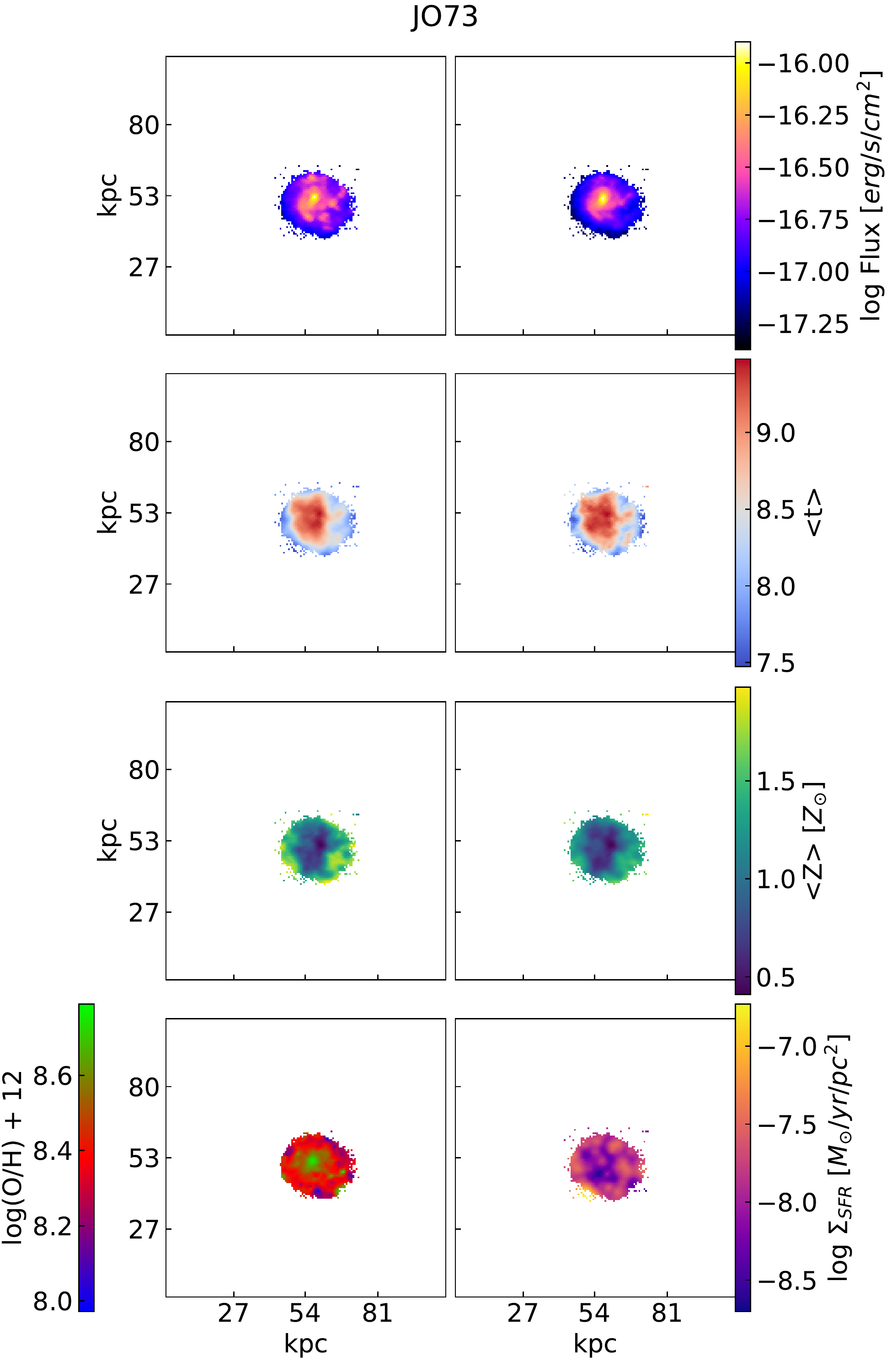}
\end{figure}

\begin{figure}
\centering
\includegraphics[width=\linewidth]{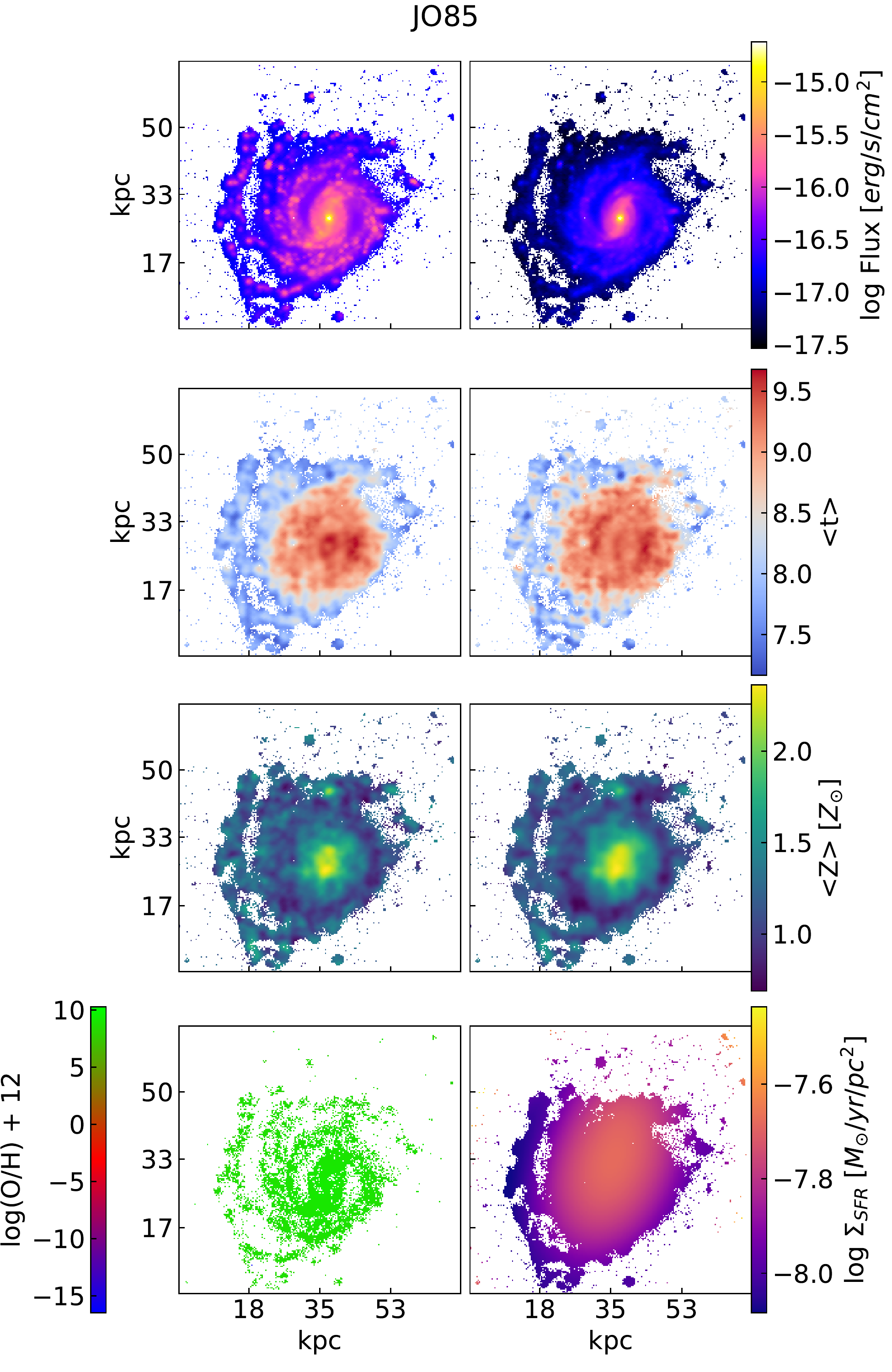}
\end{figure}

\begin{figure}
\centering
\includegraphics[width=\linewidth]{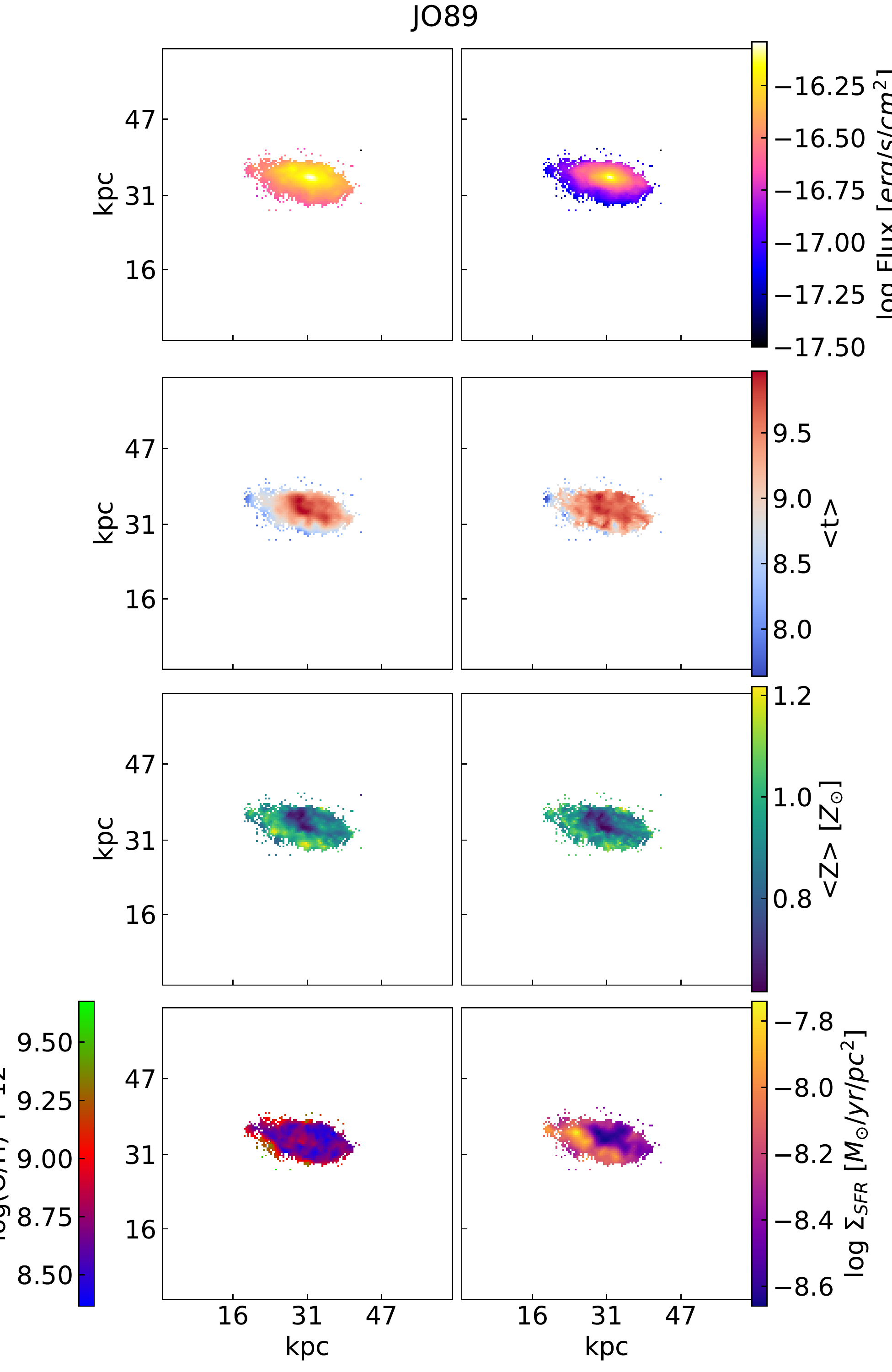}
\end{figure}

\begin{figure}
\centering
\includegraphics[width=\linewidth]{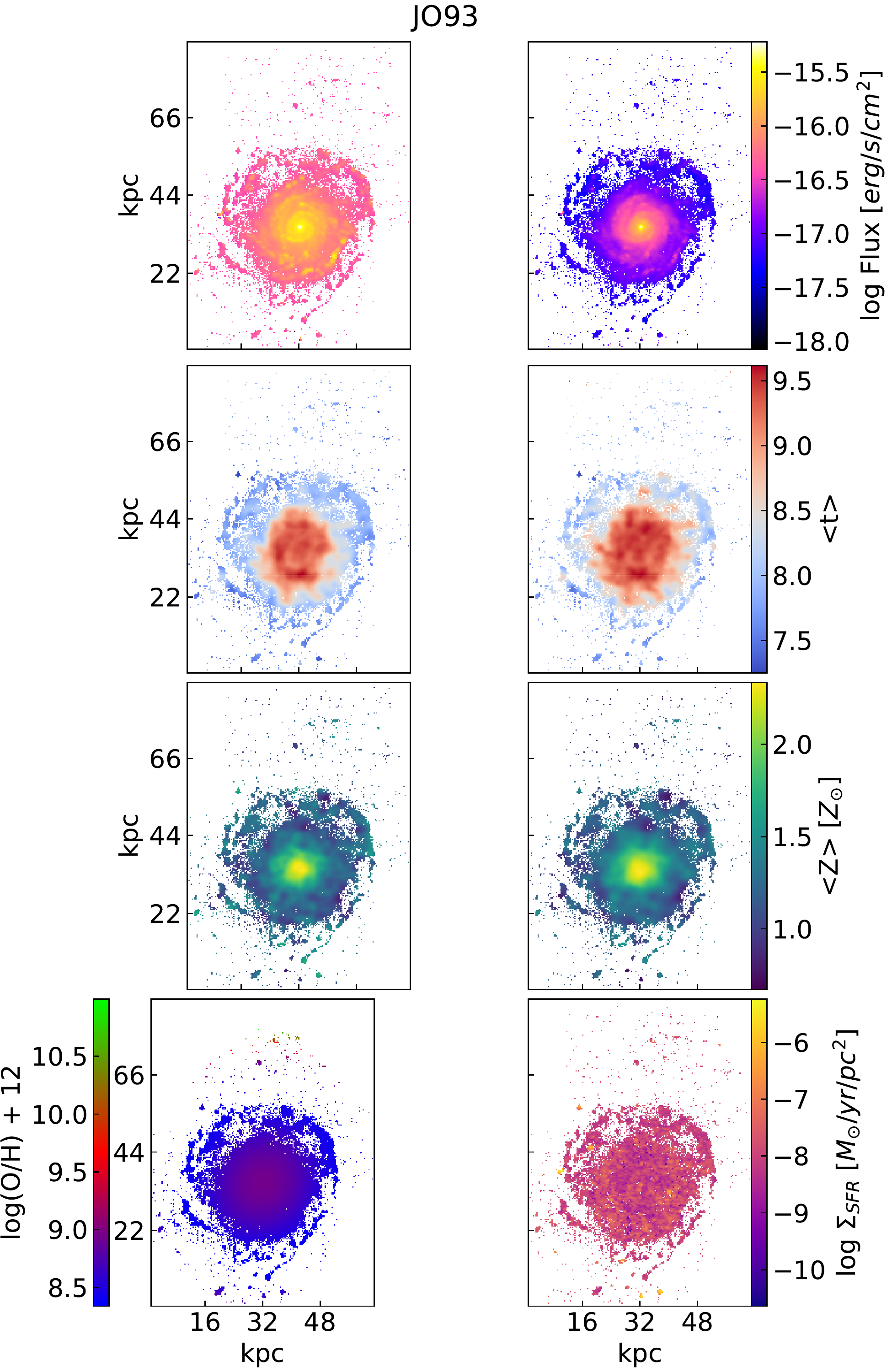}
\end{figure}

\begin{figure}
\centering
\includegraphics[width=\linewidth]{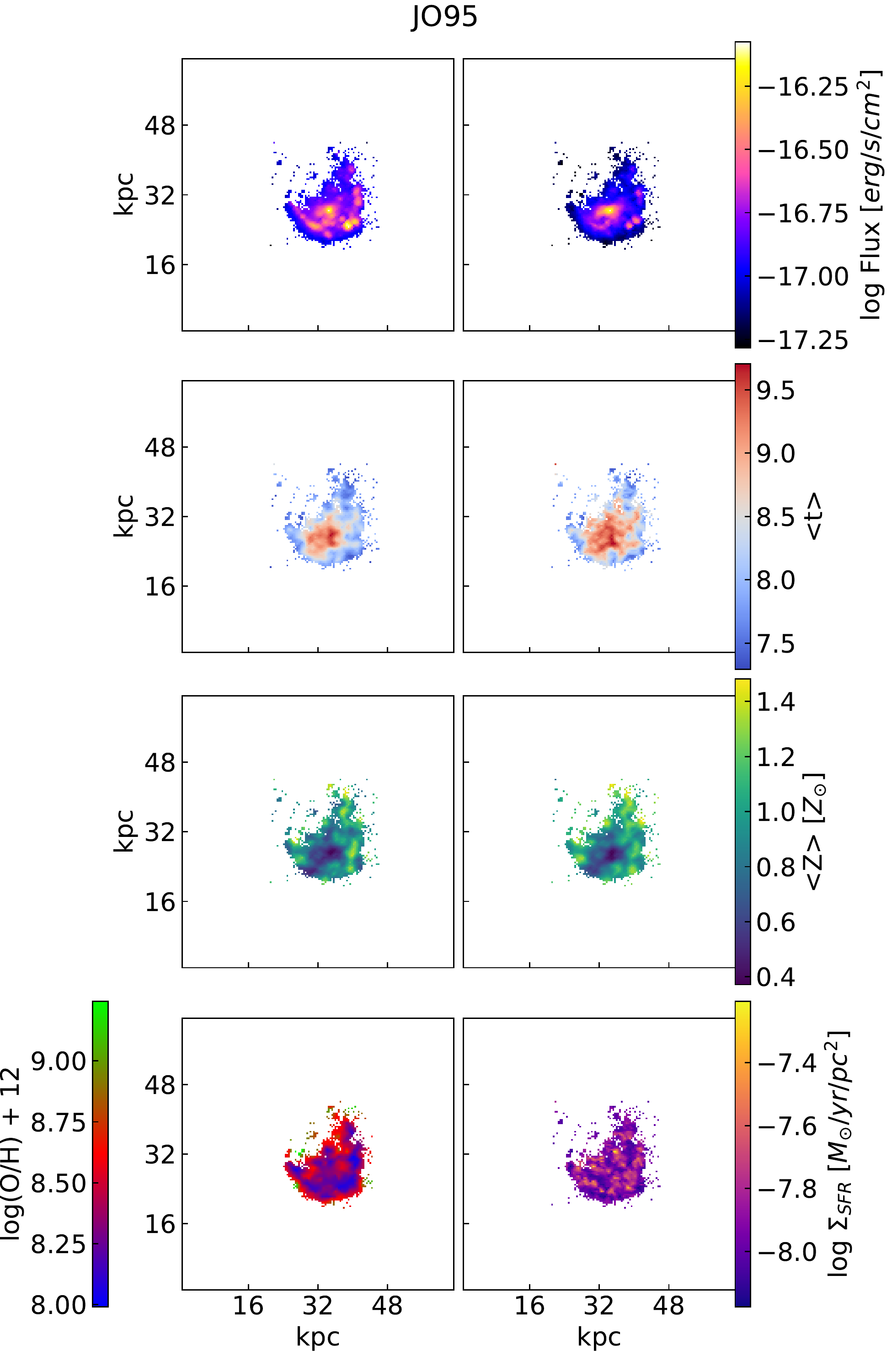}
\end{figure}

\begin{figure}
\centering
\includegraphics[width=\linewidth]{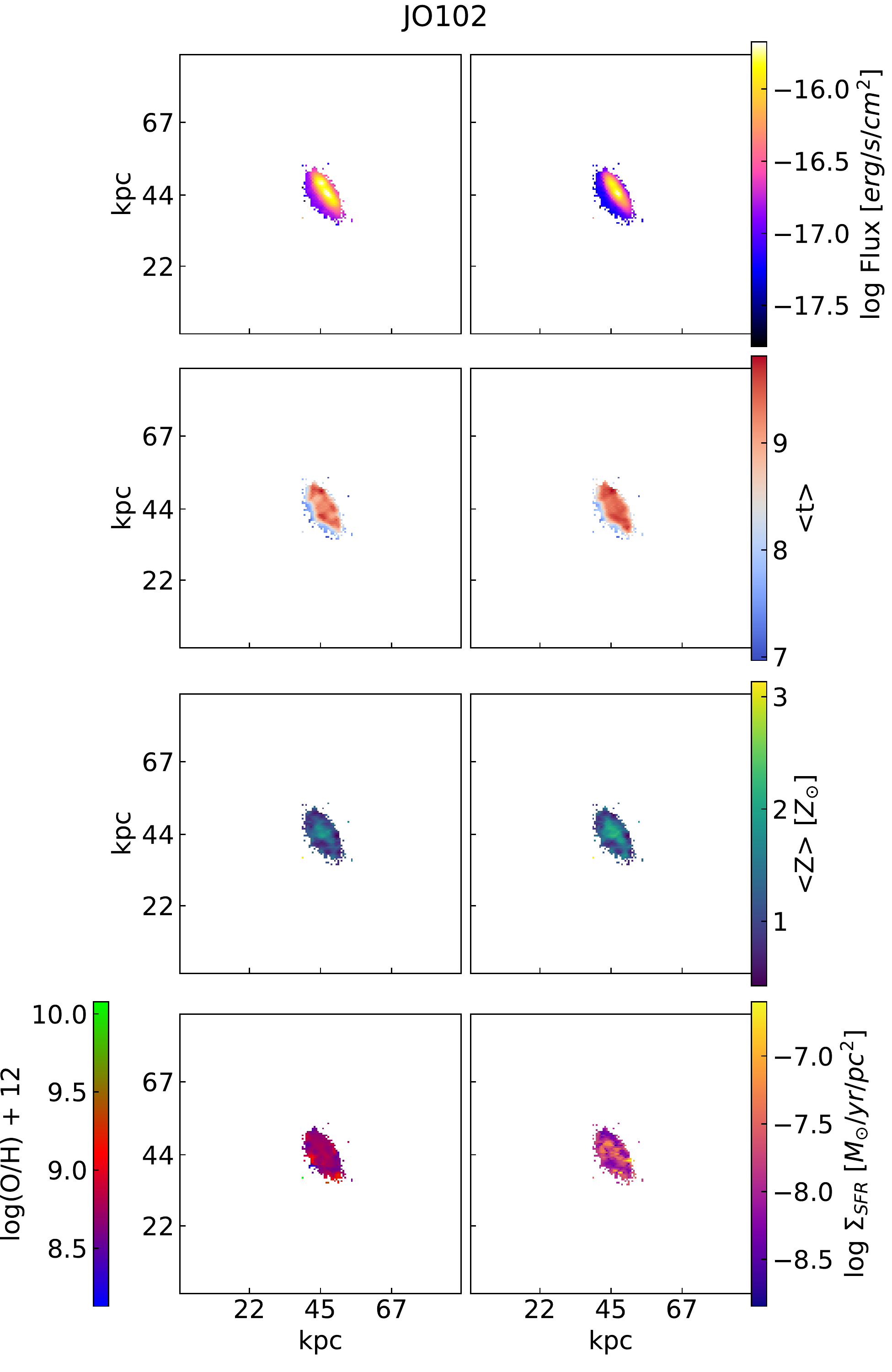}
\end{figure}
\begin{figure}
\centering
\includegraphics[width=\linewidth]{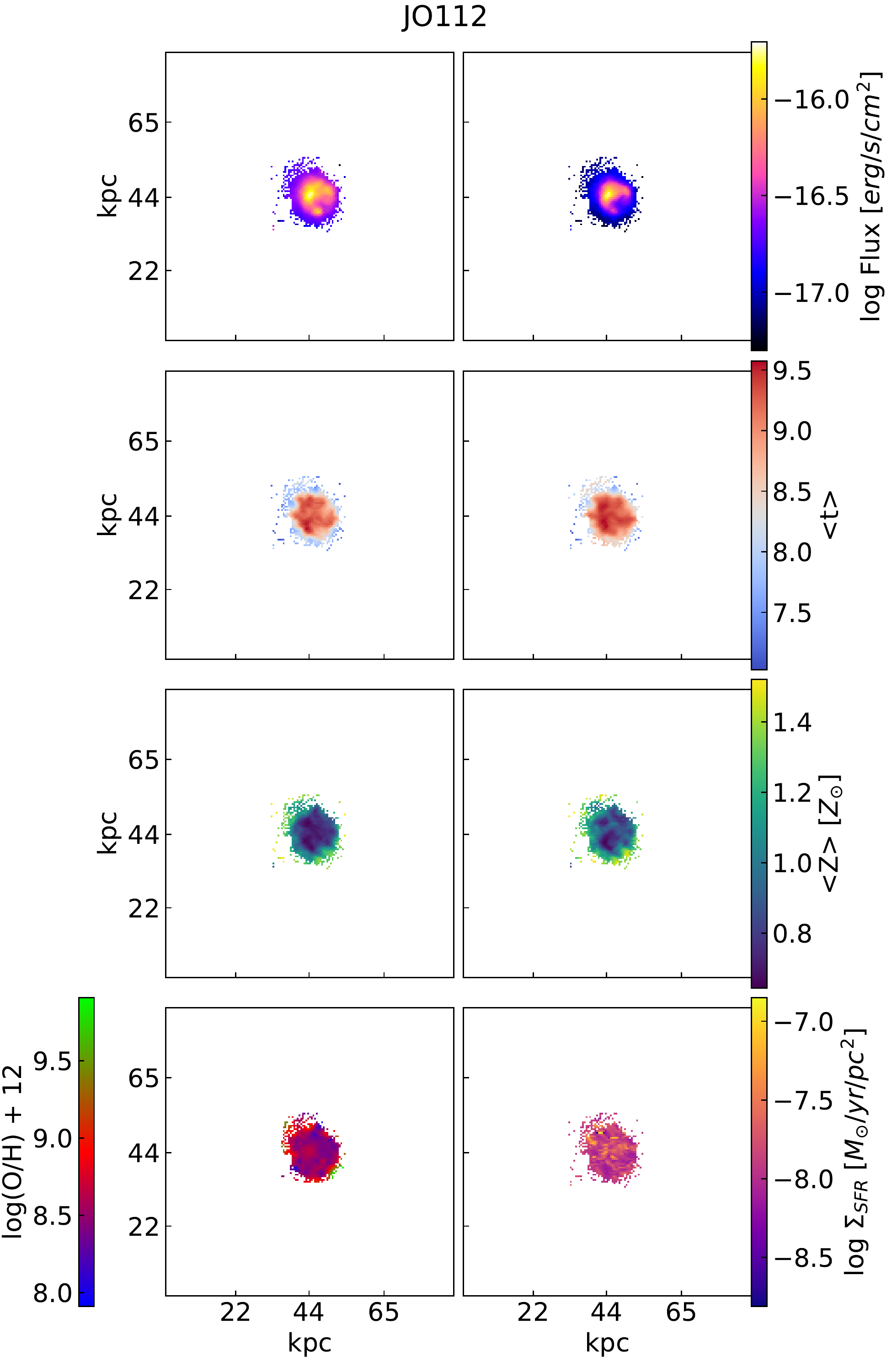}
\end{figure}

\begin{figure}
\centering
\includegraphics[width=\linewidth]{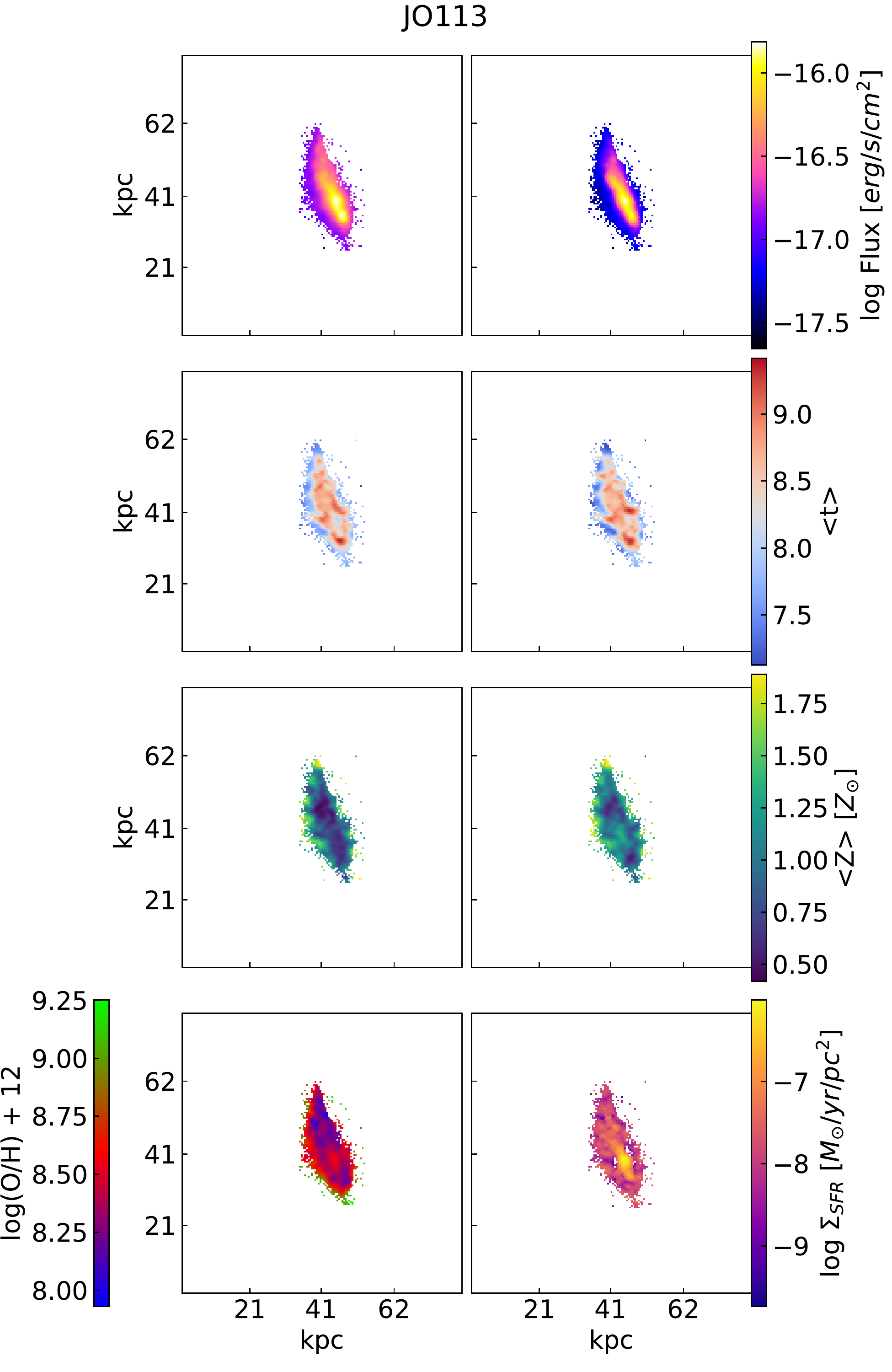}
\end{figure}

\begin{figure}
\centering
\includegraphics[width=\linewidth]{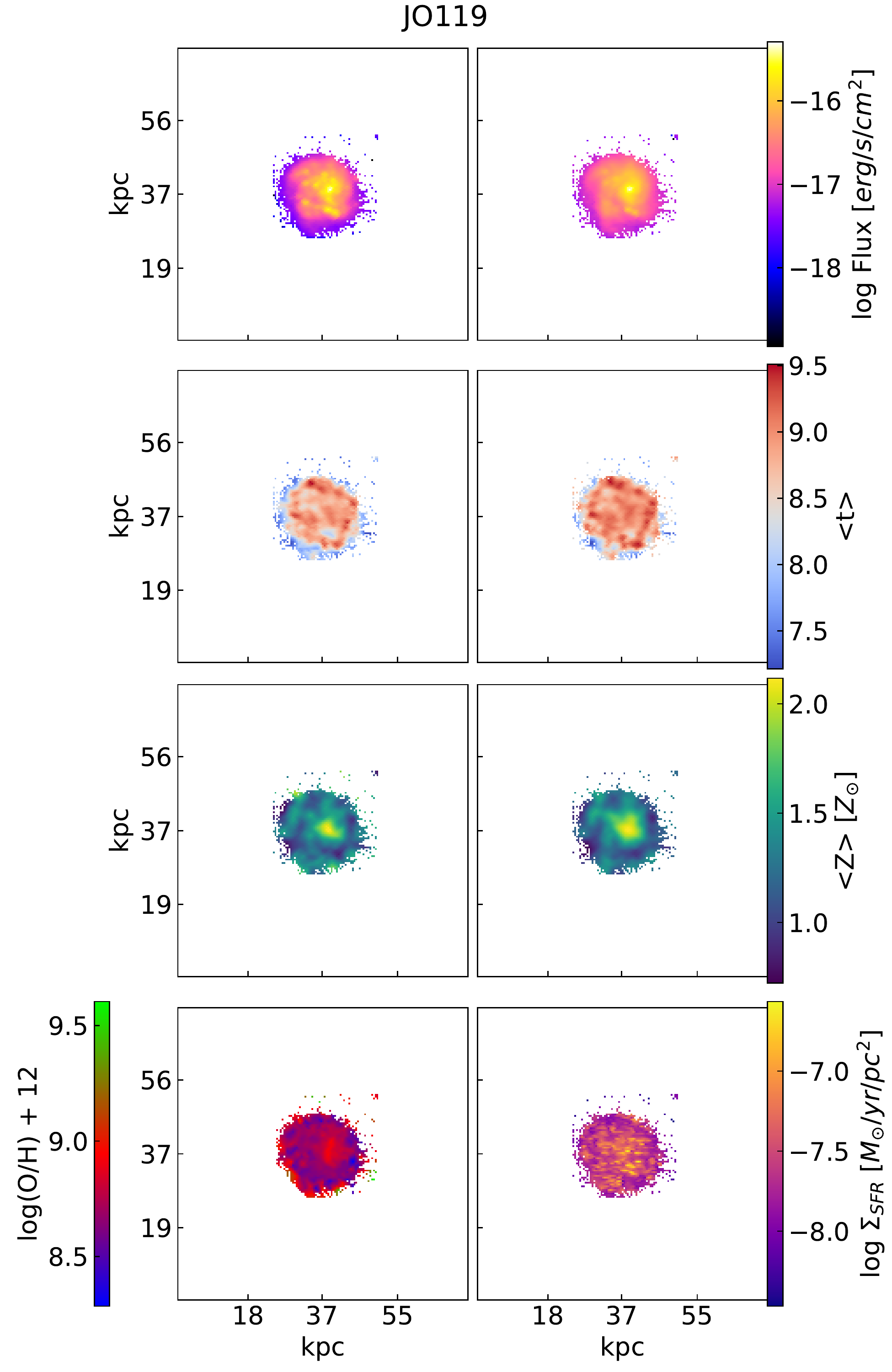}
\end{figure}

\begin{figure}
\centering
\includegraphics[width=\linewidth]{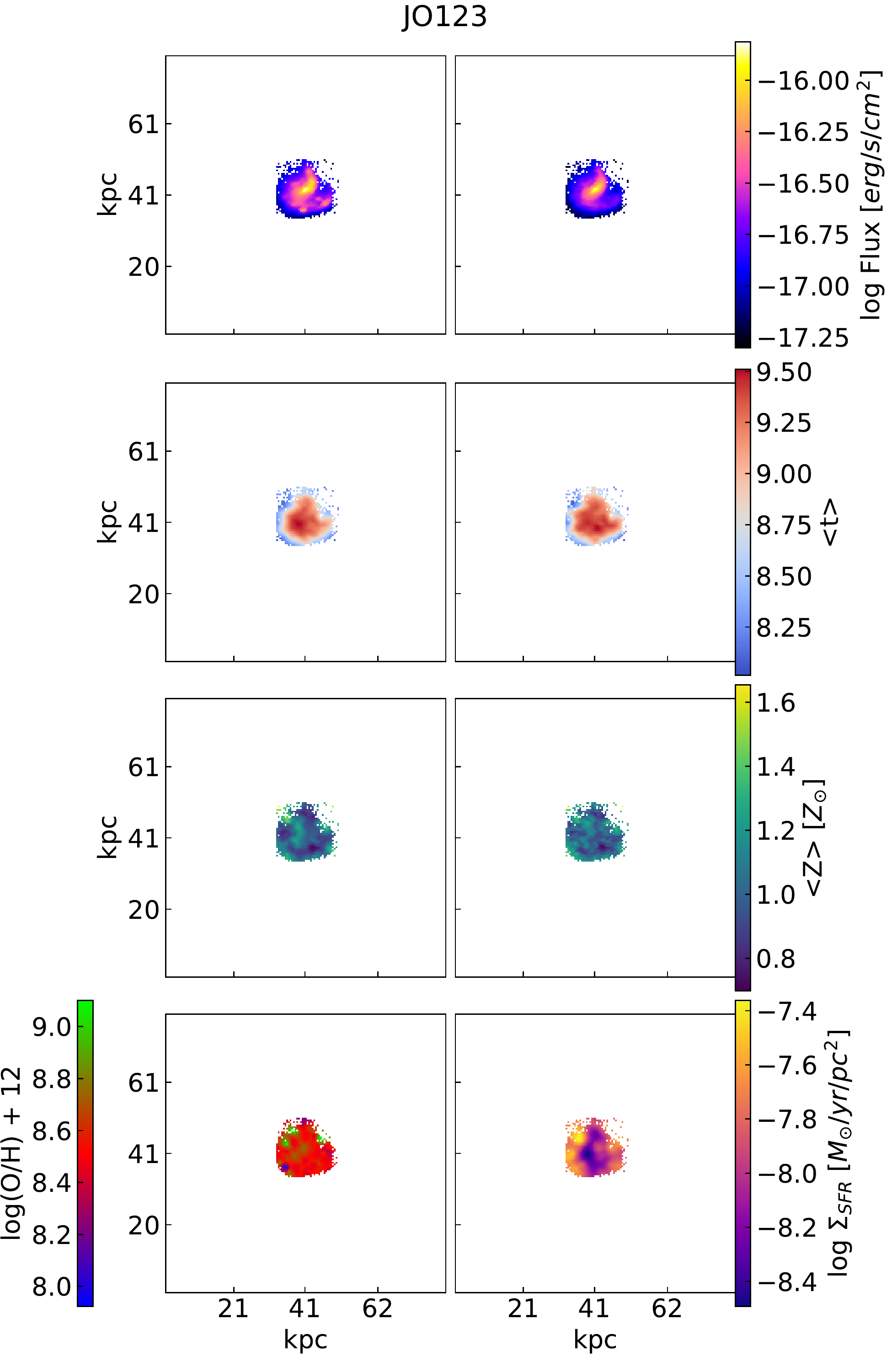}
\end{figure}

\begin{figure}
\centering
\includegraphics[width=\linewidth]{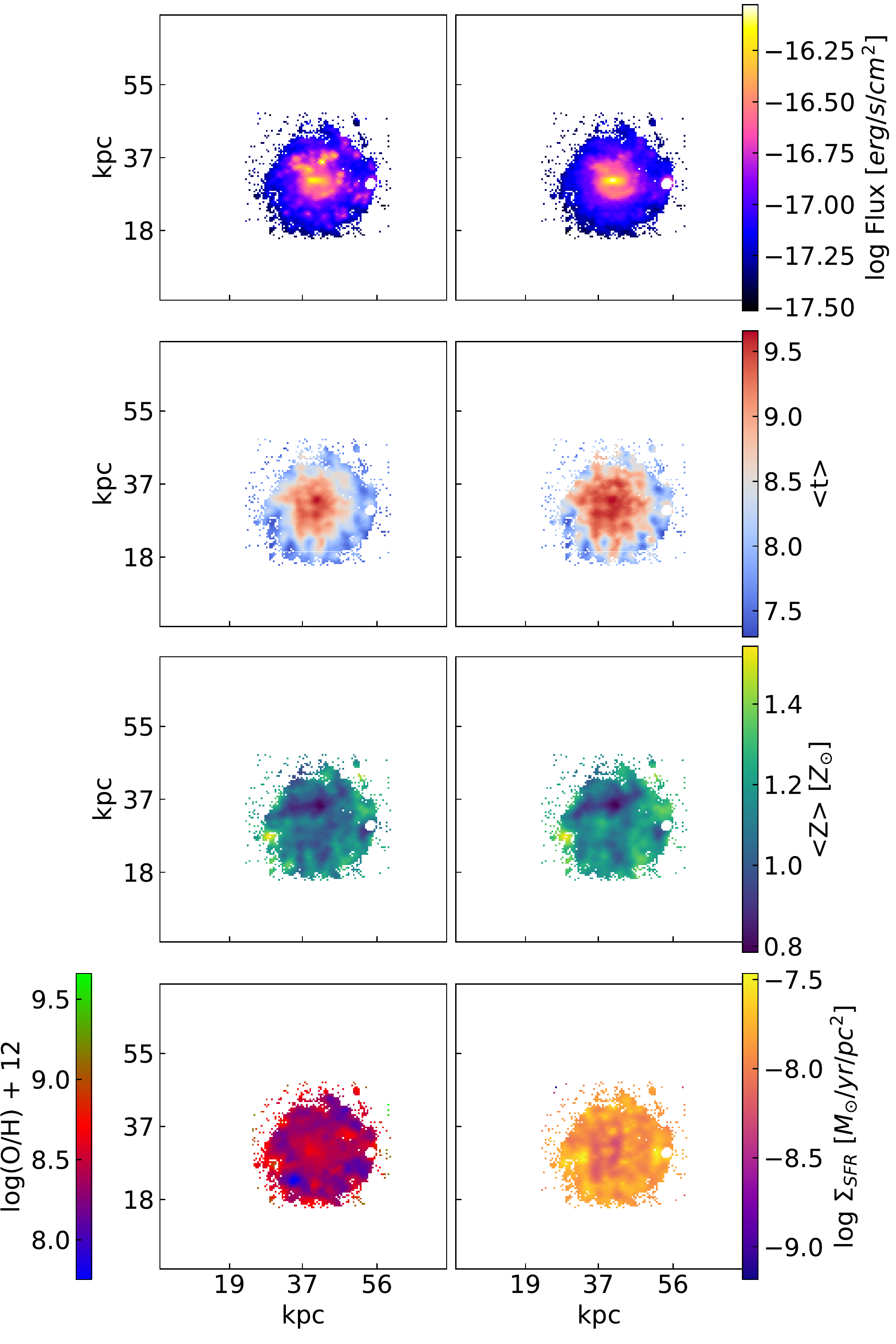}
\end{figure}

\begin{figure}
\centering
\includegraphics[width=\linewidth]{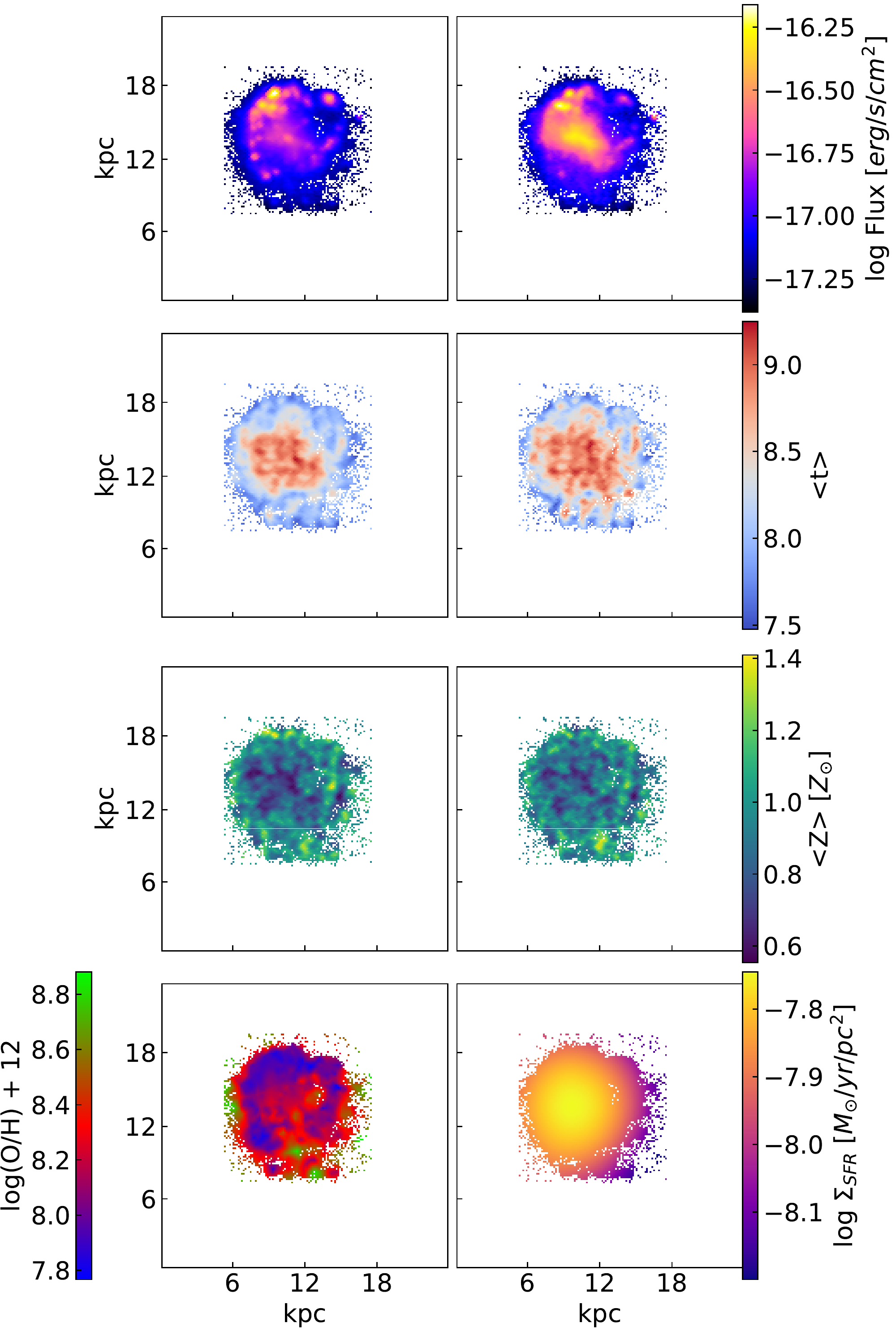}
\end{figure}

\begin{figure}
\centering
\includegraphics[width=\linewidth]{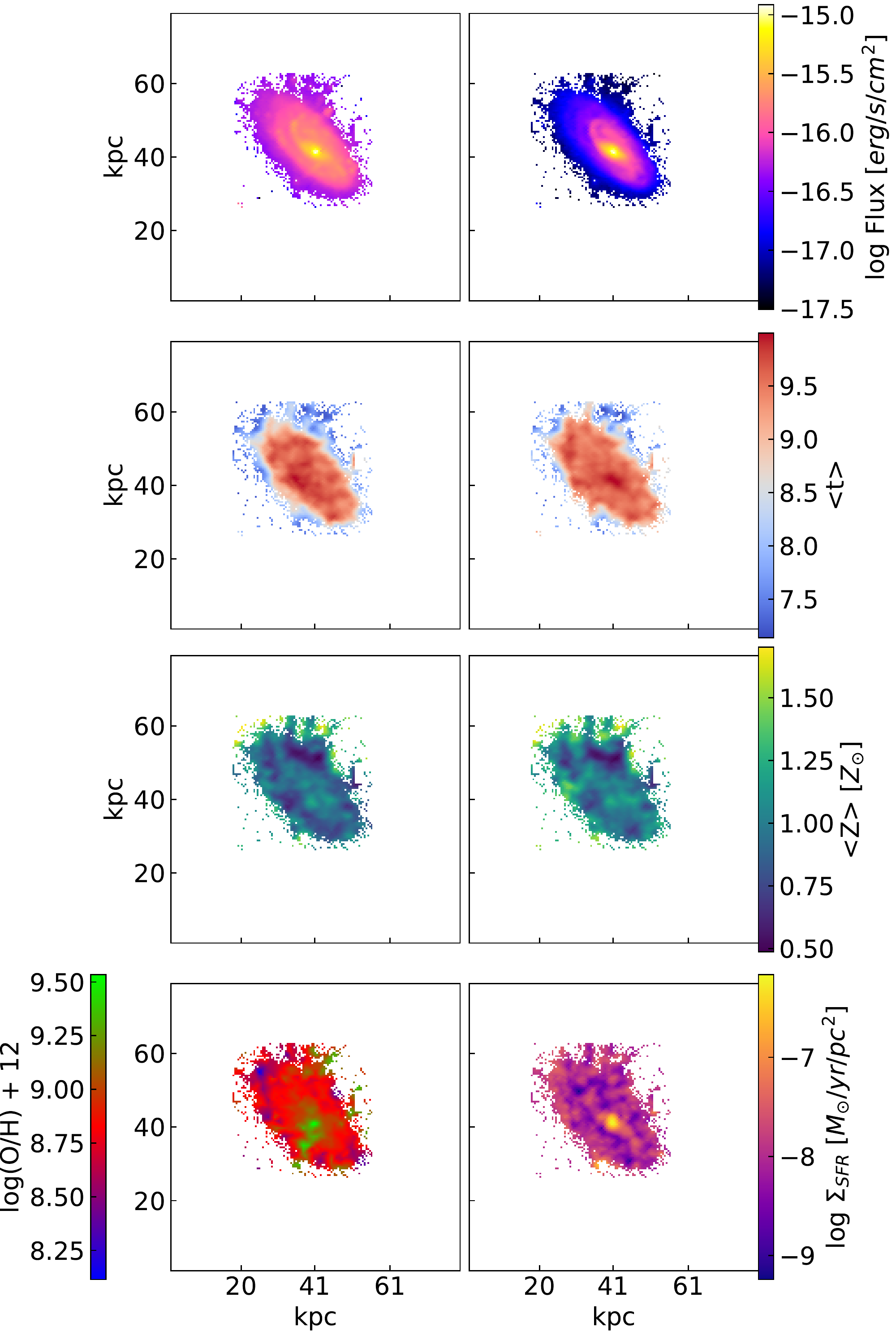}
\end{figure}

\begin{figure}
\centering
\includegraphics[width=\linewidth]{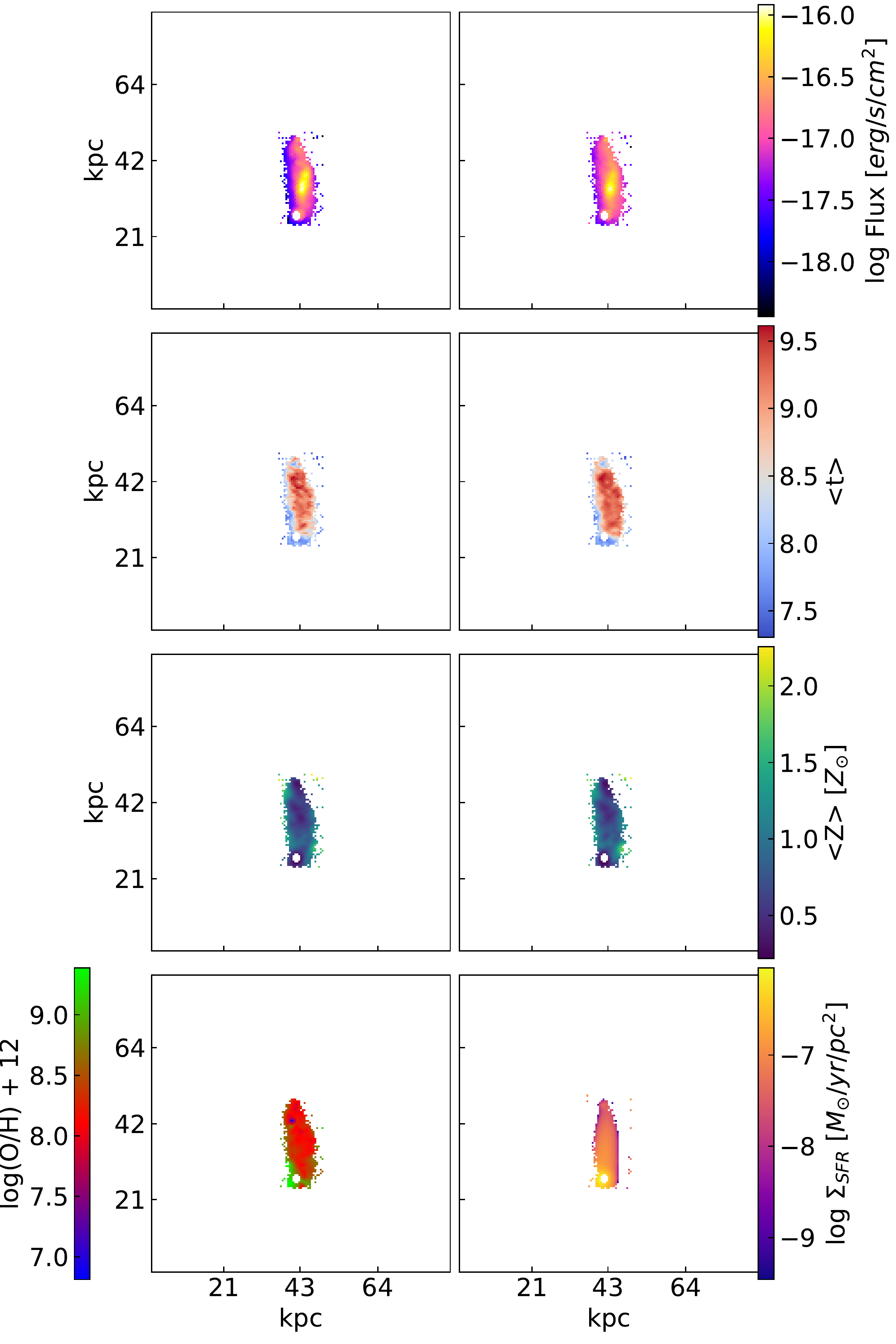}
\end{figure}

\begin{figure}
\centering
\includegraphics[width=\linewidth]{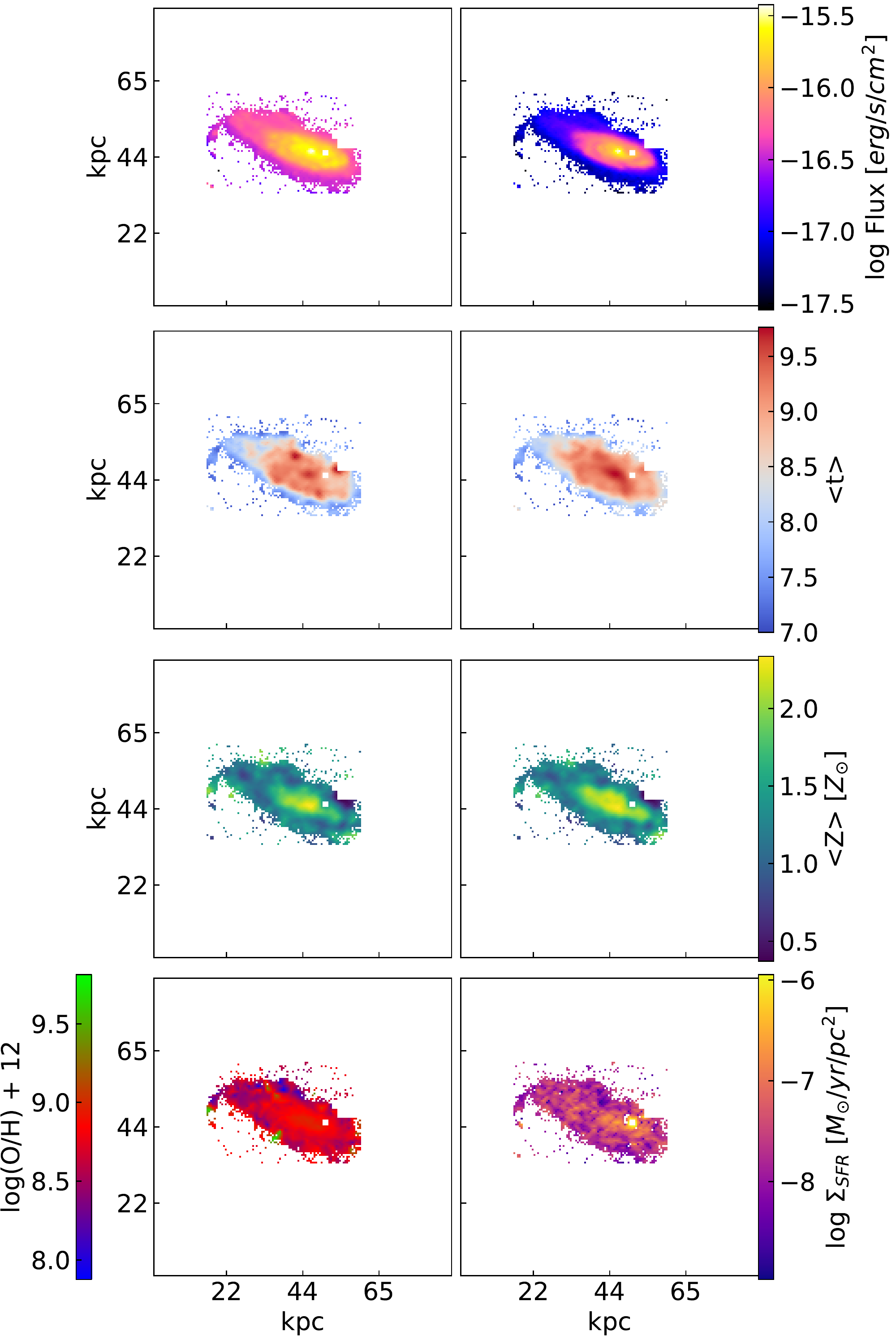}
\end{figure}

\begin{figure}
\centering
\includegraphics[width=\linewidth]{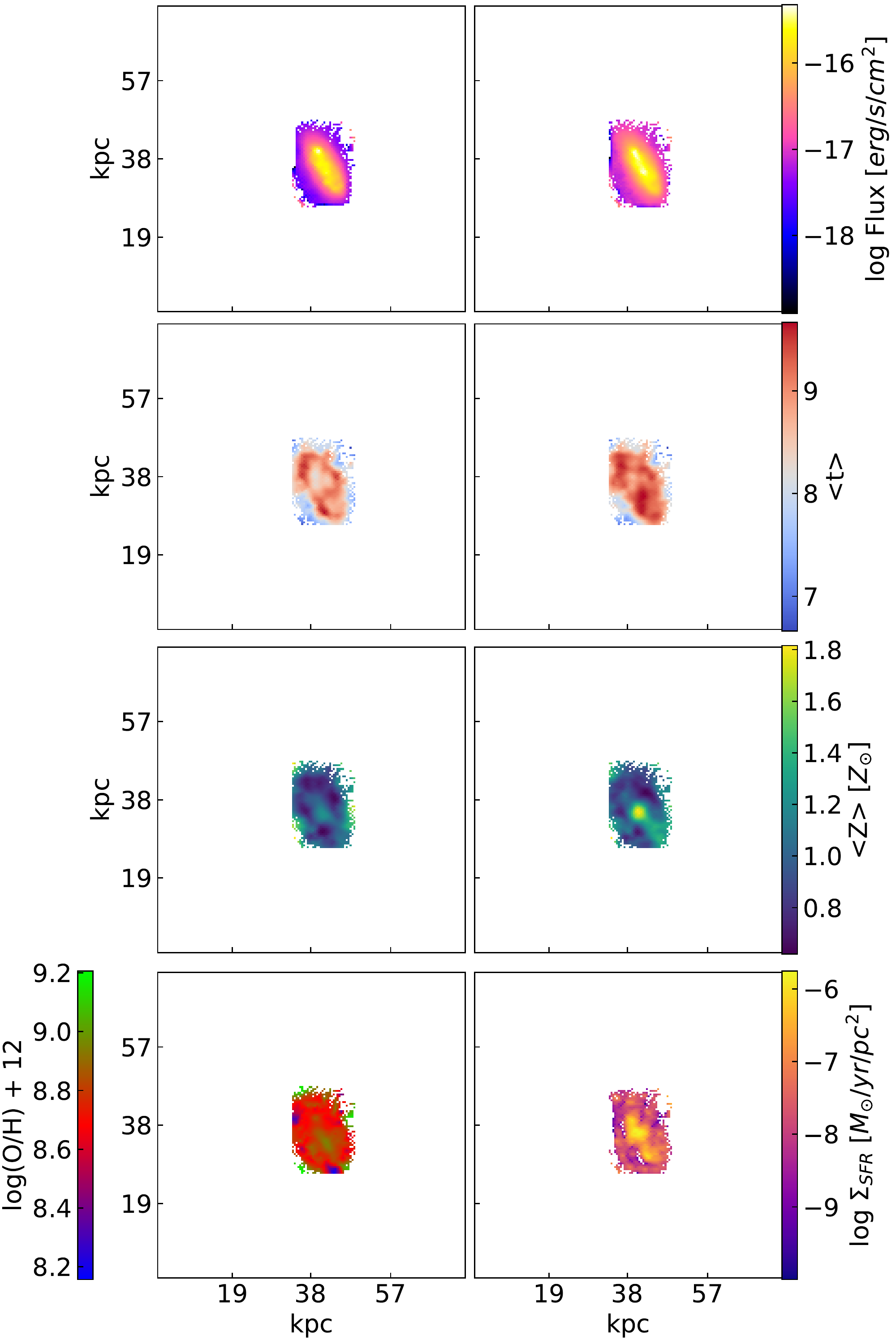}
\end{figure}

\begin{figure}
\centering
\includegraphics[width=\linewidth]{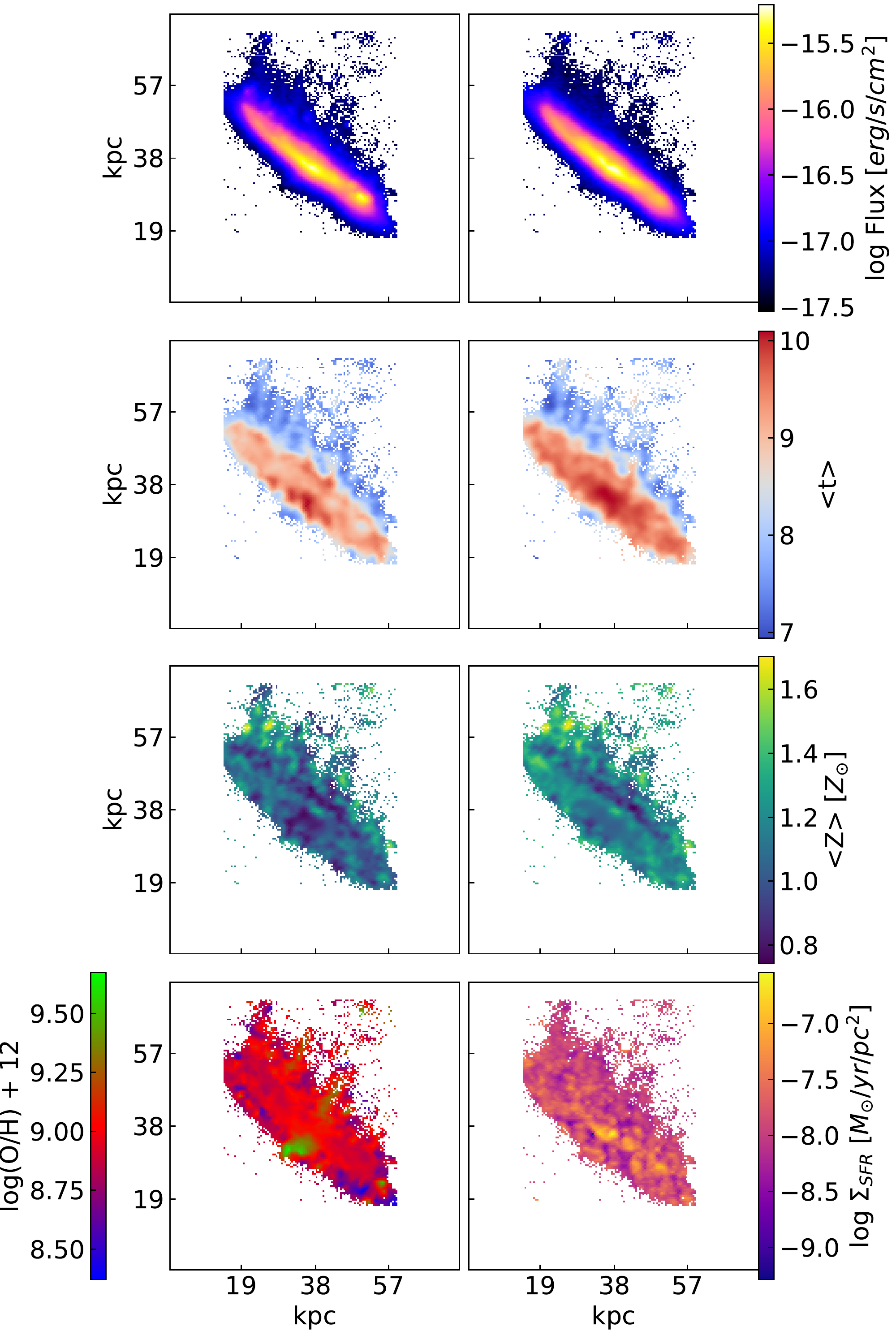}
\end{figure}

\begin{figure}
\centering
\includegraphics[width=\linewidth]{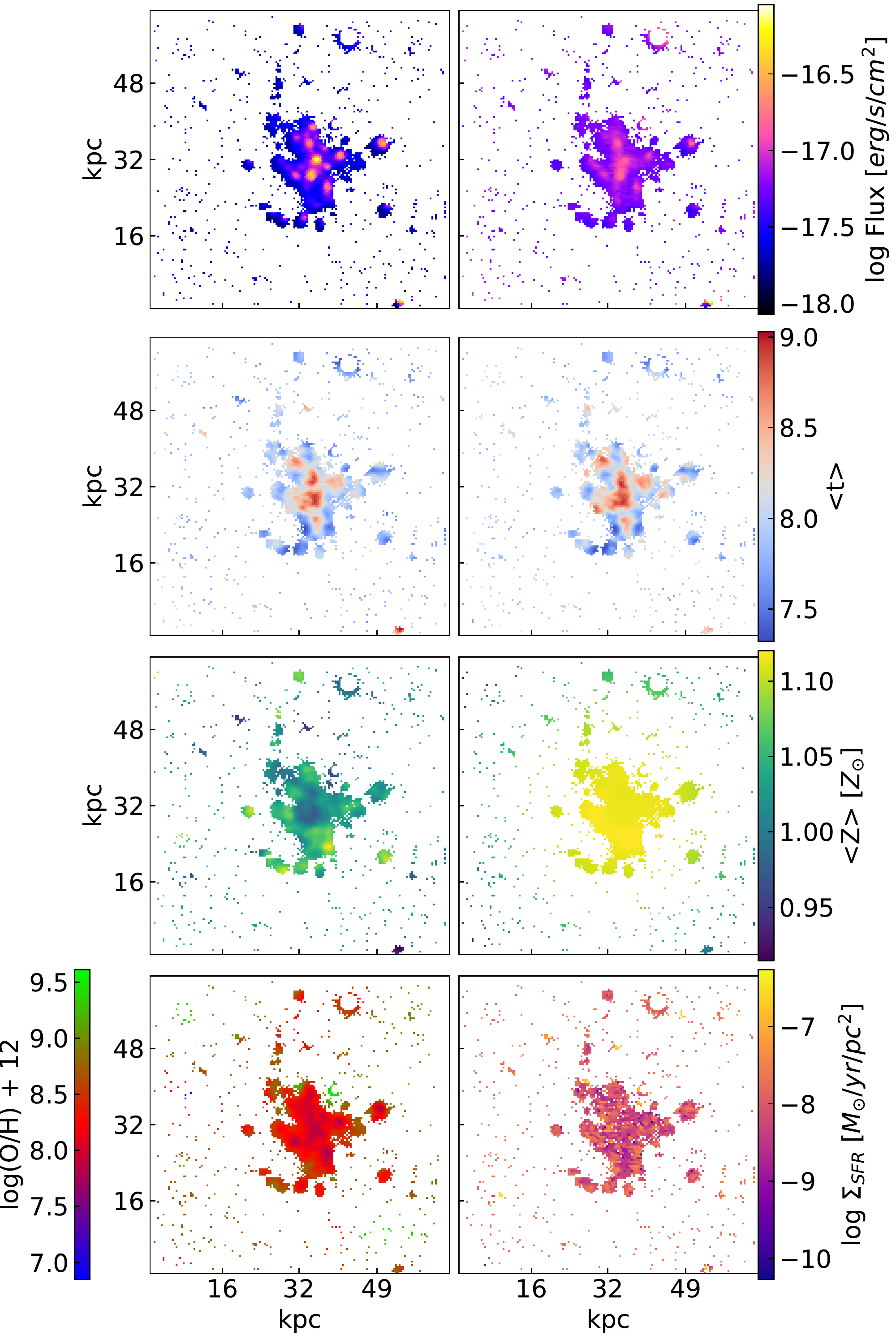}
\end{figure}
\begin{figure}
\centering
\includegraphics[width=\linewidth]{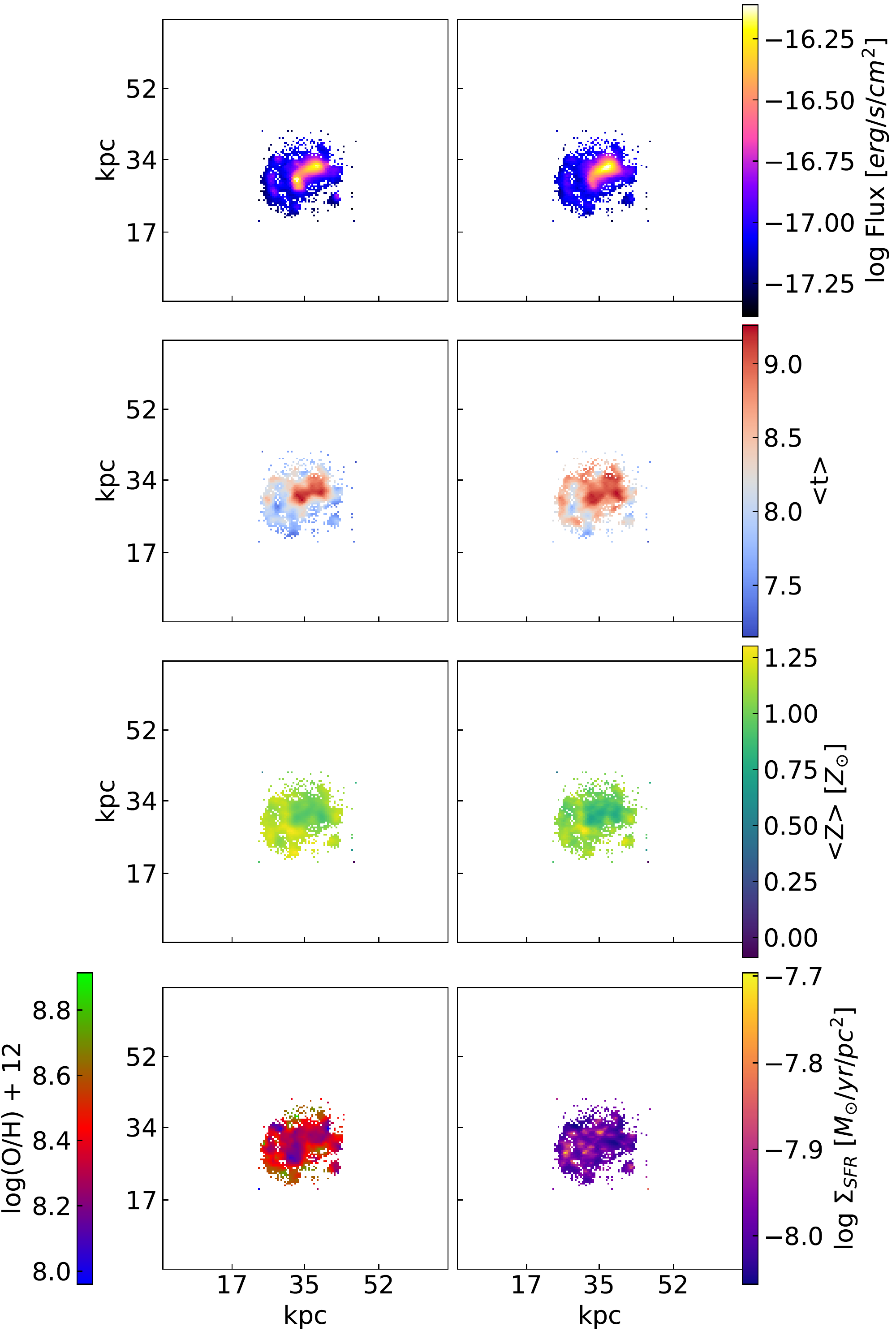}
\end{figure}

\begin{figure}
\centering
\includegraphics[width=\linewidth]{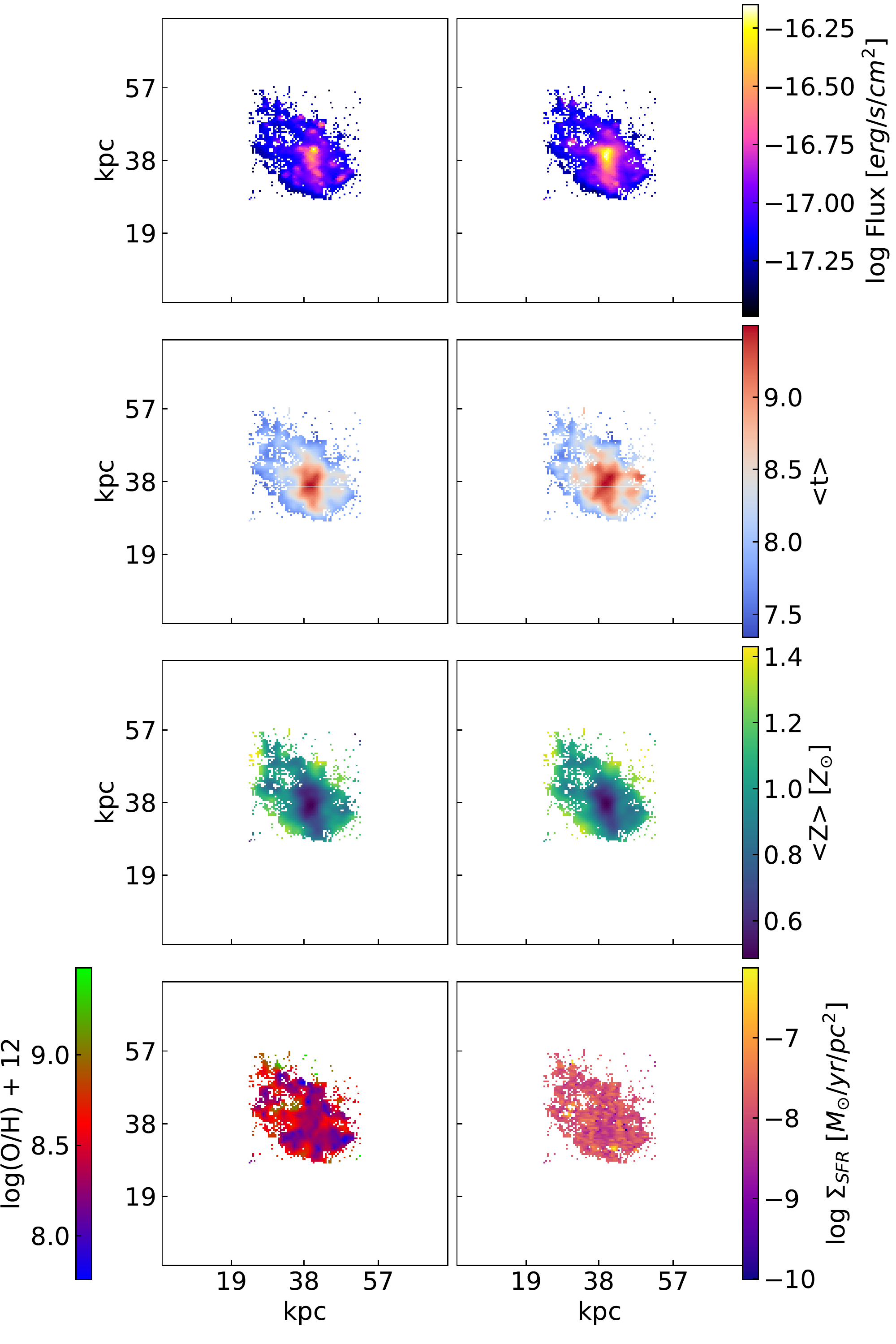}
\end{figure}

\begin{figure}
\centering
\includegraphics[width=\linewidth]{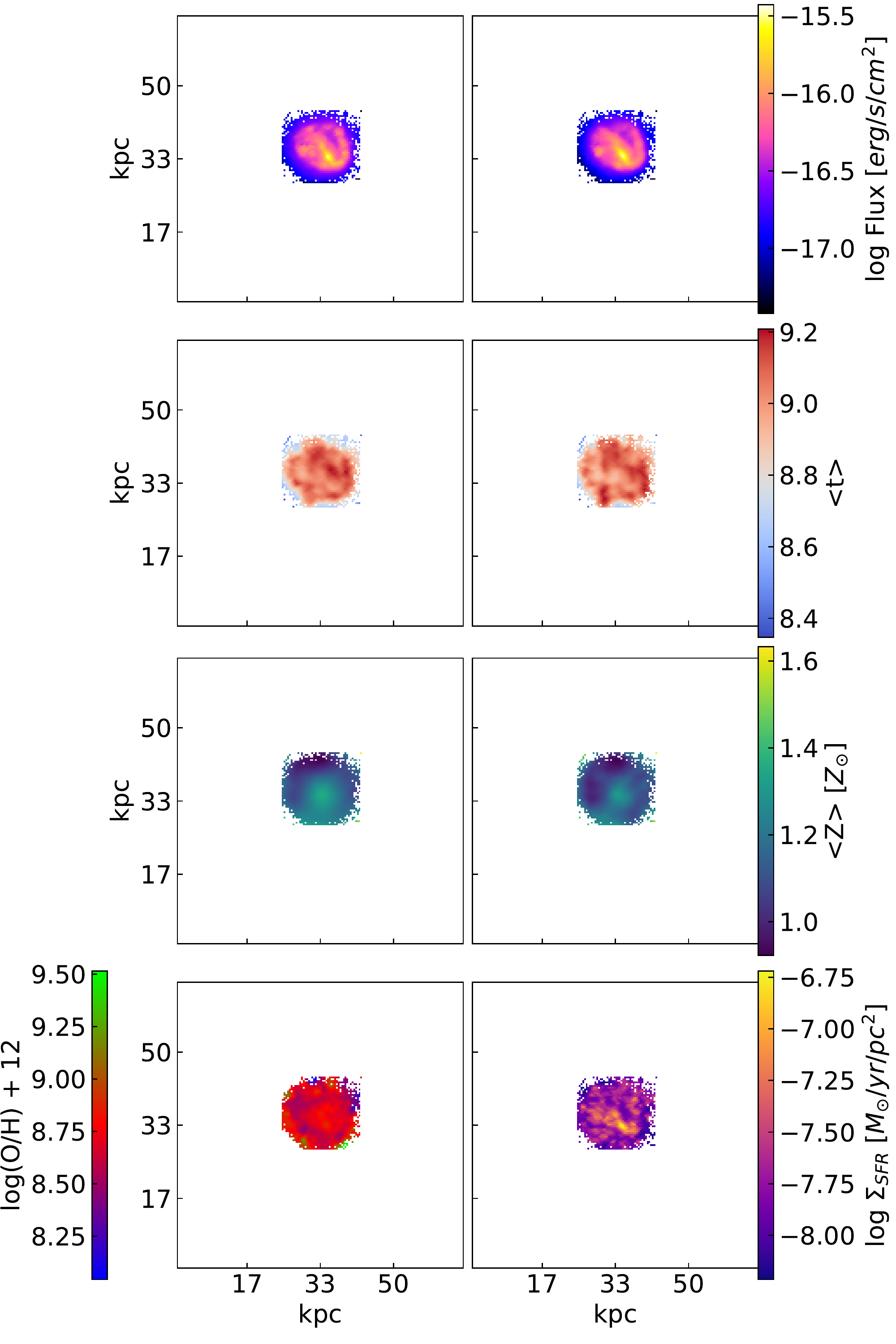}
\end{figure}

\begin{figure}
\centering
\includegraphics[width=\linewidth]{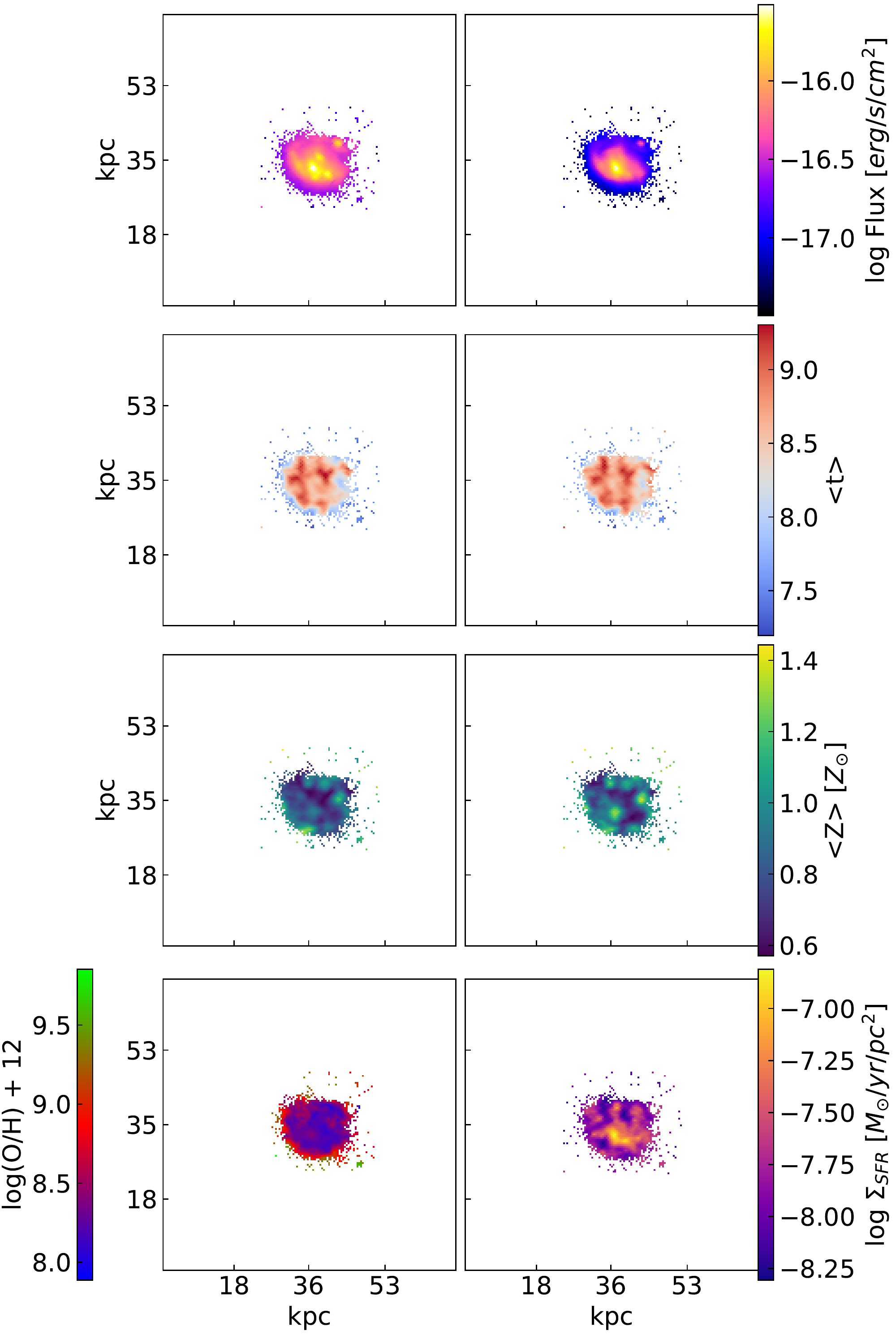}
\end{figure}

\begin{figure}
\centering
\includegraphics[width=\linewidth]{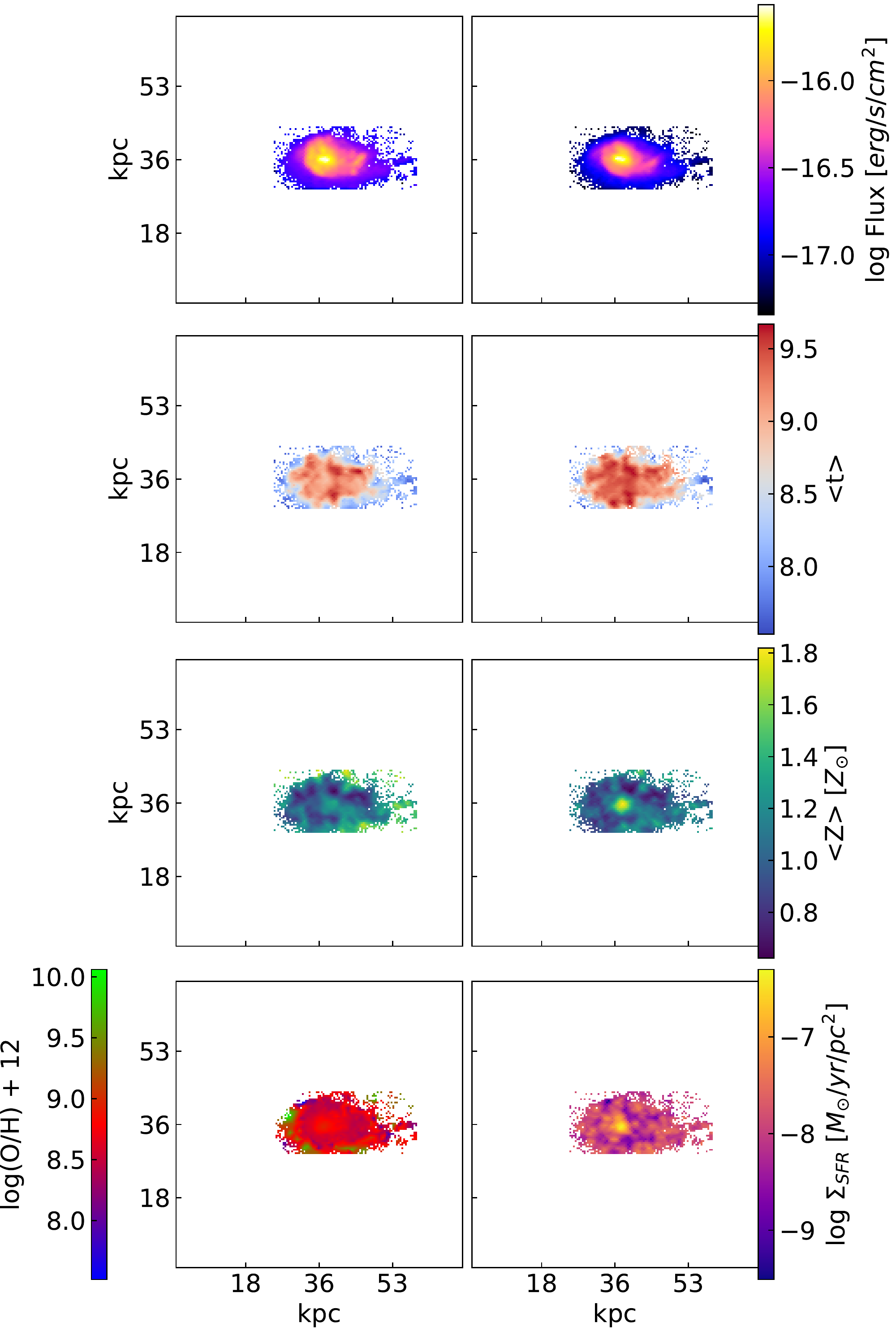}
\end{figure}

\begin{figure}
\centering
\includegraphics[width=\linewidth]{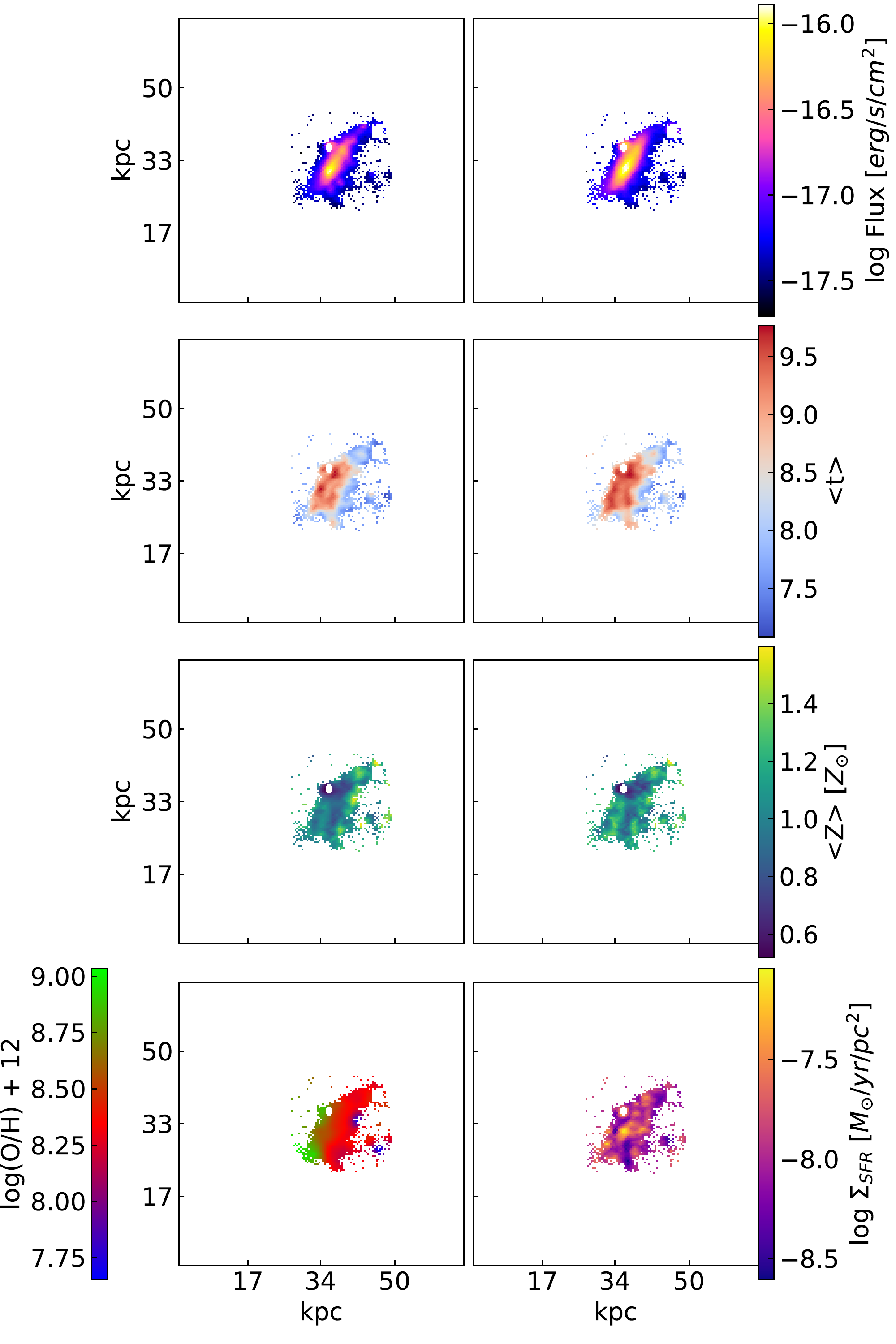}
\end{figure}

\begin{figure}
\centering
\includegraphics[width=\linewidth]{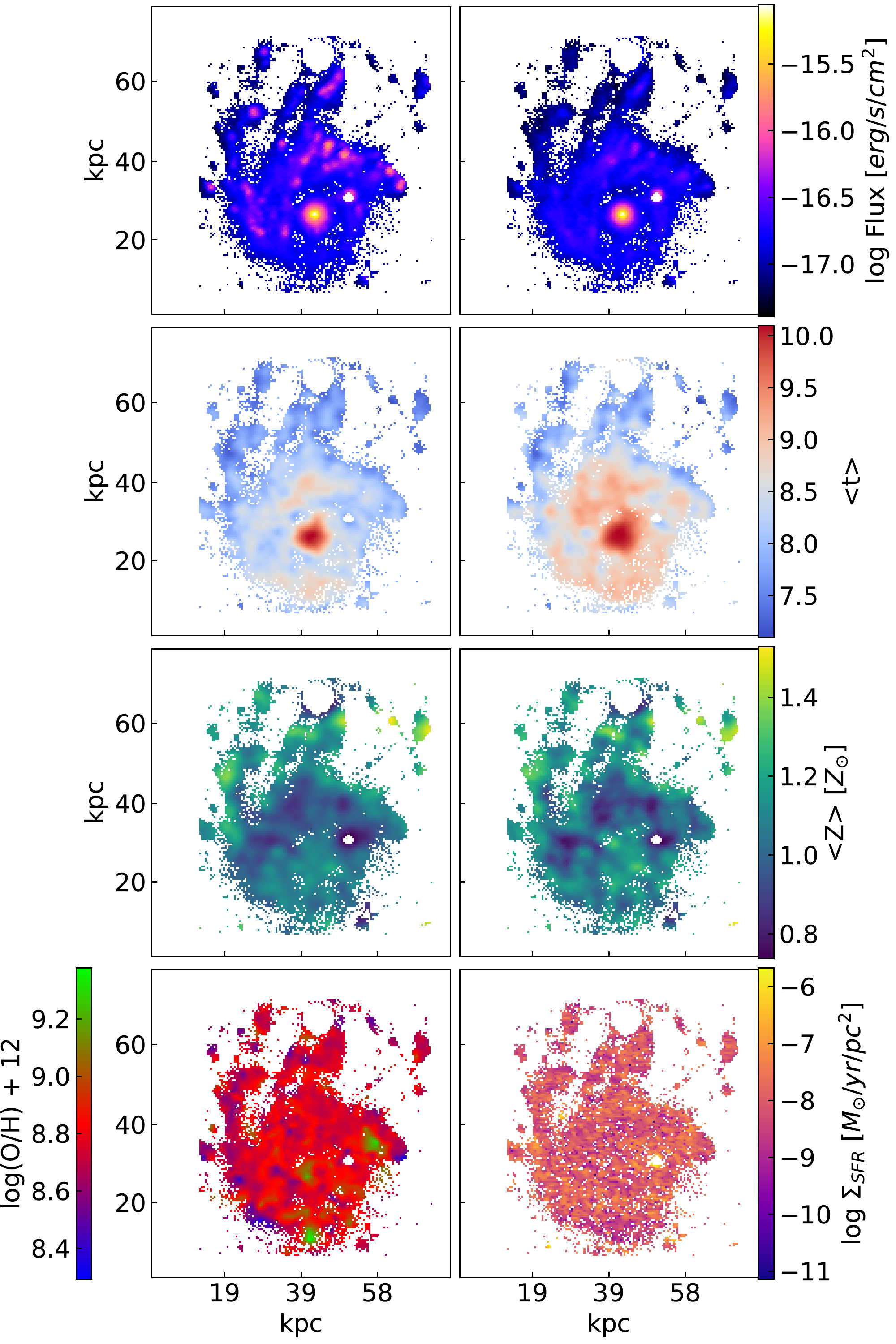}
\end{figure}

\begin{figure}
\centering
\includegraphics[width=\linewidth]{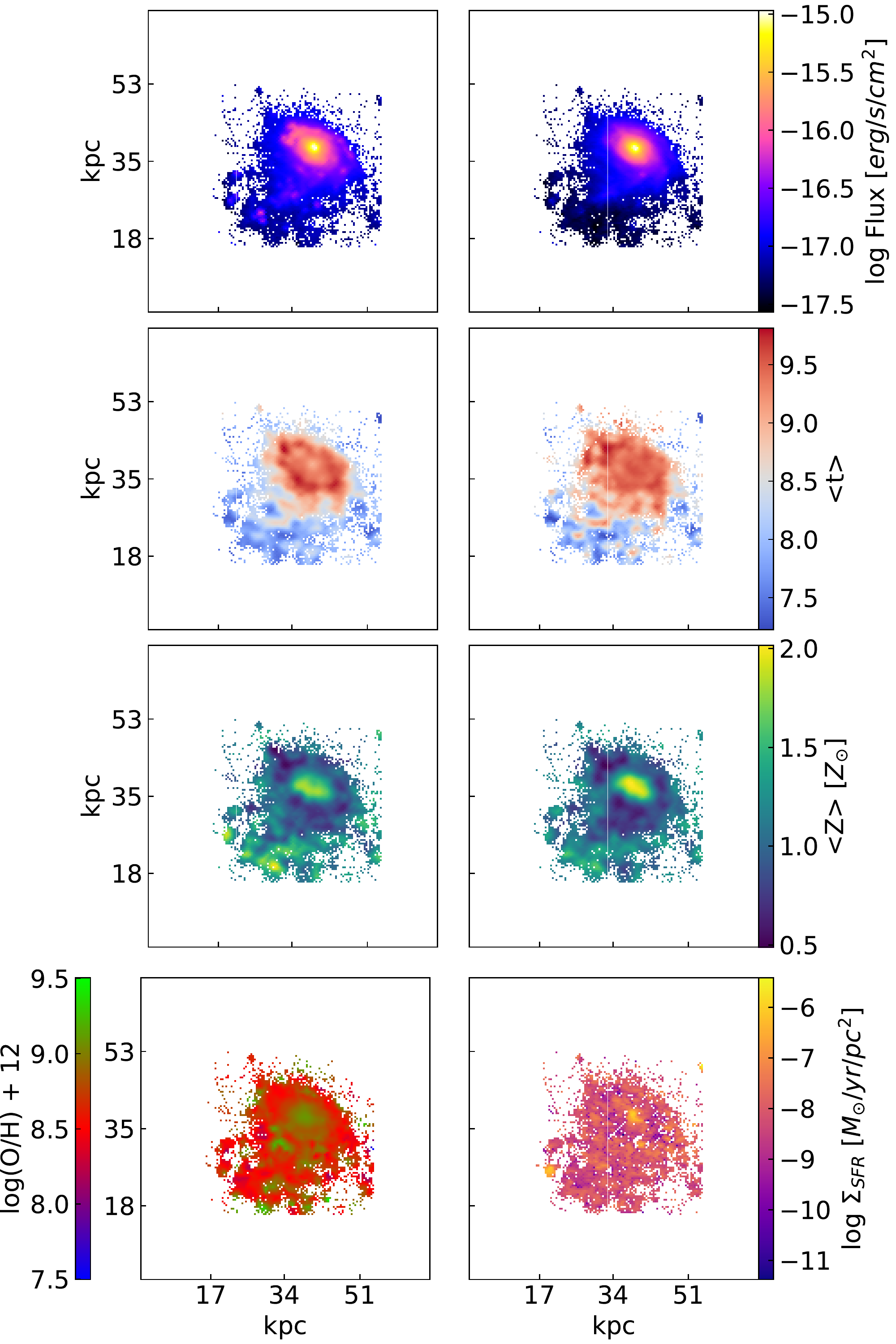}
\end{figure}

\begin{figure}
\centering
\includegraphics[width=\linewidth]{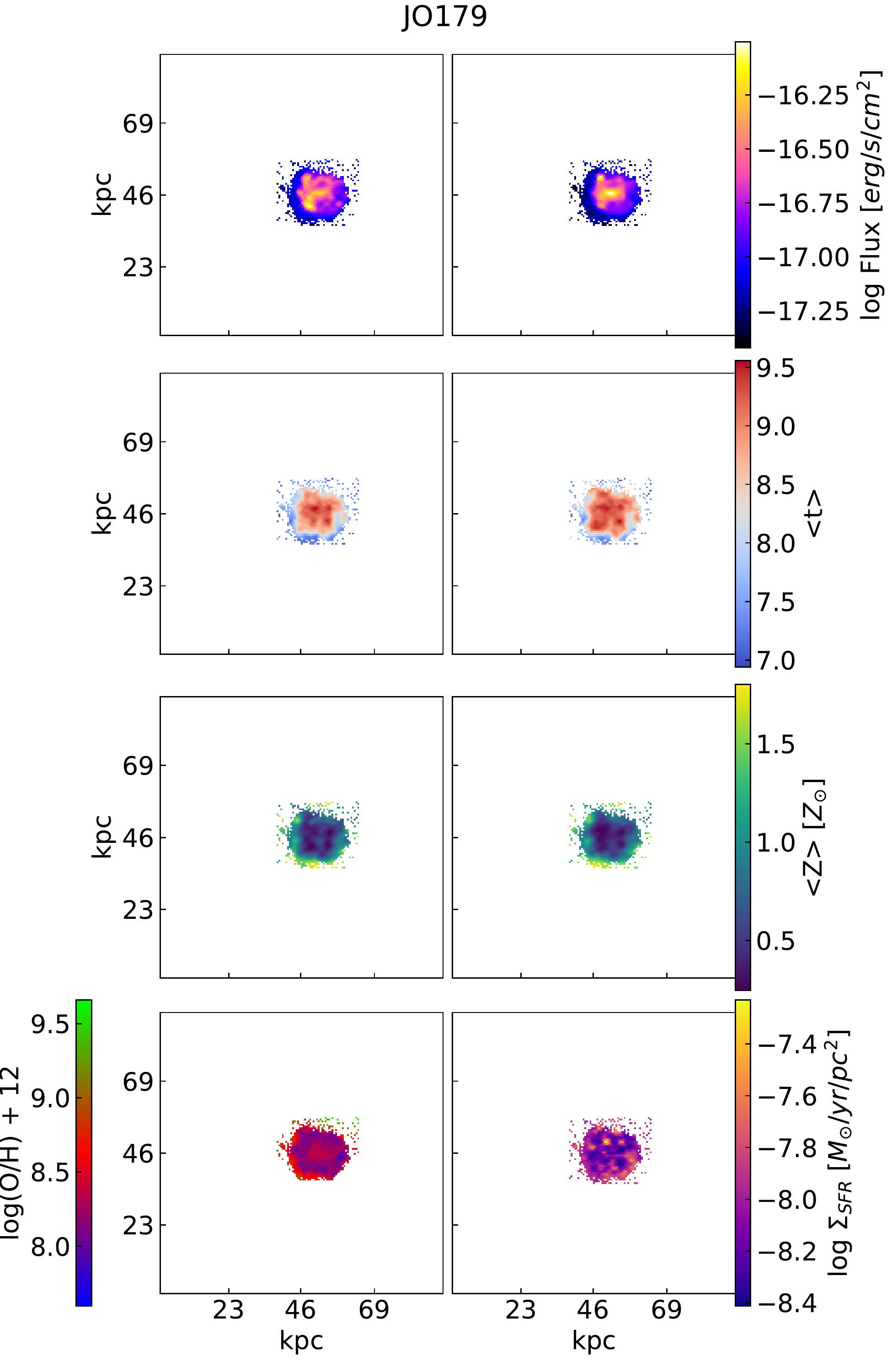}
\end{figure}

\begin{figure}
\centering
\includegraphics[width=\linewidth]{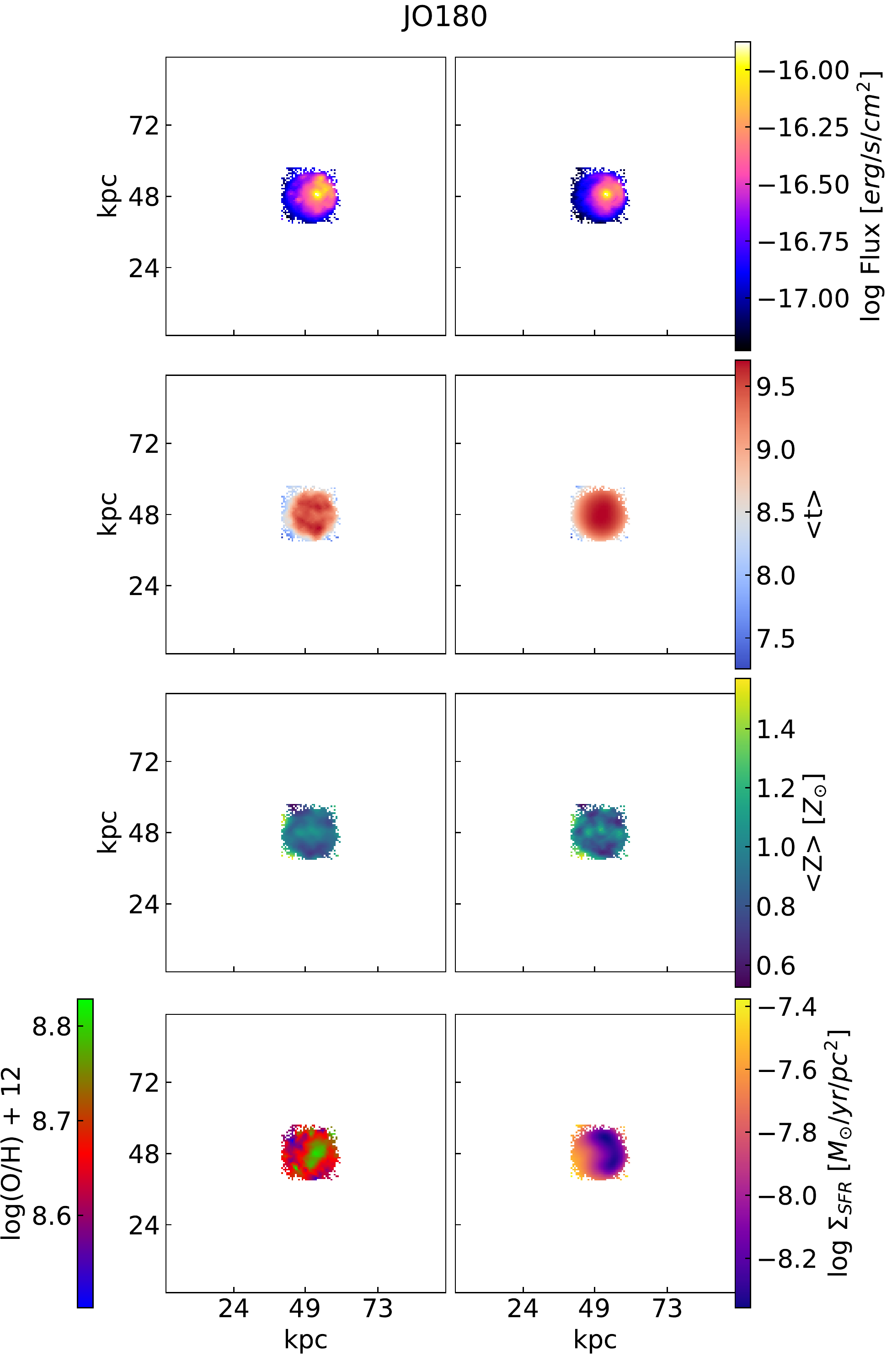}
\end{figure}

\begin{figure}
\centering
\includegraphics[width=\linewidth]{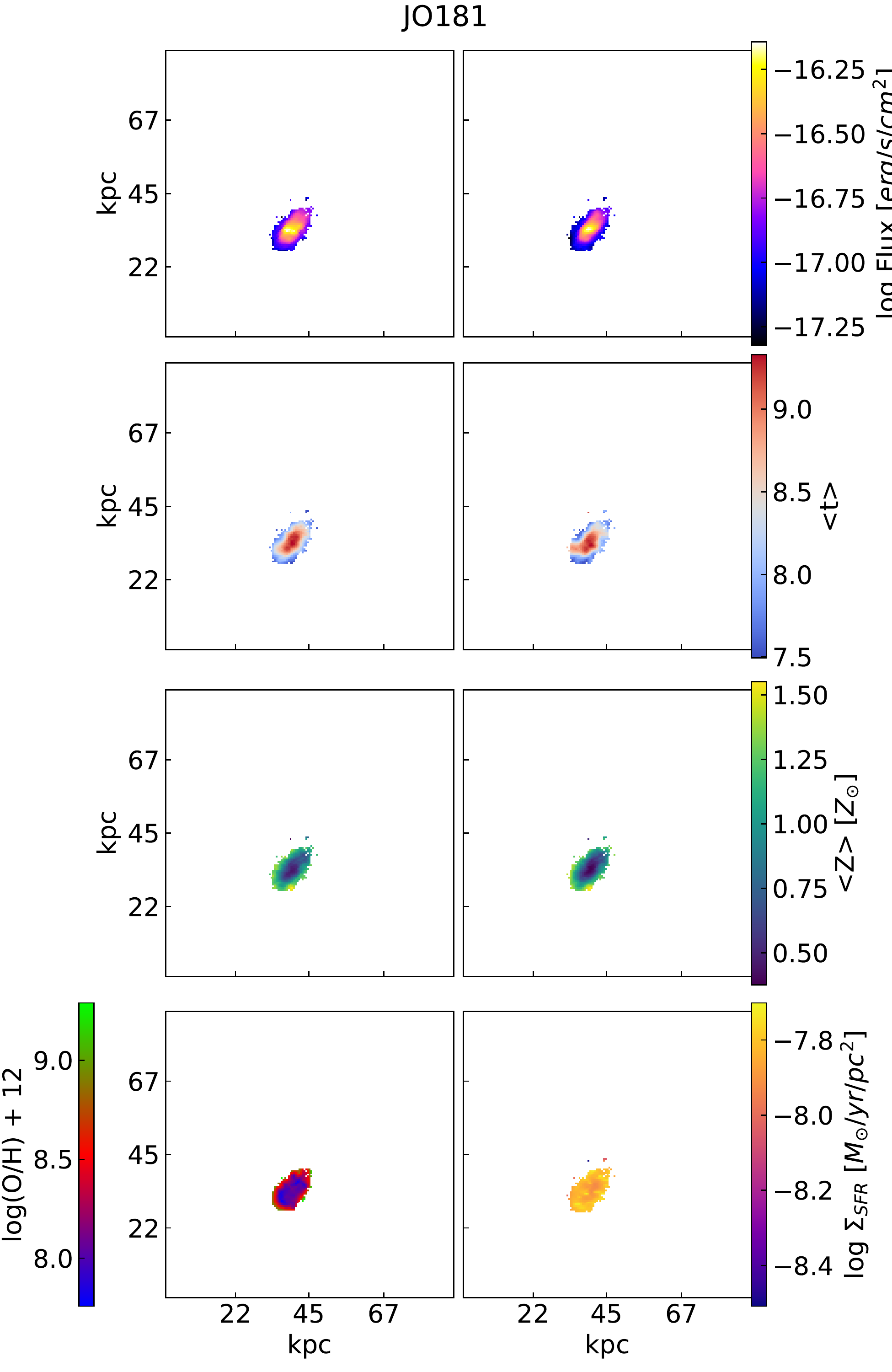}
\end{figure}

\begin{figure}
\centering
\includegraphics[width=\linewidth]{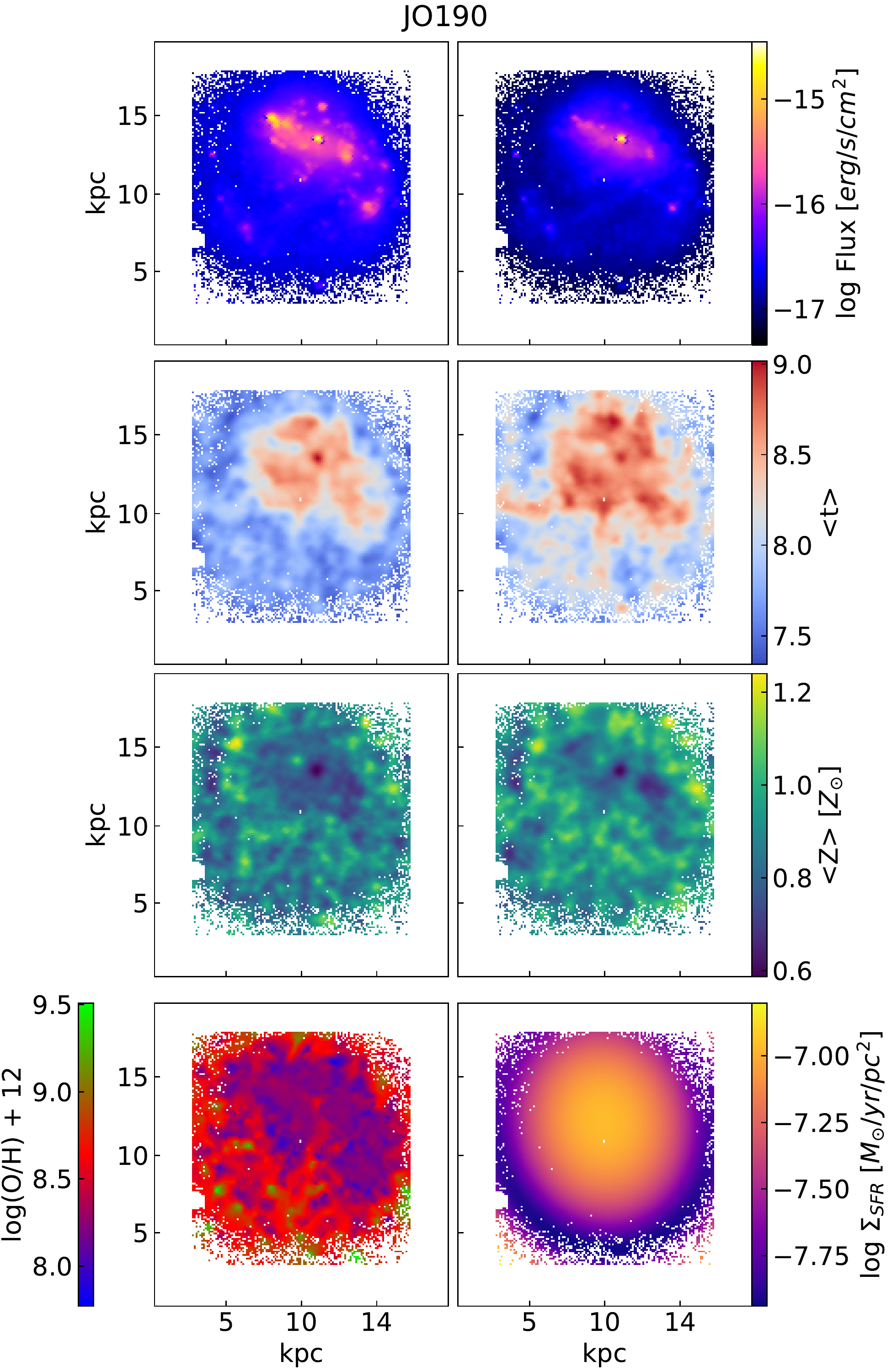}
\end{figure}
\begin{figure}
\centering
\includegraphics[width=\linewidth]{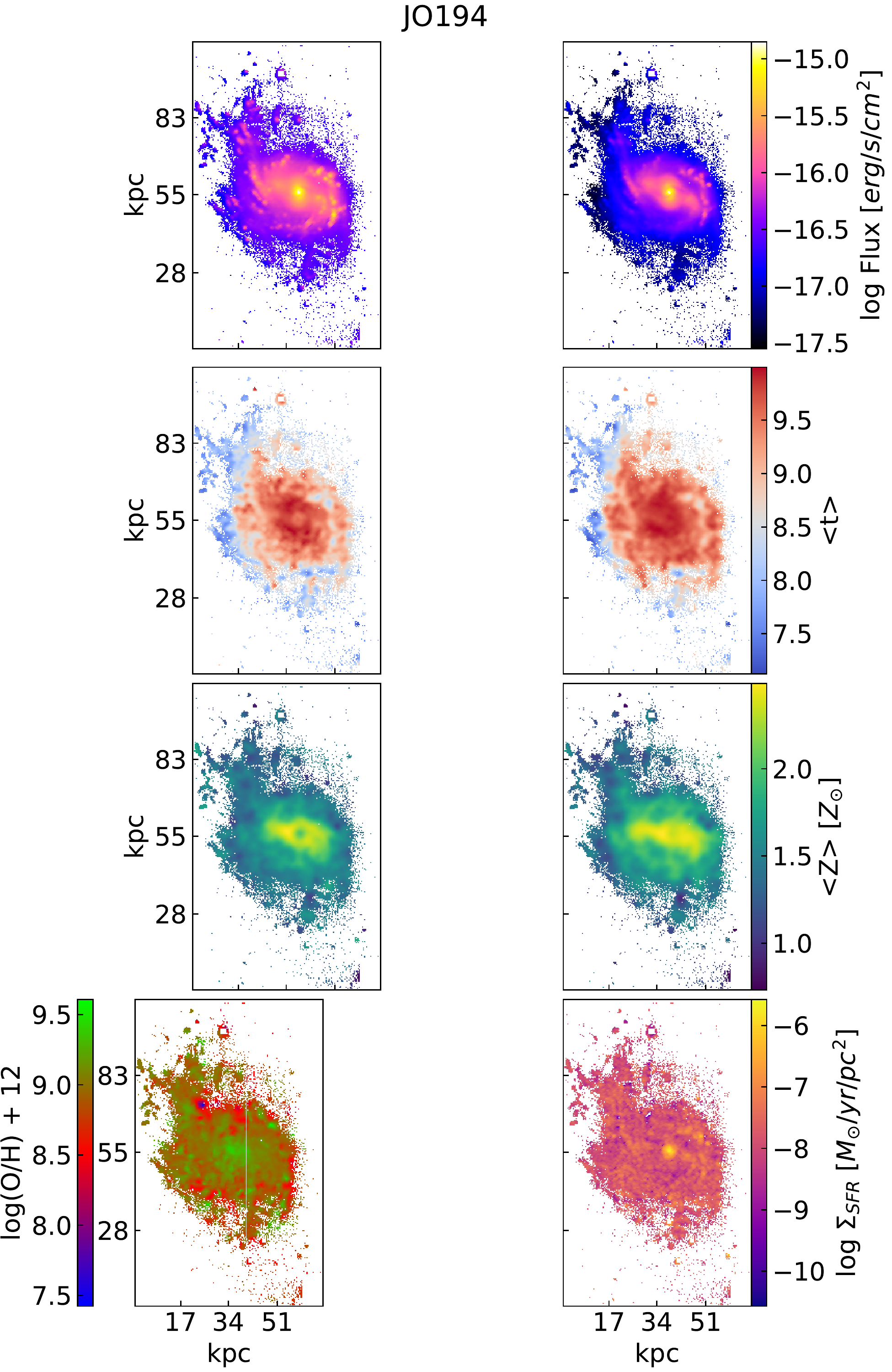}
\end{figure}

\begin{figure}
\centering
\includegraphics[width=\linewidth]{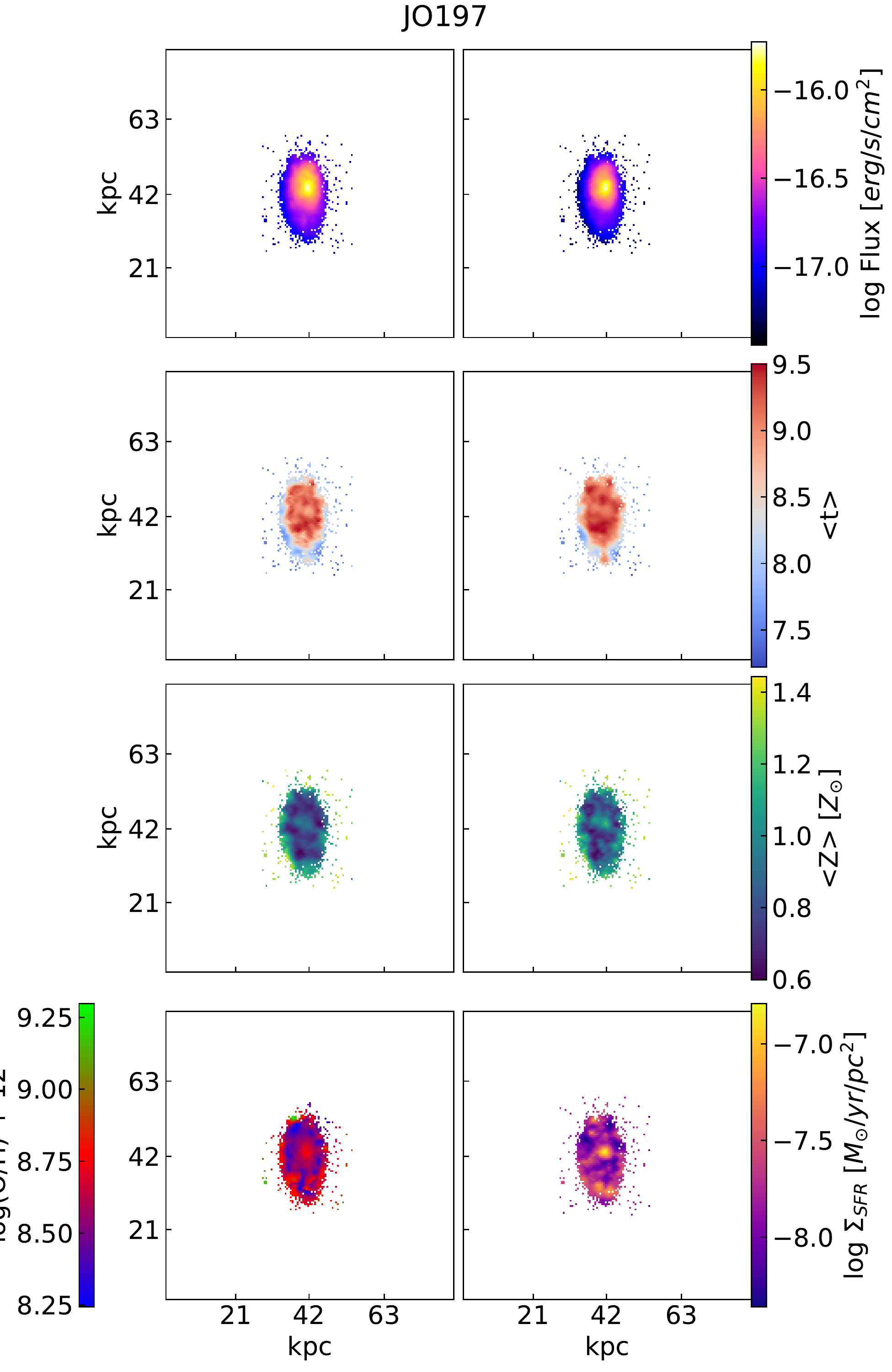}
\end{figure}

\begin{figure}
\centering
\includegraphics[width=\linewidth]{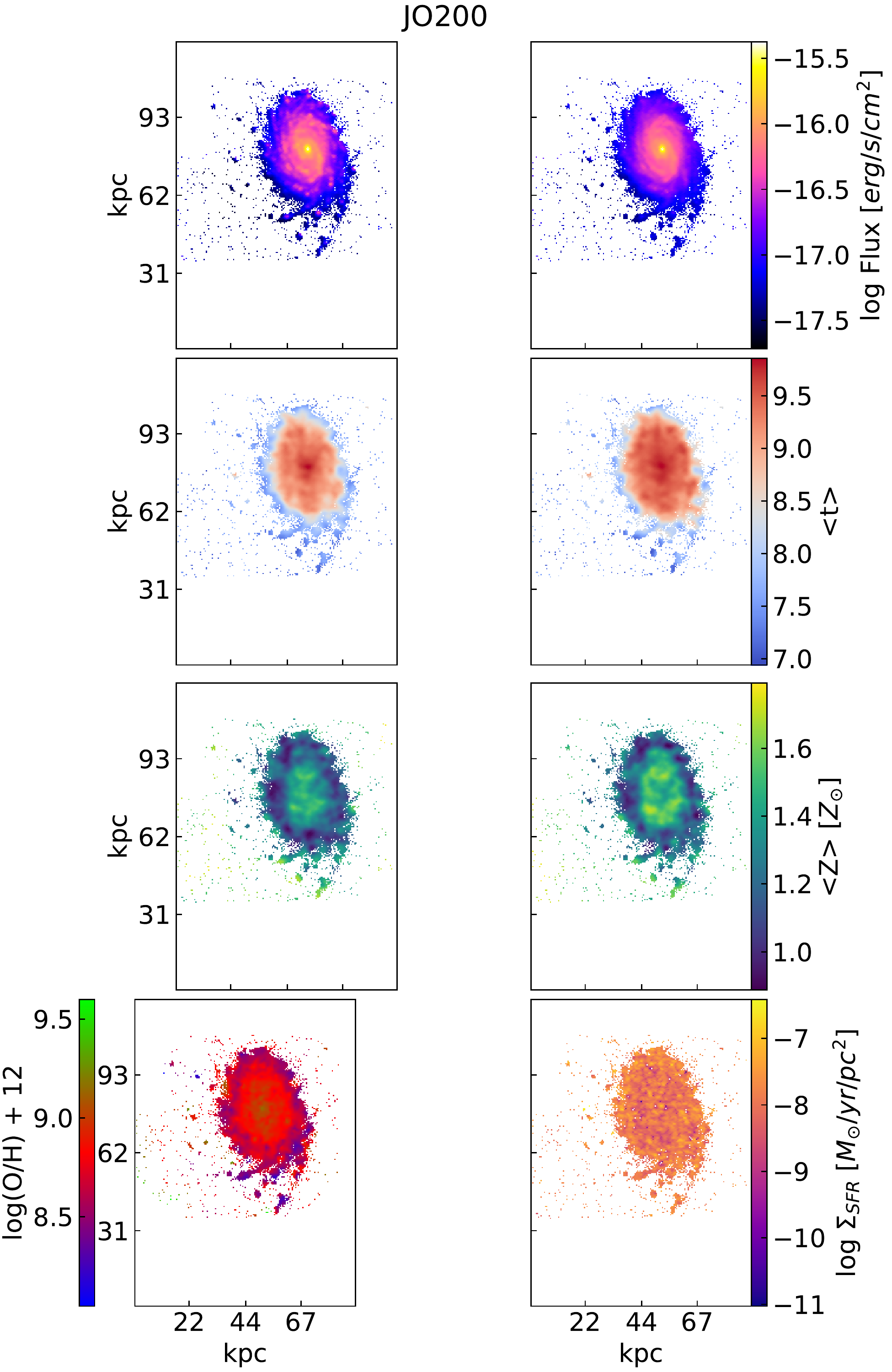}
\end{figure}

\begin{figure}
\centering
\includegraphics[width=\linewidth]{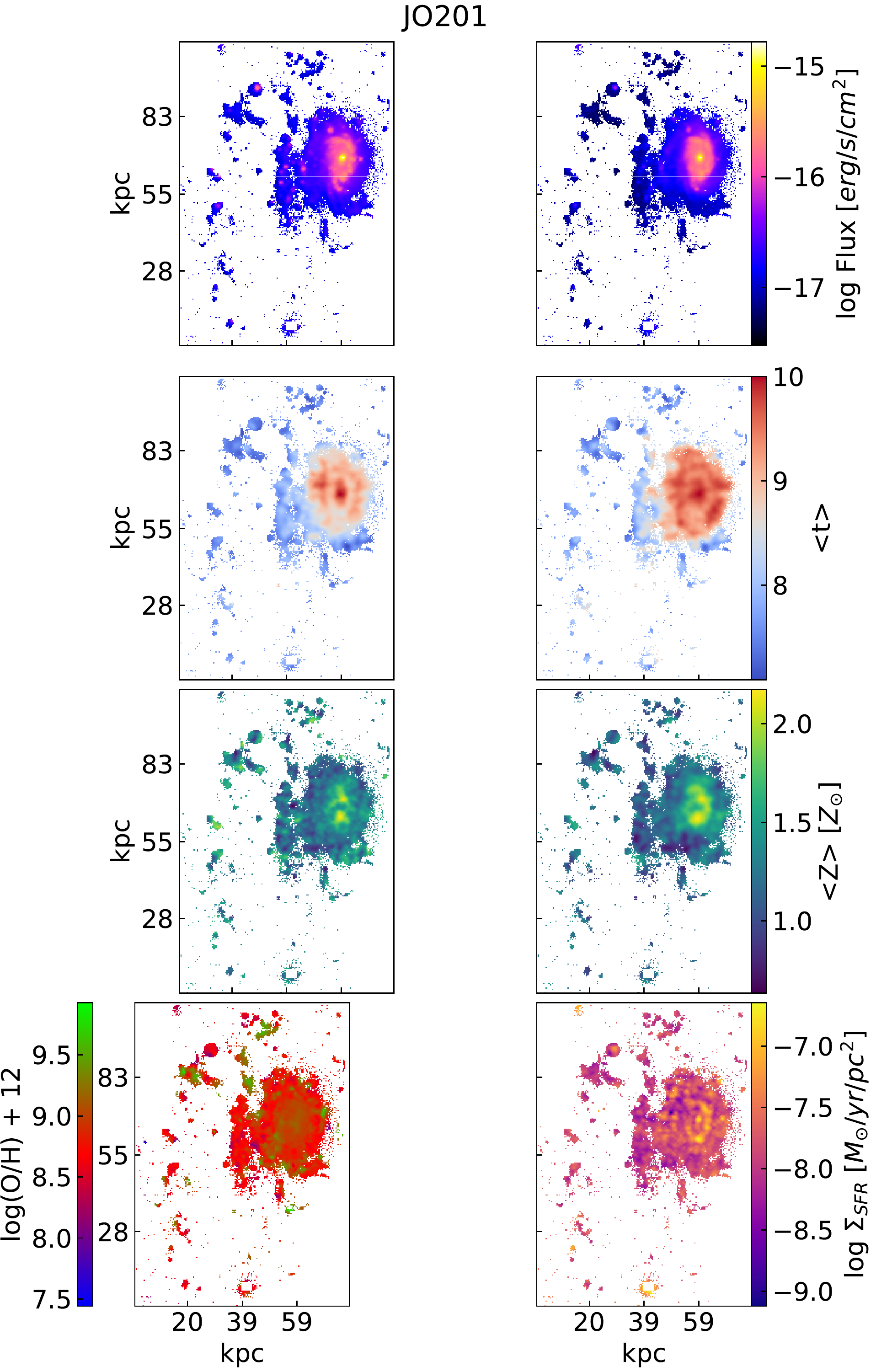}
\end{figure}

\begin{figure}
\centering
\includegraphics[width=\linewidth]{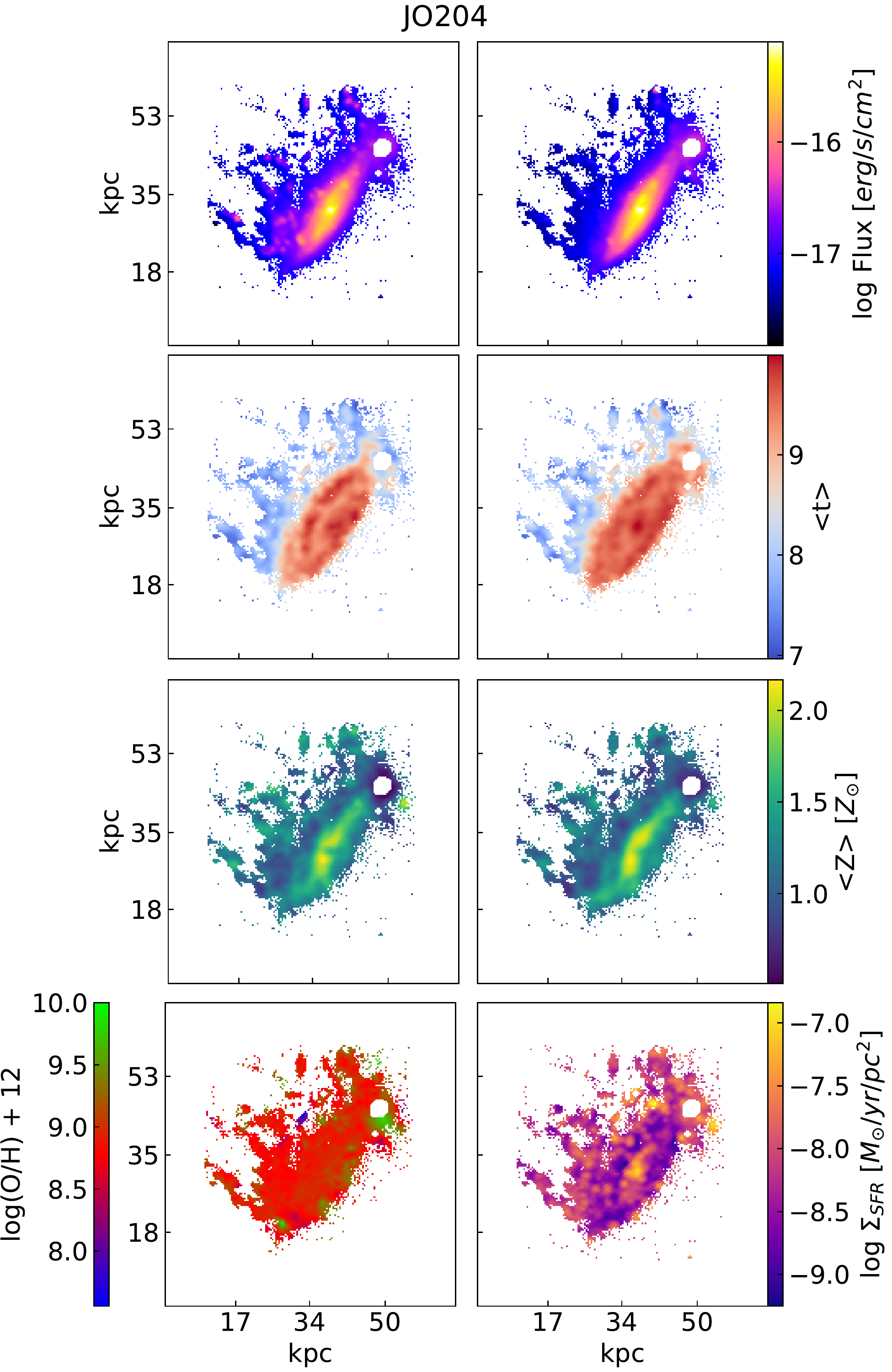}
\end{figure}

\begin{figure}
\centering
\includegraphics[width=\linewidth]{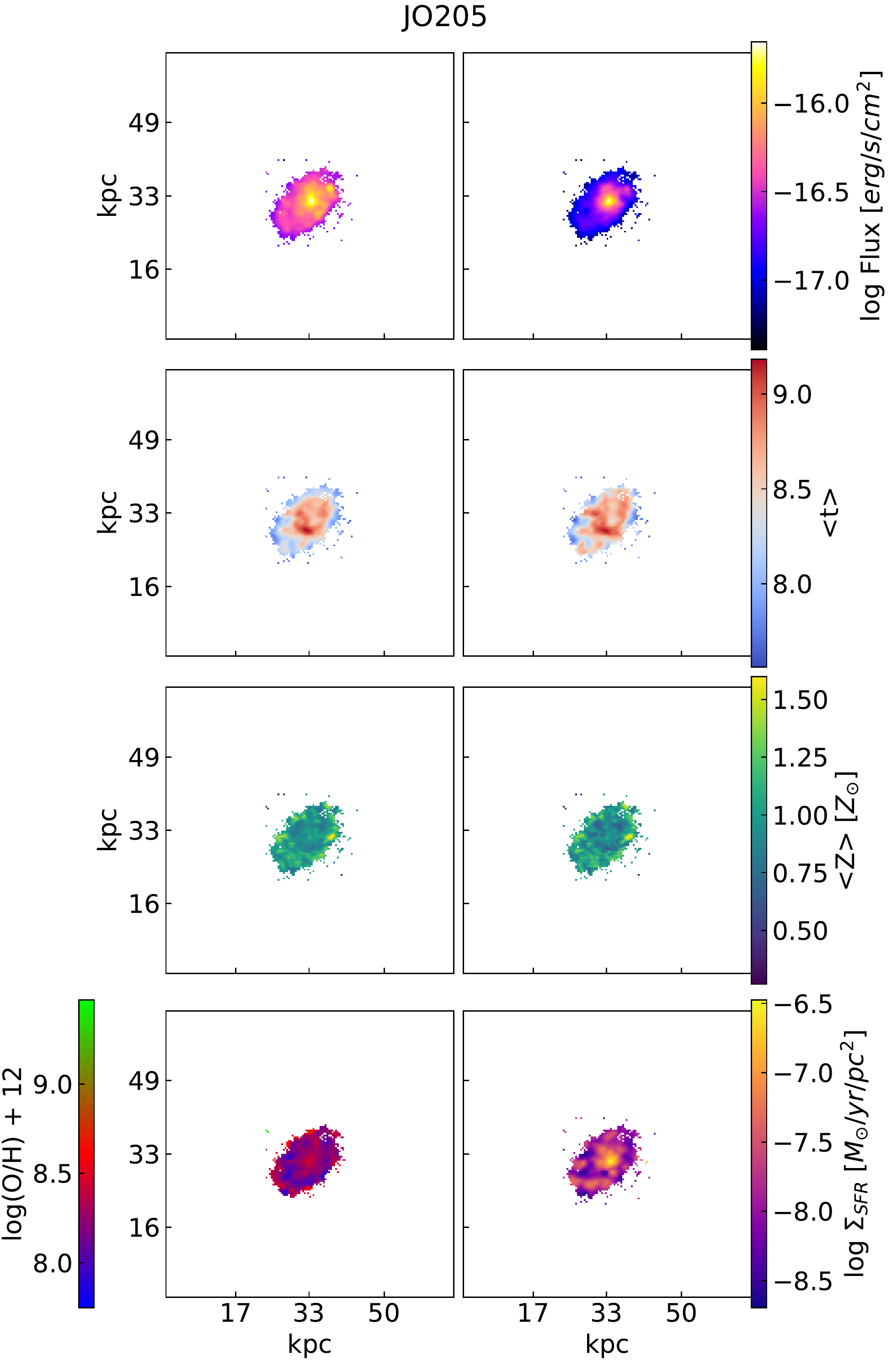}
\end{figure}

\begin{figure}
\centering
\includegraphics[width=\linewidth]{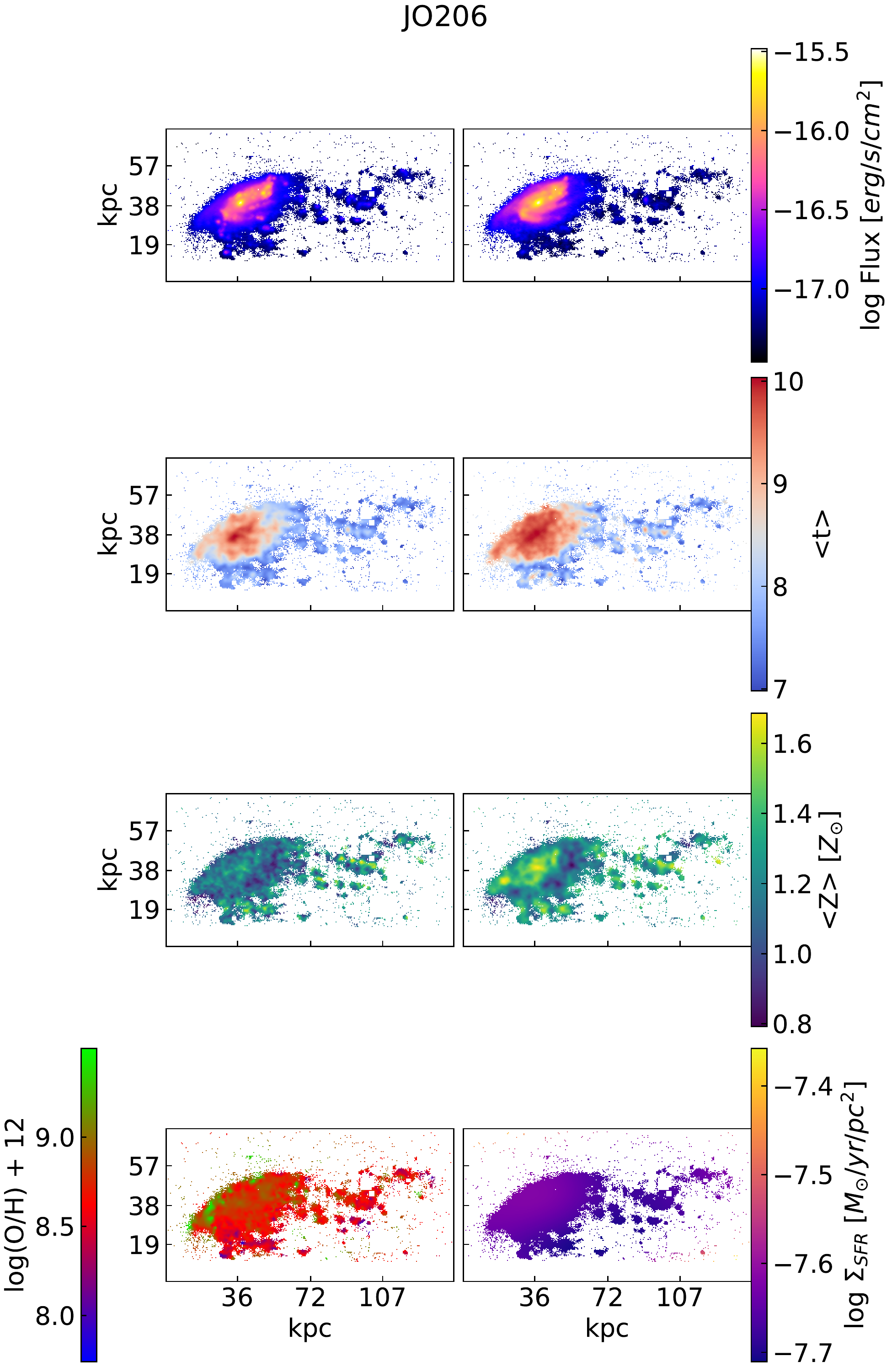}
\end{figure}

\begin{figure}
\centering
\includegraphics[width=\linewidth]{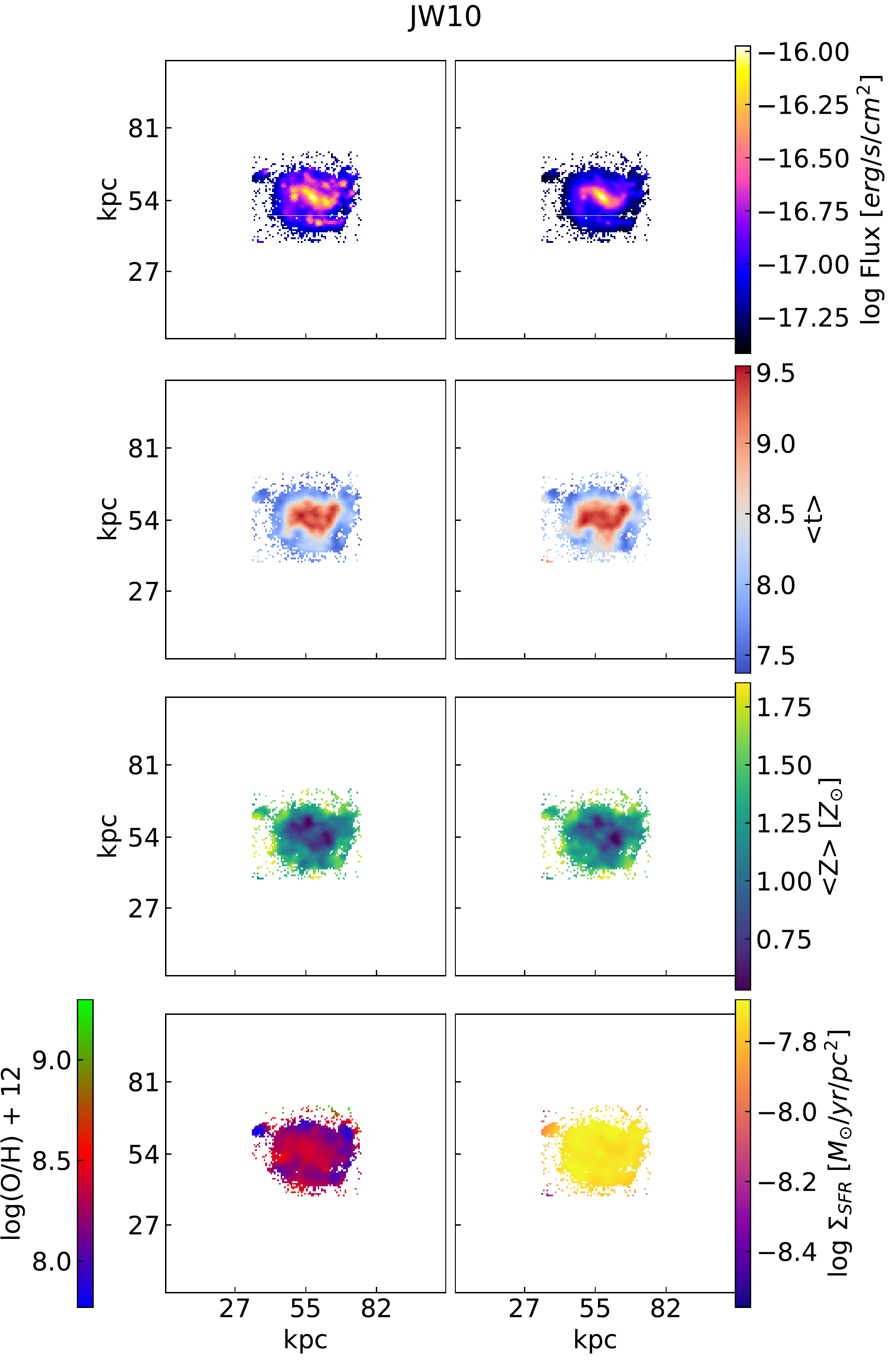}
\end{figure}

\begin{figure}
\centering
\includegraphics[width=\linewidth]{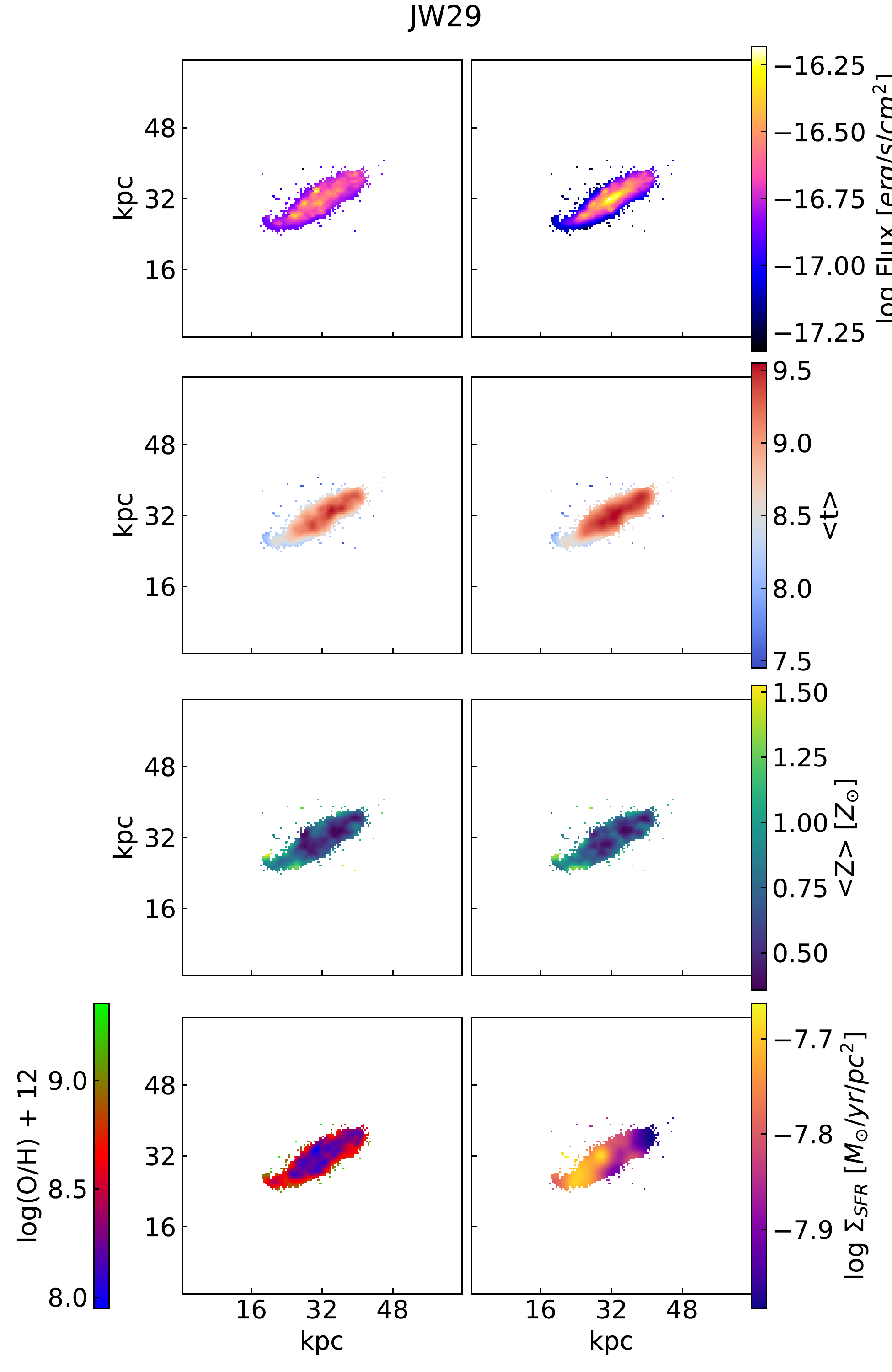}
\end{figure}

\begin{figure}
\centering
\includegraphics[width=\linewidth]{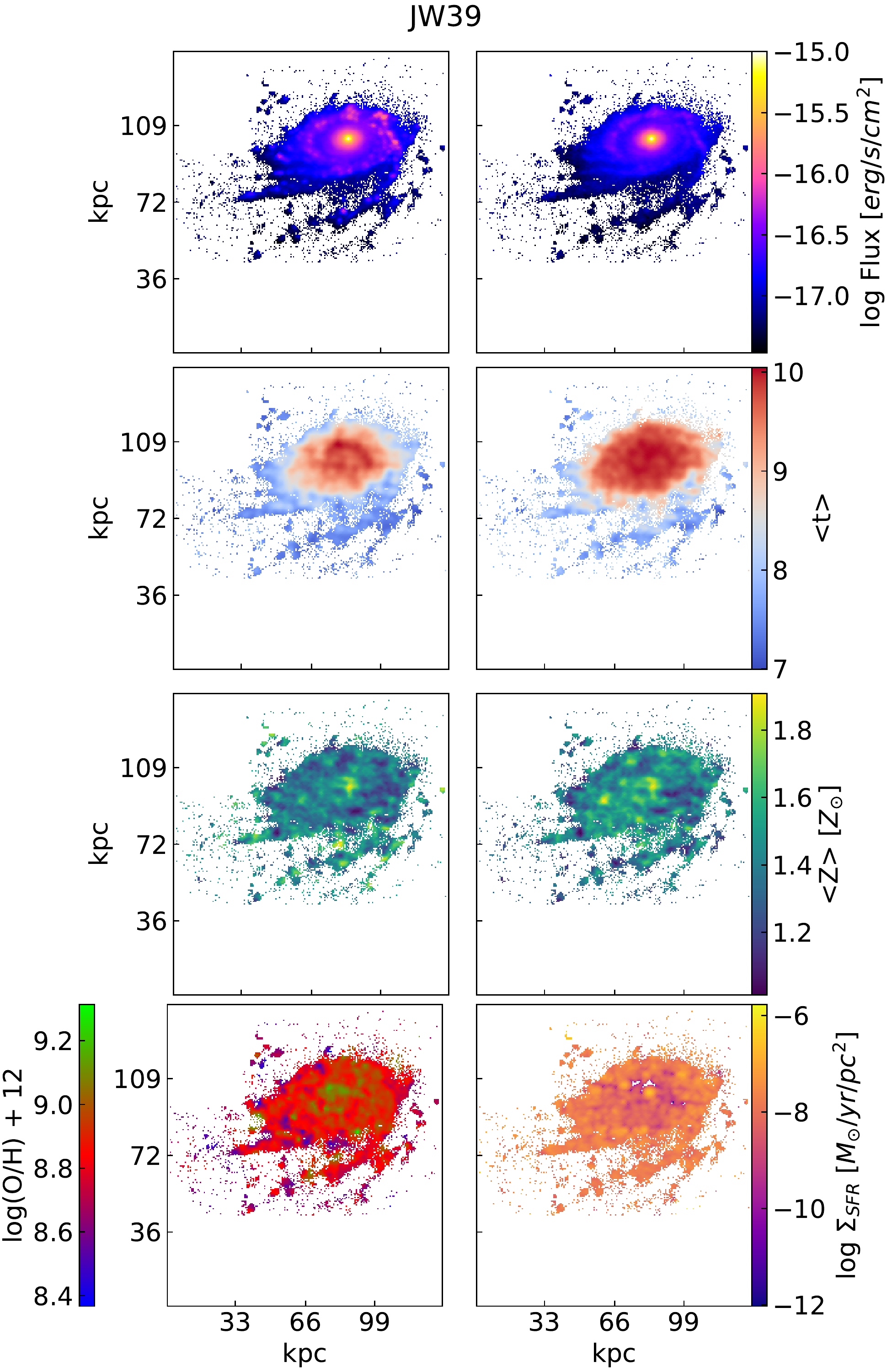}
\end{figure}

\begin{figure}
\centering
\includegraphics[width=\linewidth]{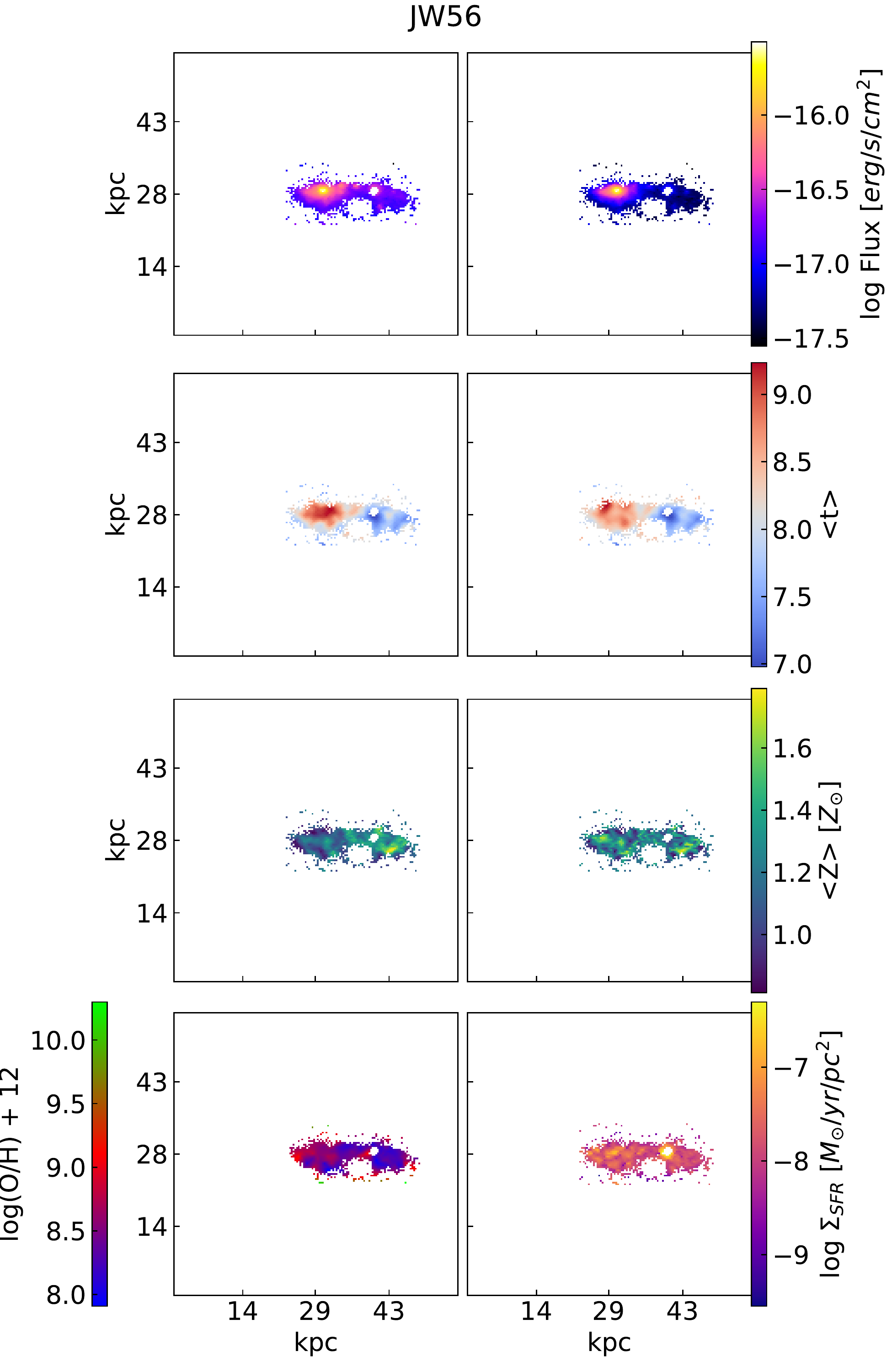}
\end{figure}

\begin{figure}
\centering
\includegraphics[width=\linewidth]{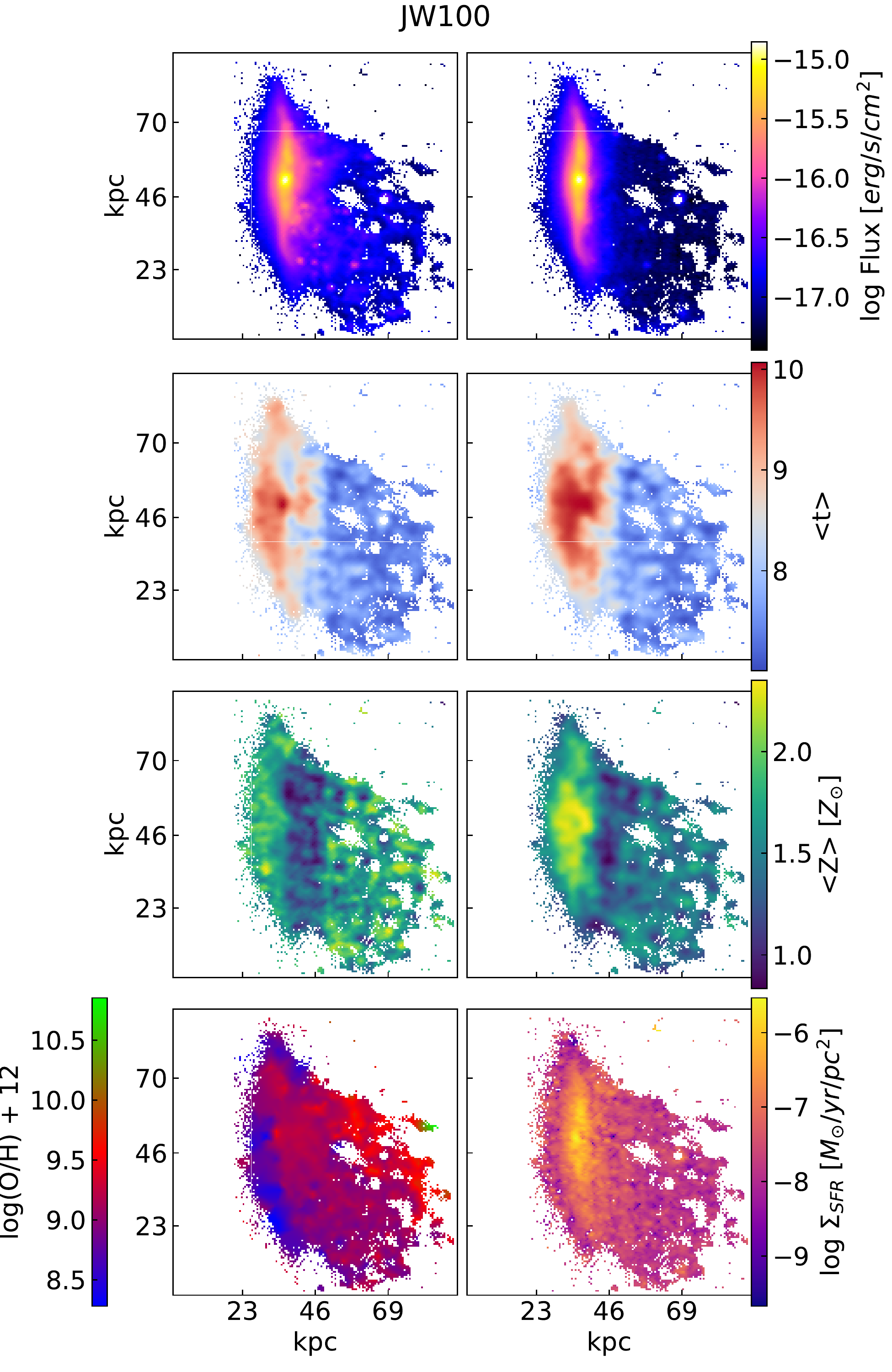}
\end{figure}
\begin{figure}
\centering
\includegraphics[width=\linewidth]{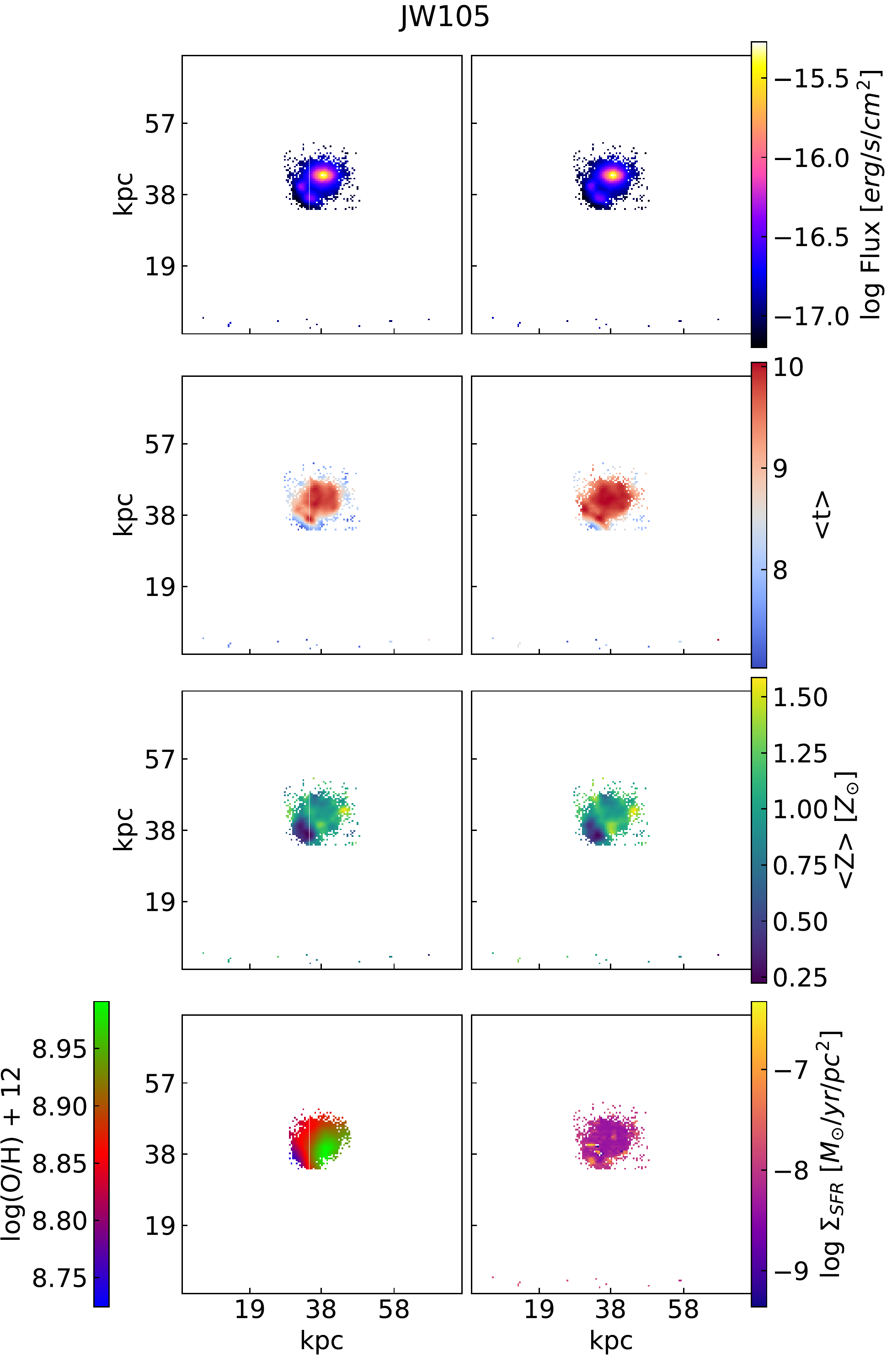}
\end{figure}

\begin{figure}
\centering
\includegraphics[width=\linewidth]{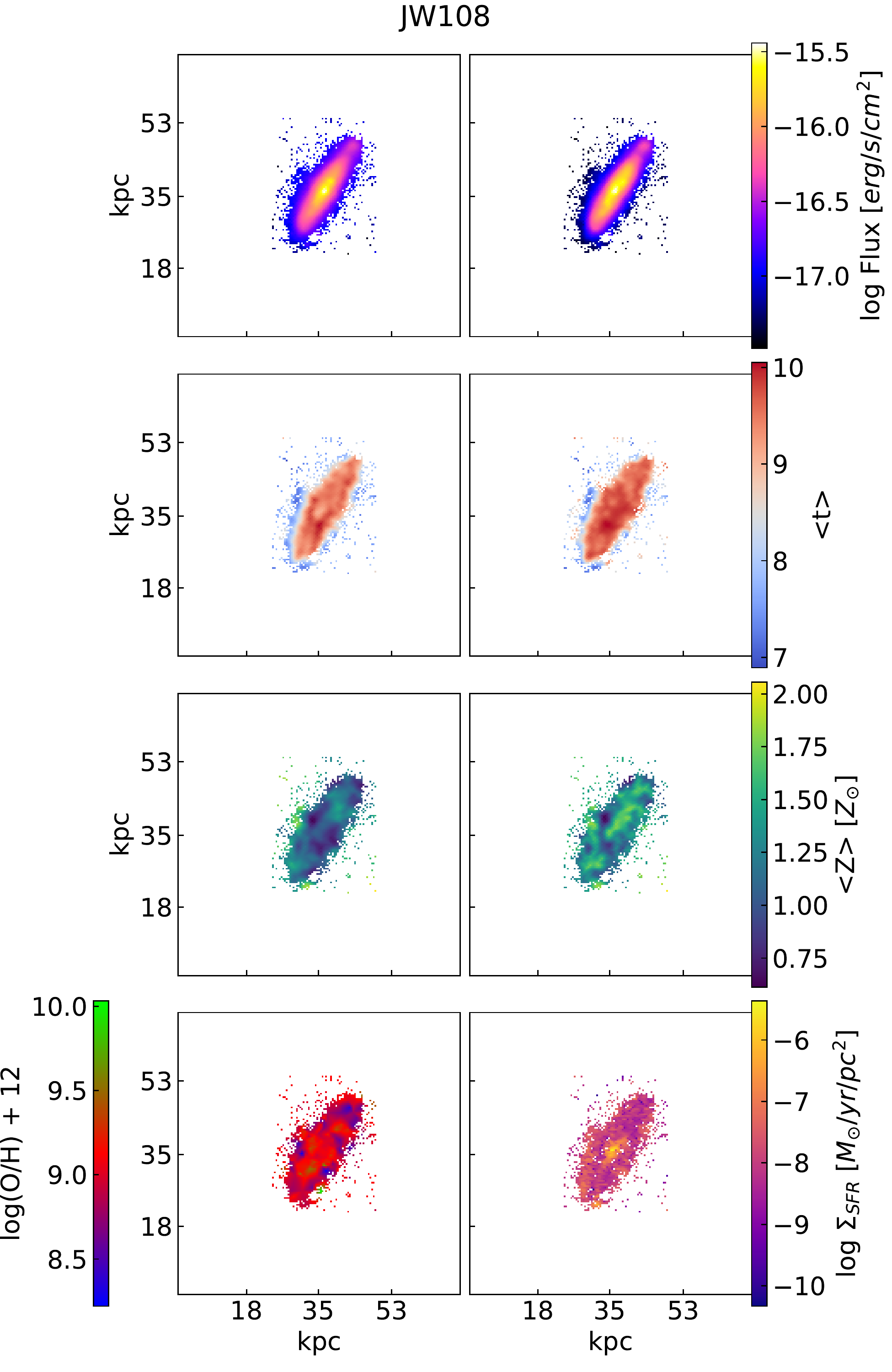}
\end{figure}

\begin{figure}
\centering
\includegraphics[width=\linewidth]{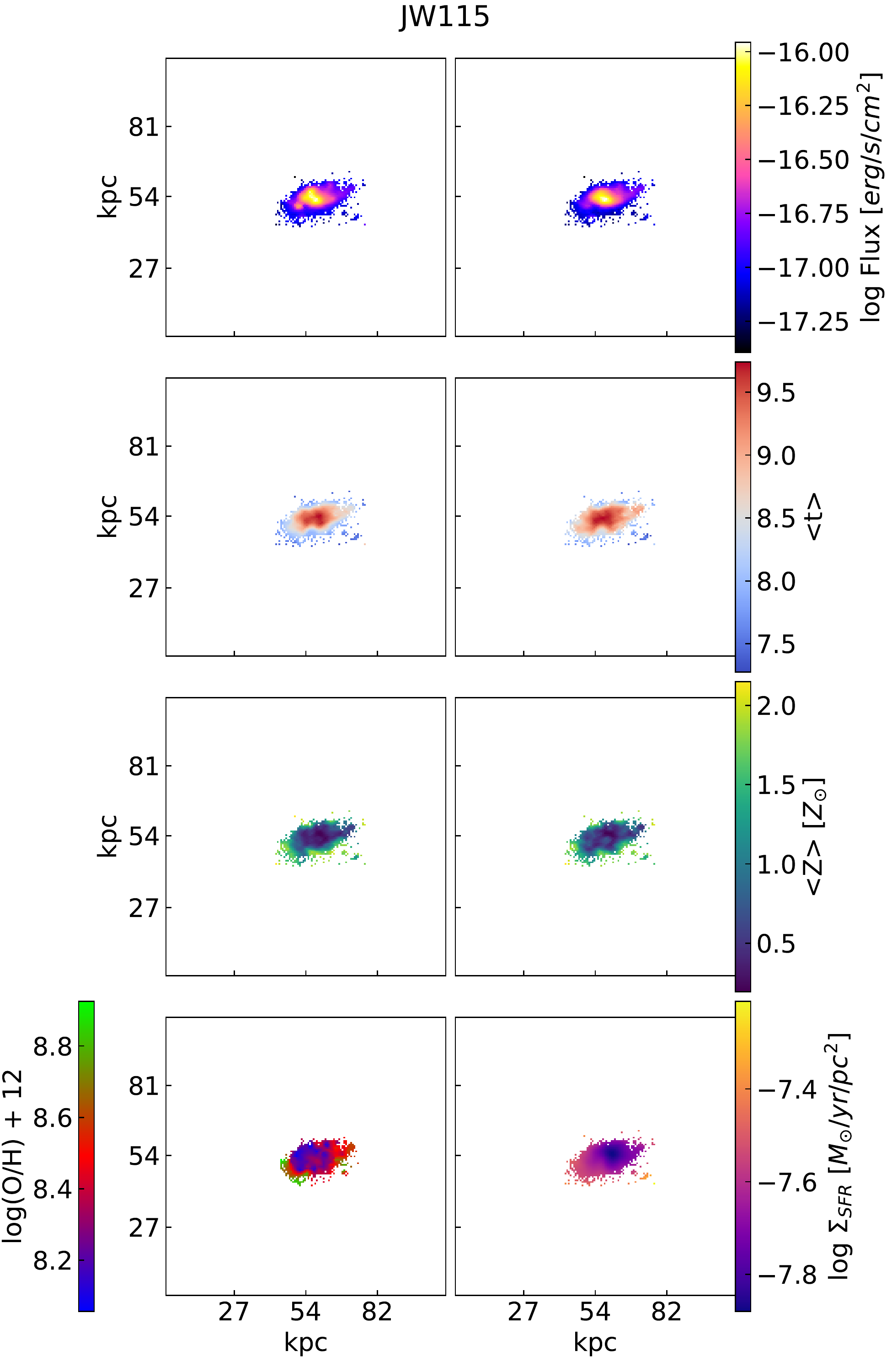}
\end{figure}

\begin{figure}
\centering
\includegraphics[width=\linewidth]{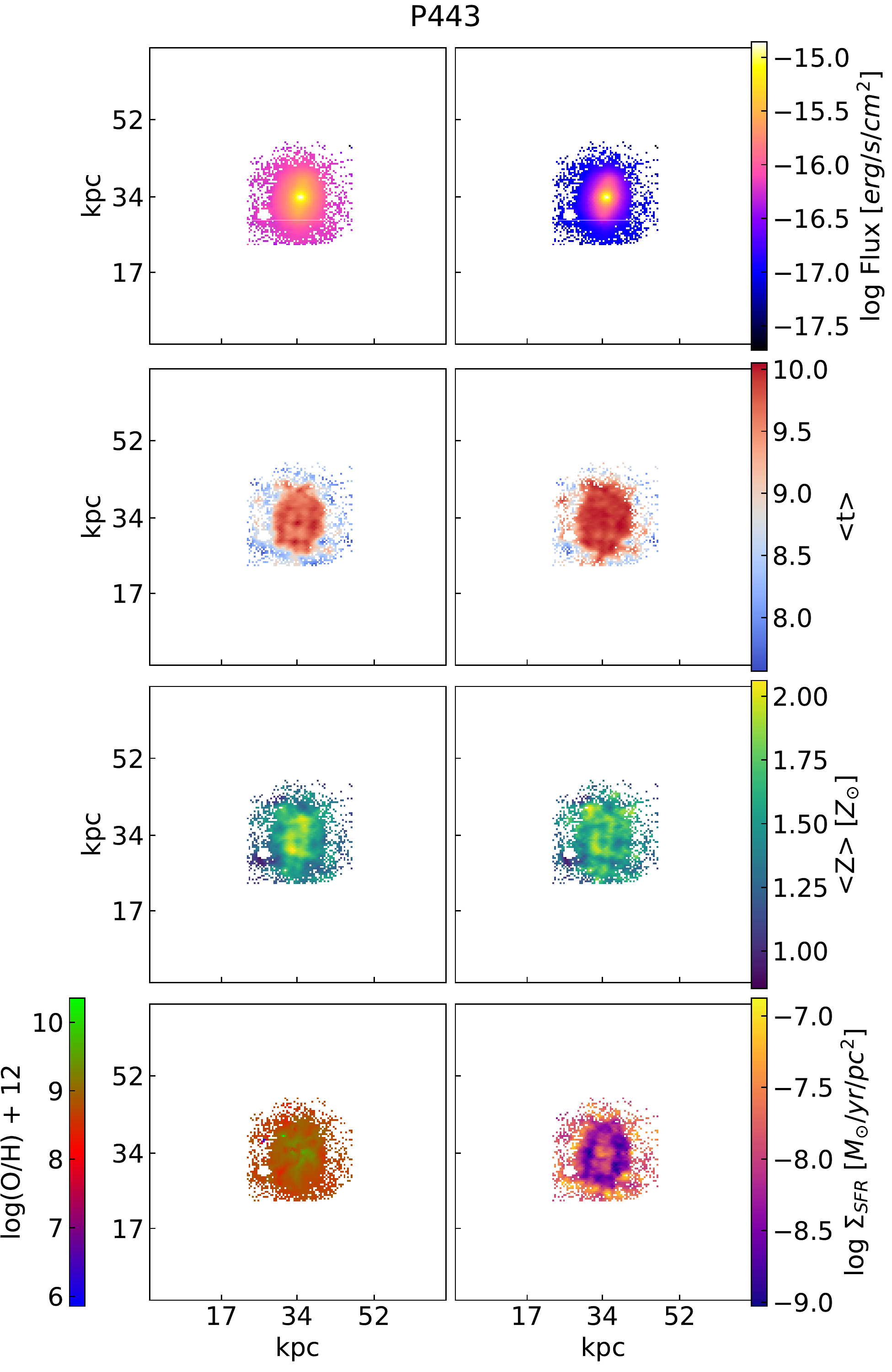}
\end{figure}

\begin{figure}
\centering
\includegraphics[width=\linewidth]{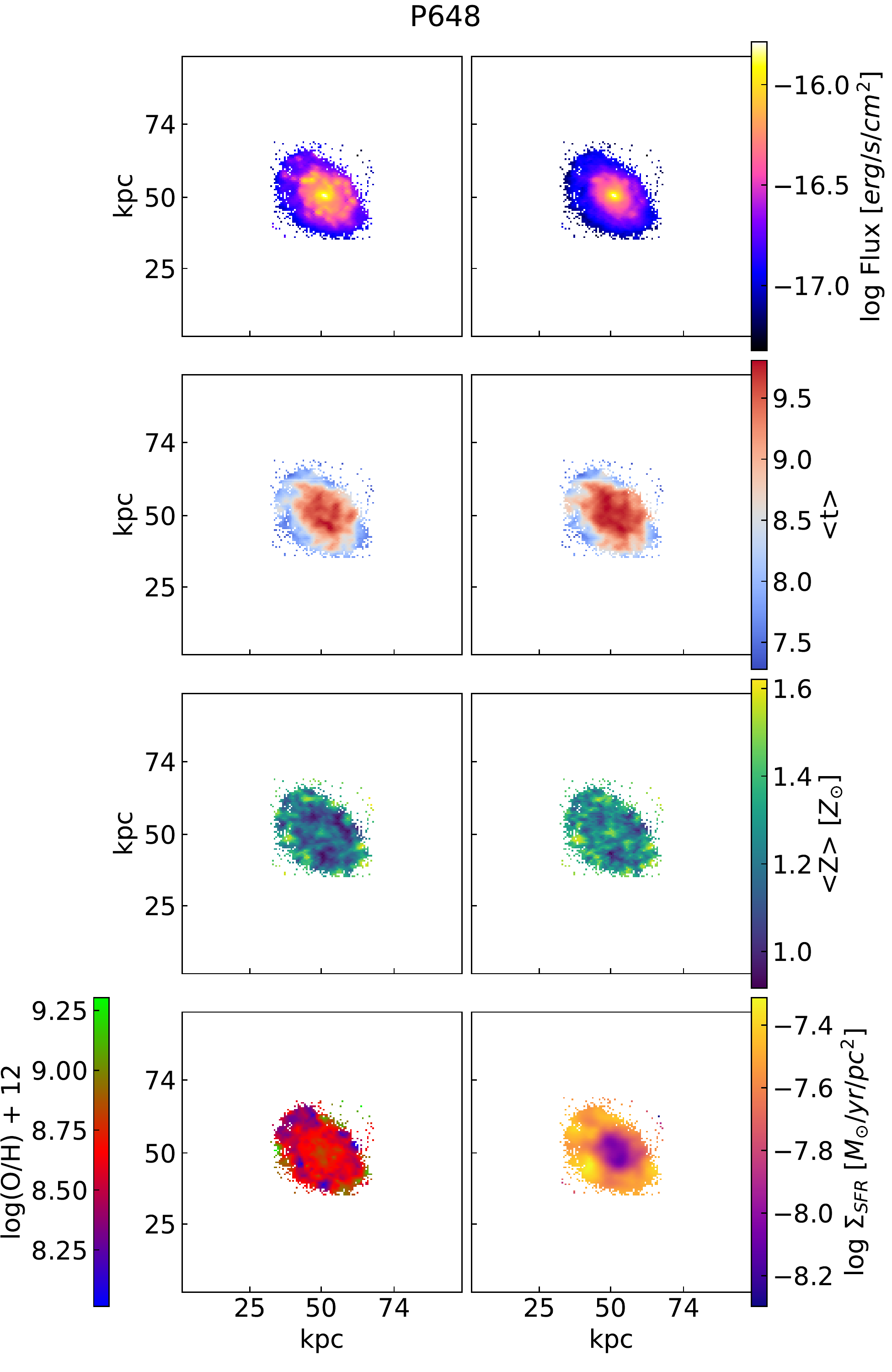}
\end{figure}

\begin{figure}
\centering
\includegraphics[width=\linewidth]{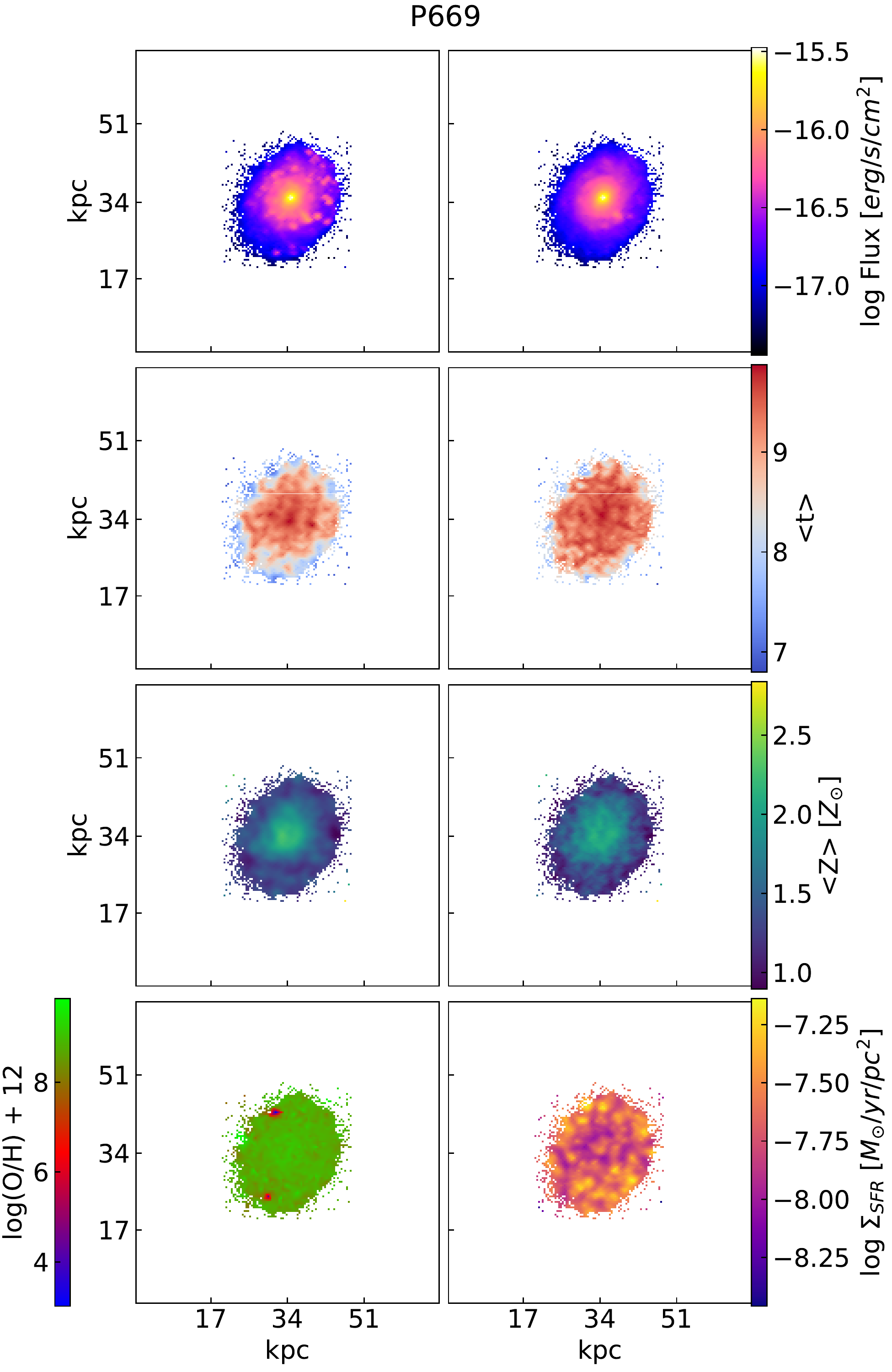}
\end{figure}

\begin{figure}
\centering
\includegraphics[width=\linewidth]{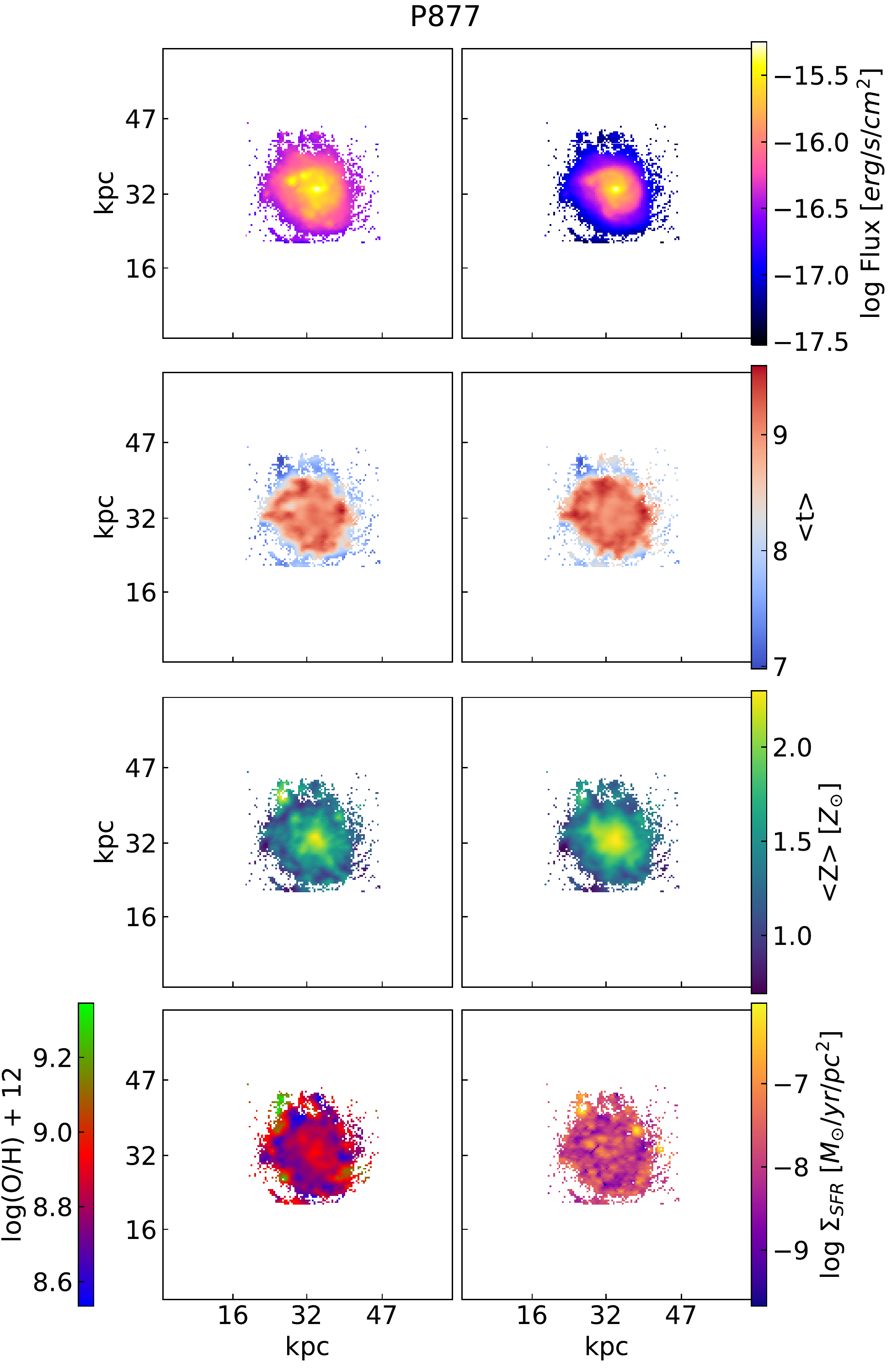}
\end{figure}

\begin{figure}
\centering
\includegraphics[width=\linewidth]{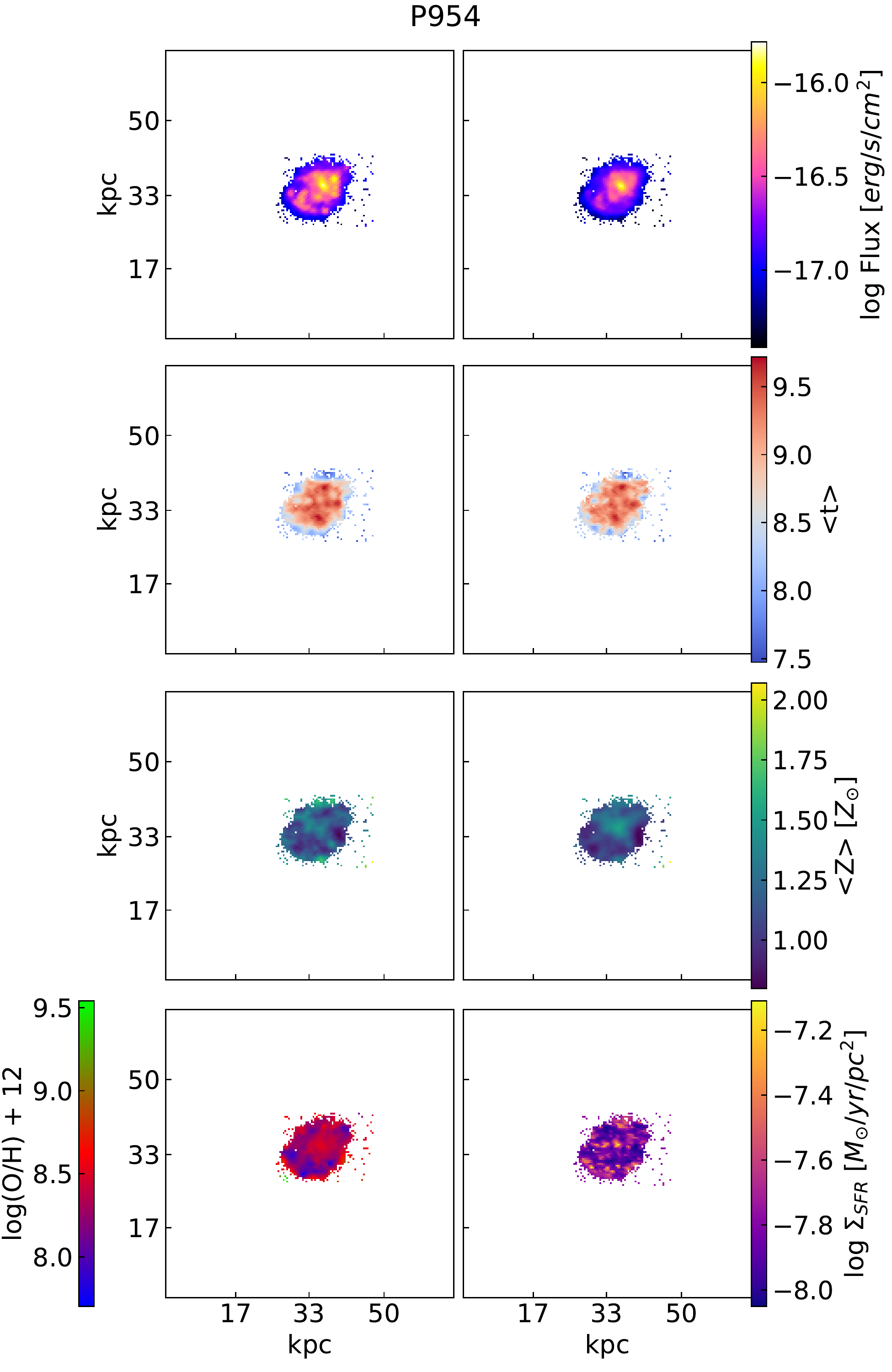}
\end{figure}

\begin{figure}
\centering
\includegraphics[width=\linewidth]{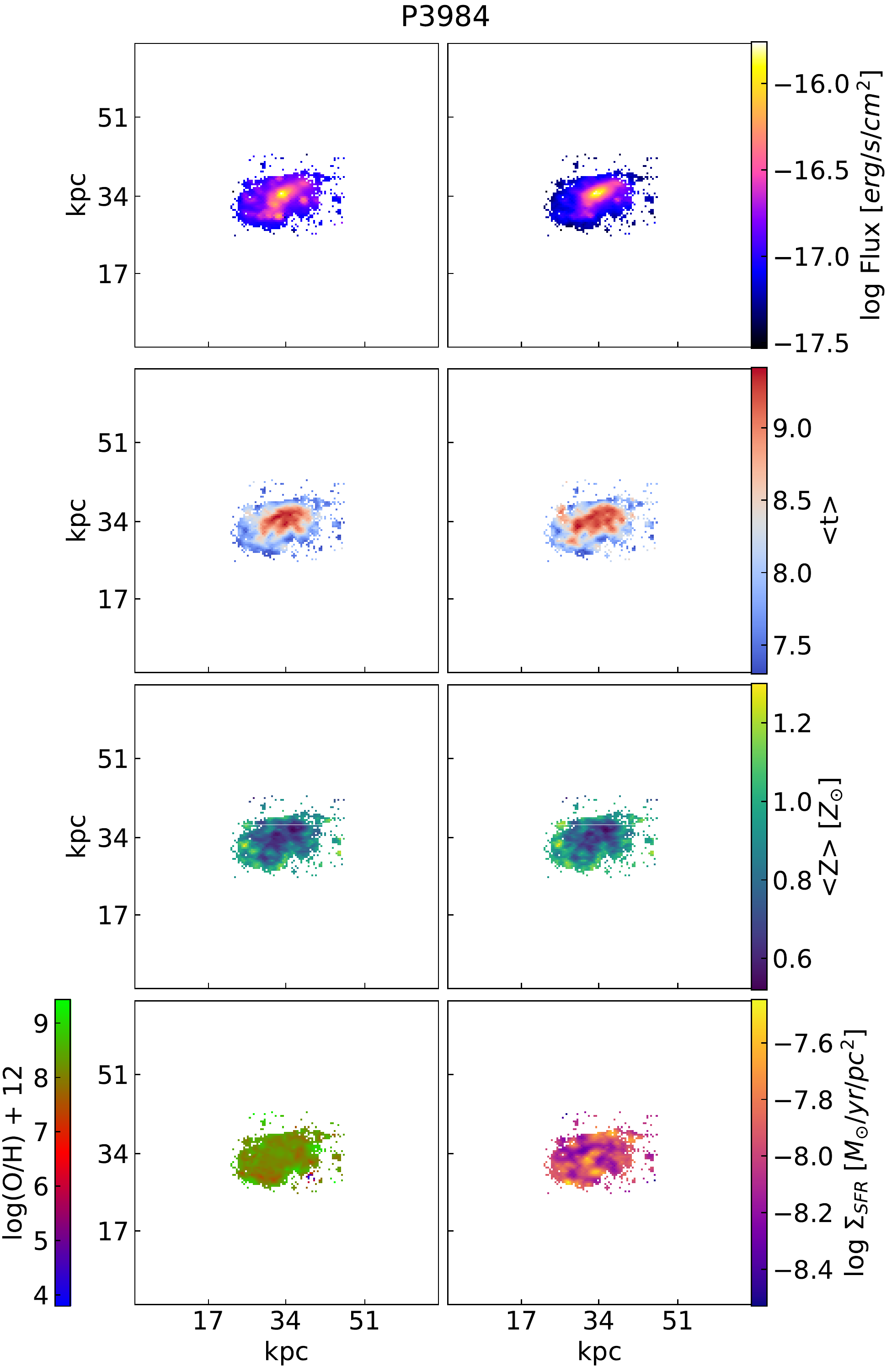}
\end{figure}

\begin{figure}
\centering
\includegraphics[width=\linewidth]{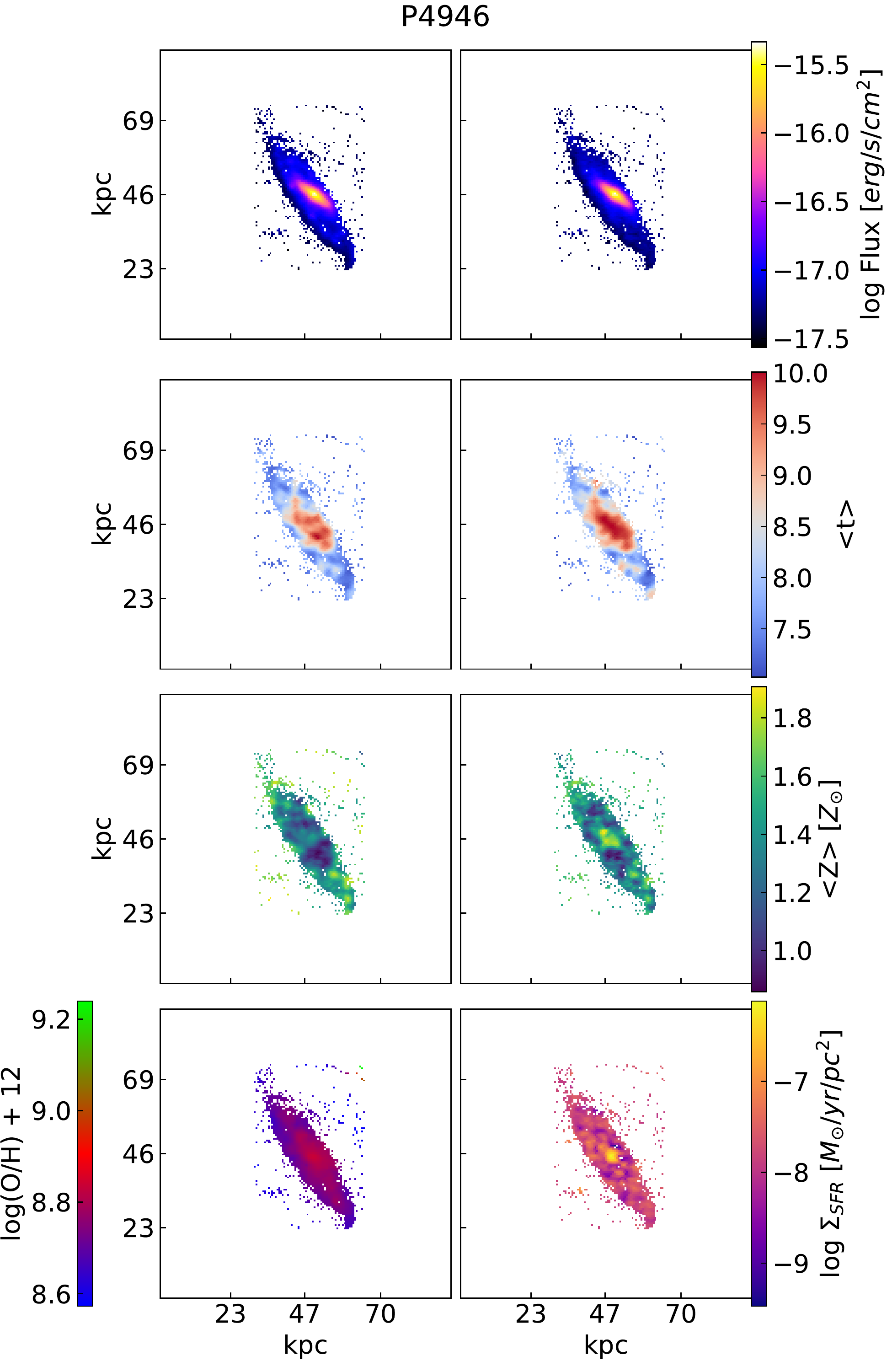}
\end{figure}

\begin{figure}
\centering
\includegraphics[width=\linewidth]{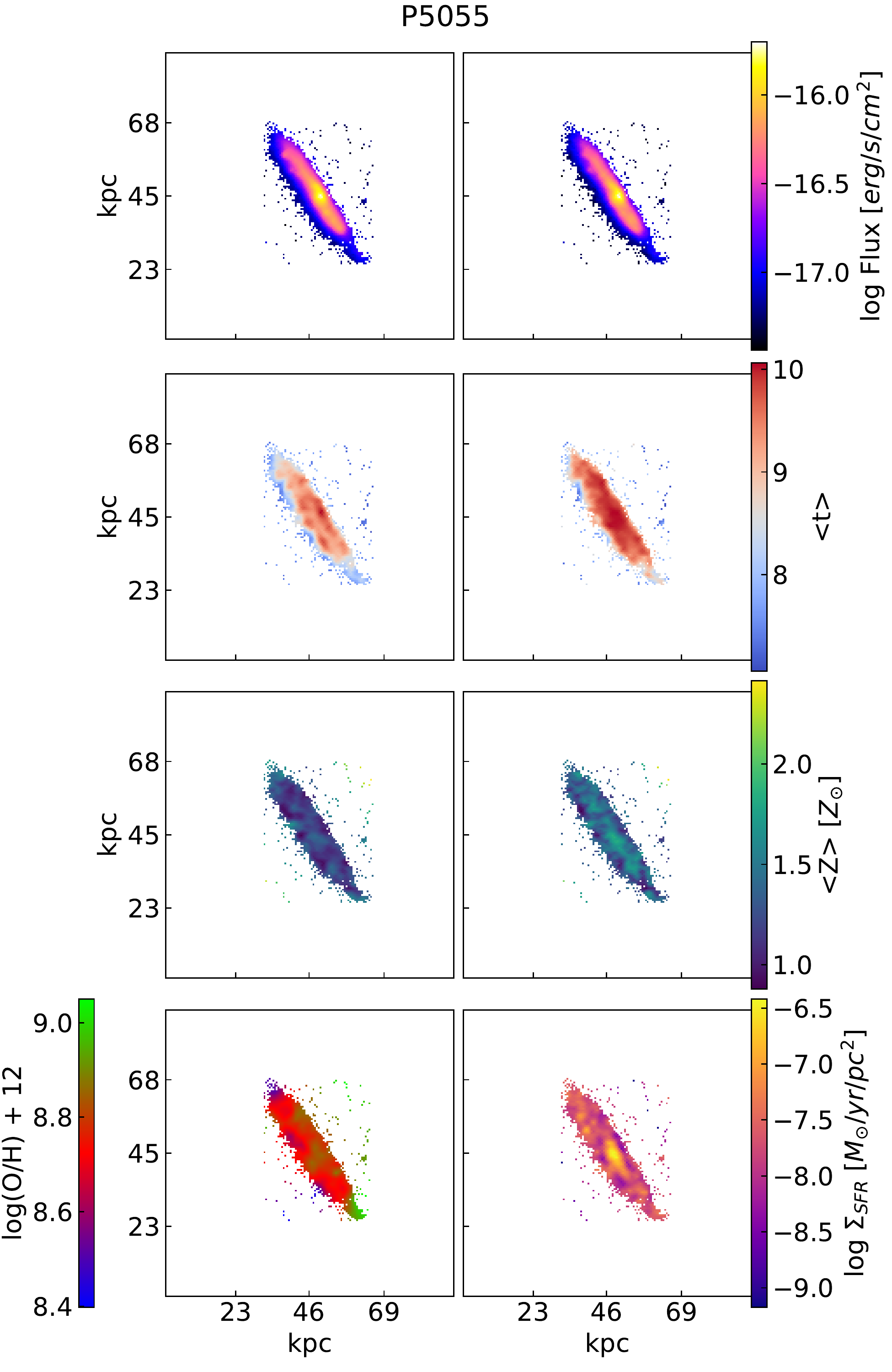}
\end{figure}

\begin{figure}
\centering
\includegraphics[width=\linewidth]{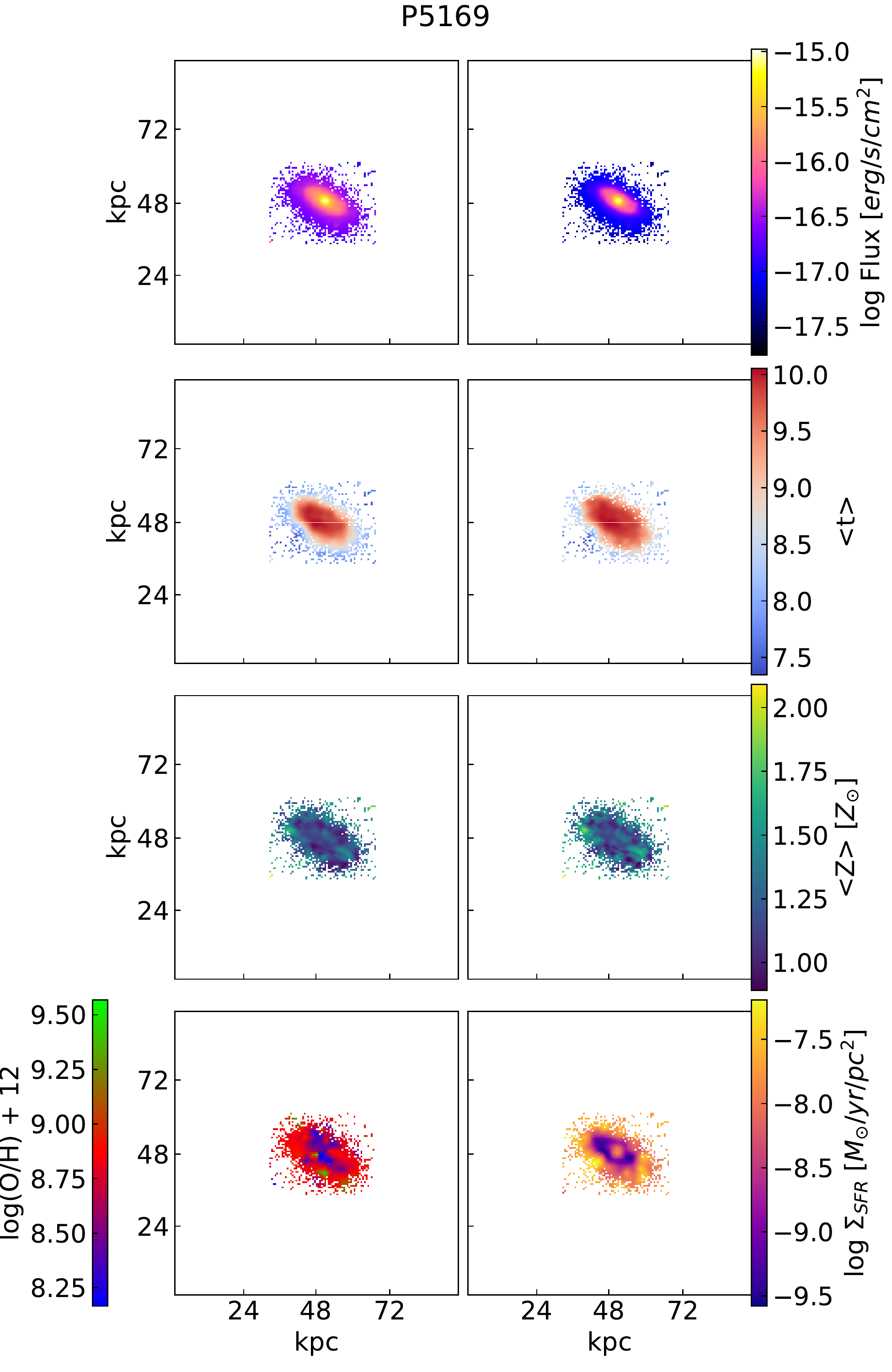}
\end{figure}
\begin{figure}
\centering
\includegraphics[width=\linewidth]{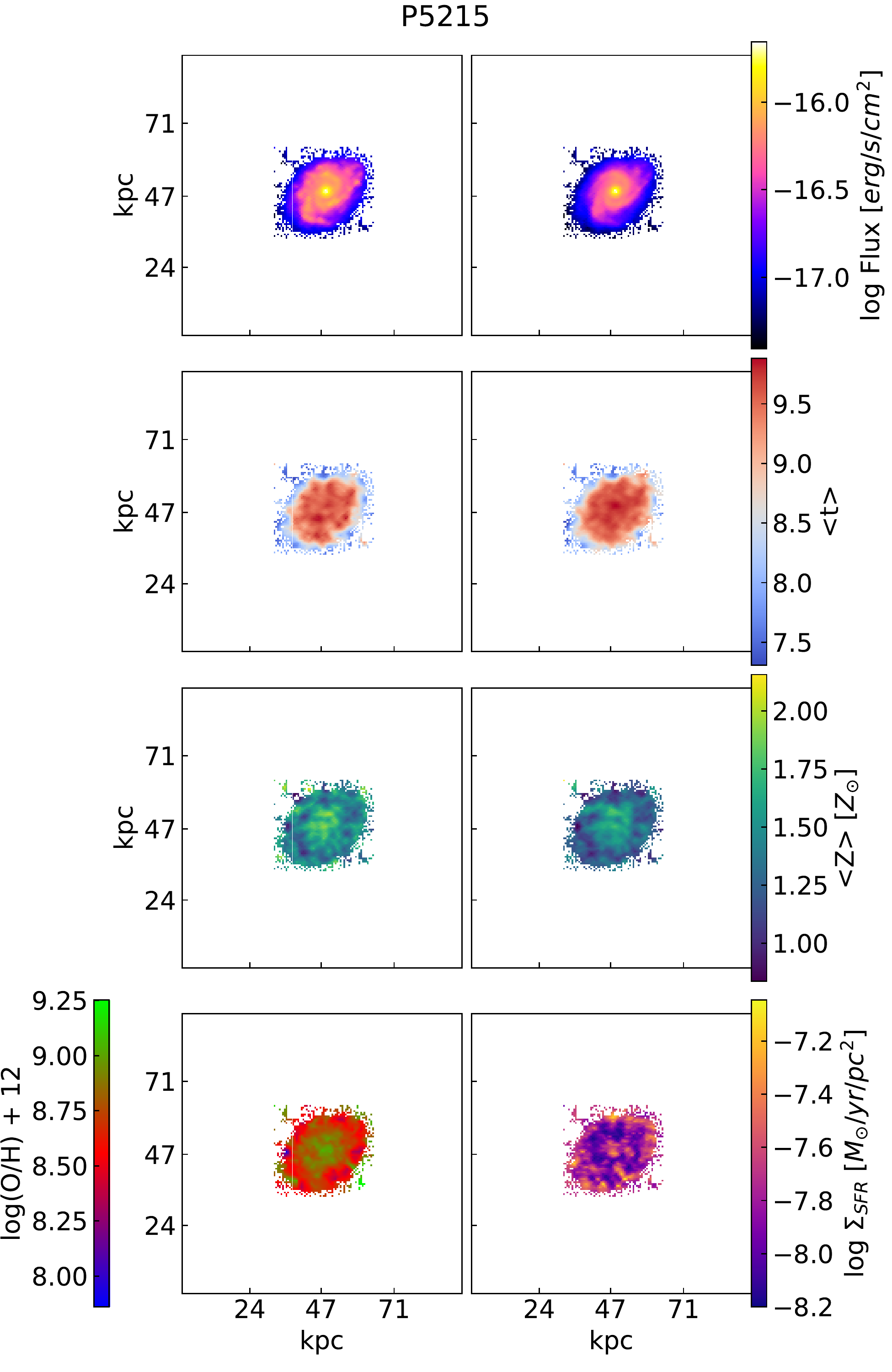}
\end{figure}

\begin{figure}
\centering
\includegraphics[width=\linewidth]{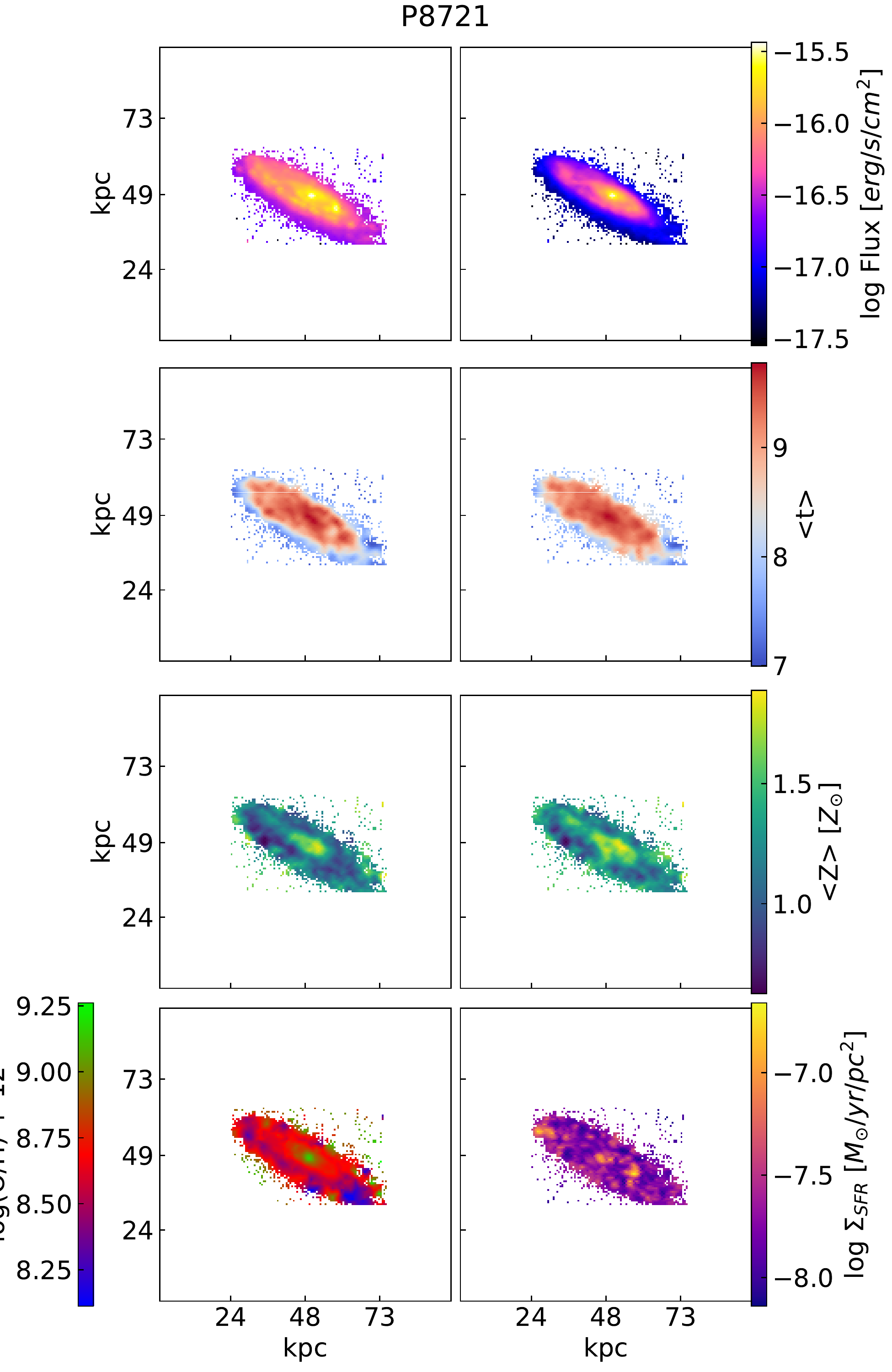}
\end{figure}

\begin{figure}
\centering
\includegraphics[width=\linewidth]{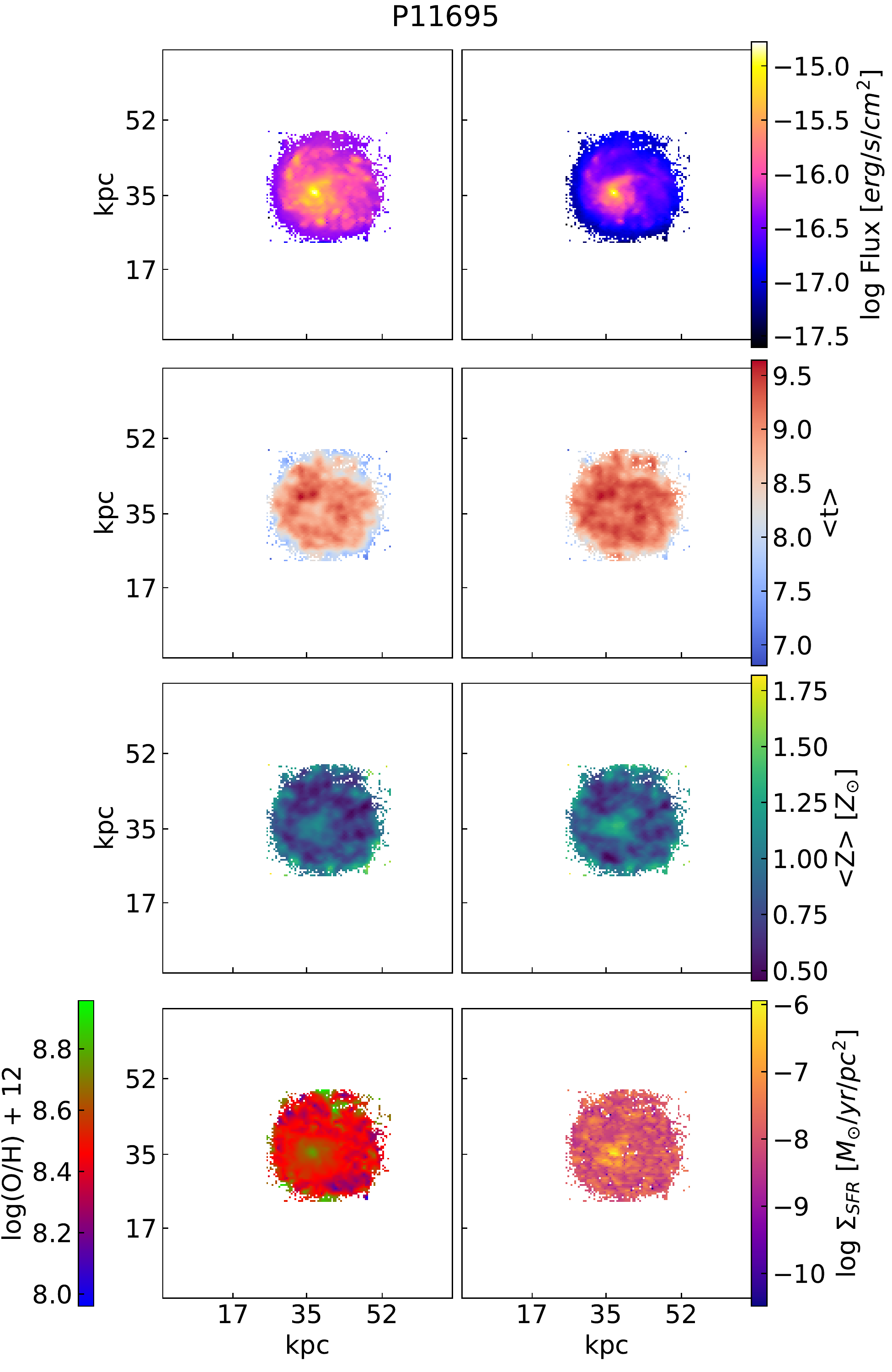}
\end{figure}

\begin{figure}
\centering
\includegraphics[width=\linewidth]{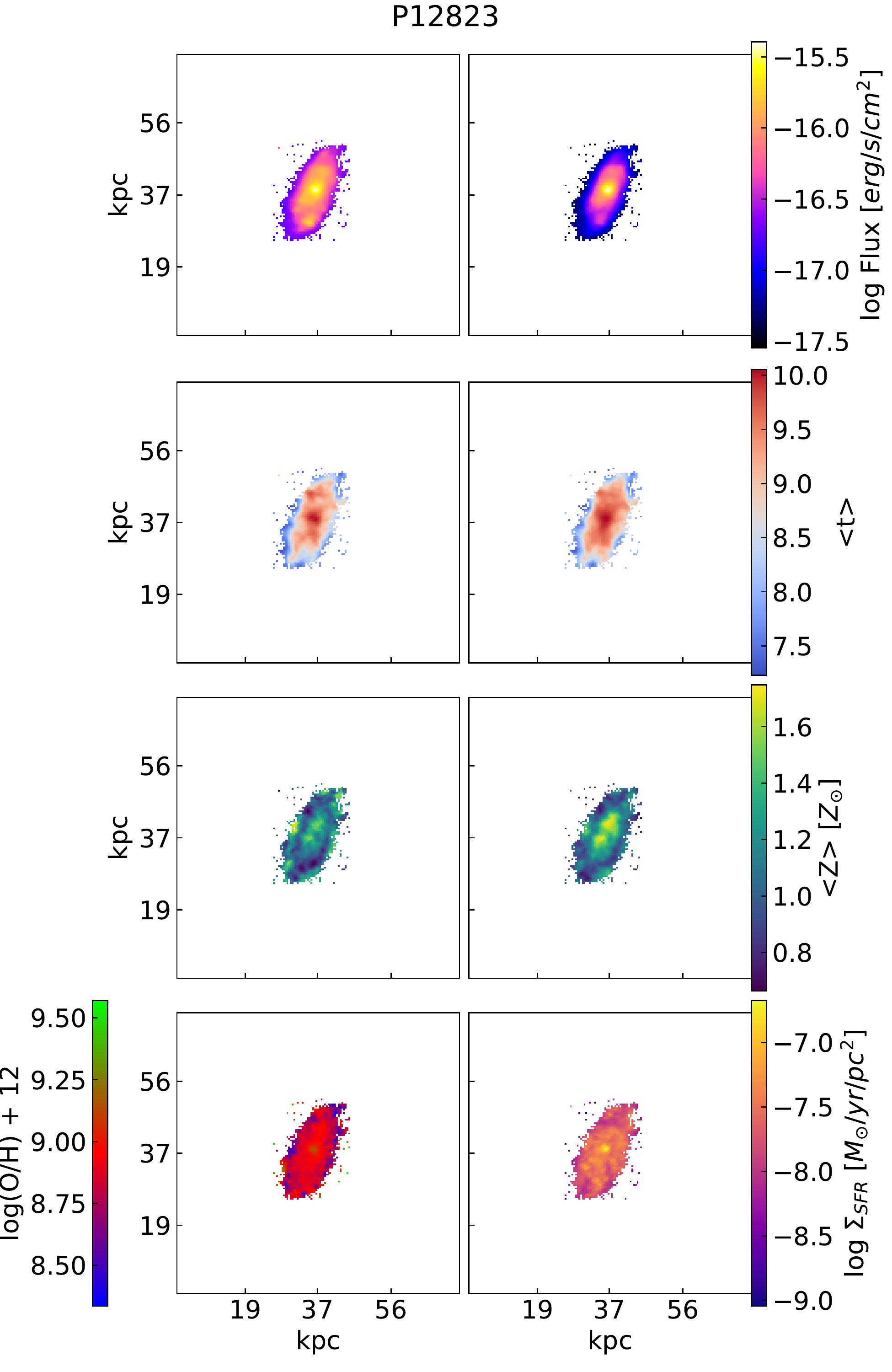}
\end{figure}

\begin{figure}
\centering
\includegraphics[width=\linewidth]{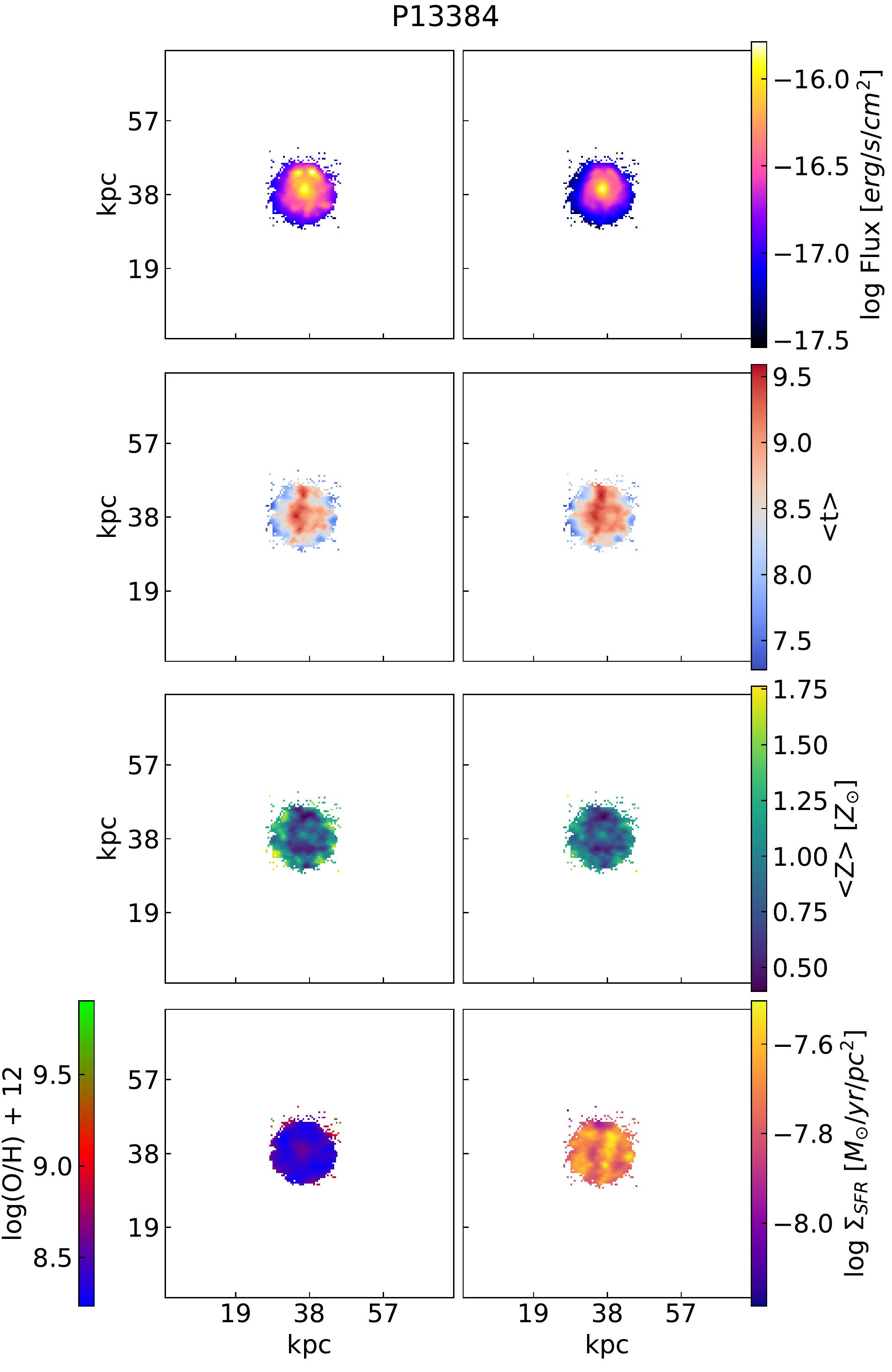}
\end{figure}

\begin{figure}
\centering
\includegraphics[width=\linewidth]{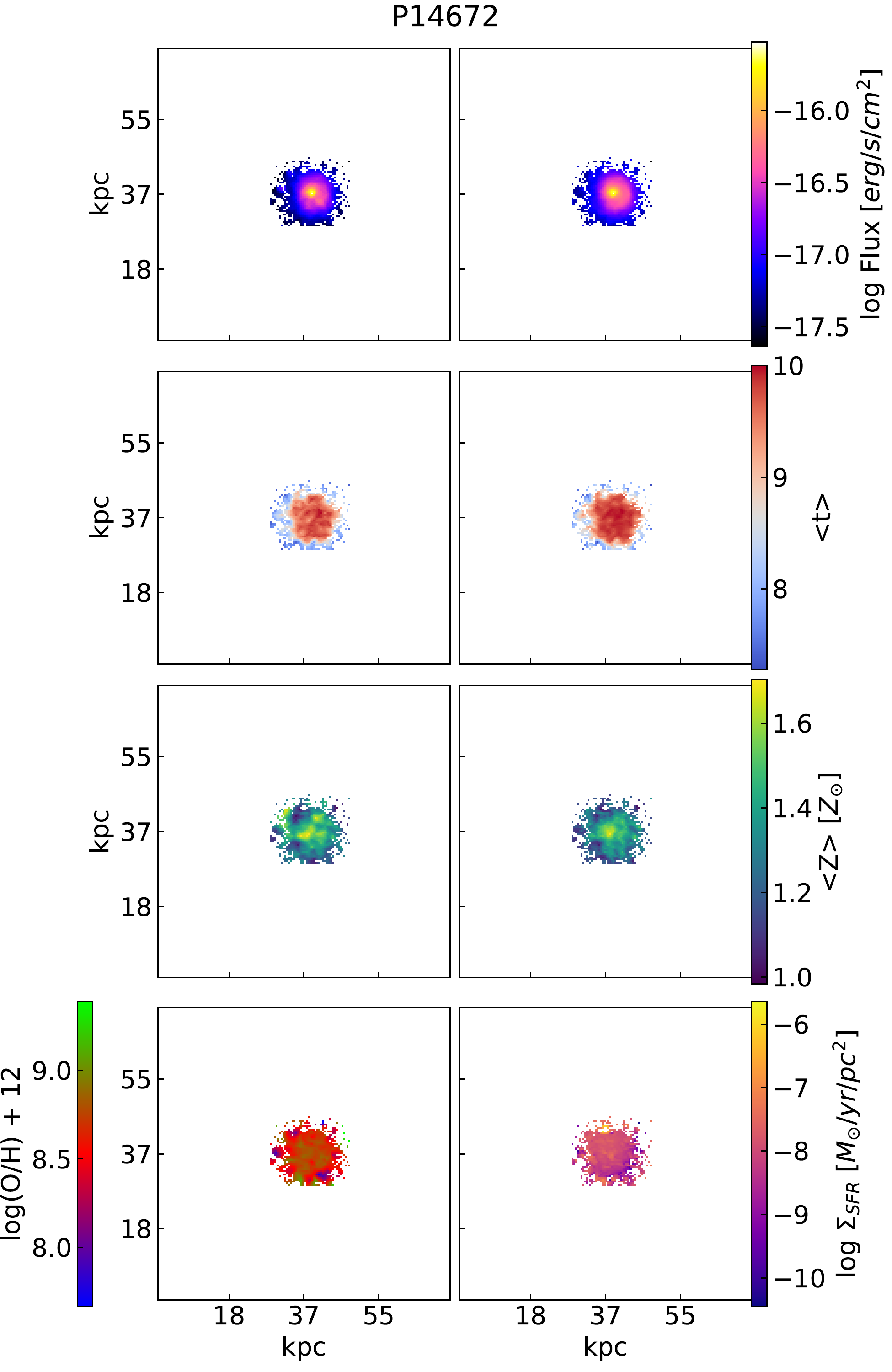}
\end{figure}

\begin{figure}
\centering
\includegraphics[width=\linewidth]{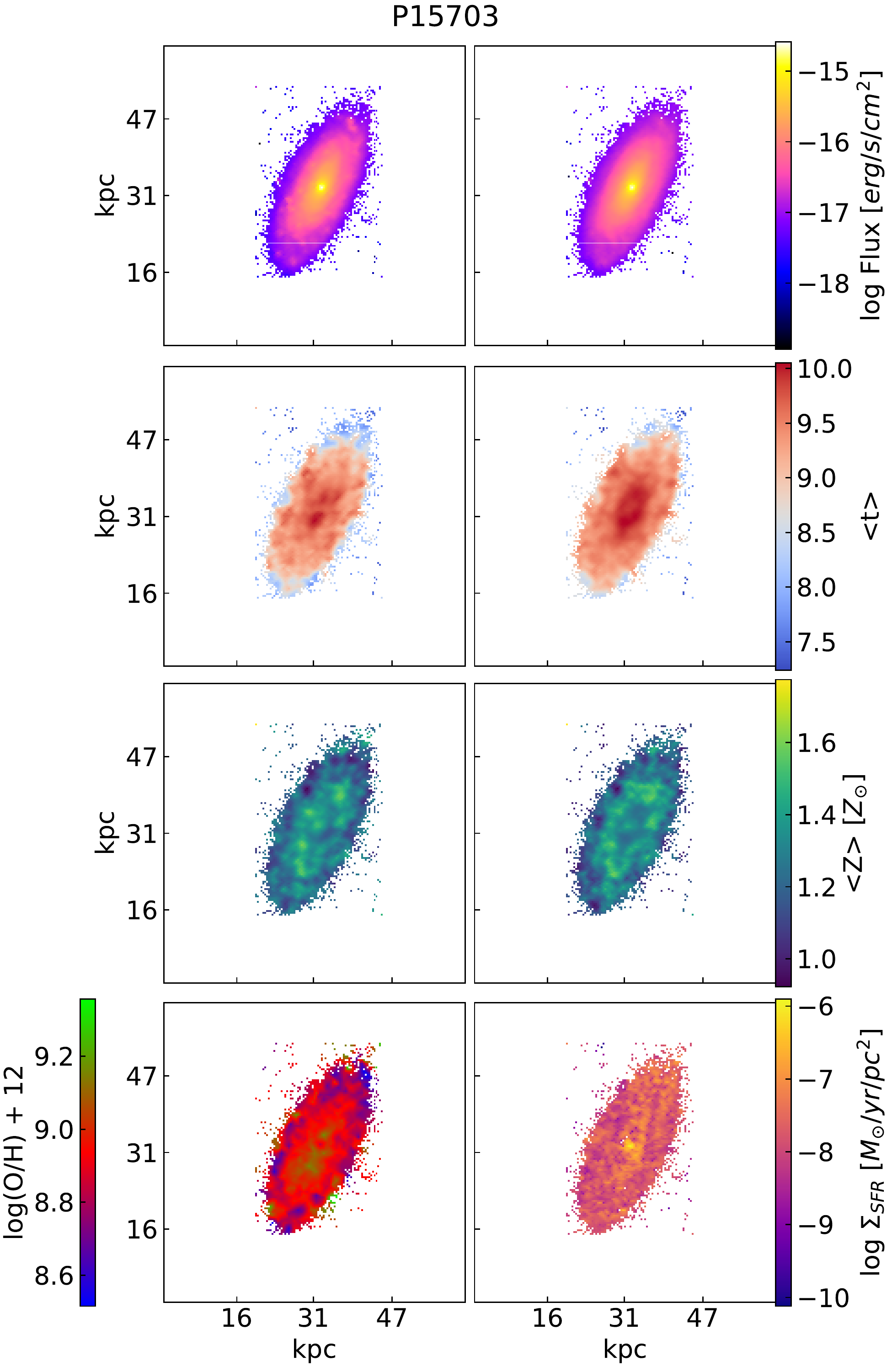}
\end{figure}

\begin{figure}
\centering
\includegraphics[width=\linewidth]{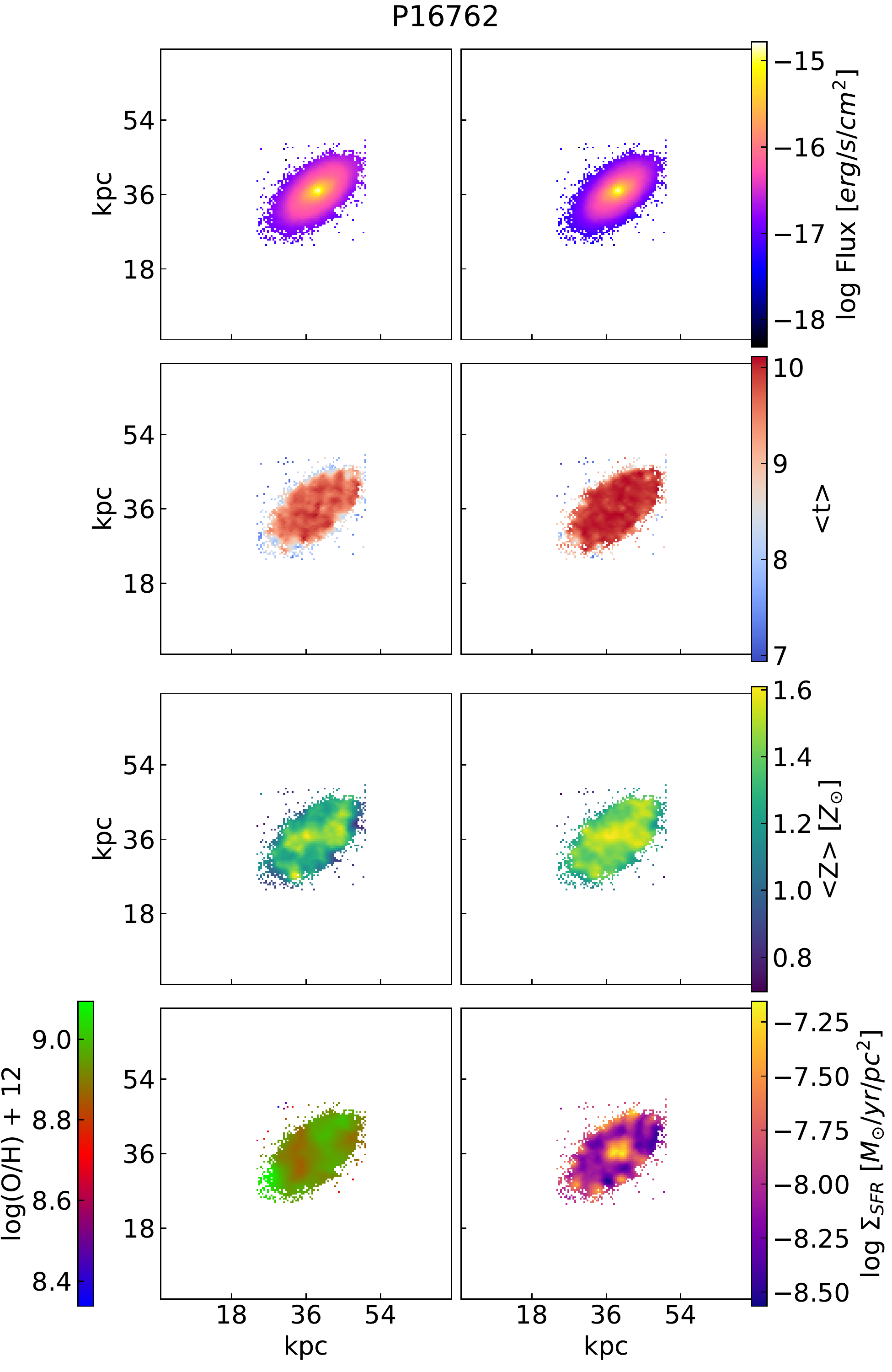}
\end{figure}

\begin{figure}
\centering
\includegraphics[width=\linewidth]{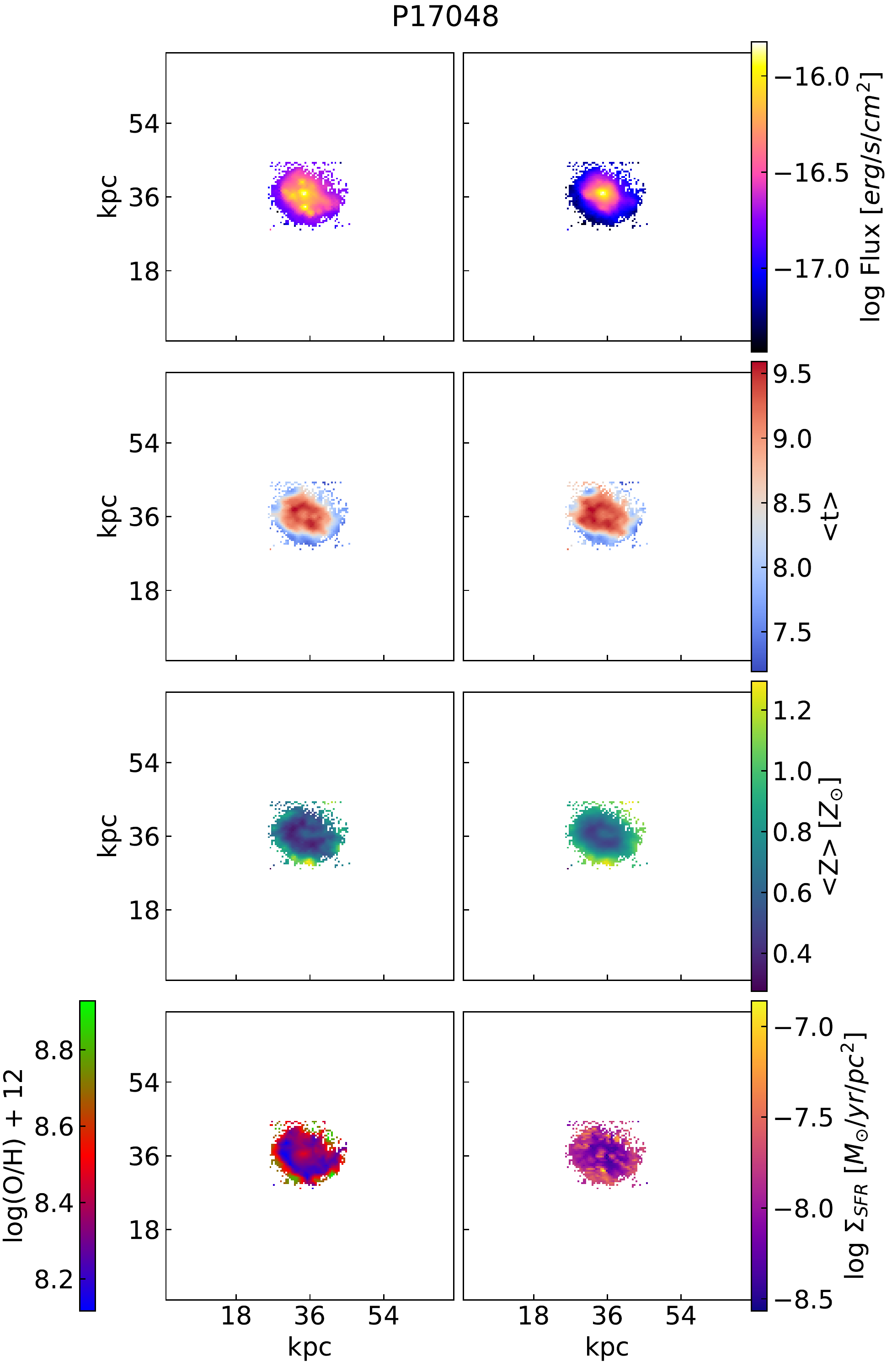}
\end{figure}

\begin{figure}
\centering
\includegraphics[width=\linewidth]{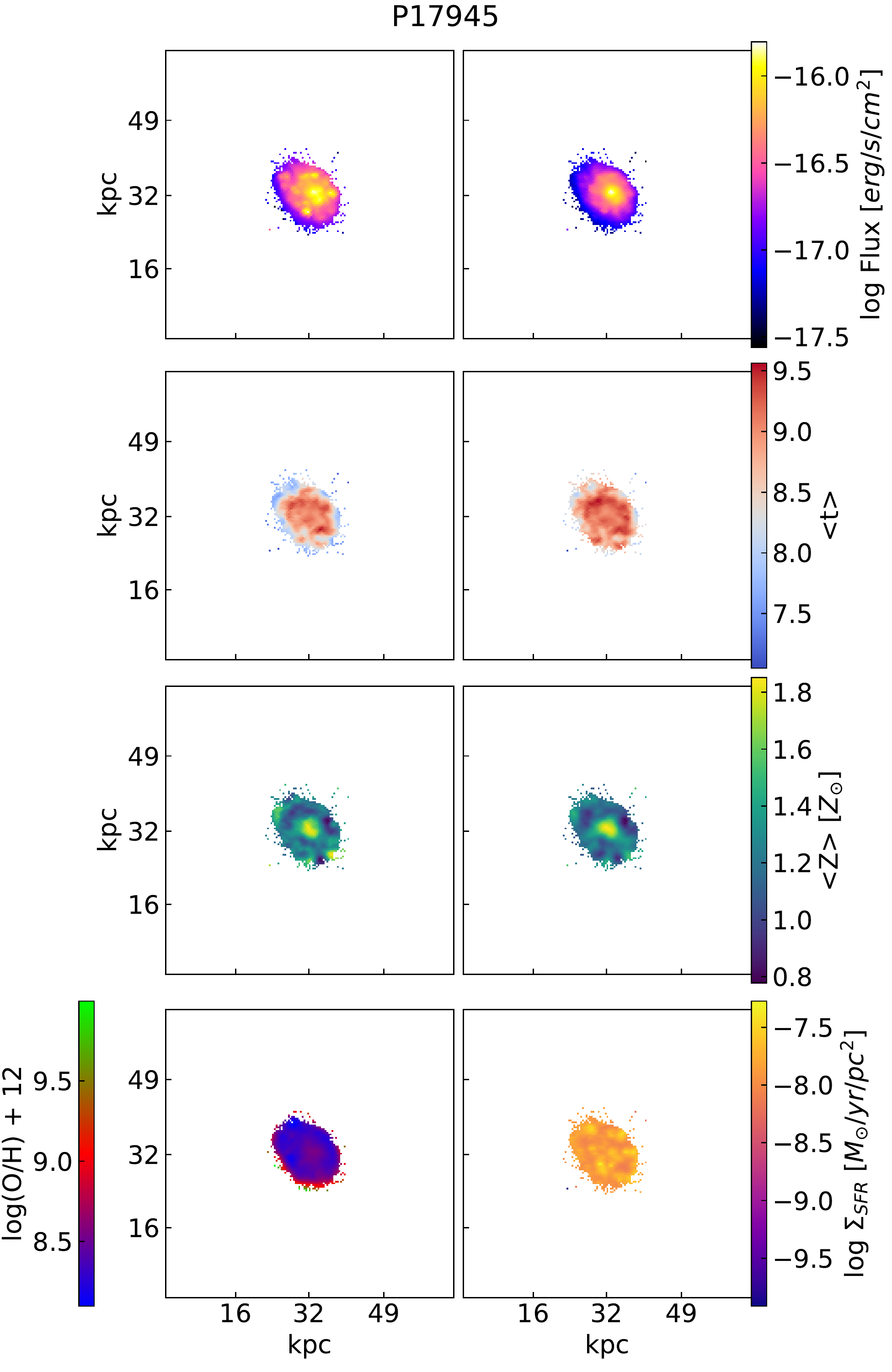}
\end{figure}

\begin{figure}
\centering
\includegraphics[width=\linewidth]{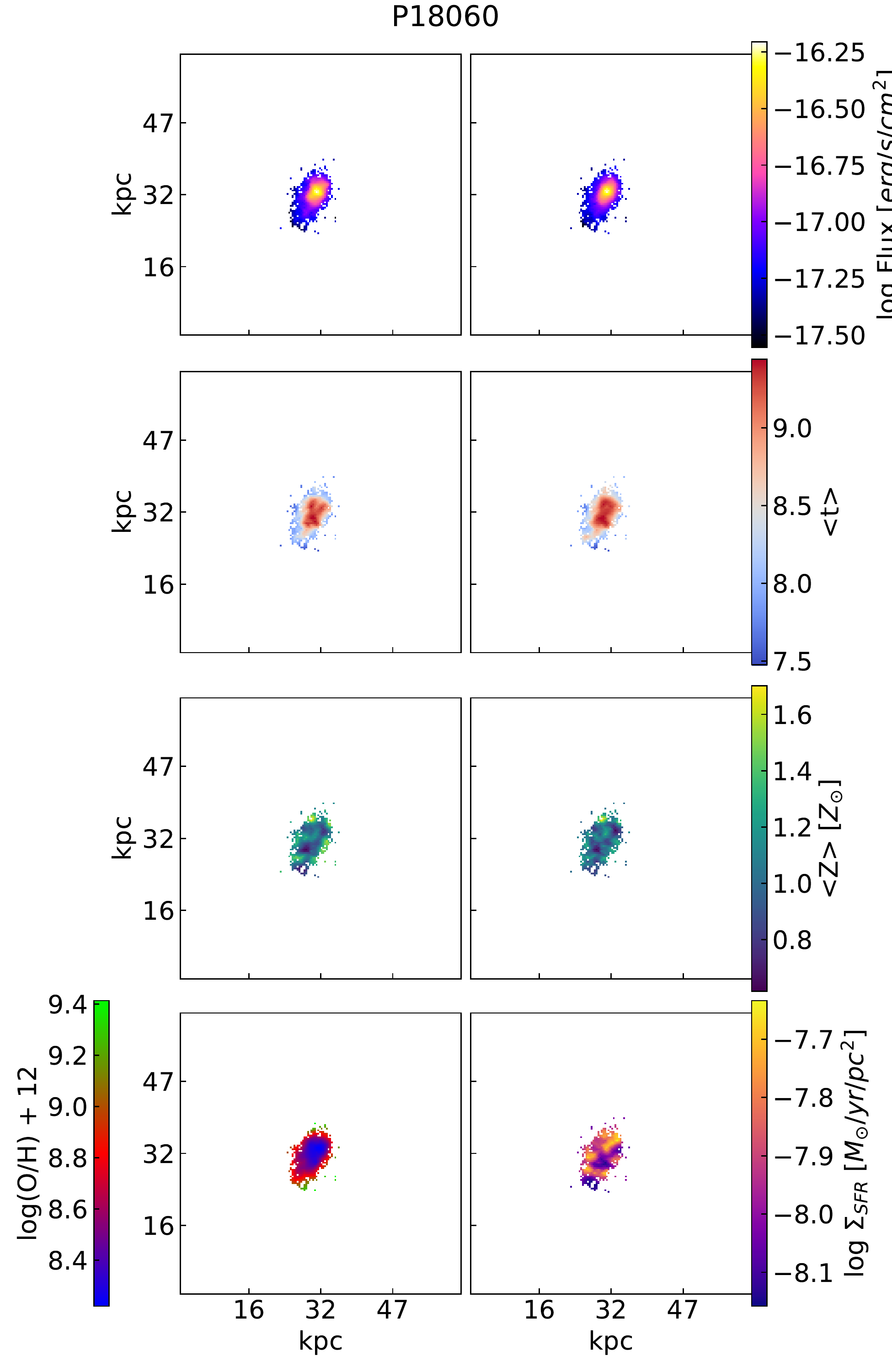}
\end{figure}

\begin{figure}
\centering
\includegraphics[width=\linewidth]{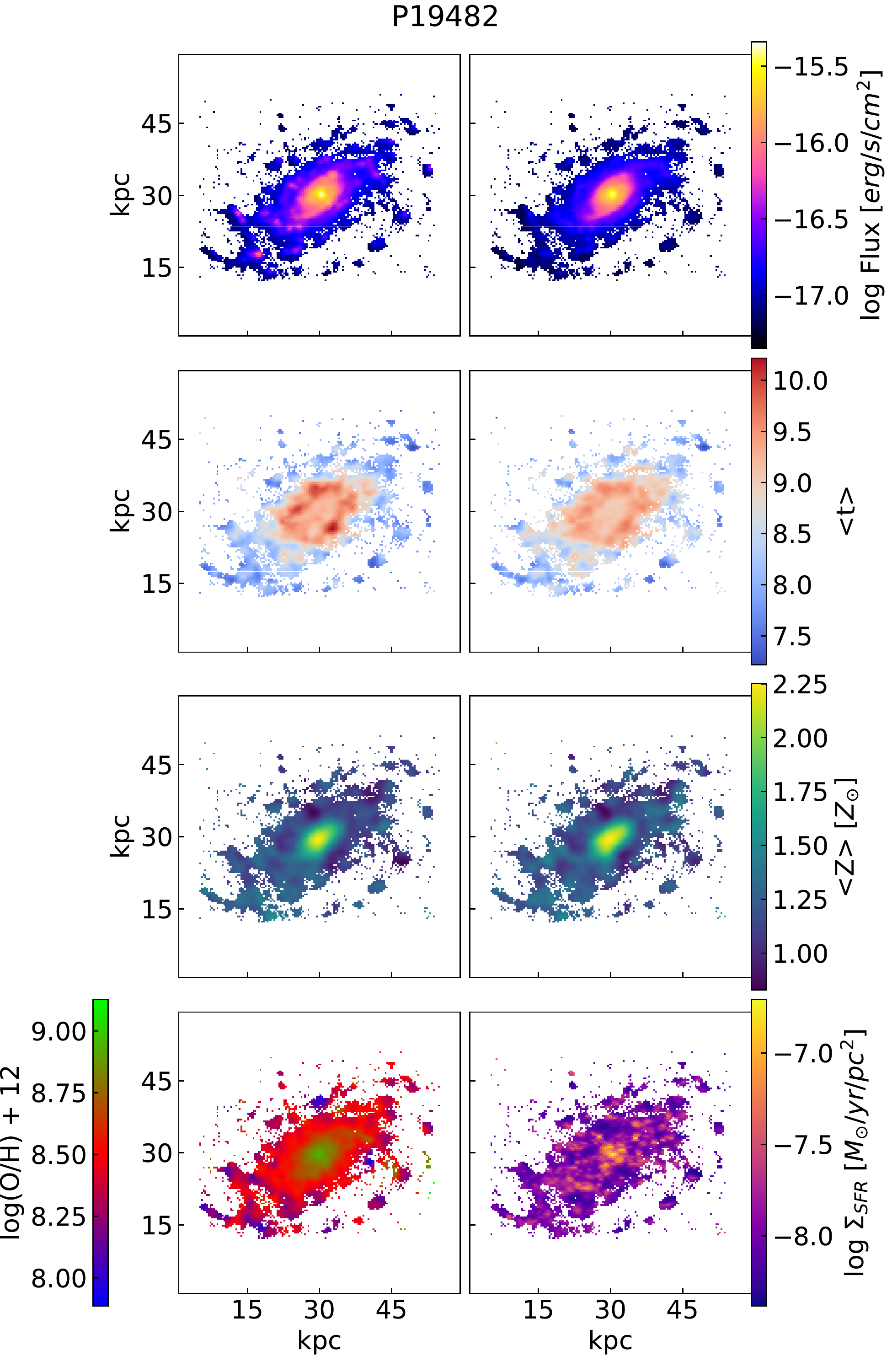}
\end{figure}
\begin{figure}
\centering
\includegraphics[width=\linewidth]{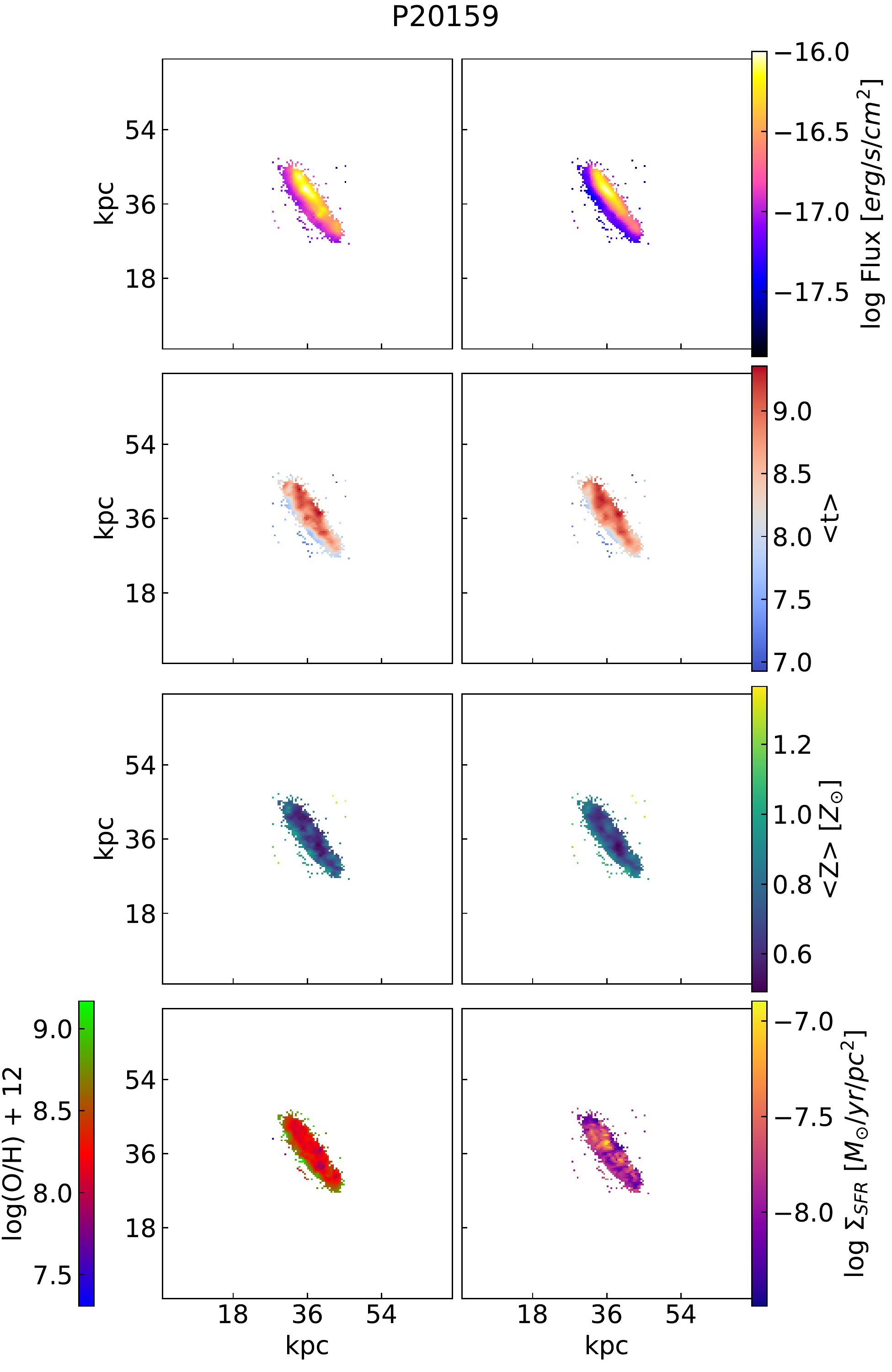}
\end{figure}

\begin{figure}
\centering
\includegraphics[width=\linewidth]{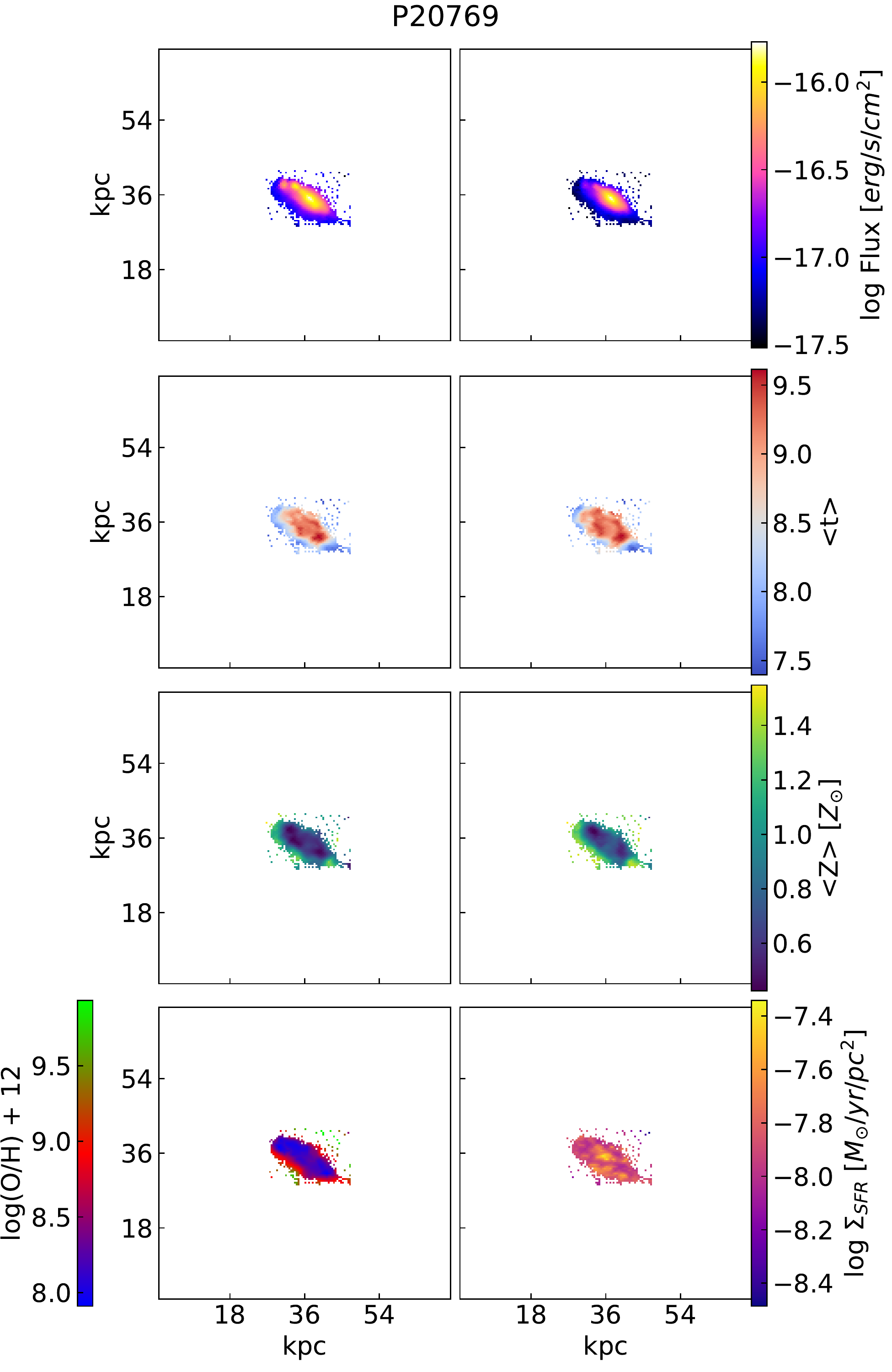}
\end{figure}

\begin{figure}
\centering
\includegraphics[width=\linewidth]{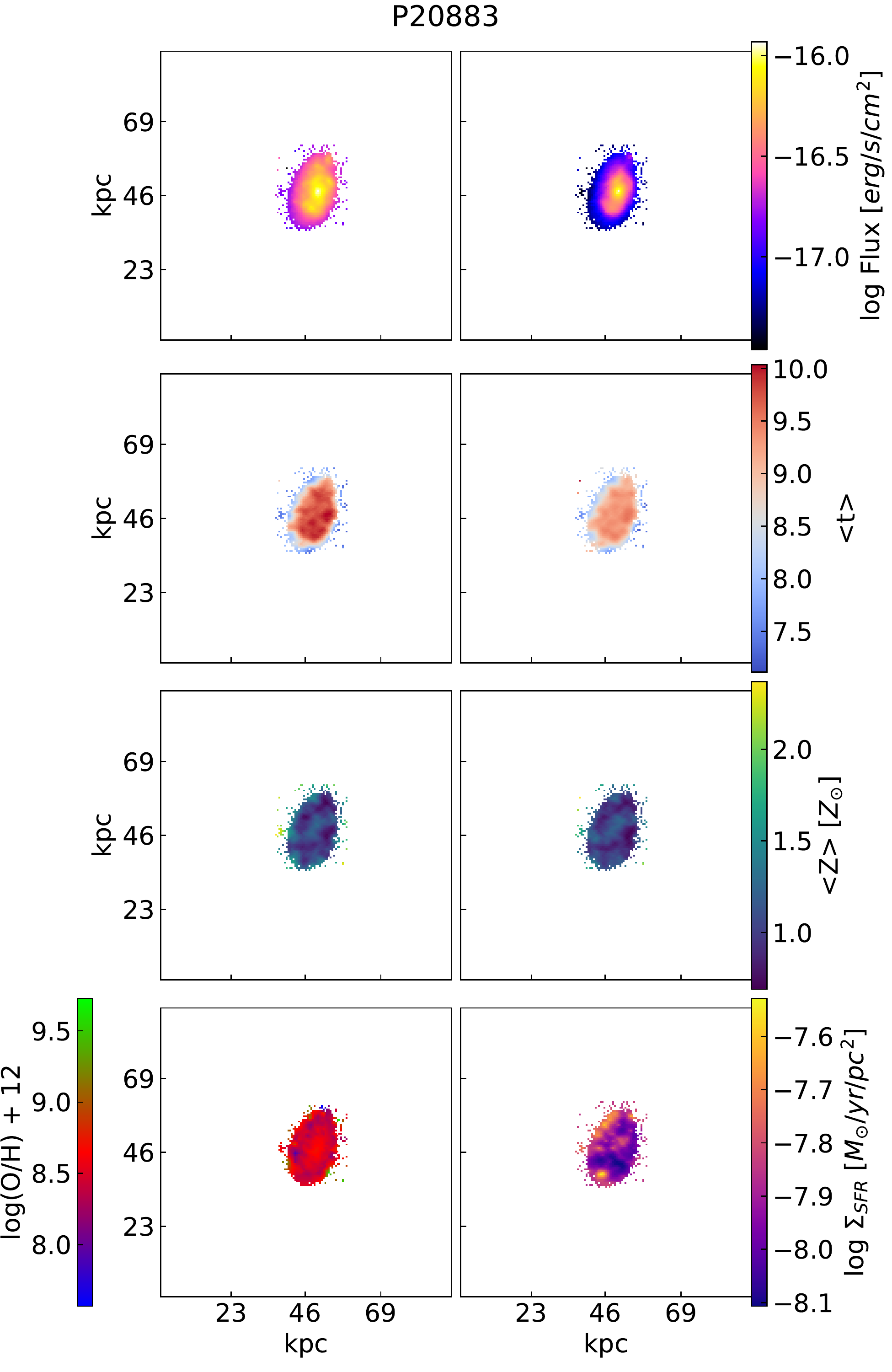}
\end{figure}

\begin{figure}
\centering
\includegraphics[width=\linewidth]{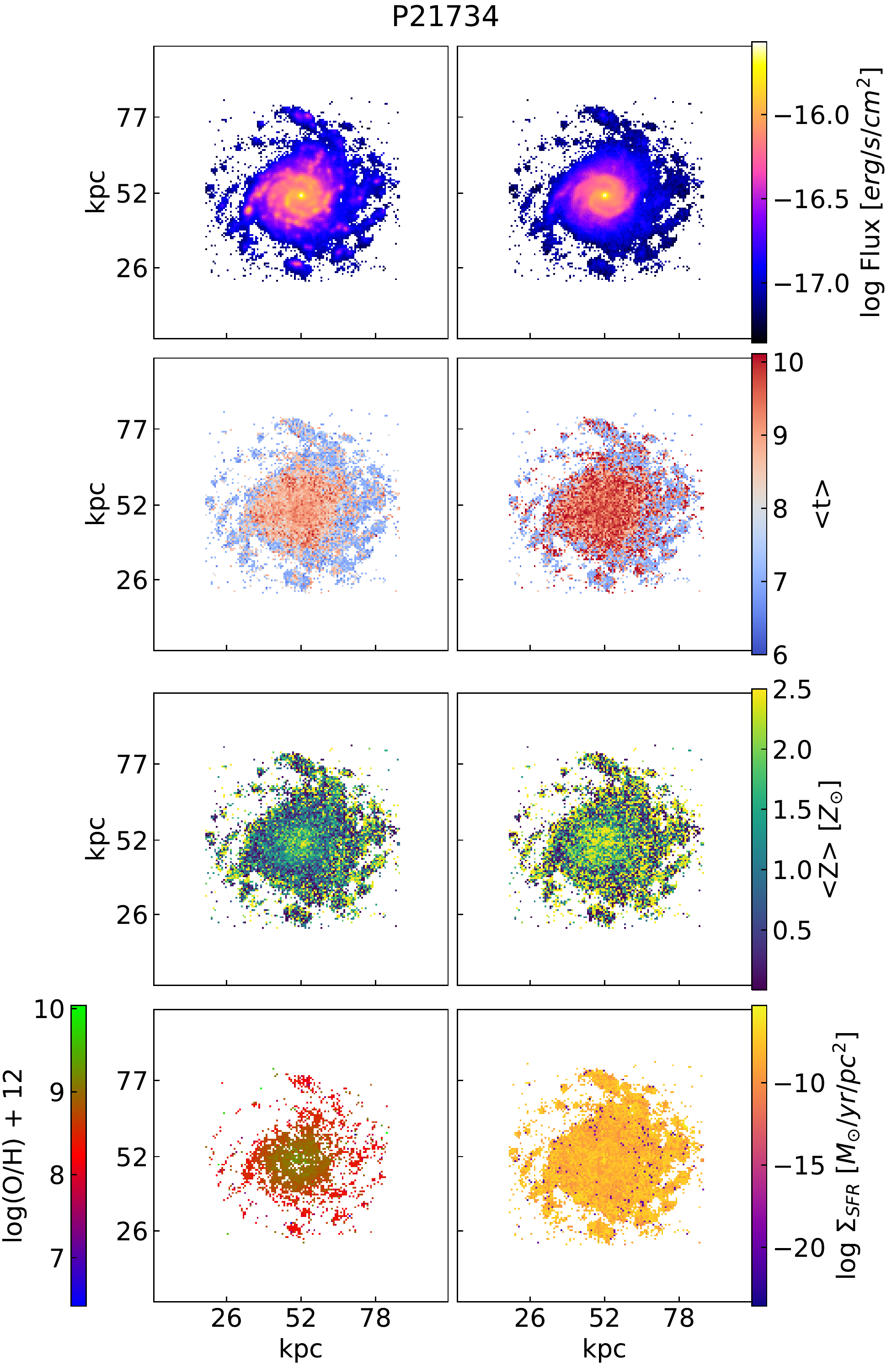}
\end{figure}

\begin{figure}
\centering
\includegraphics[width=\linewidth]{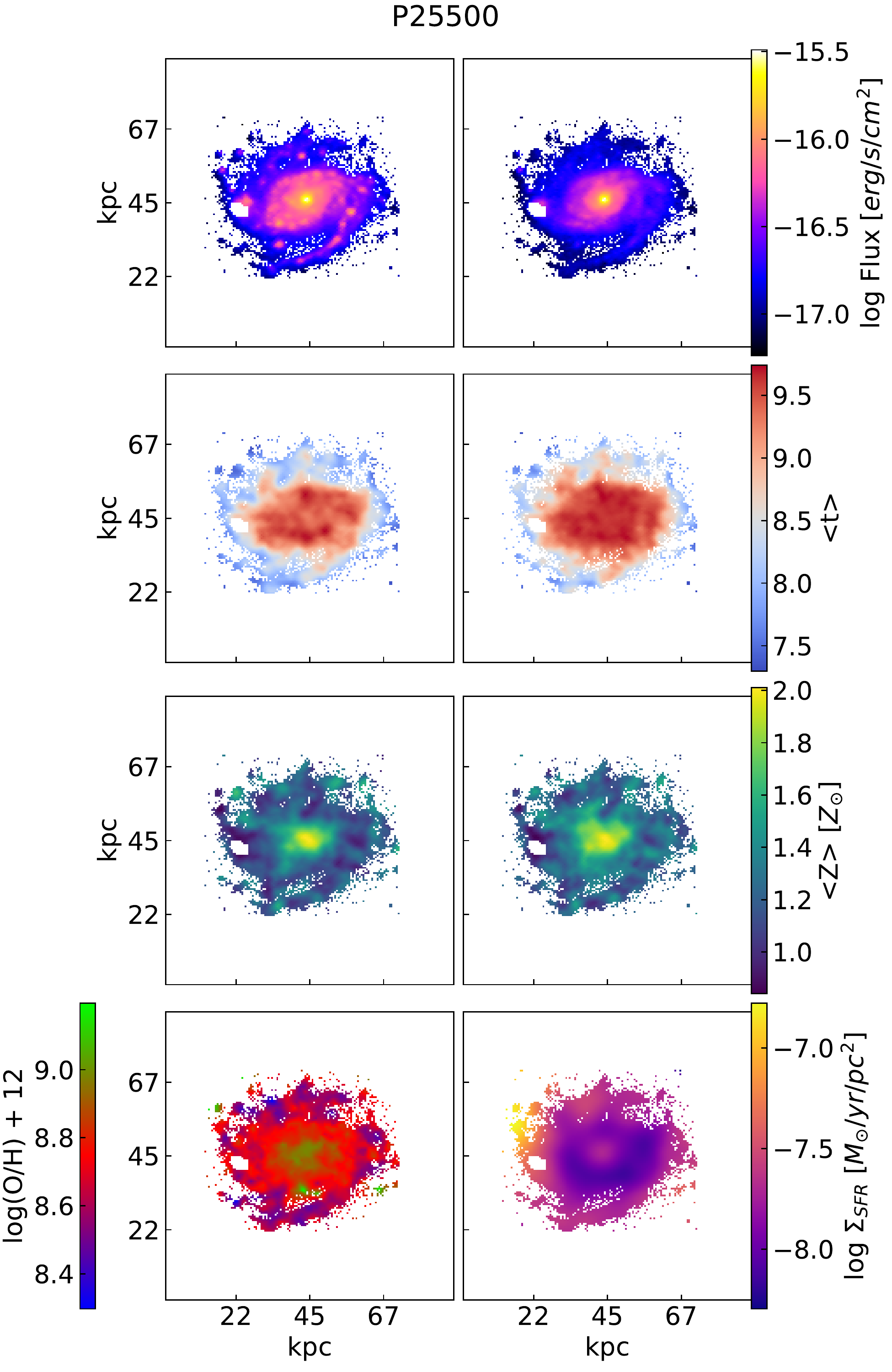}
\end{figure}

\begin{figure}
\centering
\includegraphics[width=\linewidth]{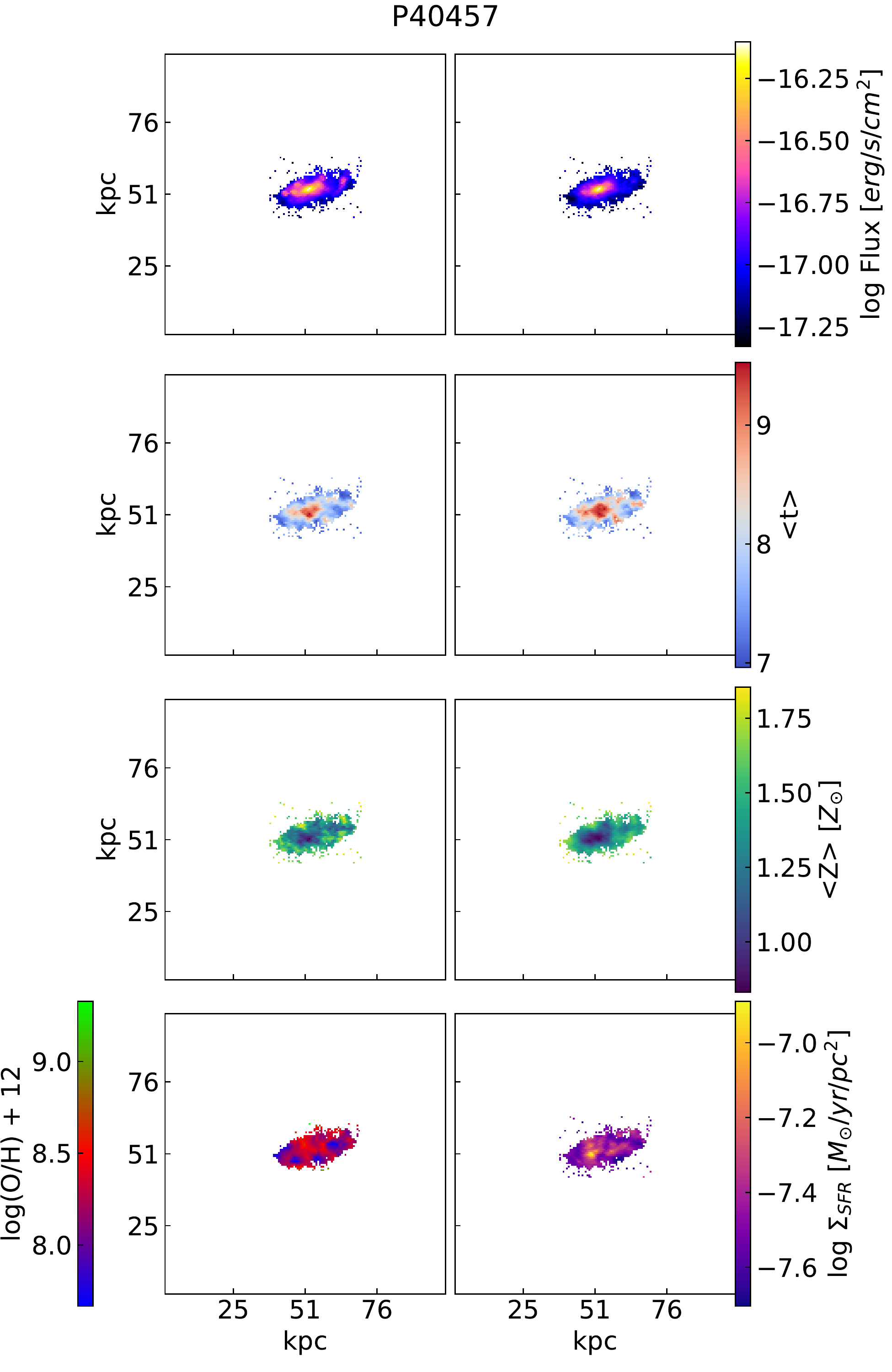}
\end{figure}

\begin{figure}
\centering
\includegraphics[width=\linewidth]{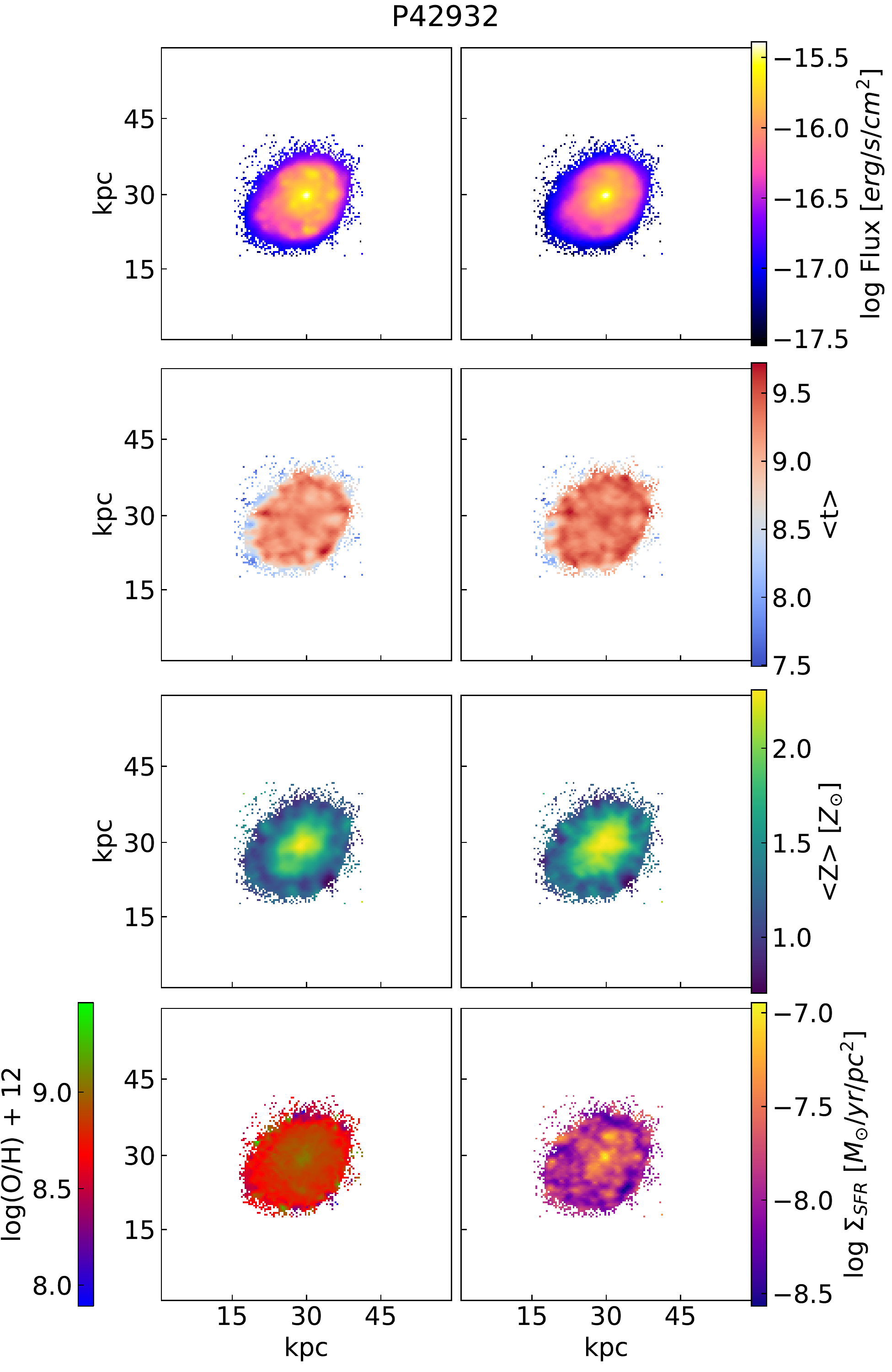}
\end{figure}

\begin{figure}
\centering
\includegraphics[width=\linewidth]{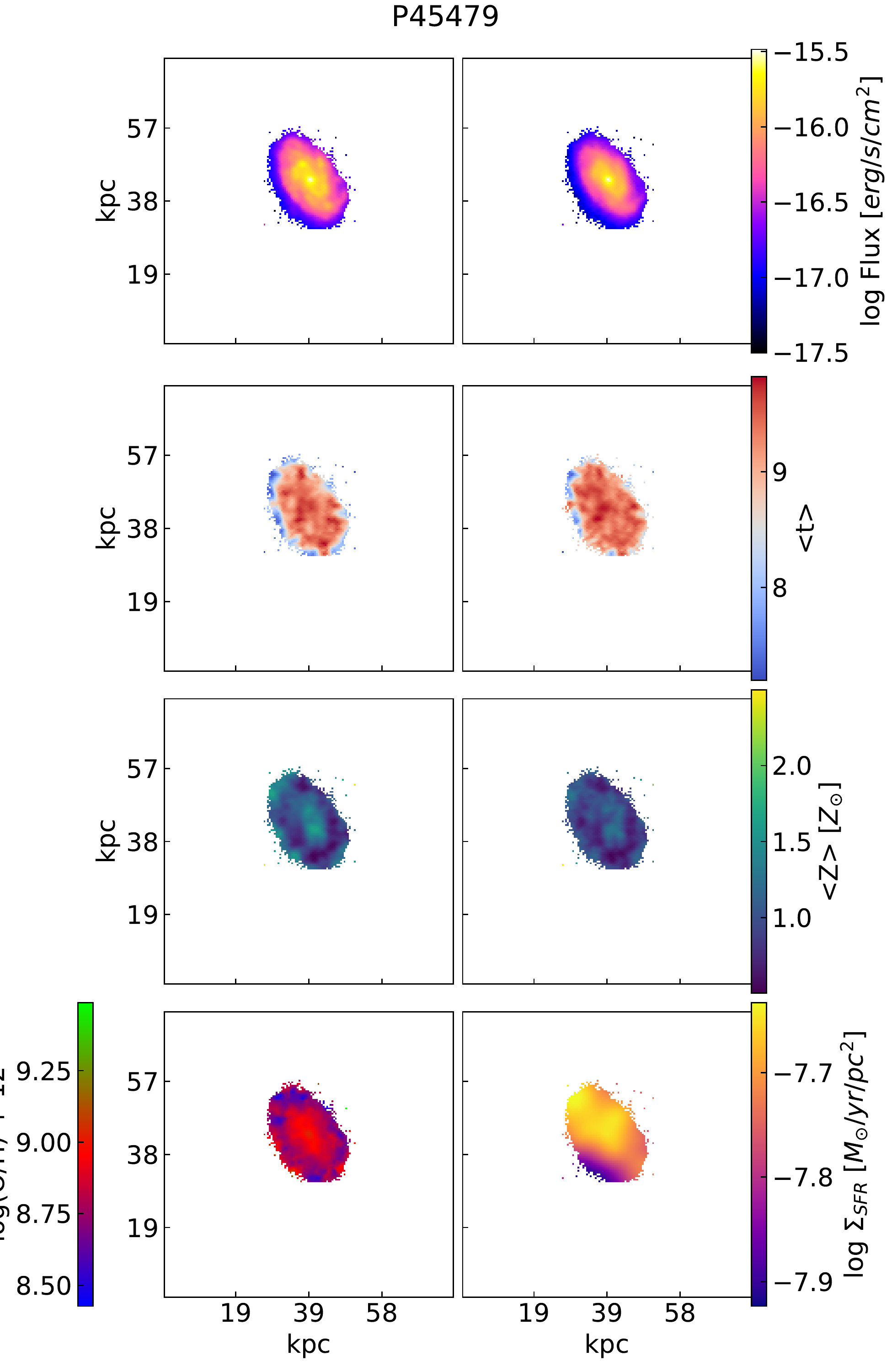}
\end{figure}

\begin{figure}
\centering
\includegraphics[width=\linewidth]{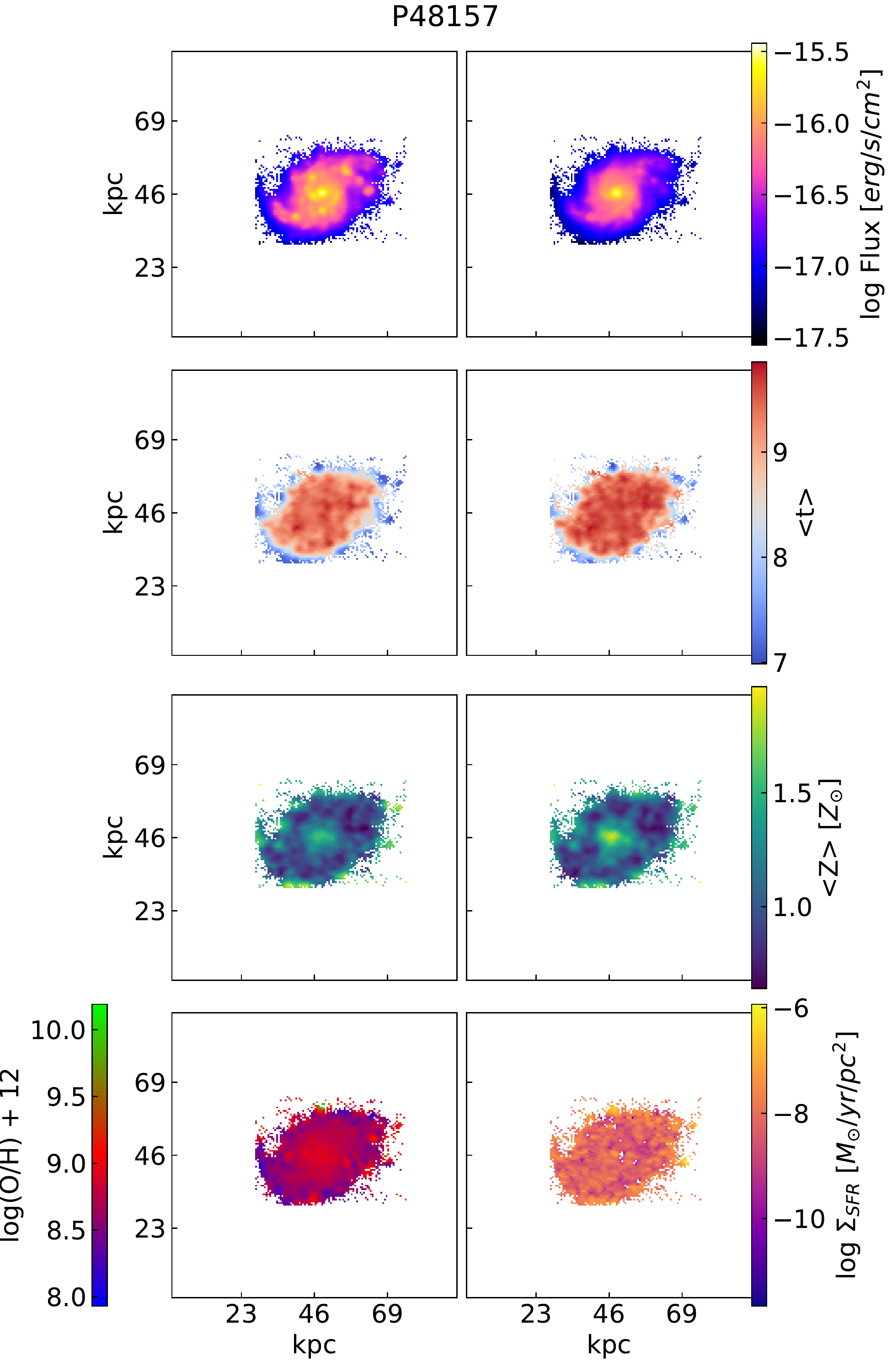}
\end{figure}

\begin{figure}
\centering
\includegraphics[width=\linewidth]{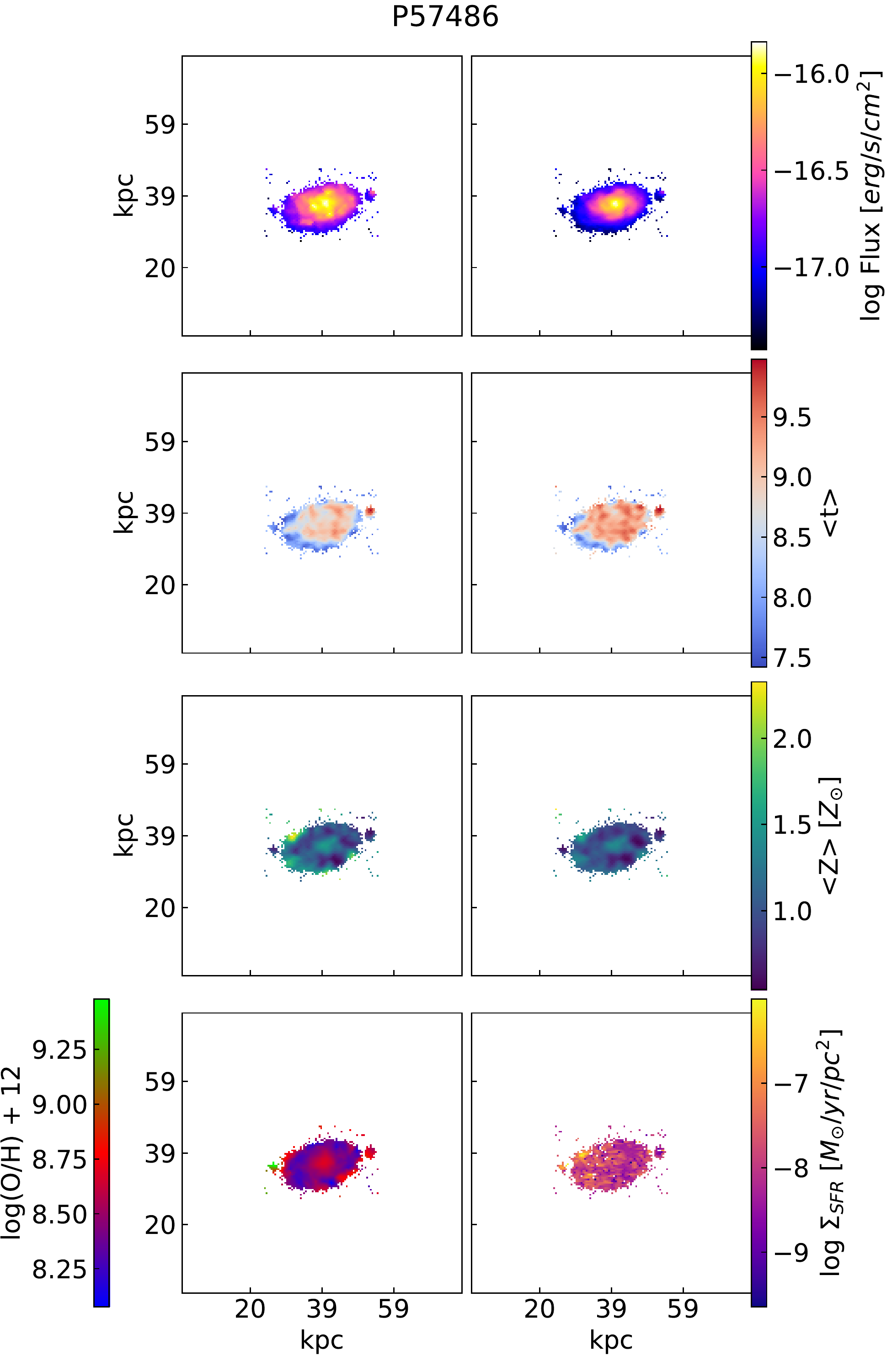}
\end{figure}

\begin{figure}
\centering
\includegraphics[width=\linewidth]{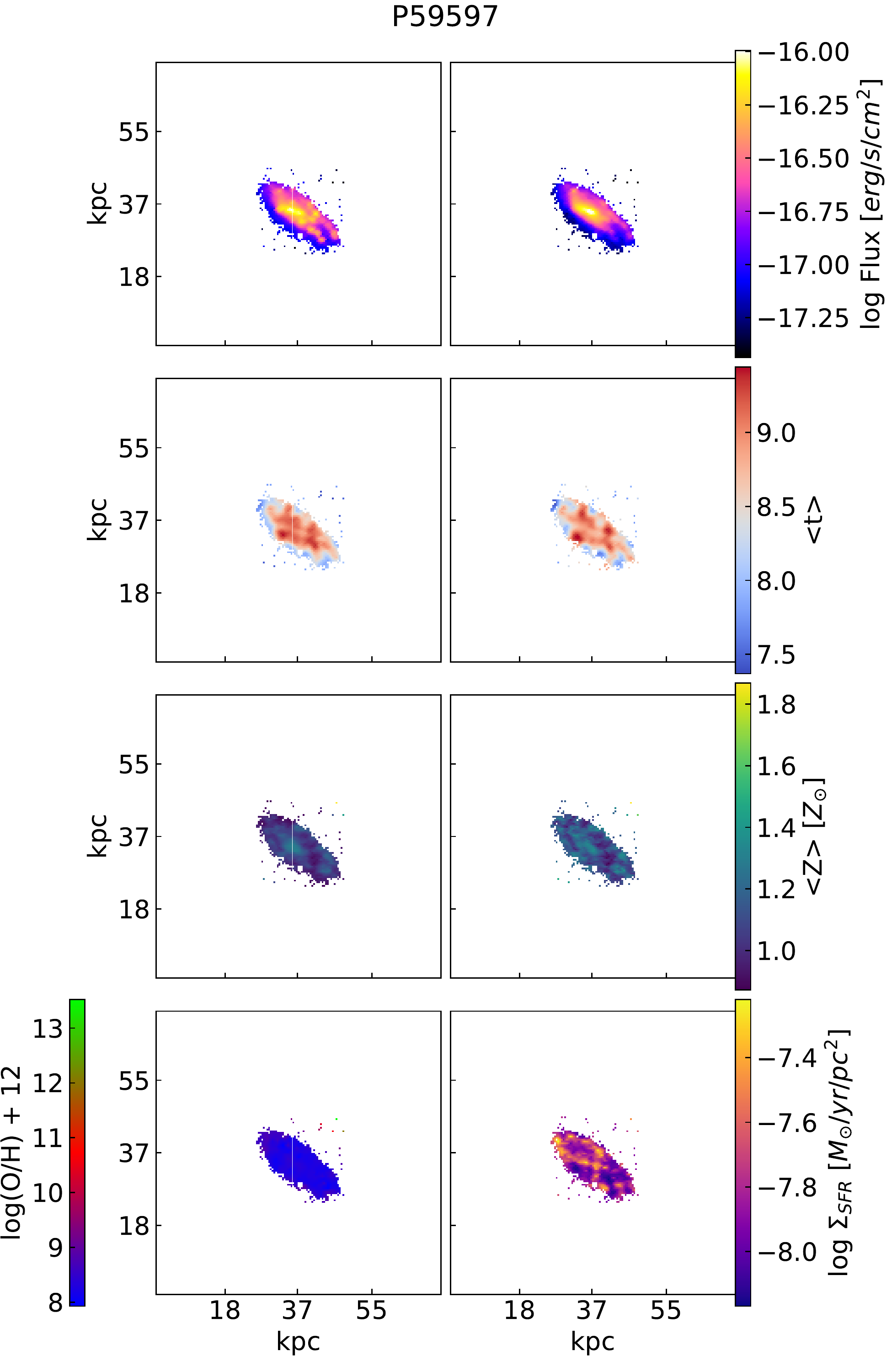}
\end{figure}

\begin{figure}
\centering
\includegraphics[width=\linewidth]{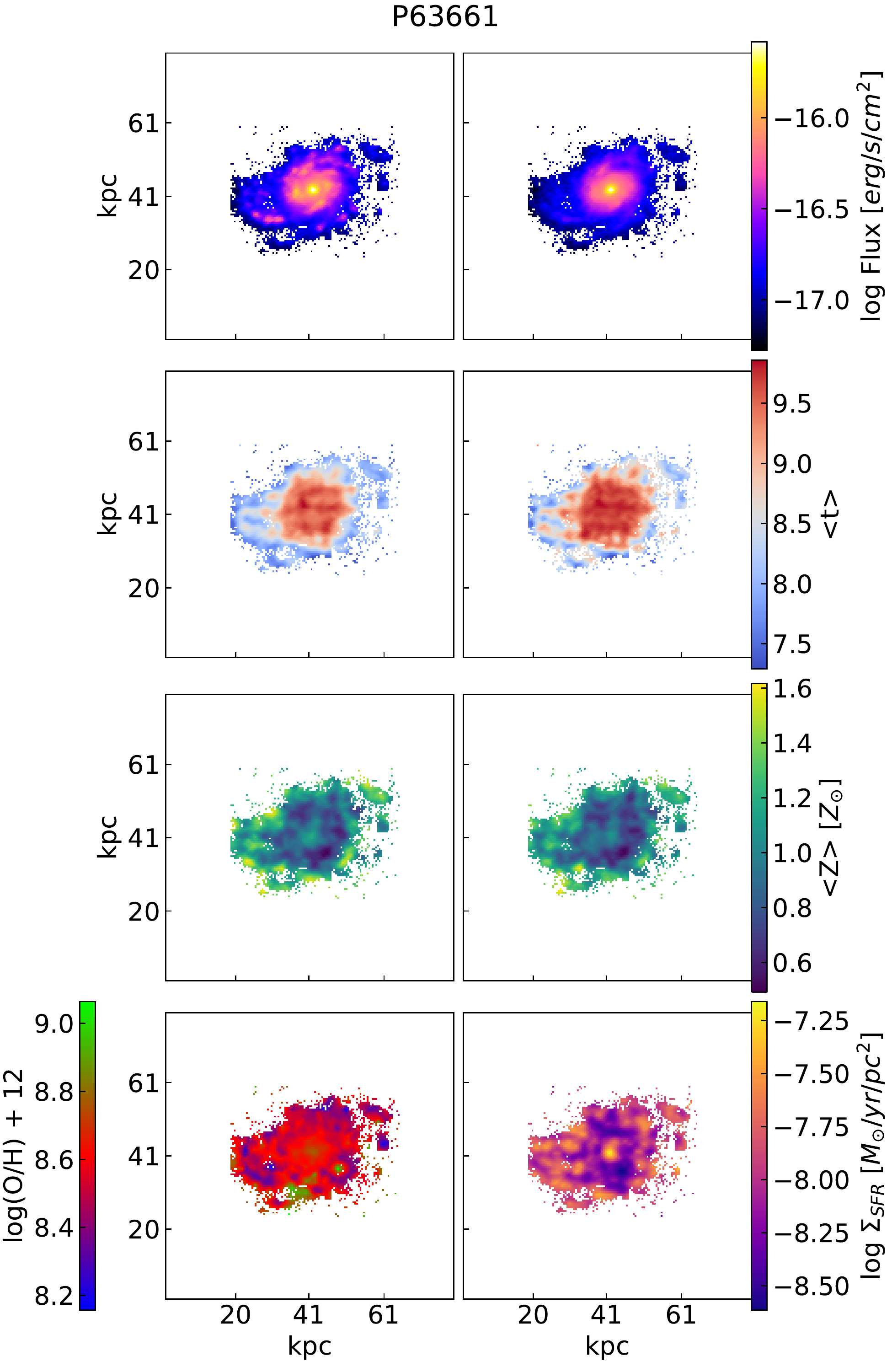}
\end{figure}
\begin{figure}
\centering
\includegraphics[width=\linewidth]{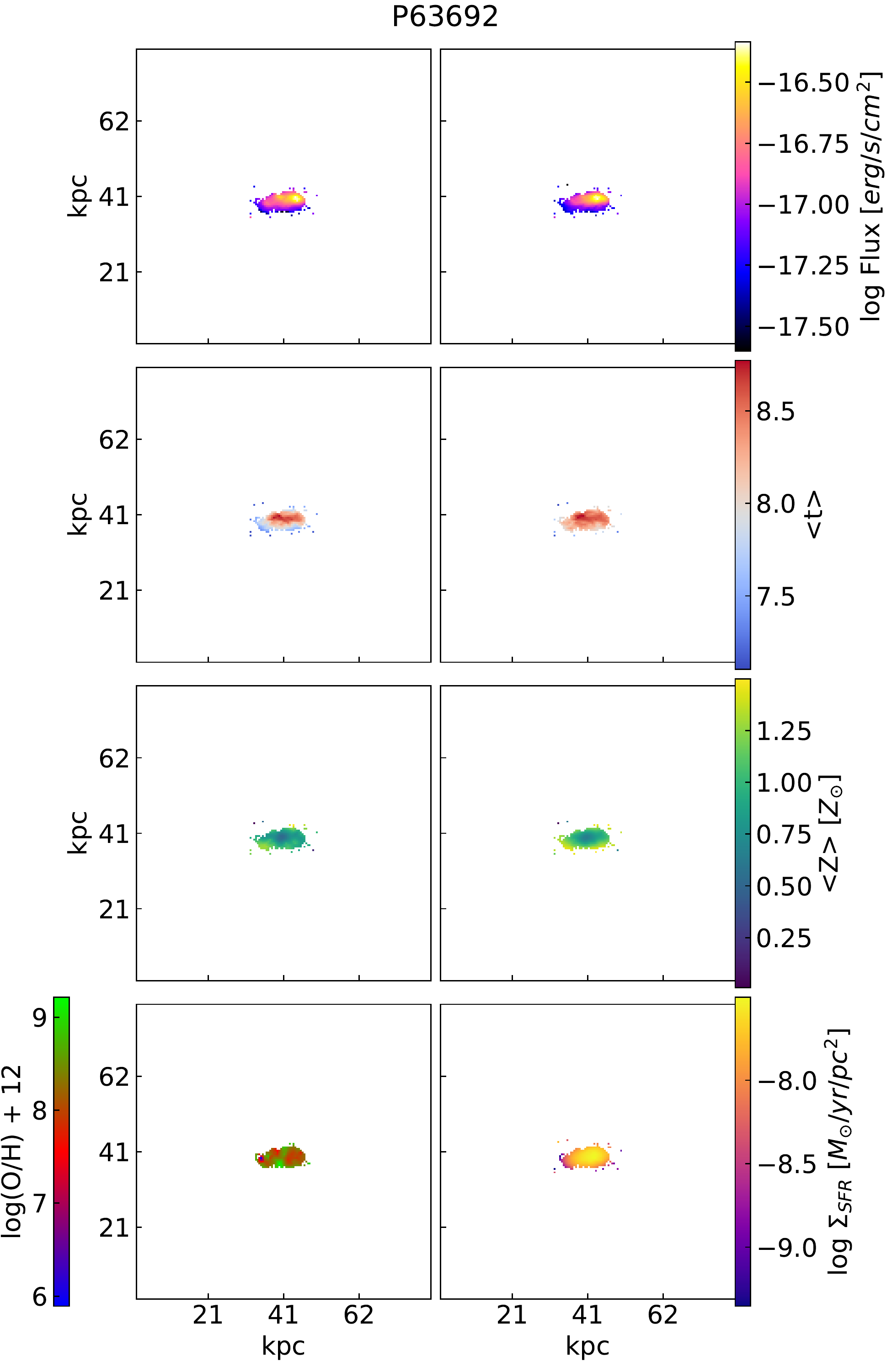}
\end{figure}

\begin{figure}
\centering
\includegraphics[width=\linewidth]{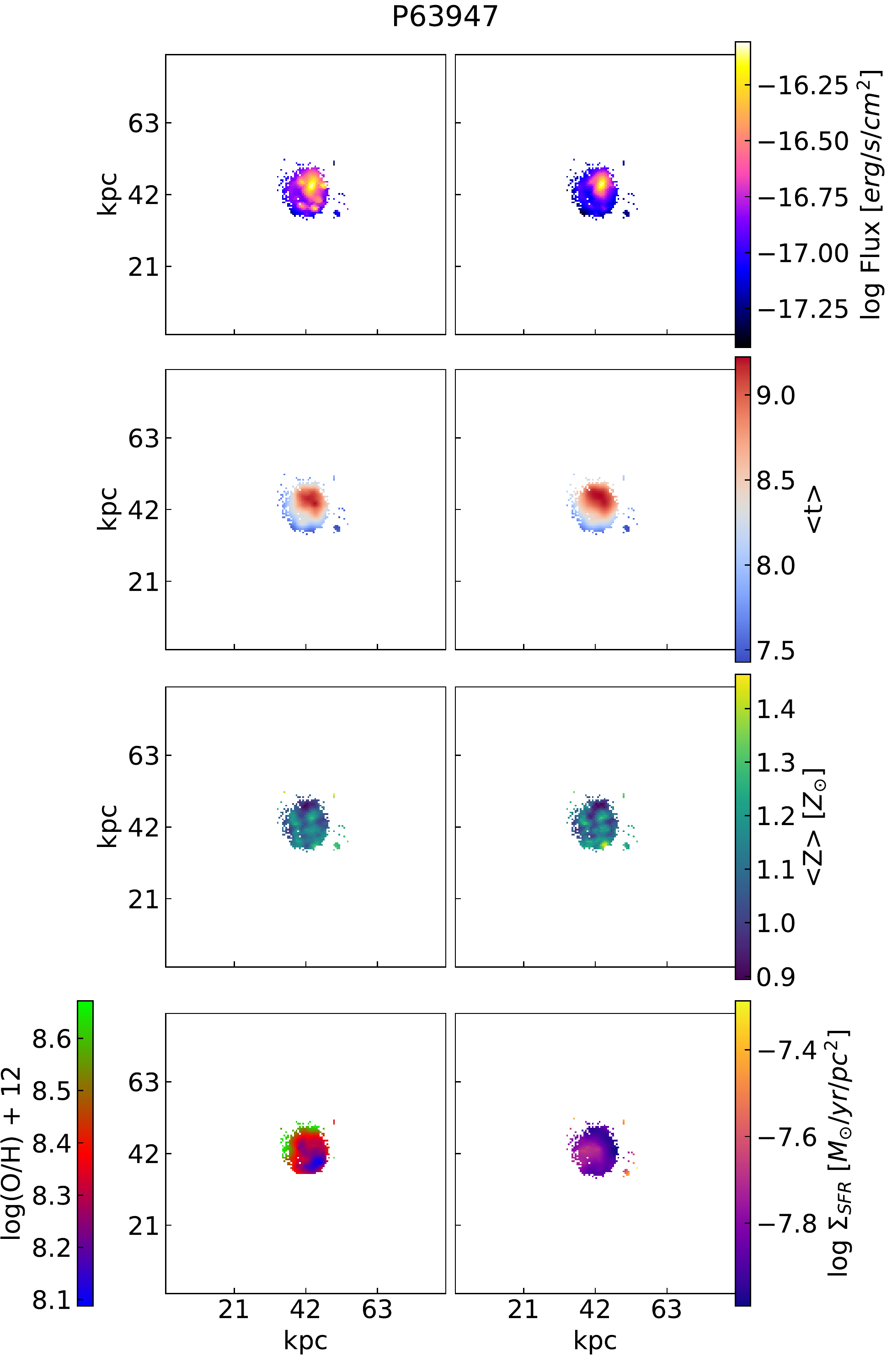}
\end{figure}

\begin{figure}
\centering
\includegraphics[width=\linewidth]{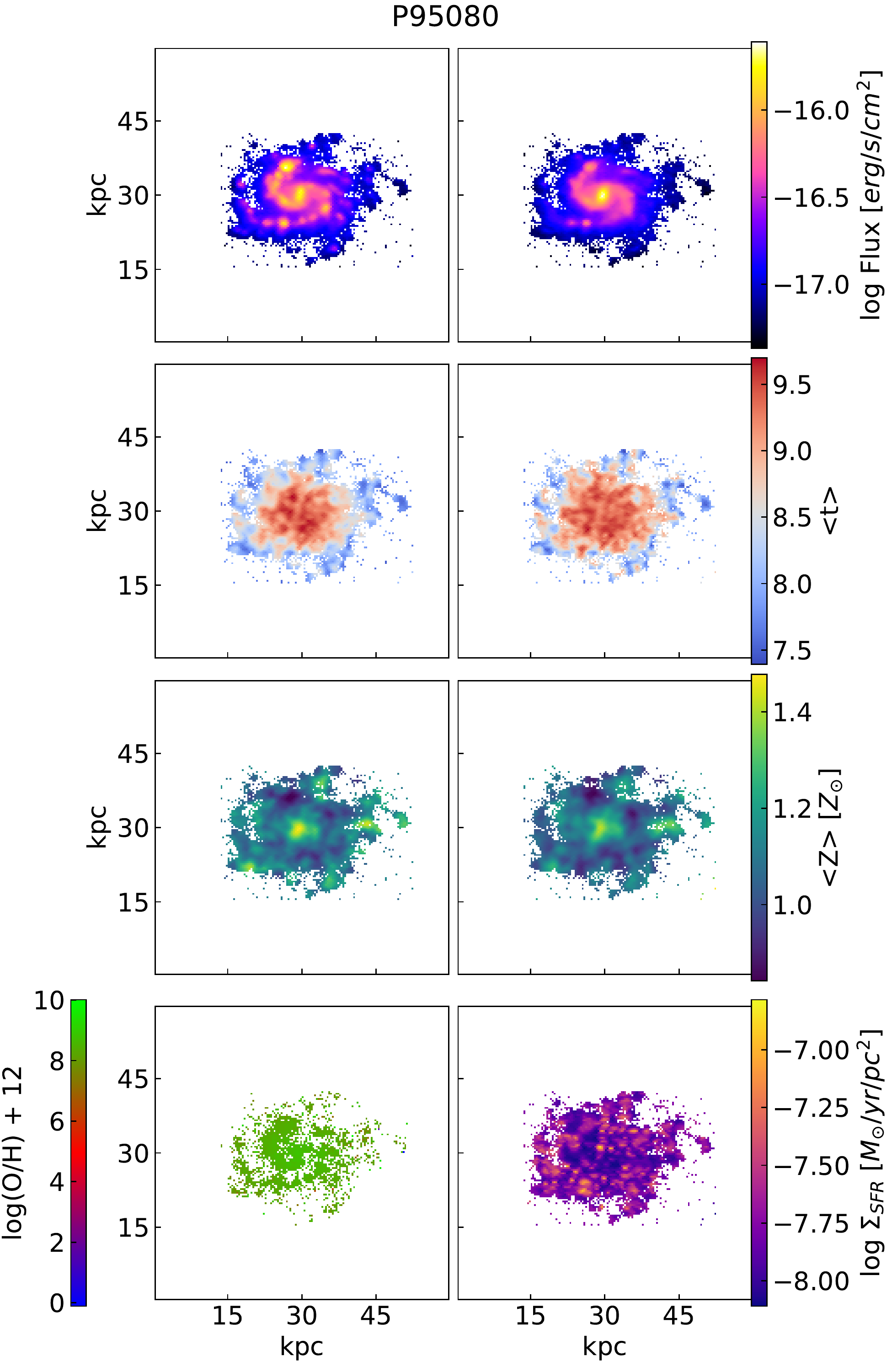}
\end{figure}

\begin{figure}
\centering
\includegraphics[width=\linewidth]{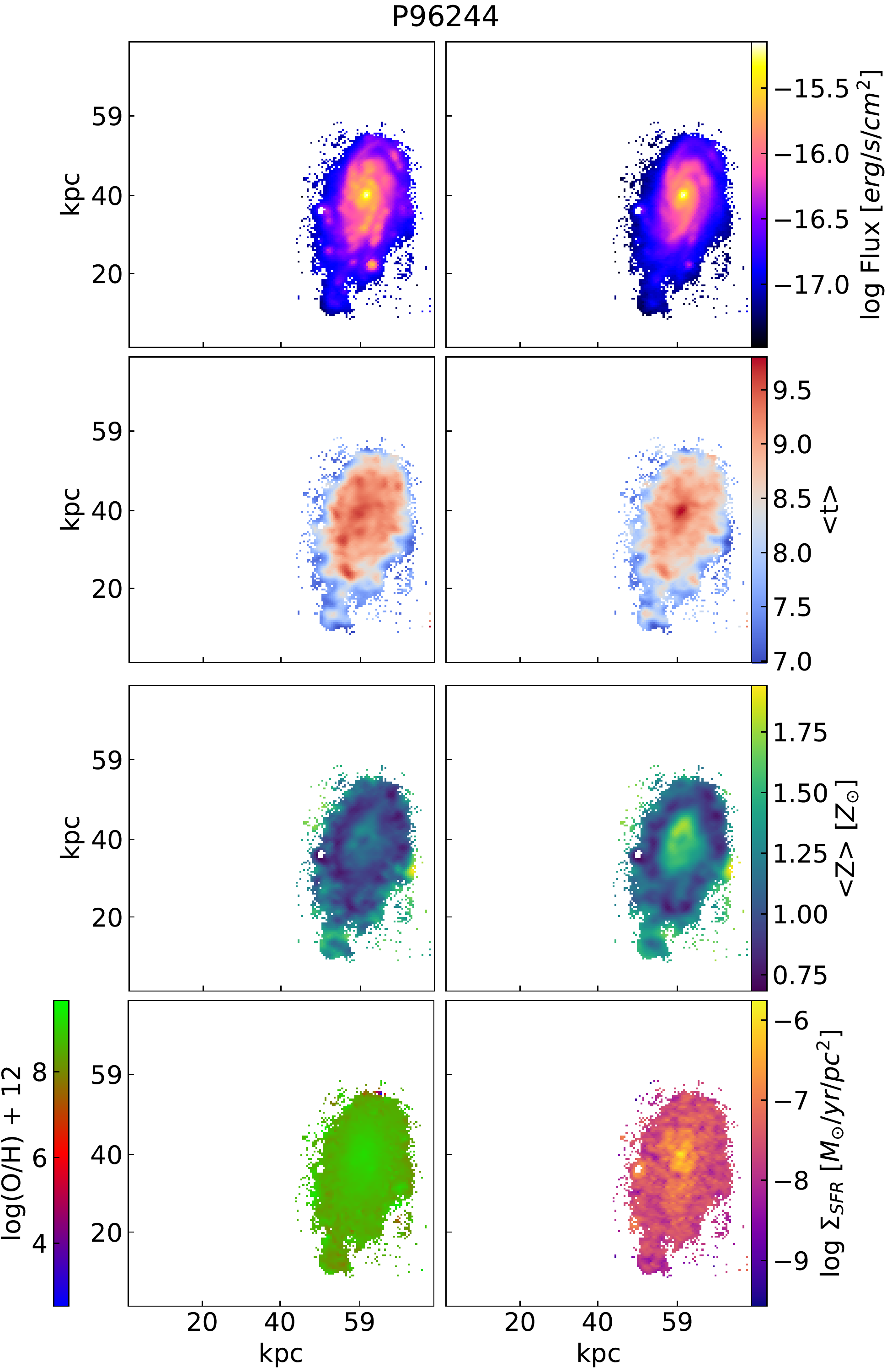}
\end{figure}

\begin{figure}
\centering
\includegraphics[width=\linewidth]{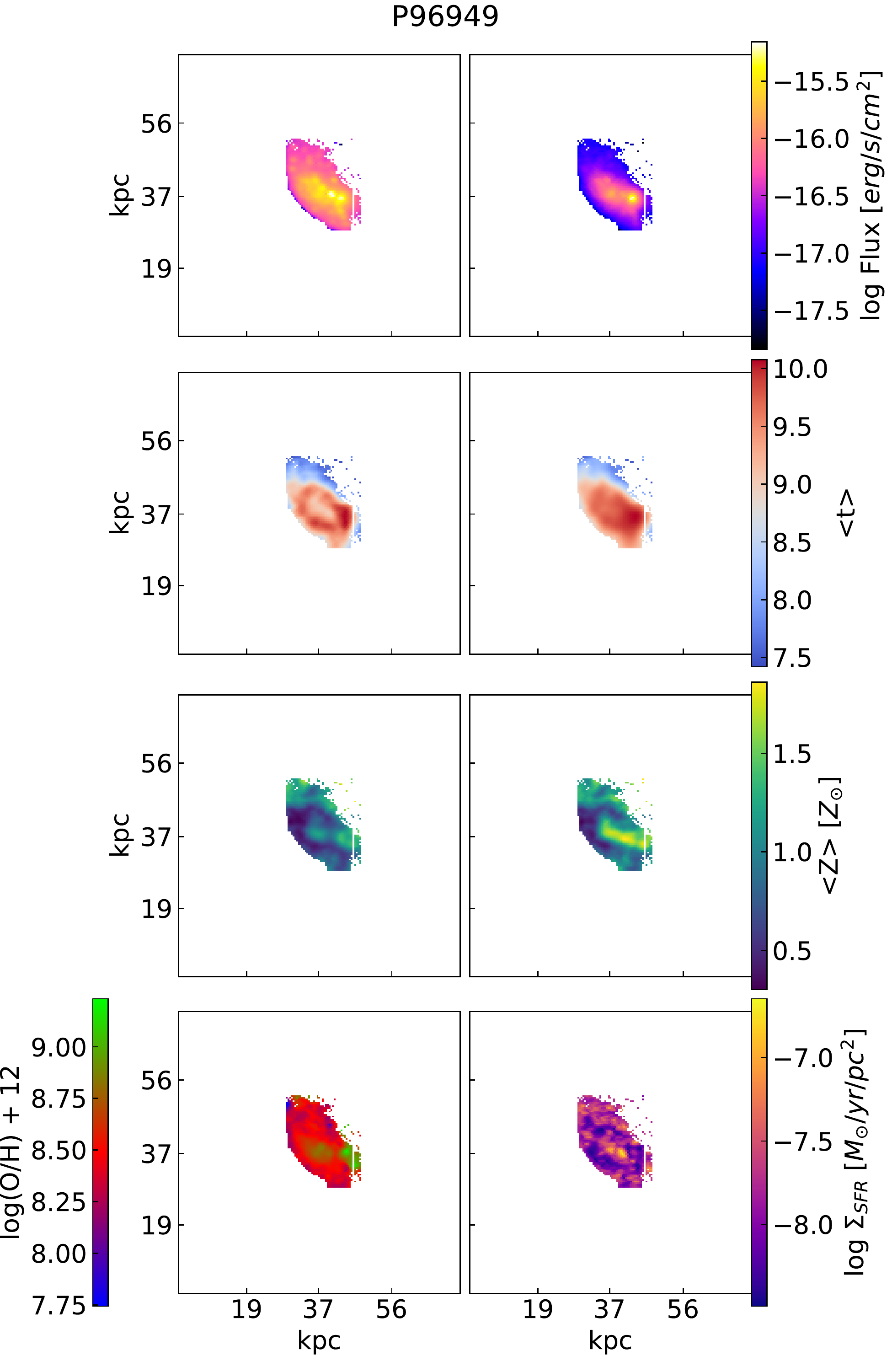}
\end{figure}

\bsp
\label{lastpage}

\end{document}